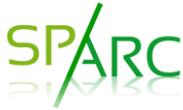
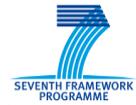

# SPARC

# ICT-258457

Deliverable D3.3

# Split Architecture for Large Scale Wide Area Networks

| Editors: | *Wolfgang John; Ericsson (EA)* |
|---|---|
| Deliverable type: | Report (R) |
| Dissemination level: (Confidentiality) | Public (PU) |
| Contractual delivery date: | M26 |
| Actual delivery date: | M27 |
| Version: | 1.0 |
| Total number of pages: | 129 |
| Keywords: | Split Architecture, OpenFlow, Carrier-Grade, Network Virtualization, Resiliency, OAM, QoS, Service Creation, Scalability, Energy-Efficient Networking, Multi-Layer Networking |

*Abstract*

This deliverable defines a carrier-grade split architecture based on requirements identified during the SPARC project. It presents the *SplitArchitecture* proposal, the SPARC concept for Software Defined Networking (SDN) introduced for large-scale wide area networks such as access/aggregation networks, and evaluates technical issues against architectural trade-offs. First we present the control and management architecture of the proposed *SplitArchitecture*. Here, we discusses a recursive control architecture consisting of hierarchically stacked control planes and provide initial considerations regarding network management integration to SDN in general and *SplitArchitecture* in particular. Next, OpenFlow extensions to support the carrier-grade *SplitArchitecture* are discussed. These are: a) Openness & Extensibility, extending OpenFlow with more advanced processing functionalities on both data and control planes; b) Virtualization, enabling a flexible way of partitioning the network into virtual networks while providing full isolation between these partitions; c) OAM, presenting a solution for both technology-specific OAM (MPLS BFD) and a novel technology agnostic flow OAM; d) Resiliency approaches for surviving link failures or failures of controller or forwarding elements; e) Bootstrapping and topology discovery issues, discussing current discovery of network devices and their interconnections; f) Service creation solutions for integration of residential customer services in the form of PPP, and business customer services in form of pseudo-wires (PWE); g) Energy-efficient networking approaches for energy savings and requirements on the OpenFlow protocol; h) QoS aspects, showing how traditional QoS tools such as packet classification, metering, coloring, policing, shaping and scheduling can be realized in an OpenFlow environment; (i) and Multilayer aspects outlining different stages of packet-optical integration. In addition, we discuss selected deployment and adoption scenarios faced by modern operator networks, such as service creation scenarios and peering aspects, i.e., how to interconnect with legacy networks. Finally, we indicate how our *SplitArchitecture* approach meets carrier grade scalability requirements in access/aggregation network scenarios.

Disclaimer

This document contains material which is the copyright of certain SPARC consortium parties and may not be reproduced or copied without permission.

*In the case of Public (PU):*
All SPARC consortium parties have agreed to full publication of this document.

*In the case of Restricted to Program (PP):*
All SPARC consortium parties have agreed to make this document available on request to other framework program participants.

*In the case of Restricted to Group (RE):*
All SPARC consortium parties have agreed to full publication of this document. However this document is written for / being used by an <organization / other project / company, etc.> as <a contribution to standardization / material for consideration in product development, etc.>.

*In the case of Consortium Confidential (CO):*
The information contained in this document is the proprietary confidential information of the SPARC consortium and may not be disclosed except in accordance with the consortium agreement.

The commercial use of any information contained in this document may require a license from the proprietor of that information.

Neither the SPARC consortium as a whole, nor any specific party of the SPARC consortium warrant that the information contained in this document is acceptable for use, nor that use of the information is free from risk, and accepts no liability for any loss or damage suffered by any person or institution using this information.

**Imprint**

| | |
|---|---|
| **[Project title]** | *Split Architecture for carrier grade networks* |
| **[Short title]** | *SPARC* |
| **[Number and title of work package]** | *WP3 – Architecture* |
| **[Document title]** | *Split Architecture for Large- Scale Wide Area Networks* |
| **[Editors]** | *Wolfgang John* |
| **[Work package leader]** | *Wolfgang John* |
| **[Task leader]** | *Wolfgang John* |

**Copyright notice**

© 2012 Participants in project SPARC (as specified in the Partner and Author list on page 7)

# Executive summary

In this deliverable, we present our final proposal for a carrier-grade split architecture, introducing Software Defined Networking (SDN) for large-scale wide area networks. Based on our conclusions from D3.1, we focus on OpenFlow as an enabling technology our proposed *SplitArchitecture*. However, earlier SPARC deliverables (D2.1 and D3.1) made it clear that current OpenFlow implementations do not fulfill carrier requirements. Thus, this deliverable presents a suitable framework for controlling carrier-grade operator networks and investigates missing features in OpenFlow as identified during the SPARC project in the context of access/aggregation network scenarios.

The overall conclusion of the SPARC project is that it is technically feasible to apply an OpenFlow-based split architecture to the carrier domain. This novel architecture paradigm promises improved network design and operation in large-scale networks with millions of customers, offering high levels of availability, flexibility and automation. The results of this project prove that these promises are valid and definitely deserve further attention by the networking industry in general and telecommunications operators in particular.

The figure below depicts a high-level representation of the main building blocks considered in our *SplitArchitecture*. In addition to the split between data and control planes, as commonly discussed in the context of SDN, we will also discuss a split of the data plane into forwarding and processing. Furthermore, we provide initial considerations of how network management-related functions can be integrated into the *SplitArchitecture*.

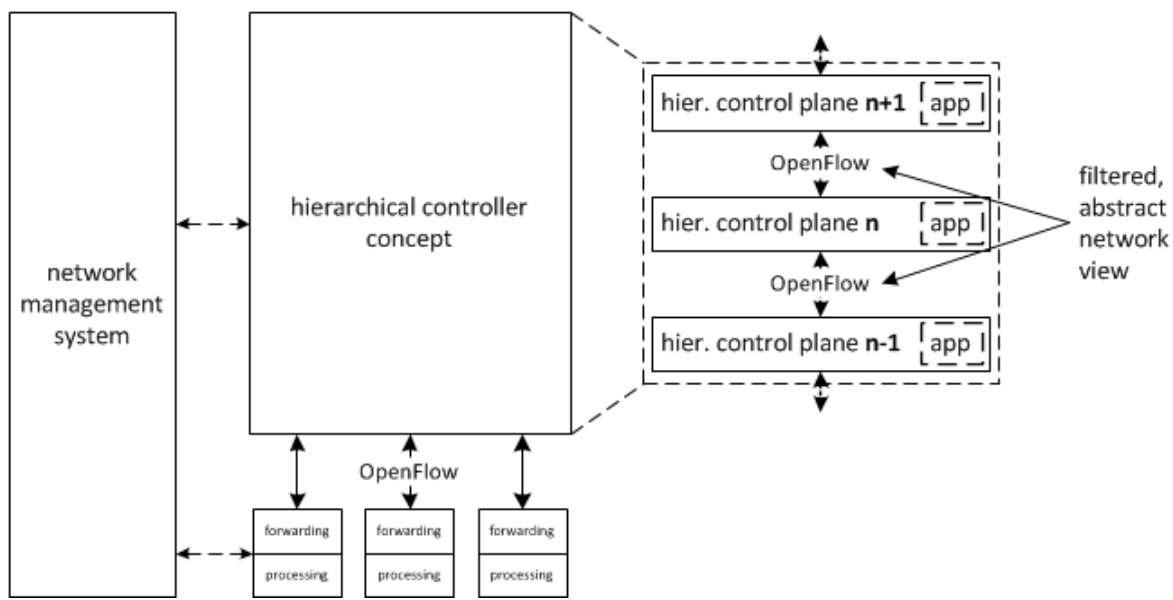

**Schematic SplitArchitecture Overview**

**SPARC Control and Management Architecture (Section 4)**

The requirements for a split architecture solution as defined in WP2, include the key aspects of support for virtualization of network resources in order to allow sharing of infrastructure among multiple operators; and increased flexibility by allowing deployment of new services in parallel to existing legacy protocol stacks. Furthermore, common carrier-grade requirements still apply, including service isolation, QoS enforcement, NMS integration, OAM provisioning, resiliency measures and scalability.

To meet this diverse set of requirements, in our *SplitArchitecture* proposal the control functions are organized as a stack of control planes, connected to each other via OpenFlow (see figure on schematic *SplitArchitecture* depicted above). Each control plane consumes services from lower control planes and acts as a controller according to the OpenFlow terminology, and, at the same time offers services to control blocks in higher planes and acts as a datapath element in this role. To facilitate this architecture, we define extensions to the OpenFlow protocol with flowspace management, which allows a control plane to express those parts of the overall flowspace it is actually willing to control. Additionally, we introduce a more generalized transport endpoint concept that maps onto the OpenFlow definition of "port" for datapath elements and supports transport endpoints on different planes in the stacked control layer.

A control plane contains several functional sub-blocks. First, an internal control block hosts the backplane control logic providing connectivity services within the sub-domain under control. Second, an external control block contains the protocol logic, which interacts with external peer control entities which manage other sub-domains. Finally, an optional interface might expose transport endpoints offered by this control entity via the OpenFlow protocol.

As mentioned, we define a recursive control layer by stacking control planes recursively. This hierarchical carrier-grade control architecture enables operators to deploy several control planes with minimal interference and assign flows dynamically based on given policies. Furthermore, it gives a network operator tight control over the level of detail of which data plane details are exposed to higher control planes or to third parties.

We provide considerations with regard to a suitable management framework flanking the controller architecture as well. When traditional network management definitions are applied to a generic SDN model with both a centralized control plane and a centralized management plane, we conclude that it is difficult to differentiate precisely between control and management functions in the context of SDN. Therefore, we present a proposal of how to integrate network management and control with the flexibility to choose whether to place network management functions (NMF) within an SDN controller or a separate network management system (NMS). Final functional assignment should be done on a by-case basis, depending on the exact scenario and use case in question. Parameters affecting such an analysis include the scale of the network (number of devices and geographical spread), existing legacy infrastructure in place, type of transport technologies in use, and type of services to be supported. Thus, in certain scenarios, either the controller based NMF or the external NMS could be designed minimalistically or even be omitted. In this deliverable, we present a way to handle the design choice of where to place network management functions. In our recommendations for carrier grade networks, we used timeliness and automatic configuration as the differentiators between control and management functions. However, depending on the specific use case and the technology used, the results of such an assessment might look quite different from case to case.

**SPARC extensions to OpenFlow and network element functions beyond forwarding (Section 5)**

In the section above, we outline a scalable, flexible control and management architecture. However, the network elements and their control interface (i.e. OpenFlow) will also require extensions, since we can conclude from our earlier deliverables (D2.1 and D3.1) that current OpenFlow-based implementations and standards do not fulfill carrier requirements. Below, we summarize the OpenFlow extensions discussed in this document. All these protocol extensions also imply additional functionality on the network elements that go beyond pure packet and flow forwarding:

- *Openness and Extensibility*: firstly, we suggest extensions required to implement the proposed recursive control plane, specifically a more flexible port management by replacing OpenFlow's current physical port model with a generalized transport endpoint model. Furthermore, we realized that more advanced processing functionalities at the data plane are desirable in many situations. OpenFlow currently provides processing of packets through state-less, lightweight actions (e.g., pushing tags, updating header fields, etc.). For statefull processing (i.e. taking proceeding packets of the flow into account), we propose using separate processing instances on a datapath element, which can be addressed by a lightweight "process" action type. Finally, we also revisit virtual ports and discuss their applicability for the use case of OAM, which requires execution of parts of the state machine directly on the datapath to meet strict timing constraints. In this final case, we specifically propose two types of virtual ports: a pre-/post-filter virtual port attached to a physical port, for cases in which we want to avoid passing OAM messages through the datapath element's forwarding engine (e.g., for connectivity checks); and a terminating virtual port, which terminates OAM messages that actually traverse the entire forwarding engine and thus also tests flow table entries (fate sharing).

- *Virtualization*: a crucial feature of future carrier-grade networks is network virtualization, enabling multi-service and multi-operator scenarios on a shared physical network infrastructure. In this deliverable, we propose improvements to the reliability, isolation and automation aspects of an OpenFlow-based network virtualization system. Regarding isolation, we identify the current lack of QoS primitives in OpenFlow as a major barrier. We also propose extensions to enable automation of virtual network setup and management.

- *OAM*: one important requirement for carrier-grade networks is the availability of proper OAM solutions. Regarding OAM integration in OpenFlow, we identified two contradicting aspects: on the one hand, integrating existing OAM tools (e.g., Ethernet or MPLS OAM) provides the desired compatibility with legacy OAM toolsets and offers well-known standard functionalities. On the other hand, integrating existing OAM tools in an OpenFlow environment requires integrating several technology-specific toolsets, which will substantially increase the complexity of datapath elements. In this deliverable, we discuss both aspects: a technology-dependent OAM solution and a novel, technology-agnostic generic flow OAM solution. As the technology-dependent OAM example, we detail how to implement an MPLS BFD based continuity check for both OpenFlow versions 1.0 and 1.1. For a technology-agnostic flow OAM, we propose a generic OAM module as a separate process on each datapath element. To still ensure fate sharing of OAM and data traffic, we suggest a virtual data packet approach that allows the OAM tool to test the entire forwarding engine.

- *Network Resiliency:* resiliency and reliability are important requirements for carrier-grade networks. In this deliverable, we first study mechanisms to ensure data plane resiliency in an OpenFlow scenario. We show how data plane resiliency can be realized through rerouting, restoration and protection. We also discuss control plane resiliency by outlining scenarios of how out-of-band and in-band control networks can be used together to achieve increased robustness in the control network.

- *Control Channel Bootstrapping and Topology Discovery*: current OpenFlow specifications do not describe how initial address assignment and control channel setup are performed. This is especially challenging in case of an in-band control network, since the datapath elements need to be able to establish IP connectivity towards the network control in the absence of a node configuration protocol. We will propose a method that facilitates automatic bootstrapping of datapath elements in such an in-band control case. Furthermore, we will present our extensions to the topology discovery module that is implemented in the NOX controller.

- *Service Creation*: we define service creation as the configuration process of network functions at service creation points within (or at the border of) the access/aggregation network to provide services to various types of customers. Besides connectivity, the configured functions include, among others, authentication, authorization and accounting (AAA) aspects. We propose integration of residential customer services in the form of PPP and business customer services in the form of pseudowires (PWE) in an OpenFlow-controlled operator network.

- *Energy-Efficient Networking*: centralized control software like OpenFlow offers additional options for reducing network energy consumption. We discuss possible energy saving approaches and functions (e.g., network topology optimization, burst mode operation, adaptive link rate). We then propose two sets of extensions for energy efficiency: one set of functions relating to port features (e.g., switching them on/off, enabling and configuring energy-efficiency functions etc.); and another set relating to configuration and monitoring of components of the switch itself (e.g., switch internal power management, switch temperature monitoring etc.).

- *QoS:* support for QoS mechanisms is generally considered a key requirement for carrier-grade networks. Furthermore, QoS plays an even more important role in a virtualized, multi-provider, multi-service operator network enabled by *SplitArchitecture*. In this deliverable, we show how traditional QoS tools such as packet classification, metering, coloring, policing, shaping and scheduling can be realized in an OpenFlow environment.

- *Multilayer Aspects:* we discuss the extension of OpenFlow toward control of configurable transport technologies, like wavelength and TDM switched networks. We used the example of circuit switched optical layers, i.e., packet-optical integration. We first discuss the additional requirements placed on a control framework by optical network equipment. We then outline possible phases for realizing packet-optical integration in an OpenFlow environment. Finally, we detail a proposal for GMPLS-aware multi-layer/multi-region extensions to OpenFlow.

**Implementation scenarios of the SPARC SplitArchitecture (Section 6)**

We differentiate between four types of how to integrate an OpenFlow-based *SplitArchitecture* into carrier grade-networks:

1. *Basic emulation of transport service*: OpenFlow data and control plane emulates and replaces legacy transport technologies (e.g., Ethernet, MPLS).

2. *Enhanced emulation of transport services:* As in 1., OpenFlow is used to provide transport services. Additional features and functions are added to both the data and control planes in order to comply with carrier-grade requirements.

3. *Service node virtualization*: In addition to transport services, OpenFlow also takes control of (distributed) service node functionalities, including service creation, authentication and authorization.

4. *All-OpenFlow network*: OpenFlow also controls other network domains, e.g., customer premises (RGWs) and the operator's core domain.

Considering that current OpenFlow 1.x is sufficient to provide integration type 1, the main focus of this deliverable is on integration type 2, i.e., studying possible extensions to OpenFlow in order to fulfill carrier-grade requirements, but we are also starting to touch upon integration type 3 in some cases (specifically in Sections 5.6 and 6.2 about service creation).

In Section 6, we present implementation scenarios of the proposed carrier-grade *SplitArchitecture* in an operator network, specifically in access/aggregation networks. Besides OpenFlow-controlled transport connectivity, we propose three evolutionary approaches for OpenFlow integration of service creation functions in access/aggregation networks, depending on which elements and functionalities are to be controlled by OpenFlow. These approaches include centralized OpenFlow control of a single service creation point at the IP edge only (e.g., BRAS), a decentralized model expanding OpenFlow control to aggregation devices (e.g., DSLAM), and finally a complete OpenFlow controlled approach including all devices in the access/aggregation domain.

Regarding a decentralized OpenFlow control model, we further discuss general implementation options for service creation. In addition to the need for legacy support for PPPoE, we identify another important requirement on OpenFlow improvements as seen from SPARC: the decision logic for authentication and authorization, which we will discuss specifically in order to outline the requirements and the potential options in more detail.

As examples of residential customer services, we outline the implementation of a "SPARC BRAS" and a "SPARC DHCP++". The "SPARC BRAS" transforms current service creation design into the SPARC design principle of the split between forwarding and processing. This discussion includes a detailed analysis of how OpenFlow fulfills the requirements for a BRAS/BNG as defined by the Broadband Forums TR-101 specification. The "SPARC DHCP++" essentially introduces a new set of protocols that needs to be supported in the OpenFlow environment.

In many scenarios, OpenFlow typically does not control all parts of the network structure. As an example, access/aggregation networks might be realized in a *SplitArchitecture* design, whereas the core network segment to which these networks connect still has legacy IP/MPLS as the predominate technology. This leads to an additional requirement: The *SplitArchitecture* domain must cooperate with legacy control planes in a suitable peering or horizontal interworking model. We discuss different options for connecting domains controlled by the SPARC control framework to legacy IP/MPLS control planes. As a result, we propose to update the controller architecture with a generalized network visor within control planes that provide a virtual router model of their part of the *SplitArchitecture* domain. An NNI protocol proxy can then use this virtual router to steer the communication with legacy control planes (e.g. the IP/MPLS), using relevant legacy protocols (OSPF-TE, LDP, BGP, etc.) that run as part of the controller.

Finally, we examined the feasibility of a *SplitArchitecture* with a numerical scalability study based on an idealized deployment model of an access/aggregation network. The results show that there are no stability concerns for static scenarios. The resulting requirements from the numerical model are in the order of magnitude or even below the capabilities of existing control and datapath devices. However, dynamic behavior (e.g., reconfiguration due to link failures) might raise scalability concerns, especially when strict time constraints exist. However, we can show that changing the connection structures through more careful network planning can increase scalability significantly even in the face of dynamic behavior.

# List of partners and authors

| Organization/Company | Authors |
|---|---|
| **DTAG** | **Mario Kind, Steffen Topp, F.-Joachim Westphal, Andreas Gladisch** |
| **EICT** | **Andreas Köpsel, Hagen Woesner** |
| **EAB** | **Wolfgang John, Zhemin Ding, Alisa Devlic** |
| **ACREO** | **Pontus Sköldström, Viktor Nordell** |
| **ETH** | **András Kern, David Jocha, Attila Takacs** |
| **IBBT** | **Dimitri Staessens, Sachin Sharma** |

# Table of contents

# List of figures and/or list of tables

# 1 Introduction

## 1.1 Project Context

The SPARC project ("Split Architecture for carrier-grade networks") is aimed at implementing a new split in the architecture of Internet components. In order to better support network design and operation in large-scale networks for millions of customers, with high automation and high reliability, the project will investigate splitting the traditionally monolithic IP router architecture into separable forwarding and control elements. The project will implement a prototype of this architecture based on the OpenFlow concept and demonstrate the functionality at selected international events with high industry awareness, e.g., the MPLS Congress.

The project, if successful, will open the field for new business opportunities by lowering the entry barriers present in current components. It will build on OpenFlow and GMPLS technology as starting points, investigating if and how the combination of the two can be extended, and study how to integrate IP capabilities into operator networks emerging from the data center with simpler and standardized technologies.

## 1.2 Relation to Other Work Packages

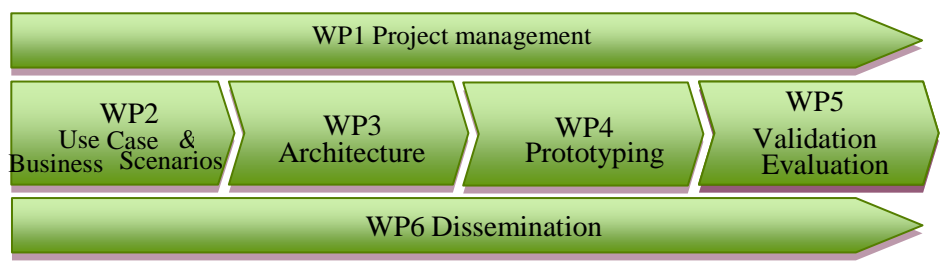

**Figure 1: Relation of SPARC work packages**

In the "workflow" of the work packages, WP3 is embedded between WP2 (Use Cases / Business Scenarios) and WP4 (Prototyping). WP3 will define the *SplitArchitecture* taking use cases and requirements of WP2 into account and will analyze technical issues with the *SplitArchitecture*. Moreover, this architecture will be evaluated against certain architectural trade-offs. WP4 will implement a selected subset of the resulting architecture, and feasibility will be validated in WP5. WP6 disseminates the result at international conferences and publications.

## 1.3 Scope of the Deliverable

In this deliverable, we present our final proposal for a split architecture for large-scale wide area networks, such as carrier-grade operator networks. Based on our conclusions from D3.1, we focus on OpenFlow as an enabling technology for the split architecture. Earlier SPARC deliverables (D2.1 and D3.1) made it clear that current OpenFlow implementations do not fulfill carrier requirements. In this deliverable, we therefore discuss a suitable framework for controlling carrier-grade operator networks and investigating missing features in OpenFlow as identified during the SPARC project in the context of access/aggregation network scenarios.

## 1.4 Report Outline

In order to set the stage, we start the deliverable in Section 2 with a summary of the use cases and requirements for a split architecture we defined and listed in SPARC deliverable D2.2. In Section 3, we then provide an overview of current software-defined networking (SDN) models, and compare them to the envisioned carrier-grade *SplitArchitecture* concept. Section 4 describes the control and management architecture of the proposed *SplitArchitecture*. Here, we discusses a recursive control architecture consisting of hierarchically stacked control planes and provide initial considerations regarding network management integration with SDN and *SplitArchitecture*. In Section 5 we propose required extensions to OpenFlow to support the envisioned carrier-grade *SplitArchitecture*. This section proposes the necessary extensions with respect to carrier-grade requirements based on missing features as identified in earlier SPARC deliverables (D2.1 and D3.1). Topics include openness and extensibility, virtualization and isolation, OAM, resiliency aspects, control channel bootstrapping and topology discovery, service creation, energy-efficient networking, QoS and multilayer aspects. Finally, in Section 6 we present selected deployment scenarios of a carrier-grade *SplitArchitecture* with OpenFlow. We show how OpenFlow-based *SplitArchitecture*s can be adopted for relevant scenarios prevalent in modern operator networks, such as service creation, general access/aggregation, network scenarios and peering aspects. Finally, we present a numerical scalability study indicating the feasibility of a split-architecture approach in access/aggregation network scenarios in terms of scalability requirements.

## 2    Review of SPARC use-case access/aggregation network

The access/aggregation network is the linking part between customers and core networks, where typically services are hosted. The general network architectures are depicted in A recursive control plane for SplitArchitecture.

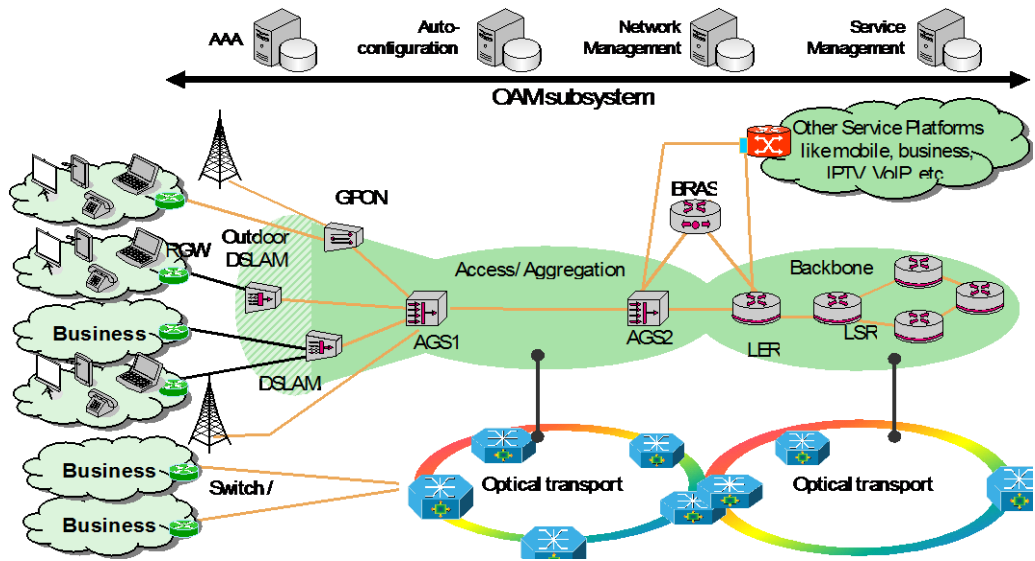

**Figure 2: The access/aggregation network**

Beside a vertical link, there exists two additional planes. First, there is a link with an optical transport network. Second, there exists a control or management plane hosting services like AAA, auto-configuration or service management.

The use case access/aggregation represents a manifold and diverse area. In SPARC, it was split in five areas (called use cases as well):

- Seamless MPLS or SDN approaches to MPLS Transport
- Multi-service/-provider environments (Service Creation)
- Mobile backhaul
- Software Defined Networking application in context of IEEE 802.11 compliant devices
- Dynamic control composition

### 2.1    Refinement of Requirements

Early in the SPARC project, we defined a set of 67 detailed requirements in WP2. In order to allow concentration on the most important ones, the total of 67 requirements was reduced in deliverable D2.1 in order to concentrate only on those requirements that are not already fulfilled with respect to existing architecture concepts and available implementations.

Four groups of general important requirements have been identified. The first group covers all required "Modifications and extensions for the data path element" or the *SplitArchitecture* itself. The other three groups deal with needed extensions of carrier-grade operation of ICT networks. The aspects related to the operation of an ICT network "authentication, authorization and auto configuration" (not to be mixed up with "AAA", as accounting is use-case-specific) are covered in a second group; "OAM" in the sense of facilitating network operation and troubleshooting in the third group; "network management, security and control" of the behavior of the network and protocols in the fourth group. Within network management, the aspects for the use of policies in network environments are included.

Additional to the work in WP2, also the assessment of existing *SplitArchitecture* approaches, such as ForCES, GMPLS/PCE, and most importantly OpenFlow, revealed a number of issues and open questions that need to be considered for future carrier-grade Split Architectures (see deliverable D3.1). The following topics have been identified that require special attention in the current architecture study:

- Requirement group "Network virtualization", e.g. to ensure strict virtual network isolation and integrity and handling of overlapping address spaces should be handled
- Requirement group "Recovery and redundancy" including open question not only with respect to the data plane of the network, but also with respect to controller and control plane failures
- Requirement group "Multilayer control" include integration of circuit switching and multilayer coordination for optimization or resiliency purposes
- Requirement group "OAM" functionalities for service management
- Requirement group "Scalability" to be considered for the data plane and proposing controller architecture

In the following architectural deliverable D3.2, the specific functions have been investigated and additional refinement of the different groups has been performed (each requirement group now effectively represents a general network function or feature):

- Requirement group "Modifications and extensions to the data path elements" has been broken down:
    - Requirement group "Openness and Extensibility" developing ways of how to extend OpenFlow to support a more complete, stateful processing on data path elements in order to enable OpenFlow support for further technologies with high relevance to carrier networks, such as PBB, VPLS, PPPoE, GRE, etc.
    - Requirement group "Multilayer" aspects is on the extension of OpenFlow in order to control non-Ethernet-based layer 1 technologies (as specified by IEEE 802.3 study group), especially the integration of circuit switched optical layers into OpenFlow (packet-optical integration).
- Requirement group "Authentication, authorization and auto configuration," is covered by the broader topic of "Service Creation"
- Requirement group "OAM" identifies a technology-dependent OAM solution (i.e., MPLS BFD) and a novel technology-agnostic Flow OAM solution
- Requirement group "Network management" was divided into several subgroups:
    - Requirement group "Network Management" covers the general framework for implementation of network management functions, fault and performance management covered by "OAM" configuration management
    - Requirement group "Quality of Service"
    - Requirement group "Resiliency" is commonly seen as one key attribute of carrier-grade networks i.e., the ability to detect and recover from incidents within a 50ms interval without impacting users
    - In order to facilitate this centralized network management operation, it was identified that automatic (Requirement group) "Control Channel Bootstrapping and Topology Discovery" is an important feature
    - Requirement group "Energy-Efficient Networking" provides functionalities to increase the energy efficiency of modern and future networks
- Requirement group "Virtualization and Isolation" enabling multiservice (within the responsibility of one operator) and multi-operator scenarios on a single set of a physical network infrastructure
- Requirement group "Scalability" is another key feature of the SPARC controller architecture [not covered in D2.1].

In this final architecture deliverable (D3.3), we add the topics of a "Recursive Control Plane" architecture (also referred to as hierarchical controller concept) and a separate discussion on "Network management" (not to be confused with the other requirement groups on QoS, resiliency, etc.).

## 2.2  Summary of Requirements

The list of harmonized requirement groups is the result of the evolving discussions during the entire project duration and has been aligned between the final deliverables in all technical work packages (WP2 to WP5). The final requirement groups have a focus on the access/aggregation use case only and are listed below:

(a) Recursive Control Plane

(b) Network Management

(c) Openness and Extensibility

(d) Virtualization and Isolation

(e) OAM (technology-specific MPLS OAM / technology-agnostic Flow OAM)

(f) Network Resiliency

(g) Control Channel Bootstrapping and Topology Discovery

(h) Service Creation

(i) Energy-Efficient Networking

(j) Quality of Service

(k) Multilayer Aspects

(l) Scalability

Note that we do not claim this list to be exhaustive. The above listed features have been chosen solely based on our own assessment of feature importance. In the continuous process of requirement specification, concepts and prototypical implementation, the scope has gradually been extended. However, in this deliverable we cover only the listed requirement groups listed above.

Overall, we conclude that the required network functions listed above have been analyzed and specified in detail where appropriate solutions could be found. With the focus on access/aggregation, it has been possible to develop a comprehensive set of solutions covering most requirements, and most solutions have been developed, implemented and tested (see deliverables of WP4 and WP5). In this deliverable, we cover requirement groups (a) and (b) in Section 4, groups (c) to (k) in Section 5, and finally assess requirement group (l) in Section 6.4.

# 3 Introduction to SplitArchitecture

## 3.1 State of the art

Today the design of network elements (i.e., switches, routers) follows a monolithic design pattern, i.e., each networking element integrates functions for control, forwarding and processing. Forwarding and processing capabilities are consolidated in the data plane where control capabilities are centralized in its own dependent control plane of a current network element. Usually it is not possible to access the interface in-between these functional blocks of today's network element, as depicted in Figure 3 (I). The upper control plane performs path computation primarily. This plane consists of network-wide distributed algorithms that enable construction of the essential forwarding information base for data plane operations. The compound of control and data plane is adjusted entirely to the provided service. The data plane performs packet forwarding, such as prefix matching, and is instructed by the control plane directly [54].

Splitting this design into functional independent planes (control and data) and opening the interface in between could be a successful way to relieve existing hardware and infrastructure of legacy architectures, thus facilitating the evolution and deployment of new protocols, technologies and architectures. Stanford's OpenFlow proposal [45] constitutes an independent control and data plane where the interface in-between is publicly accessible, as represented in Figure 3 (II). Here, OpenFlow is used to orchestrate intercommunication between separated planes. Applications (apps) are pieces of software coupled to the centralized control plane (typically called the controller). "Application" is a generic term that in this context could cover both network and service-related functions. Figure 3 II shows applications as separate entities; Figure 3 III shows network services (network-related functions) and business applications (service related-functions).

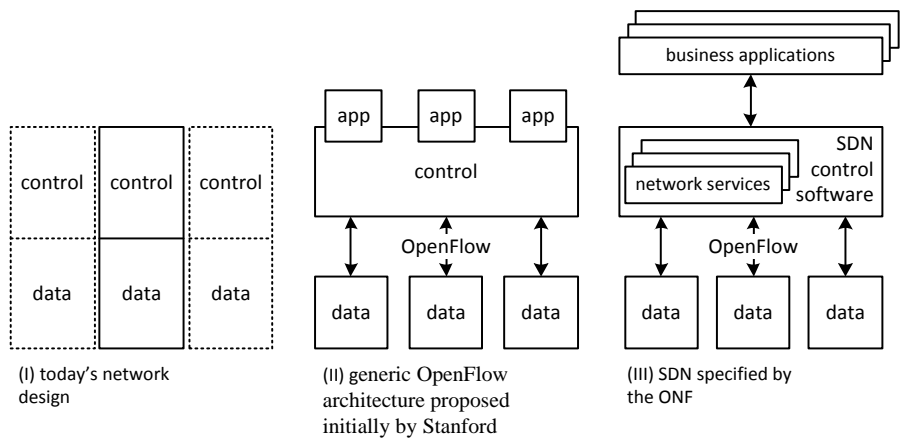

**Figure 3 Existing architectural design principles, based on research by Stanford and specified by ONF**

In this deliverable, we study how to extend the generic approach shown in Figure 3 (II) leading to our *SplitArchitecture* design proposal, in which we focus on the fulfillment of carrier-grade requirements (e.g., OAM, resiliency, network virtualization, etc.). As a result, carrier networks could benefit from the advantages of *SplitArchitecture* and enable network operators to have a disjointed evolution of data path and control mechanisms, which have the potential to pave the way toward more dynamic control of services and connectivity in carrier networks.

In deliverable D3.1, we discussed three architectural approaches that could be used to implement a split architecture. The IETF's ForCES framework (RFCs 3746 and 5810), IETF's GMPLS/PCE [69][70] and the OpenFlow approach by Stanford University [45], which is now specified by the Open Networking Foundation (ONF [43]), were considered.

While GMPLS/PCE provides a decoupling of control and data plane, most control plane functions are still distributed to each individual network element. Only the PCE architecture allows placement of parts of the control intelligence to a separate software component that can be accessed via a standardized interface, i.e. the PCEP protocol. Furthermore, GMPLS is a signaling protocol set that is useful within the control plane (e.g. for NNI); however, it does not specify the control connection between data and control planes.

Both ForCES and OpenFlow specifically target the control interface between the control and data planes. The ForCES framework initially seemed more mature in some aspects when compared to OpenFlow. ForCES provides a configuration model, allowing the specification of different technologies via libraries. OpenFlow, on the other hand, provides a more detailed, but also more rigid node model, which makes OpenFlow simpler but less flexible than ForCES. However, the strong industry support recently observed for OpenFlow, and the lack of support for ForCES in academia and industry, makes OpenFlow the more interesting and evolving technology for split architecture today.

### 3.1.1 Visions and specifications of the ONF

During 2011, the Open Networking Foundation (ONF) was founded as a non-profit consortium, consisting of network operators, equipment manufacturers, software suppliers and chip technology providers. The ONF is dedicated to developing and standardizing a software-defined network (SDN) architecture. *SDN* was coined as a new name for the concept of separating control and data plane as first proposed by Stanford when introducing OpenFlow.

The ONF adopted OpenFlow and SDN approach by Stanford and specified it as depicted in Figure 3 (III). The separated control plane is represented by the SDN control software in this architecture approach. ONF itemized applications from Figure 3 (II) into network services and business applications. Network services represent basic network functions such as routing, topology discovery and policy management. In contrast business applications are functions for service generation or a service itself. Business applications are entirely decoupled from the SDN control software (control plane) and operate on an abstracted view of the underlying network. Abstraction is provided by the interface between the applications and control software. ONF uses OpenFlow for communication between the SDN control plane and particular data plane entities, as also proposed by Stanford (Figure 3 (II)). The northbound interface between the SDN control software and business applications is currently under discussion.

The ONF defined SDN as "an emerging network architecture where network control is decoupled from forwarding and is directly programmable … Network intelligence is (logically) centralized in software-based SDN controllers, which maintain a global view of the network" [44].

According to the ONF, the key features of SDN are thus:

(a) Separation of control and data plane;

(b) Centralized control plane (or controller) with global network view; and

(c) Programmability by external software modules or applications via the controller.

The current focus in the ONF lies on the southbound controller interface, i.e., the interface between the separated control and data planes. This interface has been instantiated through the OpenFlow protocol, as depicted in Figure 3 (III). In the execution of the ONF approach SDN control software is instanced by even one or more OpenFlow-Controllers, while individual data plane entities are represented by specific, independent OpenFlow capable switches. The ONF identified OF as the unique major component of its proposed SDN concept. However, other aspects, such as commercialization and promotion, are also covered by the ONF.

The ONF is organized as independent working groups and supervised by the board of directors and a separate technical advisory group. Current WGs consider the following aspects regarding OpenFlow: Archtiecture & Framework; Extensibility; Configuration & Management; Testing & Interoperability; coexistence of conventional and OpenFlow enabled forwarding mechanisms (Hybrid switches and networks); and Market Education. The WGs with most relevance in terms of protocol standardization are the Extensibility group, driving the *OpenFlow* specification, and the Config & Management WG, driving the specification of the newly introduced SDN configuration mechanism *OF-Config* [42].

### 3.1.2 Evolution and status of the OpenFlow protocol

The OpenFlow protocol was initially specified by the universities of Stanford and Berkley. With the launch of ONF, all specification activities concerning OpenFlow have been taken over, resulting in standardization of more consistent protocol versions. The extensibility working group has driven aspects such as integration of IPv6, sophisticated matching strategies and flexibility. Version 1.2 is the first standard of the protocol evolved solely by the ONF and builds significantly on the predecessor version 1.1. In the course of enhancement to version 1.3, more and more missing aspects were addressed and added to the upcoming versions. The evolution of the protocol is outlined in the following table [41]. Table 1 assumes OpenFlow 1.0 as reference and covers major increments of the respective release.

**Table 1 Overview of OpenFlow releases**

| **OpenFlow 1.1** (Feb. 2011) | **Multiple table processing**: An OpenFlow pipeline may consist of separate flow tables which are concatenated. |
|---|---|
| | **Groups:** The abstraction of a group enables to use a set of ports as a discrete entity for forwarding packets properly, such as multicast or multipath. |
| | **MPLS and VLAN assistance:** Support of sophisticated manipulation mechanisms (i.e. add, modify, delete) of MPLS and VLAN labels. The proposed methods are not limited to single level VLAN tagging, but consider queue-in-queue tagging as well. |
| | **Virtual port concept:** Virtual ports are used to represent forwarding abstraction such as tunnels. |
| | **Connection failure:** Either switch continues to work in standalone or secure mode instead of using an emergency flow cache once connectivity with the controller is lost. |

| OpenFlow 1.2 (Dec. 2011) | **Extensibility match assistance:** Fixed structure of *ofp_match* changed to a type-length-value (TLV) structure. **IPv6 support:** Capabilities concerning IPv6 matching and header rewriting (i.e. source and destination address, protocol number, traffic class and ICMPv6 information) have been added. **Controller role change:** Switches are controlled by only one controller (master) but may maintain connectivity to a set of controllers (slaves) concurrently. This can be used for failover, where the roles of controllers change immediately on connection loss to the master. |
|---|---|
| OpenFlow 1.3 (April 2012) | **Table miss entry:** The previously assigned usage of table configuration flags is replace by a specific table-miss entry to take care of any non-matched packet. **Advanced IPv6 support:** Added ability to match the presence of several IPv6 extension headers (i.e. fragmentation, hop-by-hop). **Meters:** Meters are attached to flow entries to enable reliable measurement and control of the rate of packets per flow. |

### 3.1.3 SDN model and status of OF-Config

In parallel with the evolution of OpenFlow, the ONF configuration and management working group (WG) focussed on how to configure/monitor aspects of the datapath elements that are not associated with the OpenFlow protocol. Current versions of OpenFlow do not cover management and configuration features sufficiently (e.g., address have to be configured manually for control channel establishment). This WG thus specified OF-Config, a mechanism for configuring OpenFlow capable devices. The configuration mechanism is a first step to adding previously lacking management and configuration capabilities to SDN, which we also identified as an important carrier-grade requirement within SPARC.

OF-Config 1.0 was released in January 2012, and is based on NETCONF (RFC 4741), a transactional protocol that uses remote procedure calls on top of a secure transport channel to manage configurations on remote devices. Instead of focusing on a complete range of functions, the purpose of this first specification was to define a schema to ensure a consistent representation of configuration elements in the protocol. To specify the data model of the switch configuration, the WG made use of two specification options. As the NETCONF is XML-based, the data model used during configuration can be specified using XML schemas. However, since the XML schemas lack support for specifying behavioral constraints, the OF-Config specification was given as the YANG model (RFC 6020). The first version of the configuration protocol, OF-Config 1.0, included only a limited set of functions, such as assignment of a set of controllers, separate configuration of related resources (i.e. ports, queues) and remote manipulation. In OF-Config 1.1, additional functions such as certificate handling, capability discovery and the configuration of three basic tunnel endpoints have been added. Further functions expected in future versions include topology discovery, capability configuration, advanced tunnel configuration and instantiation and resource specification for the logical switches. The WG is also discussing whether broader OAM functions, such as fault and performance monitoring, should be part of OF-Config or should instead be discussed in separate future specifications.

To summarize, the functional scope of the OF-Config (version 1.1) protocol is the following:

1. The assignment of one or more OpenFlow controllers
2. The configuration of queues and ports
3. The ability to remotely change some aspects of ports (e.g. up/down)
4. Configuration of ceritificates for secure communication between the OpenFlow Logical Switches and OpenFlow Controllers
5. Discovery of capabilities of an OpenFlow Logical Switch
6. Configuration of a small set of tunnel types such as IP-in-GRE, NV-GRE and VxLAN

Figure 4 shows the coexistence of OF and OF-Config within the current SDN architecture as defined by ONF. In this architecture, an OpenFlow capable switch, which is a physical or virtual network element, is hosting one or more OpenFlow logical switches. The logical switches represent the actual OpenFlow network elements, which are controlled by one or more OpenFlow Controllers using the OpenFlow protocol. Network applications on top of the OF Controller use the network via the OF Controller's northbound API (NB API). Finally, an OF Configuration Point represents the service that communicates via the NETCONF-based OF-Config protocol with an OpenFlow capable switch and partitions resources among OF logical switches (such as ports and queues). Currently, the relationship between the OF Controller and OF Configuration Point is deliberately not defined by the ONF.

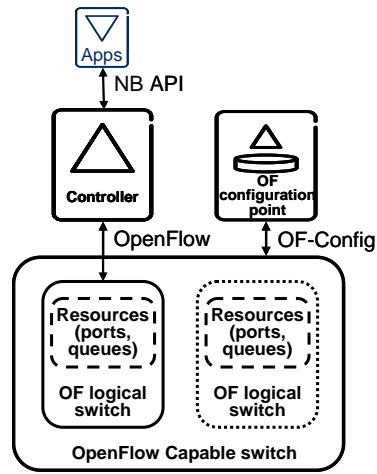

Figure 4 ONFs SDN architecture including OpenFlow and OF-Config [42]

### 3.1.4 Vision of SDN in Academia

Concurrently to the SDN model defined by the ONF, further alternative SDN architecture attempts have been made in academia. The main aspect discussed by academics is the integration of an additional element called hypervisor, responsible for virtualization and abstraction of the network. Scott Shenker, professor at the University of California, Berkley, has done the most notable work. His vision of an SDN architecture is approximated in Figure 5 (II).

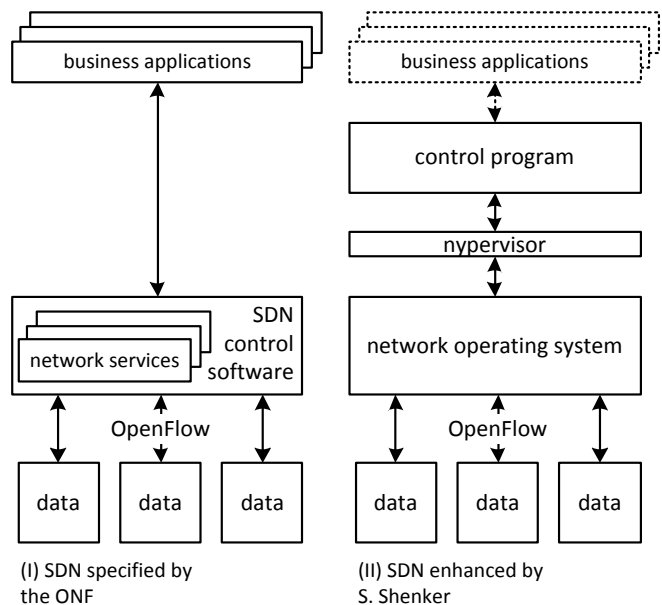

Figure 5 Enhancement to the initial OpenFlow model

Shenker separated network services as specified by the ONF (Figure 5 (I)) into control programs and services provided by the network operation system itself. In our understanding, the terms "SDN control software" as introduced by the ONF and "network operation system" as proposed by Shenker can be used interchangeable. Control programs (e.g. different kinds of routing algorithms) are network-related functions, which are not part of the essential set of functions provided by the network operation system itself (e.g. topology discovery). These programs can be implemented in different ways and substitute for one another, while functions provided by the network operation system constitute the set of fundamental functions that must be provided. The nypervisor (an amalgam of network and hypervisor) has a global view of the underlying topology and its resources (e.g. address spaces). Here the essential function of decoupling the upper layer entities from the underlying topology is organized by providing an abstracted total global view. The preferred position of the nypervisor is discussed by academics controversially and several different proposals exist in the literature. As depicted in Figure 5 (II) Shenker prefers to have the nypervisor atop the network operation system.

## 3.2　　SPARC Vision on SplitArchitecture

Based on the separation of control and data plane of current network elements and influenced by simultaneous ongoing trends, SPARC evolved the concept of *SplitArchitecture*. *SplitArchitecture* enhances the various control separation approaches integrated in different existing architectures (e.g. GMPLS or ForCES, OpenFlow Figure 3 (II)) and from the work of both the ONF and academia (cf. Figure 5).

In general, *SplitArchitecture* acknowledges the principles of SDN with a split between data and control plane as well as the introduction of a kind of network operation system. A high-level graphical representation of the concept of *SplitArchitecture* as proposed by SPARC is shown in Figure 6. There are three substantial differences to the previous concepts proposed by the ONF and academia.

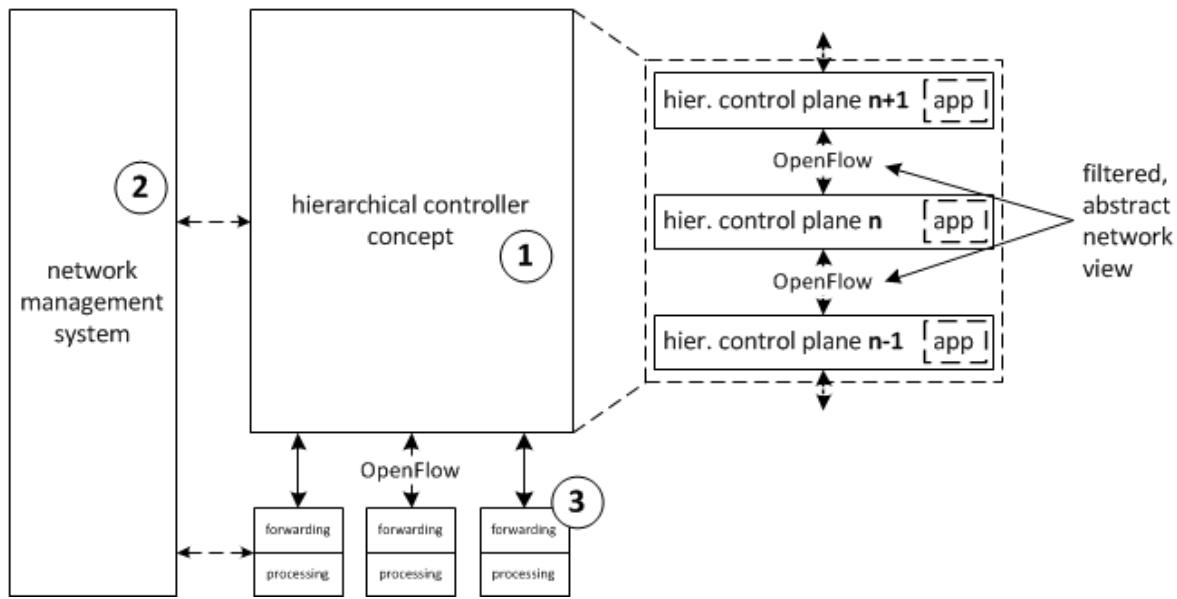

**Figure 6 SplitArchitecture defined by SPARC**

(1) First, the split between control program and network operating system (NOS) with the help of the nypervisor and two different abstractions is reduced to one abstraction (see Figure 5 II) - the filtered, abstract network view. This means that a network hypervisor is not mandatory and can be replaced by a basic set of filter function establishing a meaningful abstraction. Currently, the discussed design follows a pragmatic way of hierarchical controllers separating the network view by filtering the available addressing scheme and therefore granting access for control programs to parts of the resources only. In SPARC's concept of *SplitArchitecture*, control planes in a hierarchical controller unify the network operation system, the hypervisor functionality as proposed by Shenker (Figure 4 (II)), and control applications. The hierarchical controller concept as defined by SPARC means that several control planes are stacked upon one another recursively. In this hierarchy, each plane acts as controller toward data path elements in a lower plane and as single data plane entity toward higher planes. Further discussions about the control layer and the proposed controller architecture in SPARC can be found in Section 4.1.

(2) Secondly, we specifically discuss the role of network management in an SDN environment. Trying to apply traditional management definitions a generic SDN model we conclude that it is difficult todifferentiate precisely between control and management functions in the context of SDN, where both the control and management planes are centralized. As a result, we present a proposal network management integration with the flexibility to choose whether to place network management functions (NMF) within an SDN controller or a separate network management system (NMS). The proposed management architecture that is based on the current SDN model defined by the ONF (cf. Figure 4). In our proposal, the control plane entities includes a network management function (NMF) module consisting of an OF configuration point and an equivalent monitoring point, responsible for configuration and monitoring interfaces (OF-config and OF-mon). This NMF shares the network view of the controller, including topology information and updates to this view in terms of alarms and notifications from the data plane. In terms of functionality, the NMF can take over responsibility for configuration, fault, and performance management functions that are useful in the control layer. The controller and the NMF module can also interact with an external NMS for functions beyond the scope of the configuration and monitoring interface provided by the NMF in the controller. The network management considerations are presented in Section 4.2.

(3) The third difference is an additional split in the data plane between forwarding and the processing of related procedures. Various aspects motivate this split. Forwarding decisions are done most efficiently at the edge of the networks, but the network elements at the edge lack sophisticated processing capabilities in many cases (e.g. DSLAM). Processing capabilities are today spread around the network environment. Compared with other technologies, processing capabilities are evolving fast, so a separation of related entities could ease innovation. In addition new protocols could be integrated in existing environments more easily, for example by supporting general-purpose computing hardware. Moreover, at the end of the life cycle, legacy protocols could be phased out by moving the desired processing capabilities to other locations while keeping forwarding decisions with the same network device. Referring to the access/aggregation domain, an example could be the PPPoE recognition in a DSLAM, but moving the BRAS functions from dedicated, single-purpose devices to a general computing hardware or a data center. Section 5 discusses the necessary extensions to protocols and data path elements to fulfill carrier-grade requirements, which includes discussions on these specific topics in the Sections 5.1 on Extensibility and in 5.6 on Service Creation.

# 4      Carrier-Grade Control and Management Architecture

In this section, we will first present our architectural considerations regarding a suitable control plane for the carrier-grade *SplitArchitecture*, i.e., applying SDN principles to wide-area operator networks. Besides discussing the separation of control and forwarding, which is essential to SDN, we will also provide a discussion about how network management functions could be integrated into the *SplitArchitecture* in Subsection 4.2. Extensions required to datapath elements and the OpenFlow protocols itself will then be detailed in the subsequent Section 5.

## 4.1      A recursive control plane for SplitArchitecture

While OpenFlow is "flattening the layers" in that the datapath executes a single match for multiple packet header fields, this does not mean that the control plane has to be flat as well. Figure 6 introduced the basic concept of the SPARC *SplitArchitecture*. It shows how any controller in recursive architecture is a network application to the next lower controller.

The proposed architectural extensions to a split control plane based on OpenFlow do not dictate any constraints on control plane designers with regard to organization of the control plane. Instead, it gives the freedom of adapting the control architecture according to needs and business models. At the same time, recursive stacking, flowspace management, and advanced processing allows distributing controllers to adapt to network capabilities. The freedom to distribute or centralize controllers increases scalability, adds resiliency over a single-controller architecture and allows plugging in legacy equipment (and legacy control planes). A carrier-grade SDN-aware control architecture should also support legacy control protocol operations, as network operators require a smooth upgrading path for deploying SDN enabled devices and (sub-) domains within a network environment still based on legacy control protocols. This also includes the network management system and all OAM operations. To sum up, the main requirements discussed in SPARC deliverables D2.1 and D3.1 are:

1. Enable virtualization of network resources for sharing physical infrastructures among several operators.
2. Allow deployment of new control plane architectures and services in parallel to existing legacy protocol stacks.
3. Maintain all requirements in terms of service isolation, OAM provisioning, QoS enforcement, and NMS integration as defined in existing networks today (these topics are discussed in Sections 5.2, 5.3, 5.8 and 4.2 respectively).

The OpenFlow API defines all basic service primitives for implementing remotely accessible Service Access Points (SAP) for OpenFlow capable switches. These SAPs are different from the ones defined in ISO/OSI [73], as the architectural split enforced by OpenFlow is different from OSI. Typically, there is only the control of the data flow going through this SAP, eventhough OpenFlow also allows data frames to be relayed to the controller through that API.

The elements of the APIs, which a *SplitArchitecture* controller provides upwards, will obviously differ for network management, monitoring, and control purposes. For the latter purpose, we propose OpenFlow itself to export an SAP to client control layers for accessing a server control layer's communication services in the recursive control architecture. This OpenFlow-realized SAP also defines the information model: it models provided communication service with one forwarding device. This representation automatically hides the details of how the service is actually implemented resulting in a compact representation through the filtered network view. This design is inline with the abstract node representation provided by GMPLS [69].

We aim towards organizing a carrier-grade control plane as a stack of protocol layers, However, contrary to the monolithic controllers (such as NOX and Trema) seen today in network elements, we split the control plane into an theoretically arbitrary number of control blocks implementing layers and use OpenFlow as the base interface between these control blocks. Furthermore, splitting a protocol stack into several logical control blocks and using OpenFlow as glue among them, allows us a distribution of functional blocks across several locations and devices within the control plane, thus simplifying instantiation and attachment of new control blocks at run-time for either load balancing purposes or addition of new capabilities, processing or forwarding capacities.

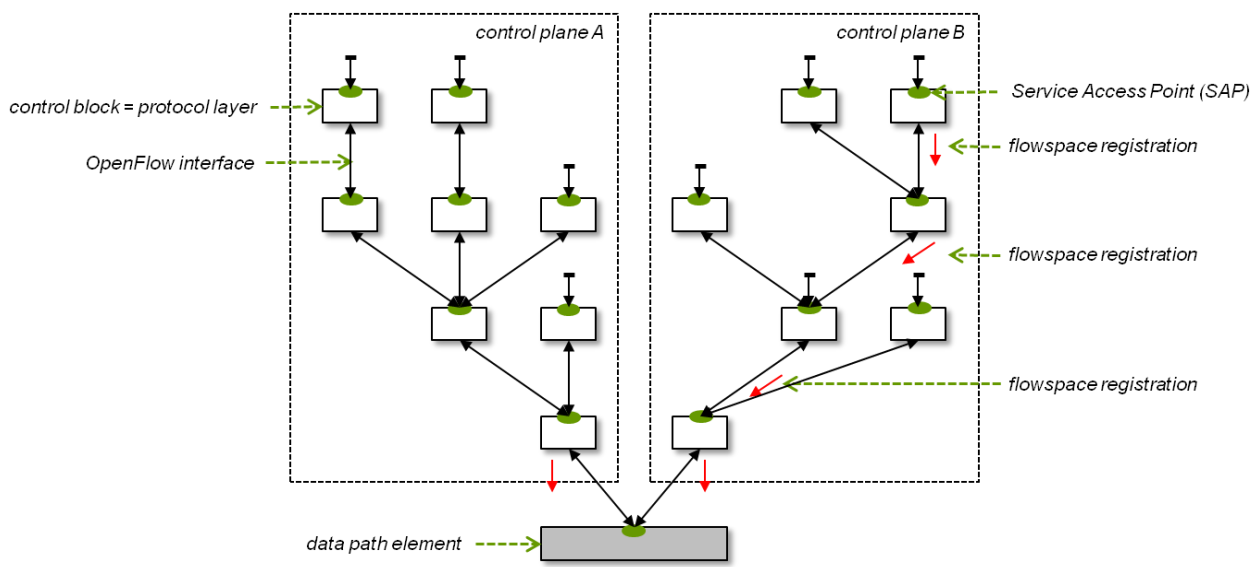

**Figure 7: Control plane organized in control blocks, OpenFlow as remote SAP**

However, OpenFlow lacks some crucial ingredients for such software based modular control plane: moving to higher layers in a protocol stack involves typically a de-multiplexing step, i.e. a Service Access Point as defined by OpenFlow can be used by several higher-layer instances in parallel. As OpenFlow defines a 1:n relationship between the control plane and datapath elements in the data plane, a single layer (n-1) entity only can attach at any time to layer (n). To solve this limitation, we introduce *flowspace management*. With flowspace registrations, a controller entity, when attaching to a datapath element (= lower-layer instance), expresses the part of the overall flowspace available at the SAP that it is willing (and able) to control. When several controller entities connect to an existing SAP, the datapath element uses the existing flowspace registrations for demultiplexing Packet-In (Flow-Removed, etc.) events and sending the event to the controller responsible for that flow.

The general properties we have introduced in this section so far are summarized below:

- We proposed OpenFlow as an open API to build a stack of control blocks. Each control block consumes services from lower-layer entities and acts here as a controller according to the OpenFlow terminology, and, at the same, offers services to higher-layer entities and acts as a datapath element in this role. A control block also behaves as an OpenFlow proxy, as it is datapath and controller at the same time, though it typically implements broader set of services rather than acting as a proxy.

- We extend the OpenFlow protocol with flowspace management that allows a controlling entity to express parts of the overall flowspace accessible at the SAP it is actually willing to control. This provides fine control of the (de-)multiplexing function of the server control block and allows several control entities to use the same OpenFlow SAP in parallel without interfering with each others.

Note that these design principles do not define or restrict a control layer's internal structure and architecture. A control plane designer is free to decompose the desired control functions in any arbitrary control hierarchy. One can collect all functions into one OpenFlow controller making the APIs between the control blocks internal to the controller. Of course, this does not preclude opening up these APIs. This paradigm of "fat controllers" was first adapted in the OpenFlow community and the most widespread controller implementations, such as NOX, follows this concept. The other extreme is when the OpenFlow-based SAP is defined between the atomic control blocks.

### 4.1.1    Controlling a single network element using SDN

Figure 7 depicts an example of how two control planes A and B attached to a single datapath element and implement a protocol stack for a single network element. Both control planes consist of a set of control blocks that are mutually connected via OpenFlow interfaces. Multiple controlling entities may register flowspaces at the same Service Access Point for parallel control of the underlying datapath element. Flowspace registrations of parallel control blocks may or may not overlap. An overlapping flowspace allows definition of "catch-all" controllers, e.g. for implementing monitoring devices or handling denial-of-service attacks.

The OpenFlow specification introduces the concept of ports. The OpenFlow port model covers physical ports only (or logical ports like trunk interfaces or simple tunnel endpoints). While physical ports do not exist at higher protocol layers, they may define specific transport endpoints for this layer. As an example, one may consider a controller block located in the Ethernet layer. To offer Ethernet transport services, the Ethernet controller will offer Ethernet transport endpoints (= endpoints with a specific MAC address assigned) and use the physical ports exposed by the underlying datapath element. A generalized understanding is that any port is a resource that an OpenFlow datapath element offers for control by one or more OpenFlow controllers. The transport service that the port is offering differs when moving up the stack of controllers, at the same time enlarging the geographic extension of the transport domain. While physical ports typically connect to one opposite port (or more for broadcast-and-select networks), learned MAC addresses define the "ports" of an Ethernet controller, eventually creating the notion of a "big switch". We extend the OpenFlow port model and introduce a more generalized one that maps to either physical ports or transport endpoints (for details see Section 5.1.1).

### 4.1.2 Controlling multiple network elements with a single controller

In non-SDN enabled networks, all network elements contain specific control stack tailored for their needs, e.g., a switch provides spanning tree and reverse learning for supporting Ethernet forwarding while a router implements IGP protocols to support IP forwarding. Introducing a split architecture concept does not necessarily change this situation. Extracting the control logic from each network element and shifting it towards the control plane still maintains the previous situation: each control block is still an autonomous entity and all these entities establish and synchronize shared state (see Figure 8). A network element makes routing decisions and programs its forwarding autonomously.

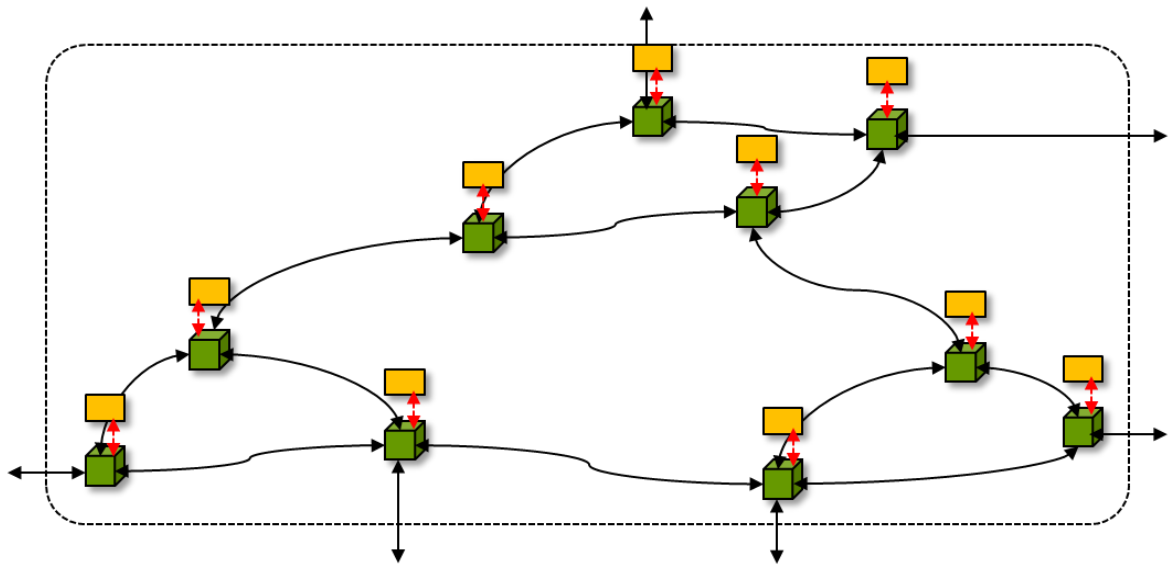

**Figure 8: A legacy network domain with distributed autonomous protocol stacks. Yellow boxes depict control blocks and green boxes datapath elements.**

The control plane designer may decide to divide the network into smaller control domains and designate a control element to be responsible for (1) supervising all control procedures performed within the control domain; and (2) establishing some form of information exchange with designated control entities for some other control domains. As part of this latter task, the control entities maintain an abstract node model of the owned control domain. As the translation between the physical sub-domain and the abstract node model is local to the controller, it can decide the amount of details it shares. This allows reduction of the amount information advertised among the control entities and thus enhances the scalability of the control plane.

In Figure 9, the network operator has split the network domain in three sub-domains. Two of them are controlled by dedicated centralized control entities; while the third is kept fully distributed. From an external perspective and with respect to its adjacent nodes, each sub-domain behaves as a single virtual node with a set of ingress/egress ports that connect to neighbor nodes not under control of the sub-domain's control block. Considering the sub-domain's internal operations, packets received by any ingress port will be either terminated within the sub-domain (e.g. at an emulated transport endpoint within the control plane) or sent via some egress port to an end system or a network domain in the next hop. Please note that two protocol layers operate in this situation: one protocol layer controls the sub-domain's internal operations and provides packet transport services in the virtual node's backplane, while the other protocol layer interacts with entities outside of the sub-domain and uses for its operation the packet transport services of the backplane controlling entity for its operation.

Note that any combination of centralized and distributed control can exist in a network. The details of the abstract model for the control entities are restricted only by the protocol they use.

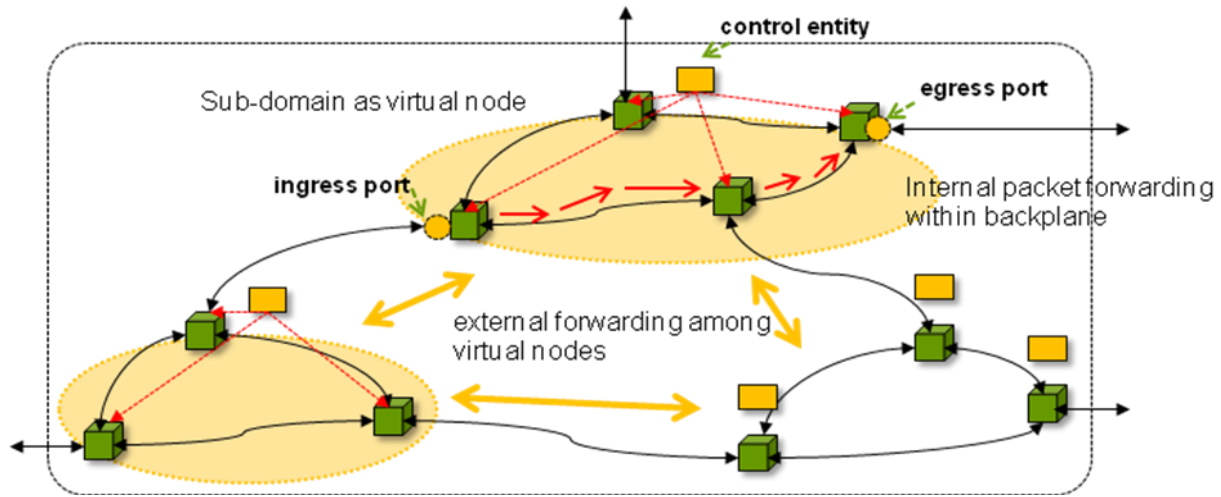

Figure 9: Split of control and data plane

The functional entities discussed above are organized into two control blocks:

1. **The internal control block** controls the sub-domain's internal operation, i.e. transport of packets across the sub-domain's internal backplane. Since a sub-domain hides all of its internal operations from the surrounding environment, the protocols and solutions adopted for managing the backplane operation are out of scope of our architecture. In principle, network designers can choose any solution that meets their requirements. However, that solution must provide packet transport services among the ingress/egress nodes of the sub-domain. This resembles the principles of an IEEE 802.11 compliant distribution system. The backplane control block must detect the ingress/egress nodes of the sub-domain through some form of topology discovery.

2. **The eexternal control block** interoperates with peering control entities. In our example discussed here, we split the network domain into two abstract nodes, which actually implement two control domains formed by three and four nodes respectively; and three physical nodes having dedicated control plane entities. In a hybrid scenario with legacy and SDN enabled network elements, the network operator would select the protocol used for interworking of the new formed abstract nodes based on the legacy protocol stacks, in order to avoid updates to the legacy network elements. For the access/aggregation use case considered in SPARC, the IP/MPLS control protocols, OSPF, LDP, RSVP-TE, BGP etc, will provide the necessary glue among the control domains. In a non-hybrid scenario, the network operator may choose either standardized protocols like in the hybrid case, or any state sharing mechanism between the SDN controllers, such as HyperFlow [72].

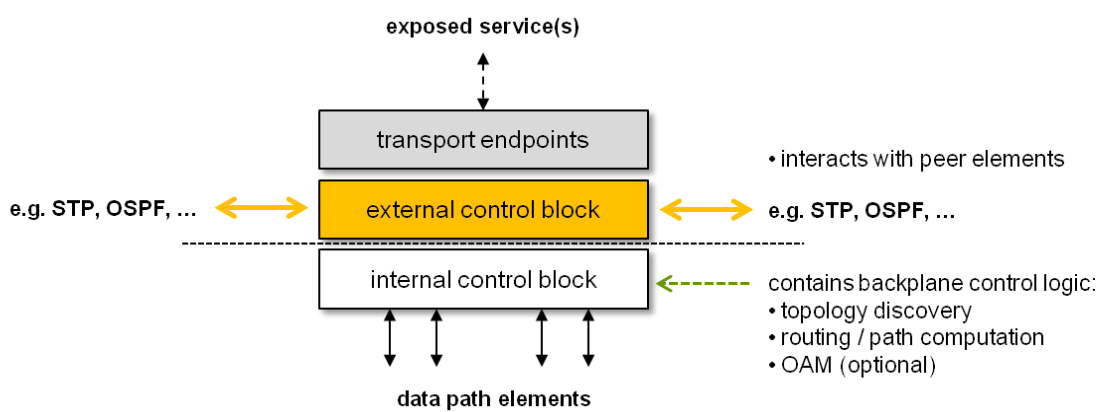

Figure 10: Functional blocks in a control block

Besides the above two control blocks, a third block is required: this module exposes transport endpoints via OpenFlow for accessing the transport services offered by this control block (optional).

The clustering of network elements in an abstract node is not limited to one control layer. Rather, we can apply this scheme recursively on top of abstract nodes again. However, for such recursive architecture, a control block must expose a datapath-like interface towards the client control entities. The depth of such a hierarchy is in principle unlimited; however, for practical reasons, it should be limited to a maximum value[1]. In Section 4.1.1 we introduced some design guidelines (OpenFlow as northbound interface, flowspace reservation) for controlling a single network element (NE) with an SDN control plane. Since a hierarchy of virtual node control entities behaves like a single node towards the control plane, we can combine both principles in a single framework (see Figure 11).

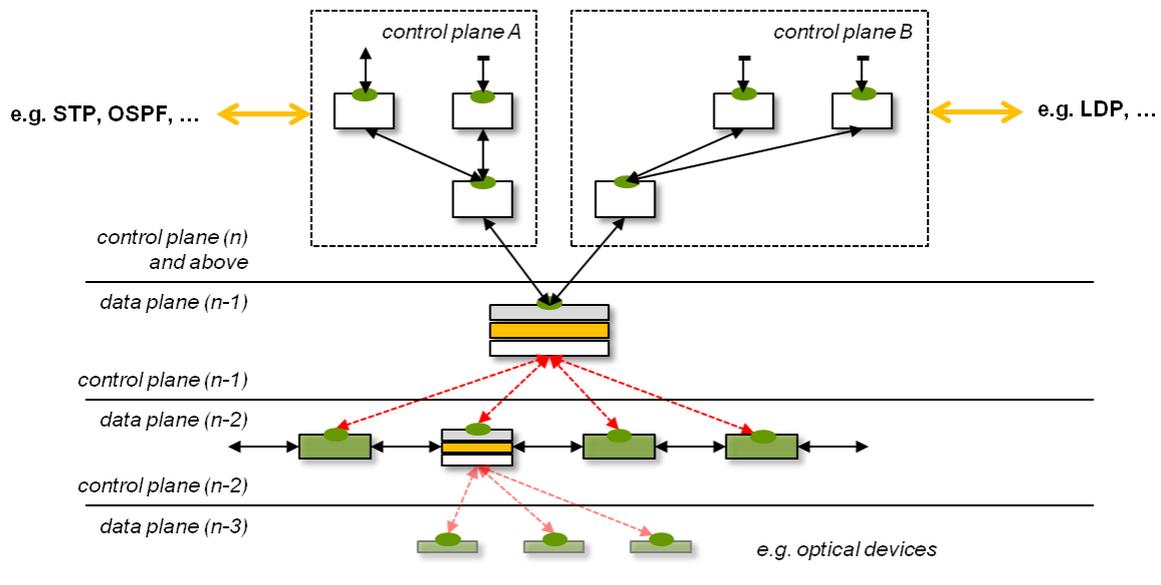

**Figure 11: Recursive Architecture of Virtual Nodes and Modular Control Plane**

### 4.1.3    Mapping the carrier-grade Architecture on the SPARC implementations

As we discussed above, the suggested control plane design does not define or restrict a control layer's internal structure and architecture. This also means that the number of such control layers and the mapping of control blocks into these layers depend only on the network operator's design decisions. Within the SPARC project, we identified two cases for developing and implementing SDN-aware control solutions for transport and service functions: an MPLS-based aggregation and core network, and a Broadband Network Gateway (BNG) service node.

For scalable control of the MPLS based aggregation, we adopted the design principle of splitting the network domain into control domains: the aggregation network is split into several control domains, while the legacy distributed control plane is retained in the core (see Figure 8). Within each aggregation control domains the control blocks required for provisioning MPLS transport services (PWE, MPLS tunnel provisioning, OAM, recovery) as well as communicating with the core control domain are grouped into one control layer, which contains the major functional groups shown in Figure 10.

As an example of service node virtualization, we implemented a control plane based on the design principles depicted in Section 4.1.1. The carrier-grade control plane consists of several control blocks and uses OpenFlow (and its protocol extensions, presented and applied in Sections 5 and 6, respectively) to create a Broadband Network Gateway (BNG) for access/aggregation domains. Each control block uses the same modularization principles internally, i.e., a control block adopts the OpenFlow interface for loading adaptation and termination functions and for exposing Ethernet and IP based transport endpoints. An initial implementation of a control plane following this design was demonstrated at the Open Networking Summit in April 2012. A detailed description of this prototype is available in SPARC deliverable D4.3.

---

[1] OpenFlow limits the maximum to 1024, as any name space will at least use one bit. In practical applications the number should of course be as low as possible, limiting the processing delay in the control stack. The actual number will be a trade-off between performance (single control block) and modularity, allowing for separation of concerns, better testing and virtualization.

### 4.1.4        Flowspace Management

In the preceding sections, we introduced control blocks as basic building blocks of a protocol stack. Each control block acts as a proxy entity, i.e. it operates as datapath element offering services to higher layers and a controller for using services from lower layers. We adopt the OpenFlow API for binding such control blocks to each other. The OpenFlow 1.x series of specifications defines a 1:n relationship between controllers and datapath elements, which means a single controller may control multiple datapath elements, but a datapath element can only have a single controller. We propose an extension to the OpenFlow framework named flowspace management that relies on the slicing of flowspaces. Before we go further, let us briefly cover some properties of flowspaces and possible options for slicing them.

Packet header information is typically structured into fields as a MAC address (source and destination), VLAN tag, IP addresses, protocol type, port, MPLS tag #1,#2, … etc. OpenFlow defines 14 of these headers in version 1.1 of the specification. The different headers correspond to multiple layers and allow the scalability of communication by reducing the number of communicating entities per layer.

It is now possible to view all these headers as one large tag that gives an identifier to an individual packet or flow. This *flattening of the namespace* is an attractive feature of OpenFlow because it reduces the total numbers of layers (which typically translate into specific boxes in a carrier-grade network). Still, however, the functions at the border of the network need to restructure this flat label into meaningful headers that can be processed.

The many-to-many relation of adjacent layers also enables the *parallel deployment of different control planes* and allows an the individual assignment of resources to one of the deployed control planes. This requires an appropriate solution for resource slicing or virtualization at any of the server control layers.

When multiple controllers share a single underlying controller or datapath element, the slicing between them requires multiplexing/demultiplexing. Controllers of the control layer (n+1) need to be addressed from the underlying controller/datapath element. This demultiplexing of messages (e.g., the Packet_In message) needs to be encoded in the flow itself, as there is no additional information than the packet header. This means that the known endpoints of layer n are exposed to the layer (n+1) controller.

Each controlling entity of a client layer must be aware of which resource slices are allocated to it. This information can be obtained from management entities, as it is done in GENI's FlowVisor and Opt-In Manager solution [15]. The controlling entity can also poll the server layer to get the usable resources. As an alternative the controlling entity may request resources it is willing to control.

A layer (n) controller exposes the known endpoints (which are addresses from flowspace n). A layer (n+1) controller then requests the slice by specifying a subset of this list. Multiple parallel layer (n+1) controllers would therefore exclusively share the set of layer (n) endpoints. For example, an Ethernet controller would pick the Ethernet ports to be controlled by it: two Ethernet controllers on one switch would be possible, and the slicing would take place on physical ports.

As a second example, one layer higher, a number of MAC addresses that is known to an Ethernet controller (through MAC learning, for instance) can be divided among multiple IPv4 controllers, corresponding to multiple routers on a single Ethernet switch. These routers would receive their own MAC addresses for the IP router ports from the layer (n), in this case the Ethernet controller.

A control layer receives two categories of configuration requests: downstream from a controlling entity (residing in a higher layer) designated to control the offered resources, and upstream from a controlled entity (residing in a lower layer) providing triggers. For example, a trigger can be a PDU from the data plane or a notification about the changes in the resources offered by the controlled entity.

To sum up: the service API between adjacent control layers shall provide a means of issuing configuration requests toward a lower control layer, receiving configuration triggers from lower control layers along with enhanced management features – such as flowspace management, virtualization, and control slice isolation. The service API will also be exposed to external content providers and thus need authentication and security features.

We define the following extensions for flowspace management:

- We add a set of messages for signaling flowspace registrations between a controller and a datapath element. Both the Fsp-Open and Fsp-Close messages carry a flowspace description. This flowspace description consists of a structure `ofp_match`, i.e. which means we use the same mechanism for describing flows as used by OpenFlow. A controller may restrict its control to a single flow or request all flows traversing a datapath element by sending an all-wildcard flowspace description.
- A controller may send multiple flowspace registrations in parallel.

- A datapath element stores all incoming flowspace registrations in a local flowspace table. The datapath element checks any event requiring a controller notification against the flowspace table. The flowspace entry with the most precise match (in terms of exact hits, wildcard hits, and priority field) wins this competition and the associated controller entity is used as destination for the event notification.

- Flowspace management can operate in either overlapping or non-overlapping mode. If multiple flowspace entries match in overlapping mode, the event is sent to all controller entities with matching entries. In non-overlapping mode a datapath element rejects a flowspace registration attempt that overlaps with an already accepted flowspace.

- For flowspace registrations, we can either adopt a soft-state or hard-state approach similar to Flow-Mod entries. When soft-state registrations expire, the datapath element automatically removes them, unless the controller refreshes the entry.

- Flowspace management does not affect the controller role model as introduced with OpenFlow version 1.2. Controllers can adopt one of the roles defined there: master, slave, or equal.

### 4.1.5 In-depth recursive controller architecture

Over the last ten years research in the area of Internet architecture has been influenced by the "clean slate" approach, largely driven by the difficulty of introducing even small changes in the services provided by - and the protocols installed in - production Internet routers. Researchers had been frustrated by the apparent block of innovation in the "real world", so a part escaped into a *Gedankenexperiment* called clean-slate research, exploring what *could be* a reasonable new architecture for the Internet, fixing the weak points of the current architecture, among others: mobility support, route oscillations, a centralized and error-prone name resolution structure, and information replication.

Clean-slate research helped in identifying basic principles of network layering (RNA, RINA), introducing new concepts like content-centric networking (CCN) and flow switching. The latter evolved into OpenFlow and promises to be the main paradigm of networking in the coming years. We also expect the previous two results to play a major role in networking over the coming years, in part enabled by OpenFlow itself. The *potential* benefits of CCN are the easy replication of much needed information in the network, reducing the overall load in the network, and - among other things - reducing the vulnerability to DDoS attacks.

The main finding of the recursive network architecture was that the layering in today's network stack follows a recursive pattern. A layer in a network according to RNA or RINA does not follow the OSI/ISO model, but in a sense is orthogonal to OSI's functional split. Analyzing the multi-layered networks of today (like IP over MPLS over Ethernet over WDM) one can observe that a certain set of functions is present in each of those layers, in a way, replicating all seven OSI layers in each of the "real" layers of a carrier network. Therefore, recursive layering splits the stack according to name spaces (the "scope") instead of functions. All functions that are required to establish communications between two or more processes must be present in each layer.

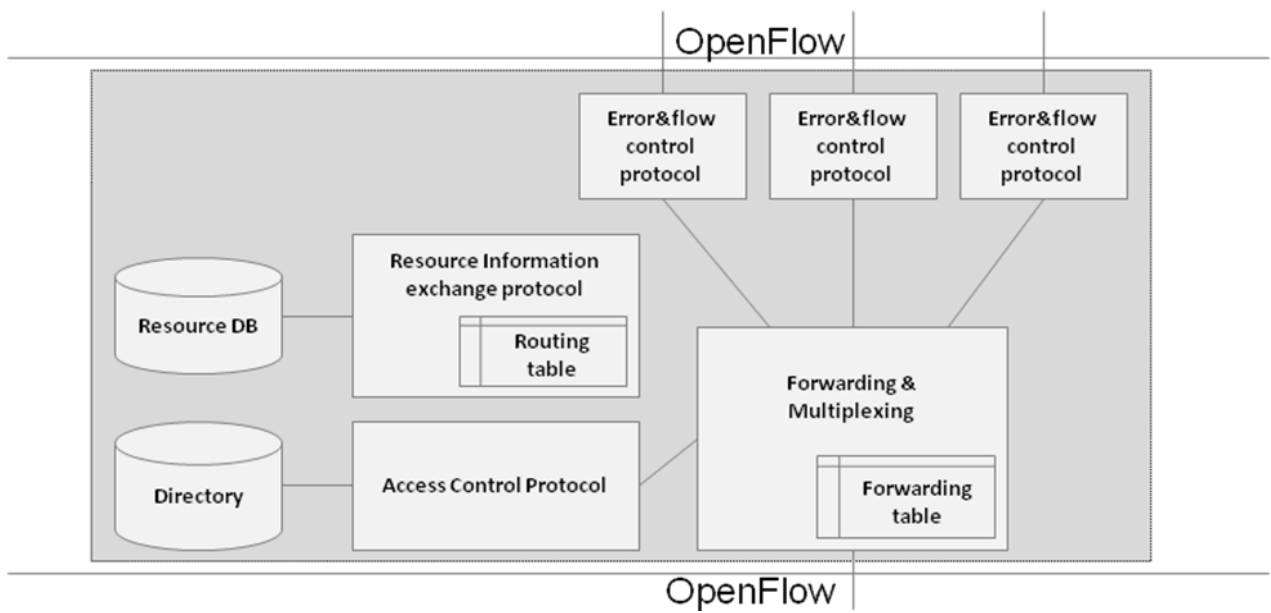

**Figure 12: Core functional blocks of a Layer.**

- Resource information exchange (protocol)
    - OSPF, Spanning Tree and LLDP all consist of three parts: a handshake for neighbour discovery, a subsequent exchange of resource information (be it link state in OSPF, distance vector in STP, or plain port identifiers in LLDP), and a subsequent calculation of a topology, and, using a CSPF, a routing table.
- Error and flow control
    - Typically, a link between endpoints of a layer is controlled by some form of error checksum and flow control. The combined error and flow control of IP is implemented in TCP: Ethernet has the option to use flow control and all layers down to the PHY assure some form of CRC.
- Forwarding and multiplexing
    - A network node has to forward and multiplex frames (or flows) from neighboring nodes. This is done by evaluating the address of the respective layer and looking up the forwarding table. The latter returns one or more points of attachment in the next lower layer, and results in the necessity to write these addresses into the frame before forwarding it. In the case of IP this means that the routing table lookup returns a next hop IP address that then needs to be resolved into the destination's lower-layer address, which is a MAC address in most cases.
- Access Control
    - Access control means that a new node is only allowed to communicate within a network layer after a certain process that assigns an address to work in the name space that defines the layer. DHCP, PPP's NCP are examples for this procedure in IP. The situation in Ethernet is different because the layer relies on fixed and a-priori assigned MAC addresses (creating ambiguities when it comes to dynamic creation of virtual machines).
- Directory (Name resolution)
    - Each layer relies on a mechanism that resolves the names of its name space to point-of-attachment addresses in the next lower layer. This mechanism is ARP for IP, or the process of MAC learning and storing the learned addresses along with the next lower name space, ports, in the forwarding table of an Ethernet switch. One of the possible upper layers of IP is nowadays predominant, and uses URL names that are resolved to IP addresses in the domain name system DNS. Other name spaces (like SIP) are used on top of IP as well, indicating that any knowledge of higher layer names in a lower layer would be potentially harmful.

## 4.2    Introducing management functions to SplitArchitecture

Before we discuss the integration of network management (NM) into a generic SDN and our *SplitArchitecture* network, we will provide some background by giving a historical perspective on network management. We start with a functional definition of network management based on ISO and ITU-T models, and a brief recap of the traditional layering of telecommunication networks (i.e. data, control and management planes). Next, we discuss the relation between the ITU TMN framework and the three planes by discussing the evolution of network management for different architectures, leading to the SDN-based *SplitArchitecture* concept as discussed in this deliverable.

### 4.2.1    Definition of Network Management

A network management model describes a set of recommendations or a framework for managing a transport network. Several network management models have been defined through the years by different standardization bodies, focusing on management of different network technologies. In order to define network management, we use the established network management model from the International Organization for Standardization (ISO) as a baseline together with the ITU-Ts Telecommunications Management Network (TMN) framework. ISO defined five functional areas of network management: Fault, Configuration, Accounting, Performance, and Security management - the so called OSI FCAPS model [46]. In this deliverable, we follow the network-centric perspective of FCAPS to divide network management functions. The functions are thus grouped as followed:

- Fault management: detection, isolation, correction and notification of faults in the network.

- Configuration management: configuration of the network devices, provision of circuits and services.

- Accounting management: collection and storage of data on network resource usage, deliver payment and accounting information.

- Performance management: collection and storage of operational statistics on resource usage for network optimization and planning.

- Security management: secure access to network elements, resources and services.

The ITU-T introduced the Telecommunications Management Network (TMN) framework [48] as a reference for how to operate and manage telecommunication networks. The TMN defines logical layers orthogonal to network management functions. There are five management layers, each providing the appropriate FCAPS functionality [47] according to the layer definition. The layers are network element layer, element management layer, network management layer, service management layer, and business management layer.

- Business management layer: functions related to business aspects, which includes rather strategically and tactical management rather than operational management, as considered in this deliverable.

- Service management layer: creation, handling, implementation and monitoring, and charging for the services build on top of the transport connectivity managed by the network management layer.

- Network management layer: distribution of network resources, configuration, control, and supervision the network consisting of network elements.

- Element management layer: handling of individual network elements or groups of network elements; including detection and handling of equipment errors, collection of statistics for accounting, and logging of event and performance data.

- Network element layer: providing an interface to the network elements, as well as instances of modules providing the required functionalities to support all FCAPS functions.

### 4.2.2　Modern view of transport networks

In this deliverable, we discuss the split between control and data planes as one of the core concepts of SDN and *SplitArchitecture*. In this section, we start to discuss a third important plane in transport networks: the management plane. Indeed, today's telecommunication networks are divided architecturally into three planes: management plane, control plane, and data plane. According to the ITU-T's generic protocol reference model for telecommunication networks [51], the user (data) plane is represented by user entities (hardware and software components) that deal with the transport of the user information ensuring switching, multiplexing, flow control and data integrity. The control plane is responsible for control related functions to establish, manage and release communications to transport information among user entities, while the management plane (as the name indicates) is responsible for management-related functions. The establishment of a communication channel is the result of cooperation between the control and management entities and the information transfer service provided. The communication channel may have different characteristics: e.g., connection-oriented, connectionless, on-demand, permanent etc.

According to the IETF GMPLS control plane framework [56], the control plane encompasses dynamic provisioning of paths, routing, path computation, signaling, traffic engineering, and path recovery. On the other hand, the fault, configuration, performance, and security management functions are placed in the management plane as defined by ITU-T [49], IETF also adds requirements for object and information models to be put in the management plane, as they are needed to manage networks and network elements [57].

With regard to the network management of packet transport networks, data, control, and management planes are defined as follows: The management plane addresses router configuration, collection of statistics, and optionally fault and performance management. The control plane exchanges connectivity and reachability information between the routers that is needed to build a routing table and computes and identifies a path between communication endpoints based on the link cost (and other types of) information. The control plane is also involved in routing of packets by identifying the outgoing interface and the next hop router to which the packet should be forwarded. The data plane receives and processes all the inbound packets, either by forwarding them to a specific interface, discarding them, or processing them specifically according to the differentiated service traffic policies.

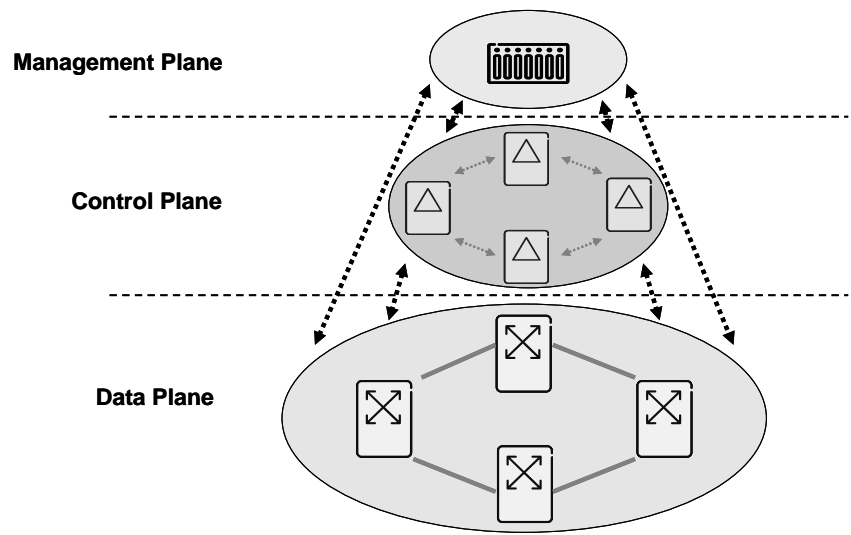

**Figure 13: Three planes of telecommunication networks: a centralized management plane, managing both the distributed control and data planes.**

The control plane is positioned between the data and management planes (see Figure 13). Network elements are controlled either by the control plane or by both the management and the control plane. The management plane configures and supervises the control plane. While the control plane can make certain decisions and control the data plane, the management plane has the ultimate control over both the control plane and the data plane entities.

### 4.2.3    Evolution of network management for different architectures

Traditionally, the network performs routing and switching functions by distributing the control and forwarding logic to all the network elements that are part of a network infrastructure. A network management system (NMS) is implemented as a centralized system on top of this distributed control plane, collecting information remotely from the network (Figure 13). In heterogeneous transport networks, a number of vendor-specific and technology-specific NM solutions have been deployed, managing a specific type of transport network elements. However, these different NM solutions are not integrated, and have proprietary interfaces towards their control plane, which have not been standardized.

To enable dynamic, policy-driven control of transport networks, the ITU-T defined an architecture for Automatically Switched Optical Networks (ASON), while the IETF was implementing a similar idea with the generalized MPLS (GMPLS) concept, focusing on control plane signaling. ASON/GMPLS defines a unified control plane for different data path technologies (i.e., different types of switching). It also includes definitions regarding management interfaces towards the control plane and the data plane. Prior to this, each data path technology had its own control plane and its own interface(s) towards the management plane.

A unified control plane enables on-demand provisioning of end-to-end services (optical circuits) to assign bandwidth dynamically on demand, thus enabling new switched services with dynamic bandwidth requirements. The optical circuits (and their bandwidth) previously provisioned statically, a priori in order to provide bandwidth guarantees for end-to-end IP traffic, and could not be modified subsequently. However, after high-demand applications appeared with strict requirements of latency, loss, bandwidth and reliability in terms of protection and resilience, the need for support of dynamic bandwidth assignment based on the actual demand (i.e., on-demand bandwidth provisioning) arose.

The ASON/GMPLS approach of addressing these provisioning issues was to add intelligence to the control plane. It integrated parts of the provisioning and configuration process into the control plane, automatically updating the network information database, all with the goal of simplifying network operations, reducing its cost, and quickly responding to failures by dynamically rerouting traffic. The GMPLS unified control plane offers a single interface to a management plane towards the control plane, while in the past the NMS had to have different interfaces for each different control plane. The functions that remain in the management plane are configuration and monitoring of the transport and control plane entities or services.

In ASON/GMPLS, we can observe the trend of moving some management functions into the control plane. The ITU-T specified the relation between ASON and the TMN architecture as the framework for ASON management in [50]. The relationship between management, control and data (or transport) plane is similar as described above, with the management plane directing both the control and the transport plane, while the control plane itself directs the data plane and reports northbound to the management plane. The control plane in this definition takes over functions that have been part of traditional TMN layers for service management, network management and element management, such as connection and call control, neighbour discovery and routing control.

However, the ASON/GMPLS control plane is still distributed (except an optionally centralized PCE), i.e. control functions are considered part of the network element layer. Thus, the datapath elements also have network knowledge and are involved in the topology discovery.

### 4.2.4      Network management for SDN

In the meantime, the SDN concept moved the network design into another direction, by making the datapath elements dumb and centralizing the control plane, e.g. by replacing distributed routing algorithms on the network elements by centralized route calculations performed in an SDN controller. Signaling in this architecture is done by the SDN controller via its southbound interface towards the datapath elements (e.g. OpenFlow). SDN has emerged as a new trend of building networks, by moving the intelligence from the data plane to the control plane. Hence, SDN is supposed to deliver cost-efficient solutions that are easier to manage, resulting in expected savings in both CAPEX and OPEX. Compared to ASON/GMPLS, the control and data planes are not only decoupled in this design, but the control plane is also logically centralized in an SDN controller.

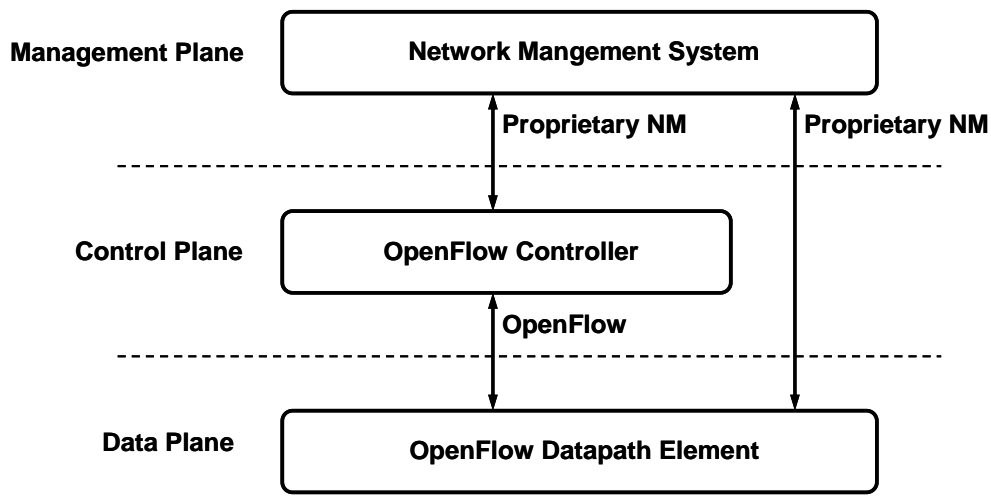

**Figure 14: Network management introduced to OpenFlow-based SDN via a fully separated, external NMS**

An obvious way to introduce a complete network management framework to an SDN network is to include a management plane on top of both the data and the control planes, according to the traditional network model (Figure 14). Network management would be fully contained within the management plane, adding additional management interface(s) to the network elements. While this might seem to be a feasible solution, we believe that this approach to the problem is not optimal, considering that both control and management planes are centralized elements in SDN.

The SDN architecture blurs the boundary between the control and the management planes. An SDN controller contains and maintains an updated network view, computes the paths, creates the forwarding rules and installs these rules on datapath elements - and in doing so provisions all the connections. Some of these functions (e.g., configuration of routers, topology discovery, service provisioning) can be traditionally seen as network management functions. The fully separated solution in Figure 14 implies that some types of events (e.g. topology updates or port state changes) would be duplicated and sent both to the controller and the NMS, and both entities would be required to keep their own central view of the network topology and state. Furthermore, certain state changes could trigger reaction from both entities. Without a certain level of synchronization between controller and NMS, this could result in an inconsistent state of the network and its elements. Finally, some notifications are only sent to the NMS. However, modern controller applications might very well require this information to react in a close-to-real-time fashion. As an example, consider device management data about power consumption, which is a main input to a power optimization-based traffic engineering policy.

An alternative way to introduce network management to SDN is to acknowledge the existence of traditional network management functions in the control plane. As a result, parts of the traditional management plane functions could be placed in the control plane. We will refer to these functions as network management functions (NMF). In the Figure 15, these functions are depicted as a separate module in the control plane. The actual realization of these functions is implementation specific, and it would be possible to integrate these functions completely into the controller software. The controller uses the southbound OpenFlow interface for control of the datapath elements. The NMF module would use open standard network management protocols and data models as southbound interfaces, such as the NETCONF-based OF-config by ONF. Due to the lack of a defined northbound interface, we assume that any API to an NMS can be implemented, similar to the fully separated solution in Figure 14.

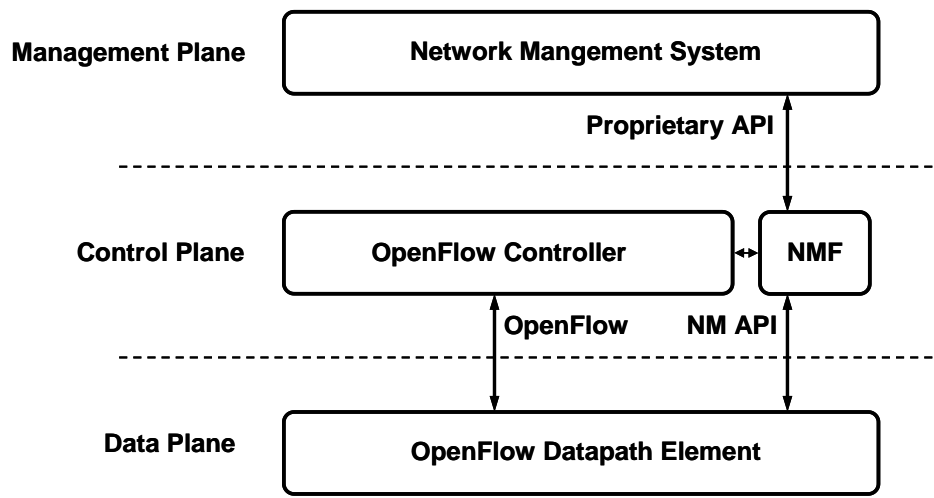

**Figure 15: Network management introduced to SDN by integration of selected network management functions (NMF) with the controller.**

The advantages of a dedicated NMF module, implementing certain network management functions in the control plane, are

- Removal of duplicate functions in two centralized elements (i.e. the controller and the NMS), thus limiting notification overhead and possible race-conditions of reactive-actions.
- Allowing for more timely reaction to management data (e.g., fault and performance) in the control plane, e.g. for TE purposes.
- Replacing proprietary management protocols and data models with open and standardized "southbound" interfaces towards the network elements.

There are however some detailed questions to be answered in a scenario as depicted above. First, we need to define the exact set of functions to be included into a NMF module within the SDN control plane. Second, we need to extend existing protocols (e.g. OpenFlow of OF-config) or define new open standards to support the southbound interaction of NMF with the datapath elements. In the next subsection, we will present our proposal of a network management architecture for SDN, which will provide some initial answers to these questions.

### 4.2.5     SPARC management integration proposal

The ONF did not yet define any complete network management framework for SDN. As described in Section 3.1.3, the recent ONF view includes the OF-Configuration point as a separate logical entity that configures datapath elements. However, no interface for exchanging the data between the OF-Configuration point and the controller has been defined. Additionally, there are no northbound interfaces defined for interaction between the applications and the OF-Configuration point, or between the applications and the controller, which could be used for management. Moreover it is not clear how network monitoring should be performed or whether a new interface needs to be defined or if the existing protocols can be extended for this purpose.

We propose a generic network management solution similar to Figure 15 for a carrier-grade split architecture transport network. The proposed architecture, depicted in Figure 16, is based on the current SDN model defined by the ONF.

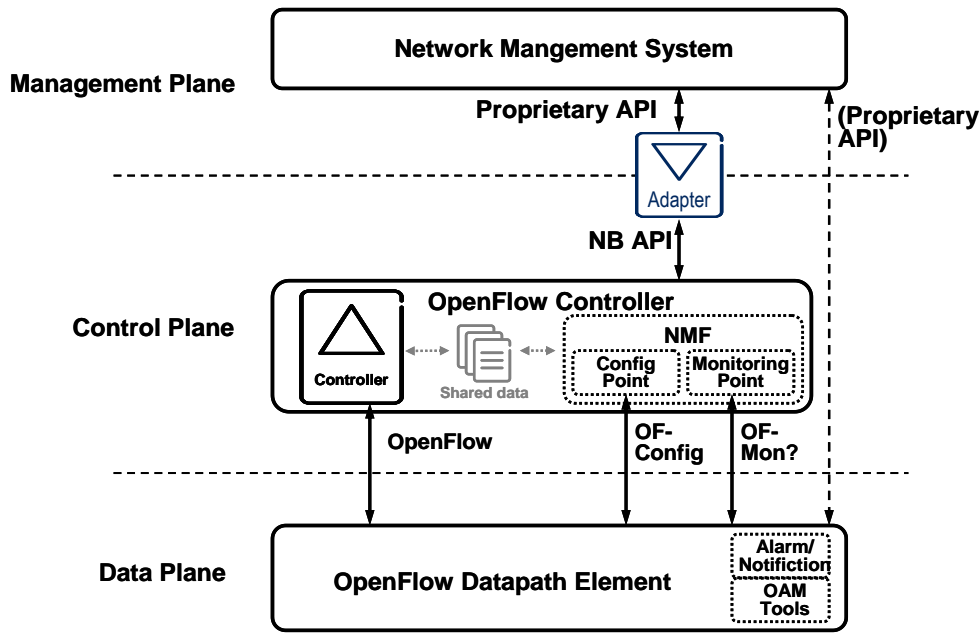

**Figure 16: Proposal for a carrier split architecture with integrated NM**

The real-time parts of a dataplane element are controlled via the OpenFlow protocol. Additionally, each datapath element needs to provide a NETCONF interface and support for the OF-config configuration scheme. In order to enable additional fault and performance management, OAM tools responsible for performing the actual measurements are required. A separate discussion on OAM tools for the OpenFlow-based *SplitArchitecture* can be found in Section 5.3. Finally, a monitoring interface is also required to report on results and alarms. In Figure 16, we call this interface OF-Mon, which could be realized as a separate interface or as an extension to OpenFlow. An example of a separate interface is the SNMP protocol, with its possibilities to allow retrieval of measurement and status data via the monitoring points, and to send alarms to the monitoring point asynchronously via traps. Alternatively, the OpenFlow protocol itself might also be suited for this purpose due to its real-time characteristics. While OpenFlow already supports retrieval of flow and port-related statistics, it lacks OAM-related reporting structures. It should be straightforward to add new message types for both asynchronous alarm messages and periodic measurement results to OpenFlow. The exact specification of these messages would be dependent on the specific alarm and OAM tool.

The SDN control plane consists in the first hand of a controller in accordance with the definition of SDN [44]. The controller maintains a global view of the network, e.g. in form of graphs with annotated nodes and edges. The controller also provides the southbound OpenFlow interface for deploying the packet forwarding rules. We propose that the OpenFlow controller includes a network management function (NMF) module for configuring as well as fault and performance monitoring. Being part of the same controller, the NMF shares the network view of the controller, including topology information and updates to this view in term of alarms or notification from the data plane. The OF configuration point is part of the NMF module, providing an OF-config interface to the datapath elements. Additionally, the NMF module comprises an equivalent monitoring point, providing the monitoring interface (OF-Mon) towards the data plane. In terms of functionality, the NMF can take over responsibilty for configuration, fault, and performance management functions that are useful in the control layer, as discussed below in Section 4.2.6. Through the controllers northbound API[2], the controller and the NMF module interact with an external NMS. These interactions include configuration of the controller, the NMF and policy input by the NMS, and updates of network state, fault and performance data towards the NMS by the controller.

The management plane is represented by an external network management system (NMS). In a typical carrier network, NMS entities will be represented by commercial NMS solutions, but could potentially also be implemented as customized NM applications or even simple CLI interfaces. In most operator environments, however, we assume that some form of NMS is already in place and could be adapted to manage the SDN domain as well as legacy equipment. To take advantage of the controllers northbound API, a lightweight NM adapter application is used to translate the controller northbound API to the proprietary API interfacing the NMS. If the management functions required for the SDN domain go beyond the scope of the configuration and monitoring interface provided by the NMF, a further device or vendor specific management interface from the NMS to the devices might be required. Such specific management functions may include many device management tasks, as identified in Table 2. In general, the external NMS is responsible for all remaining FCAPS functions not covered by the control layer, as discussed in the next section.

---

[2] The northbound interface of a SDN controller is yet undefined by subject of ongoing discussions in the ONF. Here, we assume that each controller implementation has its own common northbound API defined.

In general, this generic architectural proposal makes it possible to flexibly choose whether to place network management functions within a controller or in an NMS on a per-case basis, depending on the exact scenario and use-case in question. Parameters affecting such analyses include the scale of the network (number of devices and geographical spread), existing legacy infrastructure in place, type of transport technologies in use and type of services to be supported. Thus in certain scenarios, either the controller based NMF or the external NMS could be designed to be minimalistic or even be non-existent. In the next section, we will present a way to handle the design choice of where to place network management functions, and provide our recommendations for carrier type networks.

### 4.2.6       Analyzing the placement of Network Management functions

The key question in the proposed NM architecture is the assignment of traditional network management functions to the layers in an SDN environment (e.g. the generic architecture depicted in Figure 16). In other words, we assess which functionalities should be integrated into the SDN controller, and which should remain in an external NMS. Note that in the current SDN model by the ONF, the OpenFlow controller is already responsible for the updated network view, path computation, determination and configuration of forwarding rules and provisioning of connections. The OF configuration point supports additional configuration functions, such as controller assignment, resource configuration, certificate handling, capability discovery and basic configuration of tunnel endpoints. In Table 2, we list common network management functions of the TMN layers and asses the responsibilities for them in an SDN environment. Besides functions that are already part of the control plane, the analysis as to whether a NM function should be placed within the control plane or stay in the management plane is based on the following three questions:

Q1: Is the function already included in the ONF SDN model or the carrier-grade controller framework as proposed by SPARC? If not,

Q2: Does the information provided by the NM function help the controller framework to configure and steer the network in timely and automated fashion in order to provide carrier-grade performance? If so,

Q3: Do the southbound controller interfaces defined by the ONF allow straight-forward support for the NM function (i.e. would required extensions to the OF / OF-Config protocols be simple and keep the protocol "open", without bloating or overloading it with vendor or device specific elements)?

Table 2: Assessment of NM functions and potential for control plane integration

| NM function | FCAPS Groups | Q1 included? | Q2 timely? | Q3 open interfaces? | Proposed CP integration |
|---|---|---|---|---|---|
| *Element management functions:* | | | | | |
| Firmware management | config | no | no | no | no |
| Device monitoring (temp., etc) | performance | no | no | no | no |
| Device monitoring: Power consumption | performance | no | yes[1] | OF-mon | yes[1] |
| Control network bootstrapping | config | no | no | no | no |
| Resource and capability discovery | config | yes | no | OF, OF-config | yes |
| Logical swtich instatiation | config | yes | no | OF-config | yes |
| Control channel (addresses and credentials) | config / security | yes | no | OF-config | yes |
| Fault detection (equipment) | fault | no | no | no | no |
| Alarm management | configuration | no | yes | OF-config | yes |
| Logging of alarms | fault, accounting | no | no | no | no |
| Logging of statistical data | performance, accounting | no | no | no | no |
| Resource usage (cpu, buffer, queue-length) | performance | no | yes[2] | OF-mon | yes[2] |
| *Network management functions:* | | | | | |
| Topology discovery (creation of network view) | config | yes | yes | OF | yes |
| Path computation & setup | config | yes | yes | OF | yes |
| Flow table management | config | yes | yes | OF | yes |
| Tunnel management | config | yes | yes | OF-config | yes |
| Traffic engineering (creation of QoS paths) | config | yes | yes | OF | yes |
| Fault detection (link level) | fault | yes[3] | yes | OF-mon | yes |
| Link performance monitoring | performance | no | yes | OF-mon | yes |
| Network performance optimization | performance | no | yes | OF, OF-config | yes |
| Resiliency measures | performance/config | yes | yes | OF, OF-config | yes |
| *Service management functions:* | | | | | |
| Accounting | accounting | no | no | no | no |
| User management and AAA | accounting / security | no | no | no | no |
| Service definition and administration | config | no | no | no | no |
| Service OAM configuration | config | no | yes | OF-config | yes* |
| QoS management (service delay, loss) | performance | no | yes | OF-mon | yes* |
| SLA management | accounting | no | no | no | no |

[1] for energy-aware networking (see section 5.7)
[2] for logical switches sharing switch resources (see section 5.2.4)
[3] implemented in SPARC as BFD (see section 5.3.3)
* assuming service controller functionality in the CP, as in SPARC D4.3

Table 2 lists the most common network management function required by the SPARC use case of a carrier-grade access/aggregation network. We grouped the functions according to the TMN layers element- to service management. The second column indicates the broader function in terms of FCAPS. Columns Q1-3 refer to the three questions stated above, helping us to assess for which functions it makes sense to be integrated in an SDN controller. Finally, the last column provides our recommendation, based the results of the columns Q1-3.

#### 4.2.6.1        Element management layer functions

Functions in the element management layer are mainly related to bootstrapping, device configuration and monitoring, as well as logging of statistical data and alarms received from the OAM tools at the network element layer. Most of these functions are not time-critical in terms of controller reaction, or require hardware or vendor specific interfaces (e.g. firmware management) that might not need to be integrated into open SDN standards. As an example, bootstrapping of the control network connection could be done via existing auto-configuration mechanisms like DHCP (see Section 5.5). However, configuration of the devices is currently one of the prime tasks of OF-config. Hence, we suggest integration of device configuration in the control plane, given that the function can be supported by extending the OF-config configuration scheme and can thus be communicated via the NETCONF-based OF-config protocol.

Possible exceptions in terms of required quick controller reactions could be certain performance data of specific device parameters. As one example is energy aware networking, which we propose in Section 5.7. In this case, the controller needs updated power consumption figures for interfaces and devices in the network in order to be able to react accordingly, e.g by multilayer traffic engineering (MLTE) topology optimization, or by reconfiguring interfaces to burst- or adaptive link rate mode (see Section 5.7).

Another example is network virtualization, which is discussed in detail in Section 5.2. In a virtualized scenario, the physical resources (e.g. CPU, memory, queues) in a datapath element are shared among several tenants by slicing the physical network element into several isolated logical OpenFlow switches. To ensure the service level agreements (SLA) with these tenants, it might be necessary for the controller to react in cases when physical resources become scarce and logical switches are threatened with resource starvation. In these scenarios, the controller needs fault notifications or continuous performance data updates from the device, which allows it to react, for example, through resource reassignment or even migration of logical switches to other network elements.

#### 4.2.6.2        Network management layer functions

Functions in the network management layer involve configuration and provisioning of the network as well as steering and monitoring the traffic. To a large degree, these functions already covered by SDN controllers and communicated via OpenFlow – and, of late, via OF-Config as well. Topology discovery, for example, is most commonly done via a controller-based LLDP-like mechanism (see Section 5.5). Functions related to link, path and tunnel provisioning and configuration are the very essence of many controllers and are nearly fully supported by existing OpenFlow specifications. Fault and performance monitoring, however, are currently only supported in rudimentary fashion by ONF protocols, e.g. by polling flow, port, group and queue statistics or receiving asynchronous port status messages via OpenFlow. However, fault and performance data is essential for the controller to realize protection and restoration, as well as performance optimization. In Section 5.3 we propose more advanced carrier-grade OAM tools for more fine-grained and specific fault and performance monitoring of the network. As a monitoring channel (OF-mon), we propose either a monitoring extension to OpenFlow or, alternatively, adaption of an existing monitoring interface, e.g. SNMP.

#### 4.2.6.3        Service management layer functions

Network services, such as residential internet access or E-LINE connectivity, provided to customers are provisioned on the top of the common packet based transport infrastructure. The services are managed in the service management layer according to the TMN framework, which includes functions for service creation, implementation and monitoring. Service creation typically touches a few service nodes on the edges of the transport network, which provide the underlying end-to-end transport connectivity which is managed by the network management layer. In some SDN environments, service management might not be supported by the SDN controller, which means that related management functions would need to be performed by an external NMS.

For a carrier-grade controller framework, we assume a control plane that already includes service creation functionality. In SPARC, we proposed to organize the controller into transport and service control regions, as also realized in the SPARC prototype described in SPARC Deliverable D4.3. These regions can be implemented, for example, through a recursive control plane architecture, as proposed in Section 4.1. In this case, the services can take advantage of control plane service level fault and performance management to enable fast reaction to any type of service degradation. The fault and performance management functions need support from service OAM tools on the network elements as well as a monitoring channel between the controller and the data plane (OF-mon).

However, not all service management functions are necessarily time critical or need to share topology data with the controller. For example, service billing, SLA management and accounting do not have to be done in the control plane, and should remain in an existing NMS system. In this case, the required information is exported by the controller through the northbound API and the NMS-specific adaptor application.

#### 4.2.6.4　　Concluding the Analysis

While the analysis in this section provides initial recommendations on the placement of management functions, a final assessment needs to be done on a per-case basis, depending on the exact scenario and use case in question. According to our analysis for carrier-grade networks, typical network management layer functions are best placed within the controller framework of an SDN architecture. Additionally, the controller should have access to OAM tool information from device, network and service levels, which is critical for effective performance and fault management. On the other hand, the remaining device and service management layer functions to not require time-critical controller reaction or are not straightforward to implement with existing and planned southbound controller interfaces. An external NMS in a carrier SDN scenario might be more suitable to take over vendor and device-specific device management tasks, as well as service and business-level policy management and accounting tasks.

We started our discussion of network management by considering traditional models and definitions, which we then tried to apply to a generic SDN model. It an SDN scenario, both management and control are performed centralized, decoupled from the actual datapath elements. Based on this observation, we conclude that it is difficult to differentiate precisely between control and management in the context of SDN. Within SPARC, we used timeliness and automatic configuration (i.e. real-time reactiveness) as the differentiator between control and management functions. However, depending on the specific use-case and the technology used, the results of such an assessment might look quite different from case to case. We believe that our generic NM integration proposal (cf. Figure 16) allows enough flexibility for the placement of specific network management functions to cope with this architectural tradeoff.

### 4.2.7　　Combined SPARC network management and recursive control framework

Within SPARC, we initially had a strong focus on the control plane architecture and only started to consider network management aspects very recently. In the following paragraphs, we will provide our initial ideas on a combined control/management architecture. However, a mature network management solution requires further discussions beyond the scope of SPARC.

The network management proposal presented so far targets a rather generic SDN architecture, taking current SDN models as defined by OpenFlow and the ONF into account (cf. Figure 3 (II) and (III)). In the following paragraphs, we discuss how to integrate the proposed generic management framework with the hierarchical, recursive controller architecture discussed in Section 4.1. We will revisit the question about the role of a network management system (NMS) in a hierarchical control layer scenario. Furthermore, we will clarify the functions of additional, management-specific modules in this scenario, which includes the suggestion network management function (NMF) module in control entities as well as OAM tools within (logical) datapath elements.

We start by considering a recursive control architecture as depicted in Figure 11. In this control architecture, we stack multiple control planes, where each plane acts as controller for the lower control planes, while at the same time providing a filtered, abstracted view of its own control plane to higher layers via a virtual datapath node. The concept of virtual nodes makes it possible to (re-)use OpenFlow as the interface between the control planes. We outline this basic architecture in Figure 17.

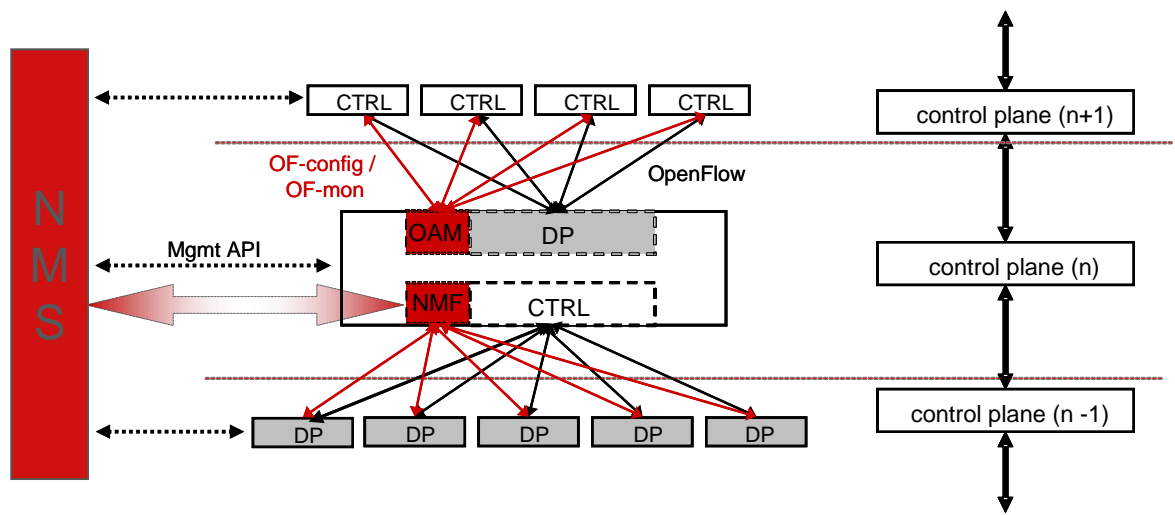

**Figure 17: Combined control and management framework in the recursive controller architecture**

We propose integrating the SPARC network management proposal (depicted in Figure 16) as highlighted in red color in Figure 17. The additional new entities added to the hierarchical controller architecture are OAM tools and the network management function (NMF) component in the control framework, as well as an external network management system (NMS)[3]. Furthermore, additional interfaces between the stacked control planes are provided through the configuration and monitoring protocols OF-config and OF-mon[4], as introduced in Section 4.2.5. We currently do not define the management interface between the individual control planes and the NMS, but we assume the control entities can adapt to a management interface supported by the NMS. Obviously, the OpenFlow-related management APIs OF-config and/or OF-mon are good candidates as well.

#### 4.2.7.1 Management interfaces

The main purpose of the network management in this scenario is configuration and monitoring of the control and data planes. Regarding configuration, the hierarchical control architecture requires additional functions besides the traditional management functions listed in Table 2. As pointed out in Section 4.1, these are flowspace management and a more flexible port management by replacing OpenFlow's current physical port model with a generalized transport endpoint model. However, given the recent introduction of OF-config, both flowspace management and endpoint (i.e. port) management could also be performed through extensions to the ONF configuration protocol. Furthermore, with regard to an external NMS, the additional functions could also be carried out centrally by using a horizontal management interface. So far, we considered a distributed method, i.e. via OpenFlow protocols from a higher control plane (n) to the lower, serving control planes (n-1). As discussed above, this partitioning of management functions needs to be done with the specific use-case and the capabilities of the network management system in mind.

Regarding monitoring, a monitoring interface (OF-mon) needs to facilitate the handling of alarms and events. This includes management of the OAM tools as well as event correlation and alarm propagation based on the output of the OAM tools. Either the NMF in the controller or the separate NMS needs to provide an event correlation function with the purpose of identifying and locating the possible cause of the potential failure or service degradation. This event correlation function can be centralized in the NMS, in which case the NMS could only control the number of events through the configuration of a collection time interval in OAM tools. Obviously, this type of centralized event correlation poses extra requirements on the NMS infrastructure in terms of memory and computing power. Another alternative is to perform this function within the NMF module in each control plane. In this case, the NMF module needs to include capabilities to suppress alarms, reducing their number and correlating them before forwarding them to higher control planes or the NMS. In this scenario, the data collection and event correlation functionality is effectively distributed along the controller hierarchy. However, this requires coordination among different control planes and the NMS, which is a study subject for future work.

#### 4.2.7.2 Distributed vs centralized network management

As pointed out earlier, it is hard to define a strict boundary between network management and control in SDN networks. Thus, we propose a flexible placement of specific network management functions in order to cope with the architectural tradeoff to be found for each use-case. This flexibility is indicated with the fat, red arrow in Figure 17. At one extreme, network management functions can be placed centrally in an external NMS, or at the other extreme be distributed completely to different control planes of the hierarchical controller. In the latter case, the element and network management functions listed in Table 2 would be placed in lower control planes, whereas service management function would naturally reside in higher planes of the control architecture.

A possible alternative is to propagate registrations, configurations, and notifications through the hierarchical stacked controller planes, where only one plane (e.g. the top or bottom plane) would be connected to the NMS via a horizontal management interface. However, the lack of direct interaction between an NMS and the lower layer controllers (including physical datapath elements) can be a shortcoming, due to the need to propagate all the messages through the whole chain of controllers. By messages, we refer to the configuration and initiation of each control entity (in terms of flowspaces, logical endpoints, and network management functions) and datapath elements (in terms of device management, performance and fault monitoring), as well as the propagation of notifications and alarms through the layered hierarchy.

---

[3] We use the term "network management system" as a generic term for the higher four layers of the ITU-T TMN model, ranging from business to element management.

[4] When the recursive control plane was discussed, the ONF has not yet specified any management related protocols: OF-Config was not yet released; and a monitoring interface is starting to be discussed in the ONF only at the time of writing this deliverable (Sept. 2012). Note that "OF-Mon" is currently only a working title used by SPARC.

These shortcomings can obviously be addressed by connecting each control plane via a horizontal management interface to the NMS. The benefit of this approach is that it enables each control plane to be configured independently of the other layers. In this case, NMS can be connected to each control plane and be used to configure, initiate, and manage all the controllers and monitoring tools. The measurement data and notifications do not need to traverse the whole hierarchy of controllers in order to reach the NMS, which can reduce the complexity of the solution. For example, there is no need to implement the forwarding of messages or keep the notification state in the controller, in case the connection with the upper controller(s) is lost. However, connecting the NMS to each control plane requires adding more interfaces, which might require some management overhead, but could result in more efficient data delivery by not needing to rely on a single channel for the interaction of all controllers with NMS. In practice, a hybrid solution might be desirable, e.g. allowing notification of both the NMS and higher control planes in case of certain events. However, this would require strict assignment of responsibilities between the centralized and distributed management functions.

# 5    OpenFlow Extensions for Carrier-Grade SplitArchitecture

In the duration of the SPARC project, we concluded that current OpenFlow-based implementations do not fulfill carrier requirements, thus protocol extensions and standardization of certain functionalities are needed. As documented in Section 2.1, the original requirements defined in deliverable D2.1 have been refined with each consecutive deliverable of WP2 and WP3, leading to the final list of requirement groups (i.e. network features) presented in Section 2.2.

In Section 4, we have covered requirement groups (a) and (b) on control and management architecture. In this section, we provide proposals for OpenFlow extensions defined to fulfill requirement groups (c) to (k) as listed in Table 3. All these protocol extensions also imply additional functionality on the network elements which go beyond pure packet and flow forwarding. For each topic, we provide an introduction and the motivation by outlining current state-of-the-art solutions. We then describe our proposed improvements to an OpenFlow-based *SplitArchitecture* in order to enable the respective feature. Specific technical details, such as extensions to OpenFlow configuration procedures or OpenFlow protocol messages, can for some topics be found in SPARC Deliverable D4.2 "OpenFlow protocol suite extensions", as indicated in the table.

**Table 3: List of study topics requiring OpenFlow extensions**

| Section | Requirement group, i.e. study topic | Extensions defined in D4.2 |
|---|---|---|
| 5.1 | *(c) Openness and Extensibility* | Yes |
| 5.2 | *(d) Virtualization and Isolation* | Yes |
| 5.3.3 | *(e) OAM: technology-specific MPLS OAM* | Yes |
| 5.3.4 | *(e) OAM: technology-agnostic Flow OAM* | No |
| 5.4 | *(f) Network Resiliency* | No |
| 5.5 | *(g) Control Channel Bootstrapping and Topology Discovery* | No |
| 5.6 | *(h) Service Creation* | Yes |
| 5.7 | *(i) Energy-Efficient Networking* | Yes |
| 5.8 | *(j) Quality of Service* | No |
| 5.9 | *(k) Multilayer Aspects: Packet-Opto integration* | No |

## 5.1    Openness and Extensibility

A closer look at the OpenFlow processing framework and its present capabilities immediately reveals one of its major deficiencies: its rather limited set of supported protocols. Initially invented in a campus-networking environment, OpenFlow 1.0 supports a basic set of common protocols and frame formats like Ethernet, VLAN, and ARP/IPv4.

The OpenFlow specification authors have continuously added new protocols to the evolving specification like MPLS in OF1.1, IPv6 in OF1.2, or PBB in OF1.3. This for sure makes OpenFlow more useful for specific use cases and environments, but a general framework capable of adding support for yet unsupported protocols seems desirable. Just refer to the access/aggregation use cases discussed in this and the accompanying documents from WP2, where protocol support for PPP and PPPoE is a mandatory requirement.

### 5.1.1    Extensions for a Recursive Architecture

OpenFlow defines in a generic manner a Service Access Point and its basic primitives: Port-Status, Port-Modify, Packet-In, Packet-Out and its more advanced version Flow-Modify. In our recursive architecture as introduced in Section 3 we use the OpenFlow API as interface between a stacked series of transport controllers (in that sense a datapath element is also a controller, a PHY-port controller). Each transport controller defines transport endpoints on its specific layer, e.g. an Ethernet controller de-multiplexes based on MAC addresses and exposes Ethernet ports to the next higher layer (it provides Ethernet transport services via these transport endpoints). An IP controller exposes IP transport endpoints to the next higher layer, thus providing IP transport services.

In the original concept of OpenFlow, a port is a well-defined entity: a physical (or logical) port with Ethernet like properties and configuration parameters. However, in our recursive architecture, a port (i.e. a protocol specific transport endpoint) may define additional or different configuration parameters, e.g. an IP port defines a local source IP address assigned to this port, a network mask, and a peer address (in case of a point-to-point link). The OpenFlow specification defines a fixed C-structure that is limited to Ethernet parameters for its Port-Status messages. We propose to use an extension to OpenFlow that utilizes a TLV-based approach for specifying port specific configuration parameters in `struct ofp_port` similar to the TLV based approach for `struct ofp_match`.

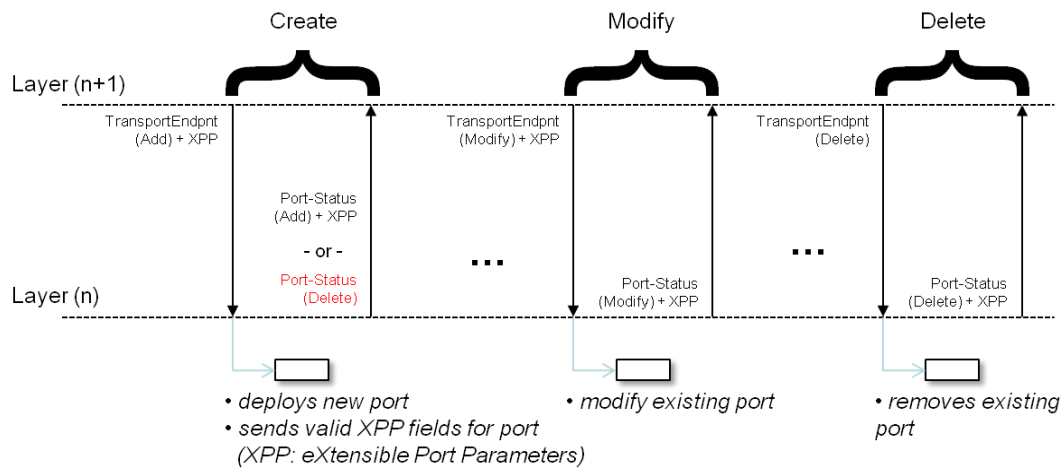

**Figure 18: Port Management Messages and Port life cycle**

In the recursive architecture, ports are de-multiplexing transport endpoints and except within a PHY-controller (i.e. an Ethernet based datapath) or an optical datapath (i,e, a fiber based datapath), these ports are logical entities used for de-multiplexing flows. We replace the physical port model defined in OpenFlow with a more generalized transport endpoint model. This includes the physical port model as used in OpenFlow so far, but adds some flexibility:

- We add an additional protocol message to OpenFlow for management of transport endpoints in layer (n) by layer (n+1). This *Transport-Endpoint* protocol message controls the CRUD[5] lifecycle of a transport endpoint.

- We replace the so far statically defined port description with a TLV based *eXtensible Port Parameter* set (XPP). A transport endpoint stores a number of layer specific configuration parameters in an XPP set.

For a controller entity on layer (n) we define a protocol extension that allows creation of transport endpoints on layer (n-1), e.g. an IP controller may request a new Ethernet transport endpoint and access to its communication services. OpenFlow defines Port-Status messages for signaling port management events from a datapath (i.e. layer (n-1)) to a controller (i.e. layer (n)), as ports in the original specification are primarily physical ports. For the recursive architecture, we propose a symmetrical Port-Status message, i.e. a controller may open a new transport endpoint (i.e. a port) in the adjacent lower layer.

Figure 23Figure 18 depicts the life cycle of a transport endpoint deployed in layer (n) and controlled by layer (n+1). All commands (Transport-Endpoint, Port-Status, and Port-Modify) adopt a TLV-based port configuration structure that replaces the current static C-structure. We call this an *eXtensible Port Parameter set* (or XPP set for short). Port-Status and Port-Modify messages are asymmetric messages in the original OpenFlow specification, i.e. OpenFlow lacks the ability to signal a result status via an acknowledgment back to the originator of the operation. Upon reception of a Transport-Endpoint message for creation of a new transport endpoint, the datapath element sends a Port-Status or Port-Modify message back with the result of the requested operation and the current port configuration.

A port creation operation might fail, e.g. when the layer (n) instance is a PHY-port controller or when a transport endpoint address is already in use by another endpoint. In addition, the set of extensible port parameters defines the type of ports exposed by a layer (n). Via a *Features Request* message, a layer (n+1) entity should obtain the set of valid port parameters for the specific layer.

---

[5] CRUD stands for "create, read, updated, delete". Note that Figure 21 visualizes "update" as "modify".

### 5.1.2　　　　The Various Processing Types

OpenFlow defines an action based processing framework and a surrounding forwarding logic for packet manipulation where individual actions define specific atomic processing operations like pushing a header tag, decrementing a field or setting some header field to a specific value. This processing model supports a number of use cases (e.g. emulating a simple IP router with decrementing TTLs, setting source MAC addresses, and so on), but it also implies a number of constraints. All actions defined in OpenFlow are lightweight and state-less in nature, i.e. an action never takes into account state from the history of preceding packets. However, there are some use cases that require a more advanced processing framework, either because they define dependencies to preceding packets, or because they need a more advanced processing logic. One example is block-chaining encryption codes that require directly results from preceding packet operations as its input. Use cases that contain encryption schemes cannot easily be realized using OpenFlow. Another example is OAM support, which is discussed in more detail in Section 5.3. OAM endpoints typically have some specific time constraints on detecting and reacting on failure conditions. Those timing constraints can only be fulfilled such when all time critical components are deployed directly in the data plane. OAM processing implies a more advanced programming logic: generating packets, setting timers, running a finite state machine, and so on. An example is our BFD implementation done in SPARC that is generating test messages and runs timers for detecting test message losses.

However, some use cases force us to define a processing logic beyond the existing packet manipulation framework. OAM is a typical use case here, where the OAM endpoint typical injects and removes test messages and runs some internal timer based logic for detecting loss of such test messages. The various OAM frameworks typically define sets of different OAM types, e.g. pure connectivity checks for testing the physical links, or OAM associations that check the proper configuration of all flow tables of datapath elements along a specific flow path. For such non-packet related processing, an additional processing framework based on virtual ports seems quite useful.

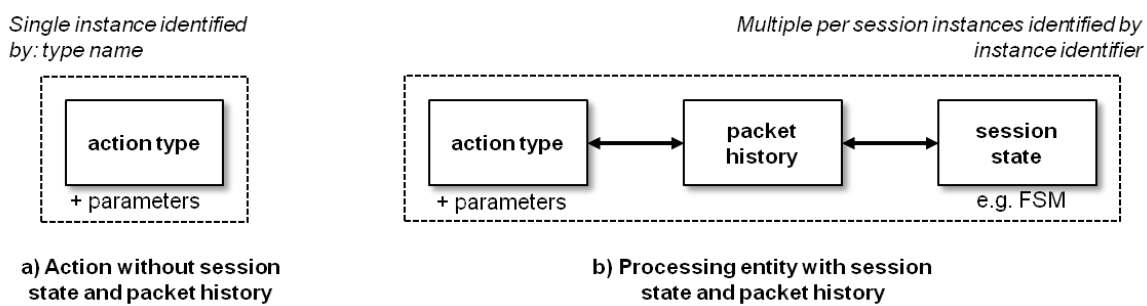

**Figure 19: Actions vs. Processing Entities**

Thus, we end up with three groups of processing requirements and means to implement these processing types:

1. State-less, lightweight packet processing (supported already since OpenFlow version 1.0)
2. State-full processing that stores and takes into account the history of preceding packets.
    → Solution approach: We propose Processing Instances and a new action named *Process*.
3. Parts of complex state machines requiring execution directly on the datapath to meet strict timing constraints.
    → Solution approach: We revisit virtual ports and discuss their applicability for the use case OAM.

We discuss state-full packet processing actions in the next subsection and re-consider the virtual port concept as non-packet processing related processing framework afterwards.

### 5.1.3　　　　State-full Packet Processing and Action Process

We propose the following extension to the OpenFlow packet-processing framework: We define processing instances that are persistent (or at least long living) entities following a CRUD approach, i.e. they are explicitly created, updated, and deleted on/from the datapath very similar to group table entries as defined by OpenFlow 1.1. A processing instance contains a specific packet processing logic and maintains state across several consecutive packets, i.e. it may store an entire flow packet history. It acts as a packet filter, i.e. it filters and processes packets. When leaving the processing instance, packets are re-injected into OpenFlow's existing action execution logic. Loading and configuring a processing instance's internal logic is out of scope of OpenFlow: a proprietary API may co-exist with the OpenFlow interface. Similar to group table entries, a processing instance obtains a unique identifier for reference purposes.

In addition to processing instances, we define a new action for OpenFlow named *ActionProcess*. An ActionProcess filters a packet through the processing instance referred to by the Processing Instance ID (ProcInstID) stored within ActionProcess.

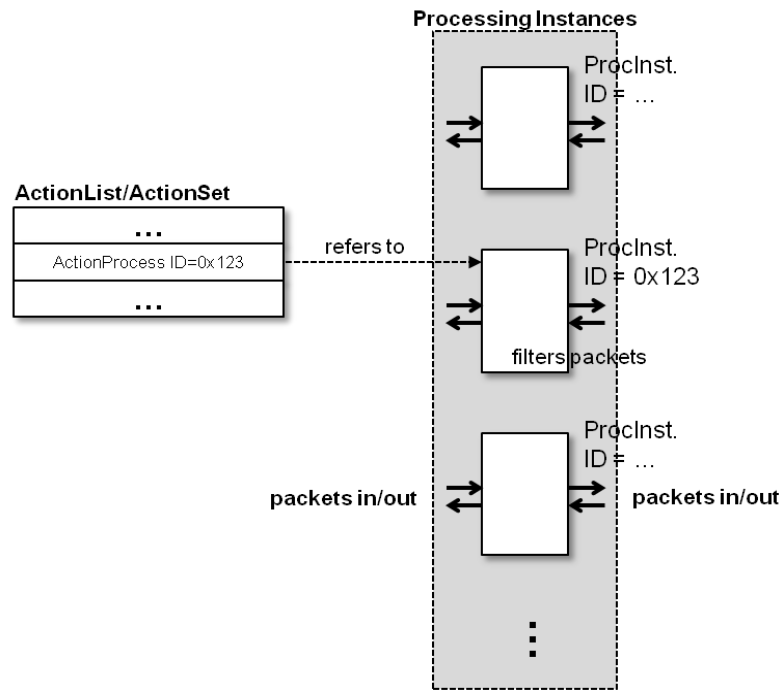

**Figure 20: ActionProcess and Processing Instances**

**Action "Process":** When executed, this action redirects packets to a specific processing entity instance. An Action "Process" decouples a Flow-Mod entry from the processing entity. All OpenFlow 1.1 action-specific constraints also apply to the Action "Process" i.e., only a single action of any type may reside within an ActionSet. OpenFlow 1.1 defines an ordered list of actions, i.e., all actions within an ActionSet must be reordered according to this ordering policy before actually executing the ActionList. All "Set" related actions (lightweight processing) are executed after decrementing TTL values (position #5) and before any QoS-related actions are applied (position #7), therefore Action "Process" should occur before or after position #6. No restrictions apply to using Action "Process" inside a group table entry.

### 5.1.4   Advanced Processing using Virtual Ports

Beyond plain packet processing, some use cases require a more advanced processing framework. As an example, consider OAM related use cases, where specific probe messages are interleaved with the existing stream of packets for testing proper operation of a specific (stitched set of) link(s). A similar problem arises in the PPPoE/PPP related use case, where LCP-ECHO request/reply exchanges monitor the state of a specific PPPoE/PPP session. Due to the strict timing constraints of most OAM schemes, we aim towards deploying all of its time critical parts on the datapath element and keep only non-critical parts in the control plane's slow path.

Consider two typical examples of OAM schemes: a connectivity check (CC) scheme that tests the physical link between two datapath elements, and an OAM association that monitors proper configuration of flow tables along a specific flow path. Figure 21 depicts both scenarios: for the connectivity check use case we want to avoid passing our test messages through the datapath element's forwarding engine. In case of a loss indication the OAM association endpoint could not determine, whether the problem is caused by the OF forwarding engine or the physical link. Thus, we propose to introduce pre-/post-filters attached to a physical in-/out-port (see part (a) in Figure 21 for details). A filter actually defines a (set of) flow-match(es) and extracts all matching packets from the packet flow. An attached virtual port consumes these packets and takes appropriate actions. A typical example for such a pre-filtering virtual port is an IEEE 802.1ag compliant OAM scheme, where we want to redirect all Ethernet frames with a specific EtherType (e.g. 0x8902) to the virtual port.

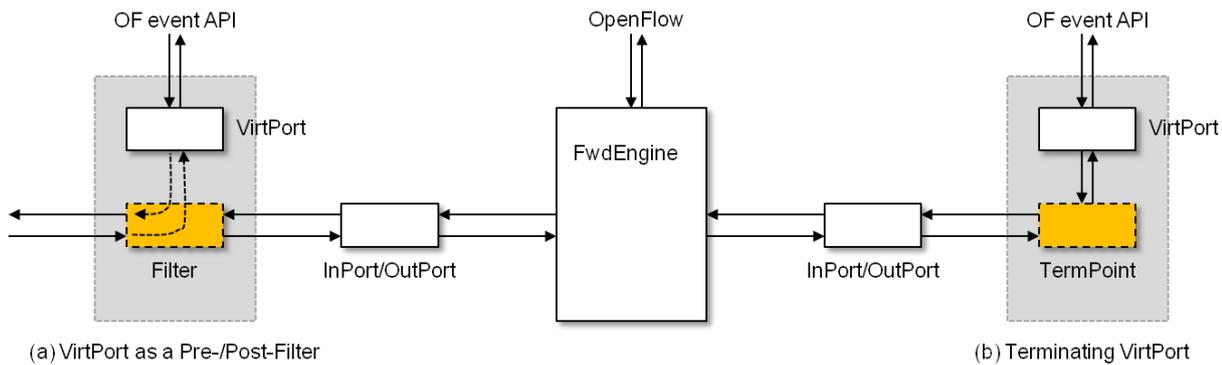

Figure 21: Virtual Port Concept

However, we might want to deploy other OAM schemes with a different scope for testing physical connectivity as well as existence of specific flow table entries along a flow path. In such a scenario, a virtual port should also stress the flow tables on the first and last datapath element. We define a terminating virtual port (see Figure 21 part (b) for details) that behaves like a physical port. All packets generated by this virtual port are actually traversing the datapath element's forwarding engine and thus, test the datapath's internal forwarding logic including flow table entries. However, for enabling true fate sharing among data and OAM packets, all OAM packets must also use the flow table entries defined for normal data packets.

### 5.1.5 Split State Machines and an Event/Action API

We have discussed briefly some constraints arising from OAM related use cases for an advanced processing framework. Due to the tight timing constraints, some monitoring must occur within the data plane. For the PPPoE/PPP example, the LCP-OAM functionality if part of a larger state machine, whose dominant part is running within the control plane while the time critical OAM part executes for each PPP session within the data plane. This example defines a split finite state machine, i.e. some parts of a state machine are executed in the data plane, while the remaining parts execute in the control plane. As both are parts of a now split state machine, we need some means to synchronize state between the two sub-state machines. This problem of synchronizing split state machines may occur either for state-full action-based processing (see Section 5.1.3) or virtual ports as discussed in Section 5.1.4.

A processing entity is in principle logically split into two halves (see Figure 22): A top handler resides in the control plane while the bottom handler is located within the datapath element. For synchronizing state both handlers exchange events and actions among each other, effectively initiating state transitions. Note that both top and bottom handlers may be NULL handlers, i.e., all processing is done entirely either in the control plane (bottom handler is empty) or no controlling instance exists at all (top handler is empty). Control-plane-only processing moves all packets of a flow into the "slow path," thus presumably degrading the flow's forwarding performance significantly.

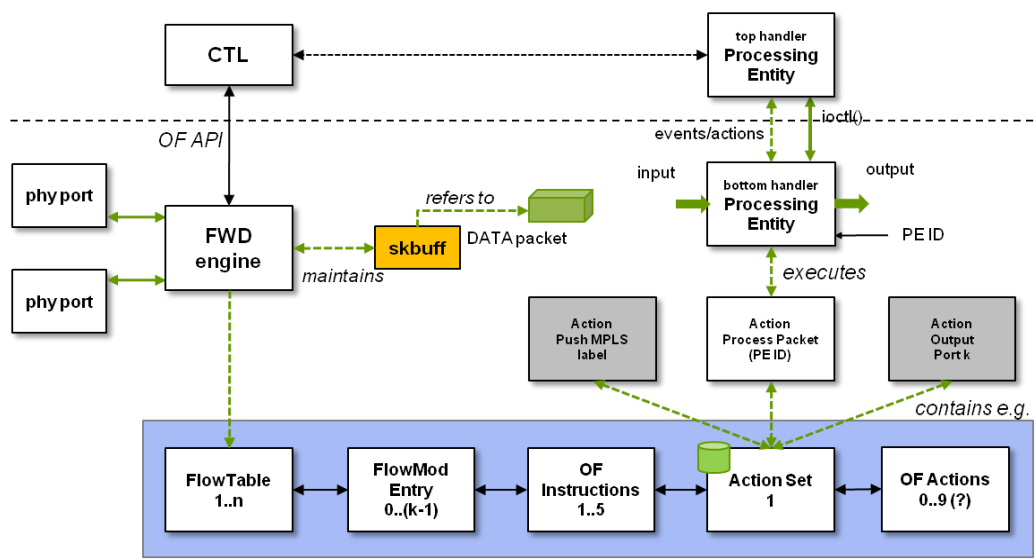

Figure 22: Proposed logical OpenFlow architecture

We propose a generic event/action API for signaling events and state transitions as well as required actions between control and data plane. The API should be usable for both virtual ports as well as processing instances. Some mandatory ingredients are:

- A namespace for identifying processing instances as well as virtual ports
- A namespace for identifying a state machines counterpart uniquely in the control plane,
- A set of OpenFlow protocol messages for sending notifications for actions, events, and state transitions
- A data model to share the same state machine between data plane and control plane
- An encoding for events, actions, and state transitions, probably based on the Specification and Description Language (SDL) [67] or some other appropriate language
- Management messages to deploy parts of a split finite state machine on a datapath and attach this to its counterpart in the control plane

We leave a final proposal for further study.

### 5.1.6    Defining New Action Types at Run-Time

Beyond a plain extension of the data structures and message formats defined by OpenFlow, adding support for a new protocol entails also changes to a number of functional elements in a datapath element, e.g. the packet parser, the matching logic within the OpenFlow pipeline, and the packet processing logic implemented by the datapath element.

OpenFlow defines an abstract view on a datapath element and its capabilities, effectively decoupling the logical from the various physical architectures adopted by hardware manufacturers. For building high performance hardware based switching/routing devices, the hardware manufacturer maps the logical architecture on the available hardware elements. ASIC-based designs impose additional constraints for extensibility, as they restrict extensibility by their internal design and the set of supported protocols that has been defined by the chip set manufacturer [17][18]. However, network processors have become more powerful in recent years and may perform advanced processing tasks at adequate speed. On the other hand, Intel is promoting chip sets for implementing fast network elements using their Intel I/O Acceleration Technology [68], so fast general purpose processing units are (or at least will be soon) available. Such programmable processing environments allow us to rethink the processing framework.

1. Since version 1.2, OpenFlow supports a TLV-based framework for defining matches as well as actions (see ActionSetField in the specification). This enables a datapath element to reveal per-table capabilities in terms of available actions and matches. The available namespace of OpenFlow extensible match fields is 23bits wide, where 16 bits contain a class identifier. Currently, exclusively assigned to ONF members, this namespace allows for a wide variety of additional header fields. While we can add new actions with means like firmware upgrades, the set of available processing capabilities remains static.

2. Exposing new non-standardized actions is a useful extension to the current OpenFlow specification. However, in the light of the generally programmable processing units entering the networking market, programming a datapath element with new, yet unknown actions and matches, seems feasible. The datapath model as defined by the ONF hides any datapath specific details. We propose to extend the OpenFlow data model with a general processing environment. While the control plane developer defines new actions and matches in the language of this virtual machine, a hardware manufacturer will map these instructions to the real hardware setup.

Based on our recursive architecture (please refer to Section 4 for details), hybrid designs may couple fast ASIC-based switching chip sets and programmable processing units within a domain. Such a domain may expose itself as a single datapath to higher layers, while its internal logic is redirecting flows based on their processing capabilities within the domain (i.e. the domain backplane). This domain controlling logic must define a constraint based routing, where the constraints are actually the processing capabilities (i.e. the supported actions and matches) of a specific datapath element within the domain.

Making a datapath programmable by the control plane is the most flexible extension framework for OpenFlow. Such an approach requires the definition of a model for parsing and matching packets and for a virtualized processing. With such a design the control plane is able to specify new actions and the virtual machine within the datapath interprets and maps these new actions/matches to the underlying forwarding and processing backend.

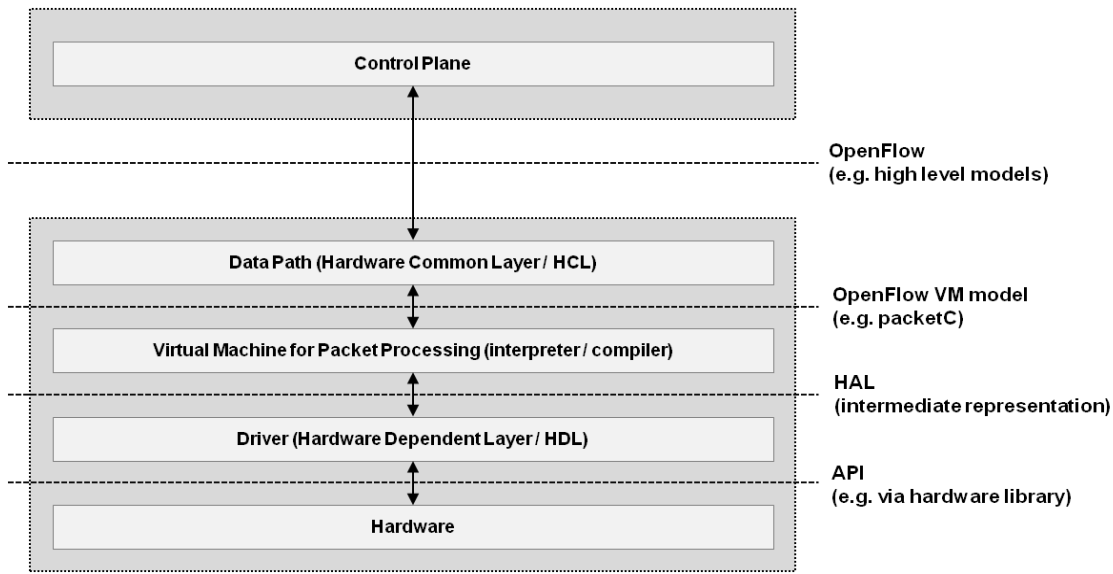

Figure 23: Programmable Datapath Model

Figure 23 depicts the logical layers in a programmable datapath element. We can distinguish four layers above the hardware forwarding and processing layer:

- The control plane designer defines a dynamic action using a programming language designed for network packet processing such as packetC or NFtables [58][66]. Some typical compound functions may reoccur frequently, like implementing a L3 or L2 routing/switching forwarding element, RIB-to-FIB mappings, etc. The datapath designer may specify a library of such compound functions with optimized hardware mapping and expose this to the control plane developers. In either case, the newly defined action results in a program based on the programming language. In Figure 23 a control module sends a code block (as either source code or in an intermediate representation) down to the datapath element.

- The OpenFlow management endpoint in the datapath receives the new action definition. Both the management endpoint and the virtual machine define a common layer, as it is agnostic with respect to the underlying forwarding and processing engine. This is called a Hardware Independent Convergence Layer (HICL).

- The virtual machine is responsible in interpreting/compiling the new action definition into a representation suitable for the hardware driver. A Hardware Abstraction Layer effectively decouples the hardware dependent driver and the virtual machine, i.e. both VM and hardware driver can be implemented and provided by separate vendors/manufacturers. The hardware driver defines the Hardware Dependent Convergence Layer (HDCL). The driver may use an arbitrary (open or proprietary) API for programming the forwarding and processing backend.

- The lowest layer is the forwarding and processing backend, e.g. ASIC; NPU, or CPU based ones.

Such architecture is currently under discussion within the ONF.

## 5.2 Virtualization and Isolation

### 5.2.1 What is network virtualization?

Network virtualization as such is not a new idea; in fact, many existing implementations exist on multiple layers in the network stack. These existing techniques are applicable in many situations, e.g., to improve network resource utilization through sharing among different tenants, to provide a logical separation of traffic between different entities, to simplify network management, and to provide secure connectivity over untrusted networks.

For example, to provide end-to-end, point-to-point or multipoint-to-multipoint connectivity, there are a number of different Virtual Private Network (VPN) techniques. There are many ways to create a VPN, for example, on top of layer 2, layer 3 or even layer 4 networks to provide the user with an interface that emulates a direct or routed/switched connection at the specific layer. For example, VPLS operates at layer 2 by creating MPLS pseudo-wire tunnels between a number of provider edge routers; these routers then provide a layer 2 interface to various customers and switch the layer 2 traffic over the tunnels. From the customer's point of view, the endpoints act as if they are connected to a standard layer 2 switch. Similar techniques exist for providing connectivity on top of many types of networks, as well as providing different kinds of connectivity. Almost all possible combinations exist: Ethernet over IP, IP over ATM, ATM over MPLS, etc. One important aspect of VPN services is that the customer or user of the service has no control over how the service is actually implemented – he is only aware of the endpoint connections.

From the provider's point of view, the techniques used to implement VPNs allow him better network utilization, since multiple customers can share network resources, including both links and routers/switches. Using, for example, Virtual LANs (VLANs) on the Ethernet layer or Virtual Routing and Forwarding (VRF) on the IP layer, it is possible to create multiple forwarding tables inside a switch or router, forwarding traffic in different manners depending on the forwarding table assignment. This capability can be used either as a mechanism for providing services such as VPNs, or as a way of simplifying management of the network by splitting it into multiple separated domains. Assigning traffic to the different forwarding tables inside a router/switch can be done in several ways, for example by adding a tag to the traffic that identifies which forwarding table should be used – this is the case with VLANs. VRF does not use any explicit tag: With VRF, traffic is assigned to forwarding tables based on other criteria such as which link it arrived on. The link may be an actual physical link, or a virtual link implemented using a tunneling protocol such as GRE, IPSEC, etc.

While there are many techniques for virtualizing parts of the network - the nodes, the links, and for creating virtual connectivity between end-points, typically they are applied separately and not as an integrated service. For example, it is common to lease virtual links for connecting different enterprise branches, but typically one has no control over routing and filtering over that link. OpenFlow, on the other hand, gives us the possibility of combining the existing techniques / concepts and provide a more comprehensive solution, namely a virtual network. A virtual network service would provide not only end-to-end connectivity, that hides all the details of how it is actually implemented, but would also give the customer complete control of how traffic is forwarded and handled inside the network, allowing him to treat it as a part of his own network.

An environment such as a multi-tenant WAN or a fixed mobile-converged network, where multiple customers and competing service providers share a single physical network through network virtualization, imposes many requirements on the virtualization system. Not only must the system make sure that the traffic is separated between customers, ideally no information should be able to leak between the virtual "partitions" (unless previously agreed otherwise); it must also enforce any established SLAs. In addition, the system should be as flexible as possible in order to avoid costly interaction and coordination between the parties involved, thus reducing operational costs.

In the following sections, we first investigate the derived requirements from Deliverable D2.1 and show how they are fulfilled by existing solutions for OpenFlow-based virtualization. We then investigate two aspects regarding virtualization – a) how to map or assign the customer's traffic into and out of the virtual networks and, b) what options are available for implementing the virtualization system itself and what changes are required to the protocol as well as the OpenFlow switch model in order to implement it in a carrier-grade manner.

### 5.2.2    Requirements for the virtualization system

In Deliverable 2.1 a number of requirements are derived and several of those apply directly to network virtualization, in particular these requirements:

> **R-2** The Split Architecture should support multiple providers.
>
> **R-3** The Split Architecture should allow sharing of a common infrastructure, to enable multiservice or multiprovider operation.
>
> **R-4** The Split Architecture should avoid interdependencies of administrative domains in a multiprovider scenario.
>
> **R-11** The Split Architecture should support best practices for QoS with four differentiated classes according to the definition documented by the Metro Ethernet Forum.
>
> **R-16** The Split Architecture must control the access to the network and specific services on an individual service provider basis
>
> **R-42** Data center virtualization must ensure high availability.

While most of these are quite general requirements, they highlight some issues with the existing implemented OpenFlow virtualization techniques: the FlowVisor and Multiple Switch Instances (a type of software-isolated virtualization). These existing solutions are described in more detail in Section 7.3-7.4 of Deliverable D3.1, but a short description is provided here.

The FlowVisor is a type of controller that acts as an OpenFlow protocol proxy, sitting between a number of controllers and switches, which forwards or denies OpenFlow commands from and to the controllers based on predefined policies. For example, the policies may restrict one controller to only send commands concerning a specific VLAN range and in that way virtualizes the network by restricting the view the different controllers have over the network. The virtual view is not only restricted to what the controller can see but also what rules they may install, or put in another way, it restricts the connected controllers to interact only with a well-defined part of the total flowspace.

Using Multiple Switch Instances, the switch runs multiple OpenFlow instances, allowing multiple controllers to connect to a single switch. Each of these OpenFlow instances is restricted through configuration to be able to use only a subset of the ports on the switch, or a number of VLANs on a particular port.

More detail about the impact of the different derived requirements on the virtualization system:

**R-2** – Multiple providers

Using a shared network, where multiple providers each have the ability of controlling part of the network via their own virtual network instance, increases the requirements for isolation in the virtualization system compared to the situation where a single provider uses virtual networks as a means to separate services. In the latter case, for example, information leakage between virtual networks or unfair bandwidth sharing between virtual networks is not a major problem since the single provider already knows the leaked information and is only "stealing" bandwidth from itself. In the first case, both of these problems are major issues and thus the virtual networks need to be strictly isolated from each other.

**R-3** – Multiple services and multiple operators on a single link

Running multiple services, each in a virtual network, does not put very heavy requirements on isolation in terms of information leakage or fair use of the control channel. However, good QoS support may be important depending on what the services are. This would include common QoS concepts like prioritization, traffic shaping, etc.

Sharing a single link for multiple operators requires some way of distinguishing packets belonging to the different operators, with more flexibility than reserving whole links per operator. Either this could involve defining strict limits on what part of the address space each operator is able to use, or somehow marking the packets with for example operator-reserved VLAN tags or MPLS labels.

**R-4** – Administrative interdependencies

When traffic enters the virtual network at the edge of the network, some mapping is necessary in order to map the incoming packets to the particular virtual network to which they should belong. This might be all traffic entering on a particular port, or traffic with some specific characteristics, such as certain VLAN tags, IP subnets, MPLS label ranges, etc. This is one area where flexibility is important in order to reduce the administrative interdependencies if, for example, two providers want to utilize or map the same VLAN range to their different virtual networks.

**R-42** – High availability

Either a robust virtualization system should be easy to duplicate or protect by other measures in order to maintain high availability, or it should be constructed in such a way that, from an availability point of view, it makes no difference if it is there or not.

Existing solutions have some problems meeting these requirements, as summarized in Table 4:

| SPARC Requirement | | FlowVisor | Multiple Switch Instances |
|---|---|---|---|
| Multiple providers | Level of support | **Poor** | **Implementation-dependent** |
| | Main reasons | OpenFlow lacks support for proper data plane isolation, good control plane isolation is possible | Data and control plane isolation may be good depending on the configuration and implementation |
| Multiple services | Level of support | **Poor** | **Poor** |
| | Main reasons | Lack of proper QoS support in OpenFlow | Lack of proper QoS support in OpenFlow |
| Multiple operators per link | Level of support | **OK** | **Poor** |
| | Main reasons | Only supports non-overlapping flowspaces | Typically separate ports per operator or non-overlapping VLAN tagging |

| Administrative interdependency | Level of support | **OK** | **OK** |
|---|---|---|---|
| | Main reasons | Flowspaces may not overlap, which requires coordination | Typically separate ports per provider or non-overlapping VLAN tagging |
| High availability | | Additional single point of failure that has to be replicated | Same as without multiple switch instances |

**Table 4: Existing virtualization solutions with OpenFlow vs. SPARC requirements**

As can be seen in this table, the current solutions do not provide adequate support for the requirements in many areas, therefore we have investigated what possibilities are available in order to create a system that fulfills the requirements to a higher degree.

### 5.2.3 Customer traffic mapping

Regardless of how the actual virtualization system is implemented, there is a problem at the edge between the customer and the provider of the virtualized network. How should traffic be mapped from the customer network onto the virtual network? At the edge nodes of the virtual network the customers may each be connected to an interface of their own, or they may share an interface through some means, for example through some tunneling protocol or simply by using different parts of the address space on some layer. Additionally, this may not be consistent for all the customers' connections to the network but may be different at various points of access. The virtualization system must be informed about how these connections are made in order to assign the incoming and outgoing traffic to the correct virtual network. The system may also require other virtualization related parameters such as the IP address, bandwidth requirements, priority of the traffic, and – depending on the virtualization system – what part of the flowspace should be reserved for each particular Virtual Network.

In the current FlowVisor implementation, these parameters are specified in a policy file on a per-switch basis, which can be a very time consuming and inflexible way of configuring the network. There is software to simplify the process, for example, the Opt-In Manager which is a web interface that allows users to define a network-wide flowspace they wish to control and to forward their request to an administrator for approval. While this simplifies the process, it still requires manual intervention for each slice.

One way of increasing the flexibility would be to allow automatic configuration either directly through the OpenFlow protocol itself or via an external protocol. Such a protocol could allow the network administrator to define some high-level SLA parameters such as maximum total bandwidth, as well as some limitations on the total flowspace the customer is allowed to use. Detailed configuration could then be left to the customer's controller, which automatically (depending on the applications running) could define the VNs he needs, within his assigned flowspace, as well as how his traffic should be mapped to these VNs.

The same protocol could serve multiple purposes, but primarily it would be used to define how traffic should be handled at the edges of the network. However, it could also be used to specify the topology of the virtual network. For example, if the customer would prefer that the virtualization system abstracts the entire physical topology into an abstract node before presenting it to him, or if he would like a certain topology to be created using virtual links it could be specified using the same protocol.

### 5.2.4 Network virtualization techniques for OpenFlow

A number of different models of how to perform virtualization in the context of OpenFlow can be imagined. All of these models contain three major parts that have to be virtualized: the control channel, the software part of the switch (the OpenFlow Instance, typically running on a general purpose CPU in the switch), and finally the hardware fast-path forwarding part of the switch. First, we will examine some different virtualization models before we delve deeper into the options available for the three parts of the system.

Most essential to the virtualization process is some kind of translation / hypervisor unit that translates values between the real physical view and the different virtual views, a unit similar to a virtual memory management unit in computer architectures and the various kinds of hypervisors used to create virtual machines. As with computer virtualization hypervisors, there are many different options for how and where to implement it, but it has to be somewhere between the application logic and the physical fast-path hardware. In Figure 24 five different models are shown:

1. The FlowVisor approach, where the translation unit resides outside of the switches and is shared by multiple switches and multiple controllers.

2. Each switch runs a translation unit that distinguishes between different connected controllers and performs the translation at the protocol level inside the OpenFlow instance on the switch.

3. Each switch runs multiple OpenFlow instances and with translation done between each OpenFlow instance and the fast-path forwarding hardware.

4. Each switch runs multiple OpenFlow instances and even the fast-path hardware has been split into several forwarding modules, one per OpenFlow instance. Translation is performed by restricting how ports are connected to the forwarding modules; the physical links may be split into multiple ports by using VLANs or MPLS labels, for example.

5. A model with multiple translation units, one responsible for virtualizing a single switch into multiple virtual ones, and another responsible for connecting multiple virtual switches and creating a virtual network representation. The first is responsible for virtualizing and isolating the switch resources, while the second connects to each virtual switch to create a virtual network topology, e.g., by presenting multiple switches as a single switch and managing tunnels that are presented as physical links similar to what is done in [29] and [30].

6.

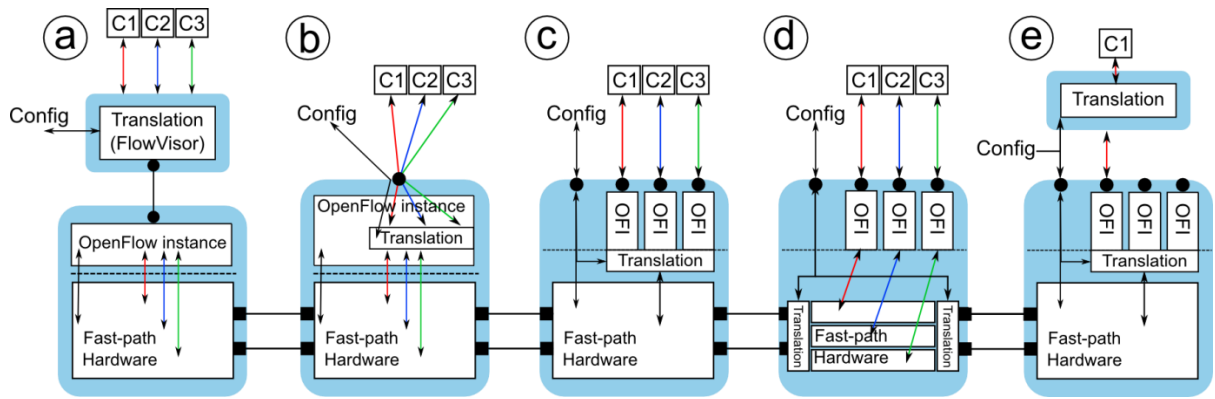

**Figure 24: Different virtualization models for OpenFlow.**

### 5.2.4.1    Control channel isolation

To ensure isolation between the control channels, the network that connects the controller to the switches (be it out-of-band or in-band) has to support some kind of rate limiting to prevent one virtual network from using all the bandwidth on the control channel, e.g., by forwarding a large amount of packets to the controller. The same is true in the other direction; the different controllers should not be able to disturb each other by transmitting large amounts of data to the switch. In the simplest scenario, with only one switch and one controller, these two problems could be taken care of in the switch and the controller respectively by limiting the amount of data they are allowed to transmit. However, if one increases the amount of switches in the system, one will reach a point where the aggregate amount of (still limited per VN per switch) traffic from all switches is enough to cause disruption.

Additionally, in both out-of-band and in-band cases, the control traffic may be competing with other traffic for network resources, for example different control traffic using the same out-of-band network, or data traffic in the case of in-band control channels. This traffic may be using its own flow control, like TCP, but this is not always necessarily true. If the only traffic on the control network is the different TCP connections between controllers and switches, the built-in flow control will react to congestion and limit the bandwidth use of each connection so that the TCP connections get an approximately equal share each. However, depending on how the different virtual networks are operating, they may likely have different bandwidth requirements for the control channel: One may rely heavily on sending data traffic to the controller for further analysis, whereas another may only be sending OpenFlow commands over the control channel. In both the out-of-band and in-band cases, network-wide QoS reservations for the control traffic can solve the problem and provide fairness in the use of control channel resources. For example, a tenant that needs to analyze data traffic in the controller will probably consume more bandwidth than one that does not have that need – TCP flow control is not enough to satisfy such requirements.

Depending on the virtualization model used, local (per VN in the controllers and switches), rate limiting may still be necessary. When using a FlowVisor, the control channels for the different VNs are multiplexed over a single TCP/SSL connection, so it is impossible for intermediate network nodes to look inside to differentiate between different flows and enforce QoS for the various controllers. On the other hand, if the switches allow the different controllers to connect and control the virtual networks using, e.g., different source or destination IP addresses or port numbers, the control traffic network can easily distinguish between the different connections and thus enforce QoS policies.

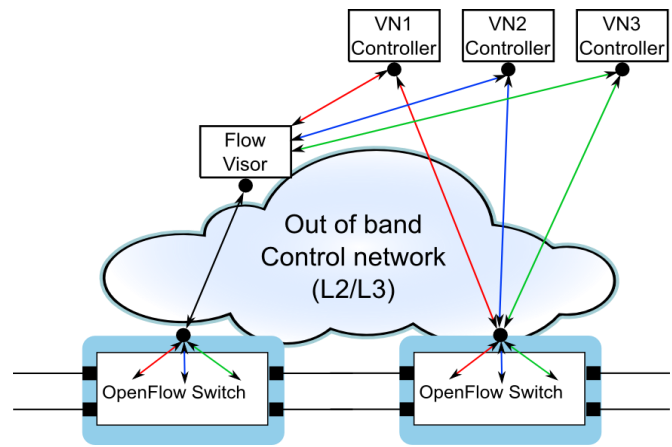

**Figure 25: A single TCP/SSL session used by a FlowVisor (left) vs one connection per controller (right).**

### 5.2.4.2 Out-of-band vs. in-band control network

An out-of-band control network has many advantages compared to the in-band counterpart in the OpenFlow approach. It is both simpler and easier to design in a reliable manner. However, out-of-band control networks might not be possible in some scenarios, for example in widely geographically distributed OLTs in access networks. Even if an out-of-band network is possible, it would be more expensive than an in-band solution due to an entire extra network and extra ports on the hardware. Given these considerations, in-band control channels are as important as out-of-band channels, if not more.

In an in-band scenario, the control channel is vulnerable to mis-configured flow table entries since all packets, including control packets, traverse them. For example, the in-band solution implemented by OpenvSwitch [31] requires the operator to pre-configure each switch with the neighboring switches' as well as its own MAC addresses and the controllers' IP addresses in order to install "invisible" FlowEntries that send control traffic to be processed by the local (non-OpenFlow) networking stack. This solution is quite fragile and configuration intensive, which could be operationally expensive in a large network. However, with a robust automated topology and controller discovery algorithm (see Section 5.5), even a quite complicated solution could be managed. It is important to note that such an automated discovery algorithm does not necessarily have to operate on all the virtual networks' control networks – these can be managed by the virtualization system itself.

In the in-band case, it is also very important that sufficient QoS support is available so that control traffic can be guaranteed priority over data traffic, in order not to lose control over switches in case of data traffic overload. This should be kept in mind when designing the QoS extensions discussed in Section 5.8. Additionally, QoS should not only be applied to control traffic entering the switch from the in-band channel, but also to outgoing traffic, without the traffic necessarily passing through the flow table when leaving the switch. This may require that an invisible / non-mutable high-priority queue is reserved for traffic from the local network stack.

### 5.2.4.3 OpenFlow state isolation

By *OpenFlow Instance* we refer to the software part of the OpenFlow implementation of a switch that is running on the normal switch CPU which is responsible for processing OpenFlow messages and performing anything else that is not handled by the hardware itself (e.g., implementing some actions, handling counters). Depending on the virtualization model and the implementation, it may also contain a translation unit that has to keep the different connected controllers separated. However, running multiple OFIs as separate processes decreases the risk of them affecting each other, e.g., if misconfigurations that cause packets to go to the wrong VN. This could still occur in the datapath or in the translation unit, but the fewer ways the different VNs share, the less likely it should be. Additionally, running them as separate processes reduces the risk of problems spreading from one VN to another through other software bugs that could cause the software to crash. Depending on the implementation, an OFI may contain memory tables that could overflow, or it may use more can than its fair share of switch CPU when processing protocol messages.. If the switch operating system supports it, this can be limited through further isolation by a virtualization system, such as the lightweight Linux Containers [32], which enables strict management of switch memory and CPU usage by the OFIs.

Local virtual ports can be used to send flows to the switch's local networking stack, e.g., to implement an in-band control network. In order to support the multiple channels corresponding to each OFI, a predetermined range of Local port numbers could be used, one for each OFI. Alternatively, if the different OFI instances are multiplexed through a single Local virtual port using their respective IP addresses, it could lead to fairness and starvation issues.

The translation unit within the OFI is responsible for mapping the virtual identifiers to their physical counterparts as well as enforcing the separation between the different virtual "identifier spaces." The complexity of the translation unit depends on the choice of the virtualization model. For example, if per-VN encapsulation is used, similar to model (d) in Figure 24, the payload of the Packet_In and Packet_Out OpenFlow messages needs to be modified based on the port and VN they are associated with, and they should have the correct QoS markings applied for fair resource allocation. In contrast, correct translation and mapping of per-VN port numbers and port counters applies to all the virtualization models discussed in Figure 13. For example, for statistics monitoring, per-VN port counters could be implemented through per-VN flow table entries.

One of the OFIs has a privileged role and has the authority to configure the translation unit as well as all the flow tables (without going through the translation unit). This process should be running at a higher priority than the others in order to increase the likelihood that the network owner can control the switch in case of issues with the other OFIs. Configuration of the flow tables can be performed through the existing OpenFlow protocol, which could be extended with, e.g., a vendor extension to also be able to define the VNs and configure the translation unit. Such an extension would contain, e.g., the VNID, the ports that should be included, which tables the VN should use and similar information.

### 5.2.4.4　　Isolation in the forwarding plane

In the forwarding plane several things should be virtualized and isolated from each other, firstly the actual entries in flow table(s), the physical links (or parts of them) and any processing modules (e.g., OAM modules as presented in Section 5.3, or other processing resources reachable by the Action *Process* presented in Section 5.1.3). The exact demarcation between the OpenFlow instance and the hardware is not very clear and depends on the implementation.

There are two approaches to keeping the virtual networks separated on the link level: partitioning and encapsulation. With *partitioning* the link is split into multiple partitions by dividing the total flowspace into non-overlapping logical chunks (called "slices"). For example one can reserve a part of the total VLAN identifier range or IP address range for each individual Virtual Network. To ensure that this separation is effective, all the virtual networks must be partitioned based on the same part of the flowspace; one cannot create some partitions based on VLAN identifiers and some based on IP addresses since it may be ambiguous as to which virtual network a packet with both a VLAN and IP header belongs to. However, these restrictions are strictly "link local," e.g., the same VLAN range could be reused for a different virtual network on another port on the same switch. Additionally one could apply translation to the field used to partition the flowspace between virtual networks. For example, if traffic belonging to two different virtual networks enters a switch from different ports but with the same VLAN, it is possible to translate one of them to a different VLAN identifier while traversing the network, and at the egress translated back to the original value.

With *encapsulation,* some type of encapsulation separates the traffic when it traverses a shared link. Before transmitting a packet, each switch adds some kind of link-local encapsulation per virtual network; the encapsulation is used at the receiver to assign the packet to a certain virtual network and then removed before the packet enters the packet processing pipeline. This leaves each virtual network with a complete flowspace that has not been sliced to create separation, thus making the system more flexible at the cost of the extra processing needed to add and remove the encapsulation at each hop. It also requires the switches to support some kind of encapsulation protocol such as IPSEC, L2TP, GRE, MPLS, or PBB. Currently the only supported encapsulation method is MPLS, which is supported with the Ericsson extensions to OpenFlow Version 1.0 and by default from OpenFlow Version 1.1 onwards.

When it comes to the flow table(s) there are multiple approaches as well, either partitioning or splitting. With partitioning the same restrictions are applied as in the case of the partitioned link sharing above. Each virtual network controller is restricted to insert flow table entries that 1) a have a match that is within the specific flowspace assigned to the particular virtual network, and 2) do not have any actions that would move the packet into a different virtual network, for example by changing the VLAN. Here again the approach has similar drawbacks as in the previous case – one needs to make sure that the partitioned flowspaces do not overlap, which causes a loss of flexibility.

Splitting divides the flow table into multiple pieces, giving each virtual network access to a specific piece. This can be done either logically through some internal logic that is specific to the particular brand of switch, or by having physically separate tables. The logical split is of course more interesting since it is more flexible and cheaper (by not using multiple hardware components). With OpenFlow Version 1.1, a logical split can be constructed by restricting the use of the multiple tables available, for example by allowing each Virtual Network access to only ten tables instead of the full range. While the OpenFlow protocol cannot address more than $2^8$ tables, this does not limit the potential number of VNs to 256. The flow table identifier has a local scope in each OpenFlow protocol session. If the switch hardware can support more than 256 flow tables, they can all be used through translation, but **not** by a single OpenFlow session.

Processing units, like virtual ports and Action *Process actions* (described in Section 5.1.3); can be shared between virtual networks if they are able to distinguish between packets from different virtual networks. For example, in the case of the BFD OAM module discussed in Section 5.3.3, the module uses a Maintenance End Point Identifier (MEP-ID) to distinguish between different BFD sessions when transmitting and receiving packets. An identical MEP-ID could exist in several virtual networks and therefore it is necessary to combine the MEP-ID with some kind of virtual network identifier to create a unique session identifier. This most likely applies to all kinds of processing units, as well as other related things, such as packet counters, which also should be kept on a per-virtual-network basis and not (only) globally within each switch.

So far, we have only discussed the ability to distinguish between different virtual networks and not mentioned enforcing fair use of resources. This could be handled through the standard QoS mechanisms (classification, metering and coloring, policing, marking, queuing and shaping) by applying them on a per-VN basis. However, support for QoS functionality is currently limited in OpenFlow, so we do not go into detail here but instead refer to Section 5.8. With the improvement suggested there, it would be possible to create guaranteed per-VN bandwidth profiles.

### 5.2.5　Improvement proposal

By combining the ideas presented above, we can construct a complete virtualization system that provides a high degree of isolation, both in the datapath and on the control level, which is flexible and requires a low degree of interaction between the operators. The system presented here is shown in Figure 26 and is based on model c) in Figure 24.

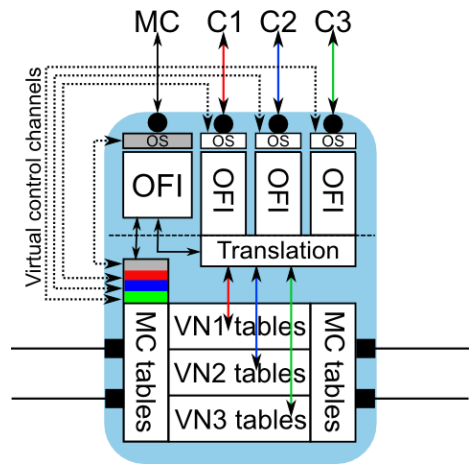

**Figure 26: A combination of an encapsulated forwarding plane with flow table splitting, an in-band Ethernet-based control network and multiple isolated OpenFlow Instances. Translation is performed between the OpenFlow Instances and the fast-path hardware, and is configurable through a privileged OpenFlow Instance.**

The datapath is based on model c) in Figure 24 to which we add encapsulation-based link separation combined with flow table partitioning. This results in a very flexible datapath virtualization system. Isolation between the VNs is handled by applying the resource allocation and QoS mechanisms discussed in Section 5.8. This can be achieved by dedicating the ingress and egress flow tables on the datapath for this purpose. Link separation can be achieved, for example, through specific, predetermined MPLS labels, where at each link they are used to encode the VNID (MPLS replaces the existing EtherType with its own, therefore each VNID needs multiple MPLS labels assigned to it, one per EtherType used).

The OFIs could be separated through the virtualization mechanisms of the OS, and they connect to the datapath hardware through the translation unit. The OFI that handles the *Master Controller* (MC), usually operated by the network owner, configures the translation unit as well as all the ingress and egress flow tables (shown as MC tables in Figure 26). This OFI should run at a higher priority on the local switch CPU so that the MC communications with the switch are given higher priority in order to handle special events such as failure scenarios. Configuration of the MC tables can be performed through the existing OpenFlow protocol, with minor restrictions enforced for each VN controller. The translation unit also needs configuration in order define the different virtual networks, and this can be achieved through non-OpenFlow channels or by OpenFlow protocol extensions containing the definition of a particular VN on a switch (e.g., the VNID, MPLS labels belonging to it, the ports belonging to it, etc.).

The control network could be implemented as an in-band IP network and separated into two parts: the master control network and the virtual control channels. The virtual control channels are managed by the MC, which is responsible for routing and establishing QoS as well as reconfiguring the control network in case of failures or topology changes. etc. The master control network, however, needs to be bootstrapped through manual configuration or via a dynamic discovery protocol.

### 5.2.6　　　　　　　　Proof-of-concept implementation

In order to test the various ideas described above a proof-of-concept implementation was constructed. However, the implementation does not exactly follow the improvement proposal since it was judged to be too difficult to implement within the timeframe of the project. Instead, the design is based on model b) of Figure 24. The major difference compared to the improvement proposal is that the translation unit is placed inside a single OFI instead of underneath multiple OFIs. The reason for this is that in the switch implementation the OFI and the "fast path" is difficult to separate. This modification does not impact the protocol extensions at all, and impacts the implementation of the translation unit very little. The proof of concept implementation consists of three parts:

1. An OpenFlow 1.1 capable software switch extended with a translation unit for virtualization

2. Vendor extensions to the OpenFlow protocol to install/remove a virtualization profile from a switch's translation unit as well as to identify the identity of the controller during the protocol connection handshake

3. A NOX application that reads a description of VNs from a file and based on the descriptions installs flow table entries and group entries to the all the involved switches as well as programming the translation unit in each switch

The overall function can be seen in Figure 27. On the left side the Master Controller can be seen connected to the physical OpenFlow switch topology, as well as the topology "seen" from the controller. The Master Controller (using a Virtualization Manager application) then installs a reduced virtual topology to the switches. The installation phase consists of programming the translation unit on each switch as well as installing a number of flow table entries and group table entries. On the right side a Virtual Controller is connected to the physical topology, however, it is only able to "see" the virtual topology since the translation unit is hiding and/or translating all relevant messages.

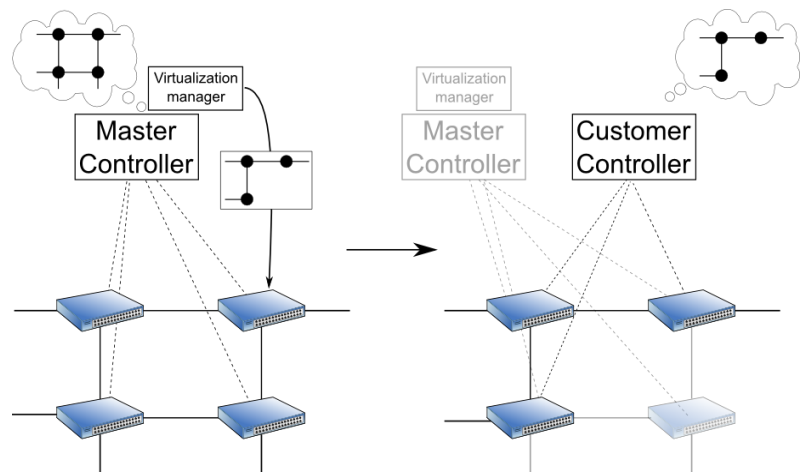

**Figure 27: The Master Controller (left) sees the full physical topology.
The Virtualization Manager running on the Master Controller configures the switches with a virtual topology. On the right side of the figure, a Customer Controller connects to the switches and is only able to see the virtual topology configured previously.**

**Dataplane virtualization**

Dataplane virtualization is performed through VLAN-tagging on shared links and by reserving a number of flow tables per virtual network inside each switch. Since OpenFlow 1.1 supports multiple VLAN-tags per packet this does not interfere with any existing customer VLANs. In Figure 28 the processing pipeline for three virtual networks within a switch is shown.

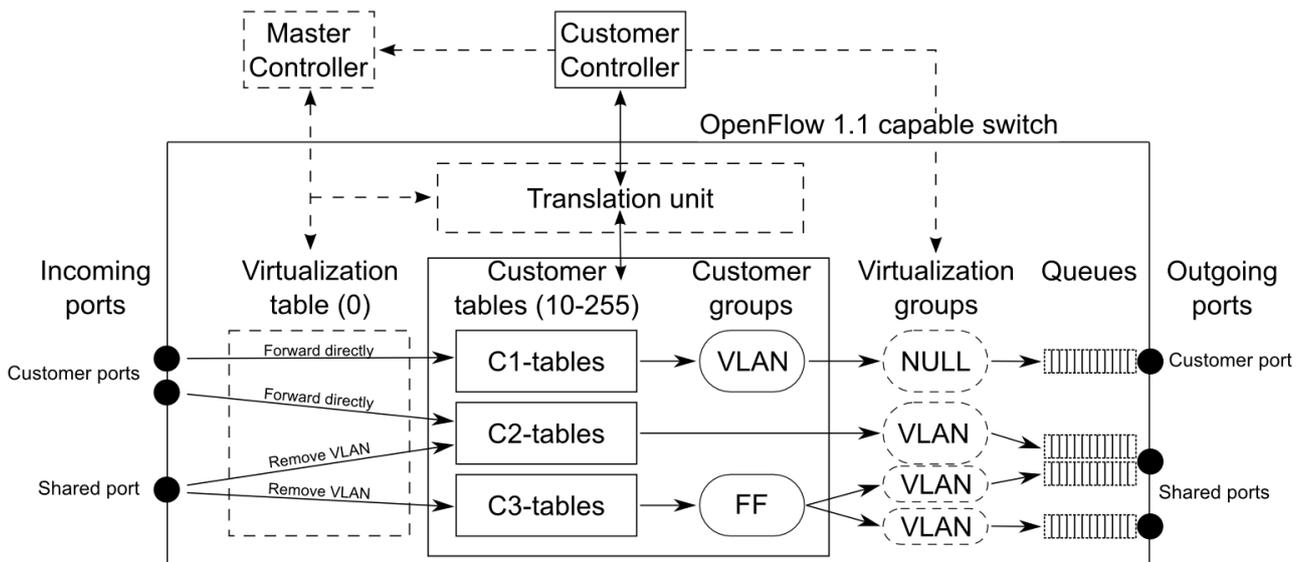

**Figure 28:** The various virtualization and customer tables and groups inside a virtualized switch. Solid lined boxes indicate regular OpenFlow datapath resources, whereas dashed lines indicate additions to realize the proposed encapsulation-based virtualization system.

When a packet first enters the switch on the left hand side of the figure, it first enters the *Virtualization table*. Here the packet is handled based on the incoming port: if it is a port shared by multiple networks the VLAN tag is examined, removed, and, based on the VLAN-ID, the packet is forwarded to the first of the tables reserved for the customer registered to that VLAN-ID. If the incoming port is dedicated to a particular customer the packet is immediately forwarded to the customer tables without any modification.

Once the packet is inside the *Customer tables* whatever the customer has programmed the tables to do will be executed. There are no restrictions on the either the types of matches nor actions that the customer can apply here. This includes modifications to the packet, sending the packet to the controller, etc.

When all customer tables have been traversed, the packet may enter one or more *Customer groups*, or it may go directly towards the outgoing port. Again, there are no restrictions placed on the customer groups, all actions and group-types are available for the customer to use.

Finally, before leaving the system, the packet goes through a *Virtualization group*. Depending on which port the virtualization group is attached to, it either does nothing, which is the case if the outgoing port is a dedicated customer port. However, if the outgoing port is a shared port, a VLAN tag is pushed onto the packet and the packet is assigned to a per-customer QoS queue. After being queued on the port, the packet may finally leave the switch.

**Rule installation and translation unit**

The Virtualization Manager application is responsible for installing the virtualization rules in the virtualization table as well as creating the virtualization groups and configuring the queues. Once these have been configured the translation unit is programmed with the relevant per VN information such as which ports are customer ports and which are system ports, which flow tables and per-port queues belong to this VN. Once this has been done, a customer controller can connect and identify itself to the switch.

The translation unit uses the information it has received in order to modify the OpenFlow messages sent to and received from the customer controller. For example, when a flow entry is installed, the translation unit modifies any "output" commands to go through the associated virtualization group instead of going to the port. The reverse process is performed when the controller requests information about a flow, if the flow refers to a virtualization group the messages to the controller is modify back to show an "output" action instead. Similar things are done for all state messages such as the one enumerating available ports, installed groups, available flow tables etc. In this fashion, only the information that the master controller has authorized can be seen by the customer controller.

## 5.3 Operations and Maintenance (OAM) Tools

OpenFlow, as an attractive element of the open split-architecture platform, aims to eliminate the weaknesses caused by proprietary hardware design and supply, i.e., the single sourced system software supply closely coupled with equipment vendors. However, this open but immature protocol still has significant drawbacks compared to the closed, traditional protocols: It only supports partial configuration of the switches. One important aspect that cannot be provisioned through the current OpenFlow solutions is OAM.

In earlier SPARC Deliverables (D3.1 and D2.1) OAM has been identified as an essential requirement for carrier-grade operator networks in order to facilitate network operation and troubleshooting. For instance, D2.1 proposes the following requirements:

*"For operation and maintenance, a number of functions are already widely defined and implemented. This includes identification of link failures, connectivity checks and loopbacks in the data plane. In addition, it should be possible to send test signals and measure the overall performance. Standards to be supported are ITU-T Y.1731, IEEE 802.3ag/ah and respective IETF standards for IP and MPLS. Interfaces of switches, routers, etc., provide additional information for monitoring and operation like link down, etc. This includes monitoring of links between interfaces as well."*

Specifically, three requirements regarding Operations and Maintenance have been derived in D2.1:

R-25    *The Split Architecture should support OAM mechanisms according to the applied data plane technologies.*

R-26    *The Split Architecture should make use of OAM functions provided by the interface.*

R-27    *The Split Architecture shall support the monitoring of links between interfaces.*

We see a few challenges with OAM in a *SplitArchitecture*:

- How to map traditional, technology-specific OAM elements to *SplitArchitecture*, i.e., how to integrate specific OAM tools (e.g., Ethernet OAM, MPLS OAM and IP OAM) into an OpenFlow-based *SplitArchitecture*.
- How to provide a generalized, technology-agnostic flow OAM for *SplitArchitecture*.
- Given that virtualization enables multi-operator scenarios, how can we support a multi-carrier service OAM and provide automatic OAM configuration in this environment?

In the following subsections, we will discuss the challenges listed above. We will start by giving an overview of the most prevailing existing packet OAM toolset, namely Ethernet Service OAM (IEEE 802.1ag, ITU-T Y.1731)) and MPLS-TP OAM (BFD, LSP-ping or ITU-T Y.1731 based). We will discuss how the traditional OAM functions and roles map to our split-architecture design. As an example of how technology-specific OAM tools can be integrated, we will then outline the MPLS BFD solution as implemented in the SPARC prototype. Finally, we will go one step further and consider the case of a novel technology-agnostic flow OAM by providing an initial architectural solution.

### 5.3.1    Background: existing OAM toolset

According to the state-of-the-art design, OAM toolsets are technology-dependent, i.e., they are bound to the specific data plane technology they have been developed for (e.g., Ethernet or MPLS(-TP)). As we will outline in the following subsections, the existing toolsets have very similar purposes, such as the detection of a connectivity loss or a violation of a delay constraint. However, the methods and architectures used to reach those goals are different, which makes the various OAM solutions incompatible.

#### 5.3.1.1    Ethernet Service OAM

Ethernet Service OAM is a key component of operation, administration and maintenance for carrier Ethernet based networks. It specifies protocols, procedures and managed objects for end-to-end fault detection, verification and isolation. Ethernet service OAM defines a hierarchy of up to eight OAM levels, allowing users, service providers and operators to run independent OAMs at their own level. It introduces the concept of a Maintenance Association (MA) that is used to monitor the integrity of a single service instance by exchanging CFM (Connectivity Fault Management) messages. The scope of a Maintenance Association is determined by the Management Domain (MD), which describes a network region where connectivity and performance is managed. Each MA associates two or more Maintenance Association Endpoints (MEP) and allows Maintenance Association Intermediate Points (MIP) to support fault detection and isolation.

The continuity check protocol is used for fault detection. Each MEP can periodically transmit connectivity check messages (CCM) and track CCMs received from other MEPs in the same maintenance association.

A unicast Loopback Message is used for fault verification. It is typically performed after fault detection. It can also confirm successful initiation or restoration of connectivity. Loopback Messages (LBMs) are transmitted by operator command. The receiving MP responds to the LBM with a unicast Loopback Reply (LBR).

A multicast Link-Trace Message (LTM) is transmitted in order to perform path discovery and fault isolation. The LTM is transmitted by operator command. The intercepting MP sends a unicast Link-Trace Reply (LTR) to the originator of the LTM. The originating MEP collects the LTRs and provides sufficient information to construct the sequence of MPs that would be traversed by a data frame sent to the target MAC address.

There are usually two types of Ethernet service performance management: Delay Measurement (DM) and Loss Measurement (LM).

DM is performed between a pair of MEPs. It can be used for on-demand measurement of frame delay and frame delay variation. An MEP maintains the timestamp at the transmission time of the ETH-DM frame.

LM is performed between a pair of MEPs, which maintain two local counters for each peer MEP and for each priority class – TxFCl: counter for data frames transmitted toward the peer MEP; RxFCl: counter for data frames received from the peer MEP. OAM frames are not counted.

### 5.3.1.2     MPLS(-TP) OAM

MPLS-TP OAM is still intensely discussed by the community, and is thus documented in IETF drafts only. Generally, there are two types of MPLS-TP OAM under discussion:

The first method is to enhance the available MPLS OAM toolset (LSP Ping and BFD) to meet the OAM requirements of MPLS-TP.

LSP Ping provides diagnostic tools for connectivity checks of MPLS tunnels, testing both data and control plane aspects. LSP Ping can be run periodically or in on-demand fashion. Essentially, LSP ping verifies the status of a complete LSP by inserting "echo" requests into the MPLS tunnel with a specific IP destination address. Usage of a dedicated address prevents the packet from being routed further at the egress LSR of the MPLS tunnel. On reception of an "echo" request, the designation LSR responds by sending an "echo" reply back to the originator of the request. For MPLS-TP, LSP Ping should be extended with tracing functionality (traceroute i.e., link-trace). Furthermore, LSP Ping should support point-to-multipoint LSPs and should be able to run without an underlying IP.

BFD (Bidirectional Forwarding Detection) is used for very fast proactive detection of data plane failures by connectivity checks. BFD is realized by "hello packets" exchanged between neighboring LSRs in regular, configurable intervals. If hello packets are not received as expected, a connectivity failure with the particular neighbor is detected. BFD packets can also be used to provide loopback functionality with replied received packets at the neighboring node. For working with MPLS-TP, BFD will be extended, e.g., for working without reliance on IP/UDP functionality.

The second method is to develop MPLS-TP OAM based on Y.1731 Ethernet service OAM. The basic idea is that Y.1731 specifies a set of OAM procedures and related packet data unit (PDU) formats that meet the transport network requirements for OAM. The actual PDU formats are technology agnostic and could be carried over different encapsulations, e.g., MPLS Generic Associated Channel [59].

### 5.3.2     Mapping OAM element roles to SplitArchitecture

Extending SDN enabled network domains with OAM functions raises two core questions: An OAM solution may aim towards monitoring and supervising the internal integrity of an SDN domain without interacting with external networking domains. However, interworking among SDN domains and legacy domains may occur frequently, at least in migration phases moving legacy infrastructures towards SDN. As discussed previously, we face a wide variety of OAM toolsets in legacy domains tailored to providing services on their individual layer, be it Ethernet, MPLS, IP, etc. We investigate potential solutions for enabling interworking among OpenFlow based and legacy networking domains first. Beyond interworking with legacy OAM, a network operator may want to monitor and supervise internal operation and health of his SDN enabled domain. Here, the need for a dedicated SDN based OAM toolset arises. We investigate potential solutions for providing OAM functionality within an SDN domain in a following section.

While the OpenFlow protocol itself does not support any OAM solutions, it does not prohibit adopting technology-specific toolsets. For example, to support Ethernet OAM, Ethernet OAM modules could be configured in every node considering the control and management architecture presented in Section 3, as in the following figure:

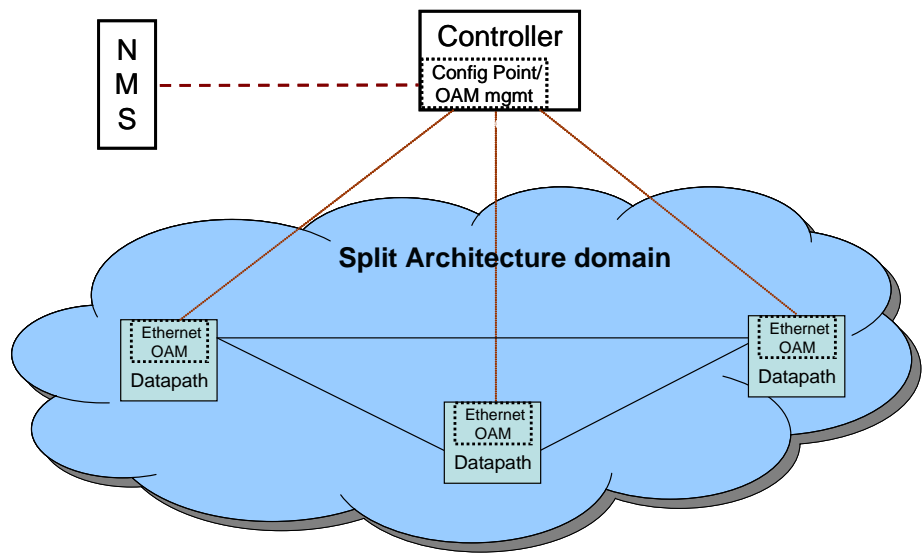

**Figure 29: SplitArchitecture OAM Configuration**

This configuration works well with Ethernet traffic flows. However, a main feature of an OpenFlow-based *SplitArchitecture* (e.g., OpenFlow) is the support of various traffic flows such as Ethernet, MPLS or IP. Thus, in the case of MPLS traffic flows, the Ethernet OAM module would become ineffective and an additional MPLS OAM module would have to be added to the datapath elements. As a result, datapath elements would contain different OAM configuration modules and OAM generator/parsing modules for different traffic flows. This contradicts to the original notion of *SplitArchitecture*s having simple datapath elements, and this type of OAM integration would significantly increase the complexity of data forwarding devices. An additional problem with multiple OAM modules for different flow technologies is their different configuration models, which further complicates both datapath and control elements. Furthermore, OpenFlow allows routing and switching of packet traffic that does not strictly follow Ethernet switching, MPLS forwarding, or IP forwarding. The OpenFlow switch allows "flow switching," i.e., switching is done based on arbitrary combinations of bits in the flowspace, which currently mainly consists of the Ethernet, MPLS and IP header fields. Finally, a general problem for OAM integration into current OpenFlow scenarios is the lack of an explicit OAM indication, which allows dispatching of OAM packets without interfering with OpenFlow match structures.

This section is devoted to proposing improvements to the OpenFlow switch model to implement OAM feature sets. During the design phase, we were faced with two contradictory aspects:

- Compatibility with legacy OAM toolsets.
- Avoiding the need of developing and running dozens of toolsets at the same switch.

The first aspect is relevant when such logical connections are monitored that span OpenFlow and non-OpenFlow domains, as done in the integration of IEEE 802.1ag OAM to OpenFlow by SARA Computing & Networking Service [20]. In such cases, the data plane switch must implement all technology-specific OAM features of the data layer. Considering the monitoring of OpenFlow domain internal connectivity only, this former aspect becomes irrelevant. As the datapath in OpenFlow domain is layering-agnostic, the goal is to provide a unified flow OAM using only a single OAM module in order to monitor all flows in this specific domain.

In the upcoming two subsections we will present improvement proposal methods for both, technology-dependent and technology-agnostic OAM toolsets. As an example of a technology-dependent solution, we present MPLS BFD for OpenFlow, which has also been implemented as part of the SPARC demonstrator. The second, more novel solution is technology-agnostic and provides a unified, generic OAM module for all flows. This solution is outlined only as an architectural concept, and many details relevant for implementation are still the subject of future work.

### 5.3.3　　MPLS BFD-based Continuity Check for OpenFlow

OpenFlow MPLS BFD-based continuity check (CC) uses the virtual ports concept. A virtual port is identified with a port number within the port table exactly like a physical port, but additional actions not part of the standard OpenFlow ActionSet may be performed on the packet. Virtual ports can be chained. A virtual port chain on the input side resides between the input port where the packet enters and the first flow table, whereas a chain on the output side starts after the flow table and ends at a physical port where the packet exits. The output side chains can be addressed by entries of any flow tables. OpenFlow 1.0 included a few built-in virtual ports for actions such as forwarding a packet to the controller. But those virtual ports are reserved for specific purposes and cannot be configured through the OpenFlow interface. An extension is defined for OpenFlow 1.0 that allows the dynamic creation and configuration of virtual ports in order to implement MPLS-related packet manipulation.

To reduce the amount of resources (e.g., timers) needed to generate monitoring packets, the OpenFlow MPLS protection switching makes use of the group concept to perform multicast. The protection switching also uses this feature to replicate single monitoring packets and then send them, after being modified to identify a single BFD session, out through multiple ports. Since the standard OpenFlow 1.0 neither supports the group concept nor includes any other multicast method, the group concept has been added to the MPLS-enabled OpenFlow 1.0 switch implementation.

### 5.3.3.1 Ingress-side BFD control packet insertion into the MPLS G-ACh control channel

Figure 30 illustrates how BFD control packets are generated and inserted into LSPs on the ingress side. A virtual port represents a BFD packet generator. The BFD packet generator generates a BFD packet template with a configured interval, for example each second, each millisecond, etc (see ① in Figure 30). The BFD template packet does not contain any information related to a particular BFD session; however, the packet generator fills in the timing fields of the packet. The BFD packet template also contains the required associated channels (ACH) TLVs (with empty values) and the generic associated channel label (GAL) (see [59] for further details) on top. The switch input source port is set to the BFD packet generator representing the virtual port number upon input to the flow table.

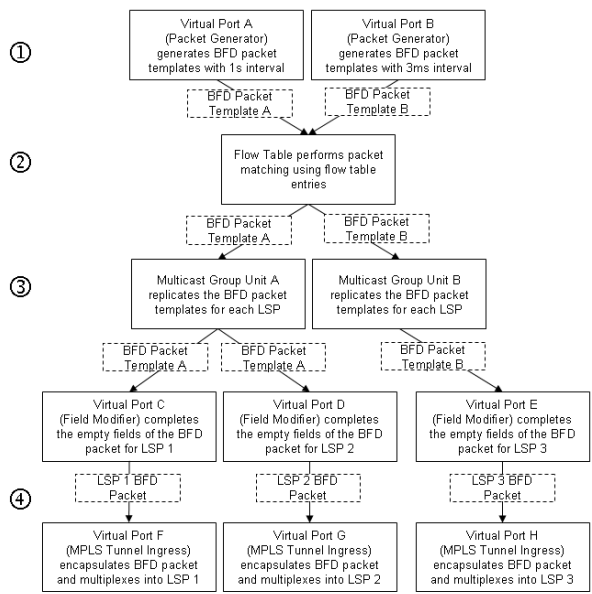

**Figure 30: Ingress-side OAM signaling generation**

The template packet is sent to the flow table for matching (see ② in Figure 30). The flow table has been programmed with a flow rule that matches the incoming BFD packet template, specifically the "In Port" field matches the virtual port number of the BFD packet generator, and the MPLS field matches the GAL label number (13). The flow action forwards the packet to a multicast group that handles packets with the same generation interval as the BFD packet generator. If there are two packet generators with different intervals, the packets generated by them are forwarded to different BFD multicast groups (see ③ in Figure 30) based on the differing input port numbers.

When the multicast group receives a packet, it replicates it and forwards it to its configured output ports. These virtual ports are field-modifier virtual ports (see ④ in Figure 30).

The field-modifier virtual port fills in the missing fields currently empty in the packet, i.e., the BFD packet fields for the particular BFD session: the timer values, the MEP-ID and the state variables, such as remote defect indication (RDI) [60]. Thus, this virtual port represents the source functions of the MEP (source MEP). The MEP-ID is used to do a lookup in an associative data structure that contains all the BFD session information (state variables, descriptors, output virtual port number, etc.) for all the BFD sessions on the egress side of the LSP. Once the BFD packet has been completed, the packet is forwarded to an MPLS tunnel ingress virtual port, where the BFD control packet is treated in the same fashion as incoming user data traffic: The MPLS label for the LSP is pushed onto the label stack and the packet multiplexed into the MPLS LSP through the output physical port.

#### 5.3.3.2        Egress-side BFD control packet processing

Figure 31 illustrates how BFD packets are processed on the egress side of the LSP. Incoming BFD packets enter the switch like any other packet on the same MPLS LSP (see ① in Figure 31). An incoming MPLS packet is matched in the flow table, and the packet is forwarded to an MPLS tunnel egress virtual port (see ② in Figure 31). This port is configured to pop the top label and examine any labels still on the stack. If the new top label is the GAL label, the packet is removed from the regular packet processing pipeline, but passed to the G-ACh module within the virtual port. The G-ACh module examines the channel type of the packet and the packet is sent to a Link Failure Detection module depending on the channel type (see ③ in Figure 31). The Link Failure Detection module performs a lookup in the BFD sessions' associated data structure storage using the MEP-ID found in the BFD packet as the key (see ④ in Figure 31). If a session is found, the associated BFD session data and failure timer for the session are updated appropriately.

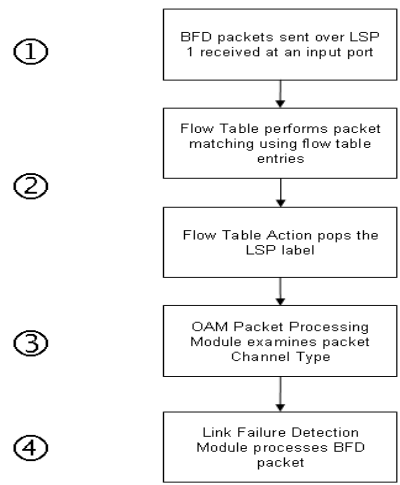

**Figure 31 : Egress-side OAM signaling reception**

#### 5.3.3.3        BFD transmission timer updates

During its lifetime a BFD session uses at least two different transmission timer intervals, one before the session has been established (longer than 1 second) and another (depending on the failure detection resolution) once the BFD session has been established. This requires that a BFD session is able to move between different BFD packet generators. To reiterate, each BFD session common timer value is represented by a BFD template packet generator and a corresponding Multicast group per timer value, and each particular BFD session is represented by a field modifier virtual port. In order to move a BFD session, when the timer changes, the multicast groups has to be reconfigured during operation in order to change the actual rate of template packets going to the field modifier ports. This design might seem overly complicated but it was done in order to reduce the number of individual high-resolution timers required by sharing them between BFD sessions with the same timing requirements.

#### 5.3.3.1        Updated design for Openflow version 1.1

The implementation described above was originally designed for the 1.0 version of the OpenFlow protocol and datapath model. With the release of OpenFlow version 1.1, the BFD implementation was updated to make use of the new mechanisms available in the new version. OpenFlow version 1.1 introduced the concept of "Fast Failover" groups. These groups contain buckets with references to either other groups or ports, when a group receives a packet it will forward the packet to the first of these buckets that is *alive*. The liveness of a port or group is determined by a liveness flag that is set by an external entity.

The new Fast Failover mechanism allowed for a less invasive design than the previous, since it is no longer necessary to modify the flowtable itself in order to perform the protection switching. This allowed us to design a more generic model, which required fewer modifications of the data plane as well as to the protocol itself. With the failover mechanism already in place we can move a large part of the BFD functionality into an external module and put a more generic OAM structure in place that could be reused for other OAM protocols apart from BFD.

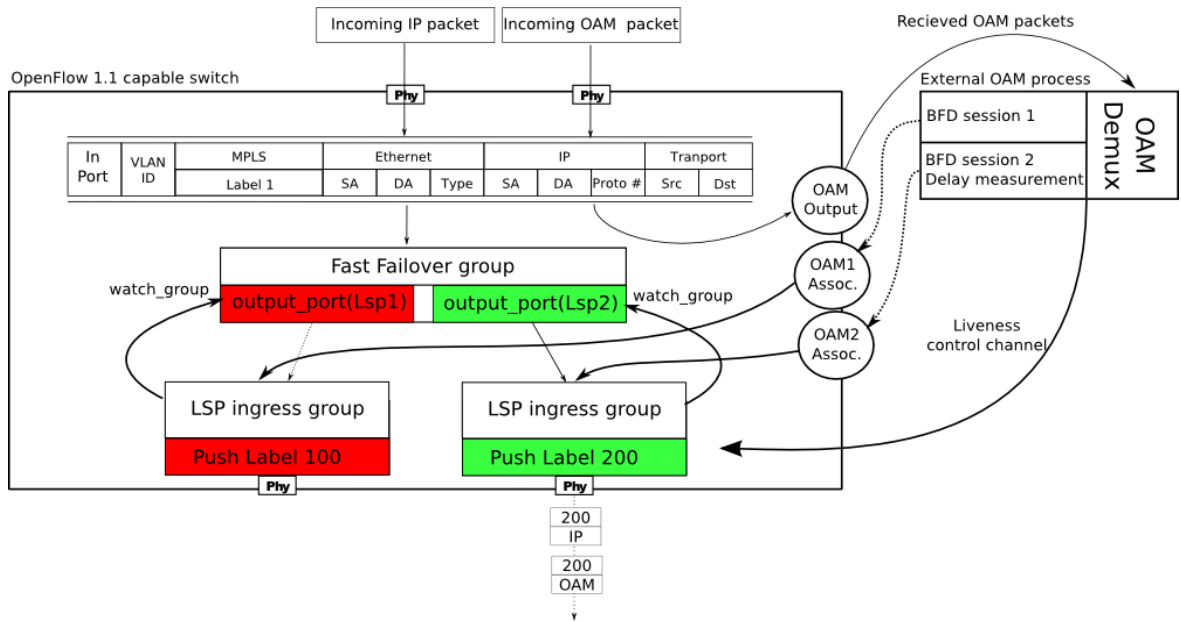

**Figure 32: BFD implementation for OpenFlow version 1.1.** Normal incoming packets are processed in the flowtable and forwarded to the Fast Failover group, which uses the first working LSP to forward the packets. Incoming OAM packets are sent through a channel for processing in the external module. The external module generates OAM packets and injects them directly in the LSP groups, whose liveness the module controls via a control channel.

The updated design can be seen in Figure 32, in this design the OAM packets (in this case: BFD packets) are processed entirely by an external OAM process. The external process creates and sends them to the OpenFlow software switch using some type of data channel, in our case, we are using UNIX sockets as a low-latency and simple means to transport the packets between the processes but other channels such as UDP/IP or shared memory could be used. When an OAM packet arrives on a data channel, the switch forwards the packet directly into a group for processing, bypassing the normal packet pipeline that does packet classification and flowtable processing. When an OAM packet enters a group it is treated exactly like any packet that has gone through the normal pipeline and from that point shares fate with the normal data packets.

When the switch receives OAM packets the same updated MPLS processing scheme that was used in the 1.0 version is applied, but instead of sending the packet for internal processing in the software switch itself, the packet is forwarded to the external OAM module with is responsible for decoding it and associating it with a particular OAM session.

In case of failure the external OAM module uses a control channel to update the liveness of the affected group, which will cause traffic to switch to a different path in the Fast Failover group (in case one is available). In case of a failure, the same notification mechanism is used as in the previous version; however, since all OAM session information is in the external module, the notification to the controller only contains the new liveness state and the group number.

### 5.3.4　　　　　　Technology-agnostic flow OAM

Technology-specific OAM solutions, as described in the MPLS BFD example above, have the advantage of straightforward integration with legacy domains in order to provide end-to-end monitoring of connections spanning multiple domains (including OpenFlow domains). However, technology-specific OAM solutions do not take the diversity of the OpenFlow flowspace into account, and target only specific flow types (e.g., MPLS flows in the BFD example). In order to support OAM functionalities for the entire flowspace, this approach would require a number of separate OAM tools (e.g., Ethernet OAM monitors Ethernet Traffic; MPLS OAM monitors MPLS traffic; IP OAM monitors IP traffic), which might lead to an unacceptable increase of complexity in the datapath elements.

Examining the different OAM technologies, we realized that they all have similar goals for the fault or failure in the network – detect, locate and report. In a *SplitArchitecture* domain, with decoupled control and data forwarding plane, different traffic flows (Ethernet, MPLS or IP) are all treated equally by the data forwarding devices – and differences are only relevant to the controlling elements. Then the interesting question arises: In an OpenFlow-based *SplitArchitecture* environment, can we decouple the OAM monitoring traffic from the actual flow traffic (Ethernet, MPLS or IP)?

There are a few issues with decoupling OAM from regular flow traffic that need to be considered for a proposed solution. First, even if OAM traffic is decoupled, fate sharing needs to be ensured. Fate sharing means that OAM traffic needs to take exactly the same path as the actual flow traffic, i.e., it must go through the same link, the same node and use the same filtering table as the data traffic to be monitored. Secondly, there needs to be a way to explicitly mark OAM packets in order to enable the datapath elements to detect and handle OAM packets accordingly. Finally, some OAM may have technology-specific OAM requirements.

### 5.3.4.1　　Independent OAM module

We propose an independent OAM module that is not associated with data traffic. This independent OAM module only generates or parses OAM PDUs (Protocol Data Units). To create a "fate sharing" path between the actual data traffic flow and OAM flow, we propose a Flow ID encapsulation/decapsulation module. This Flow ID module is associated with actual data traffic flow to be monitored so that the OAM traffic will have the same Flow ID and pass through the same link and the same node as the data traffic flow.

This general split-architecture OAM proposal has three advantages when compared to other solutions:

- Ubiquity: One OAM module supporting different traffic flows.
- Granularity: Supporting many services from a single carrier or many services from multiple carriers.
- Uniformity: Simplified and standard configuration and provisioning process.

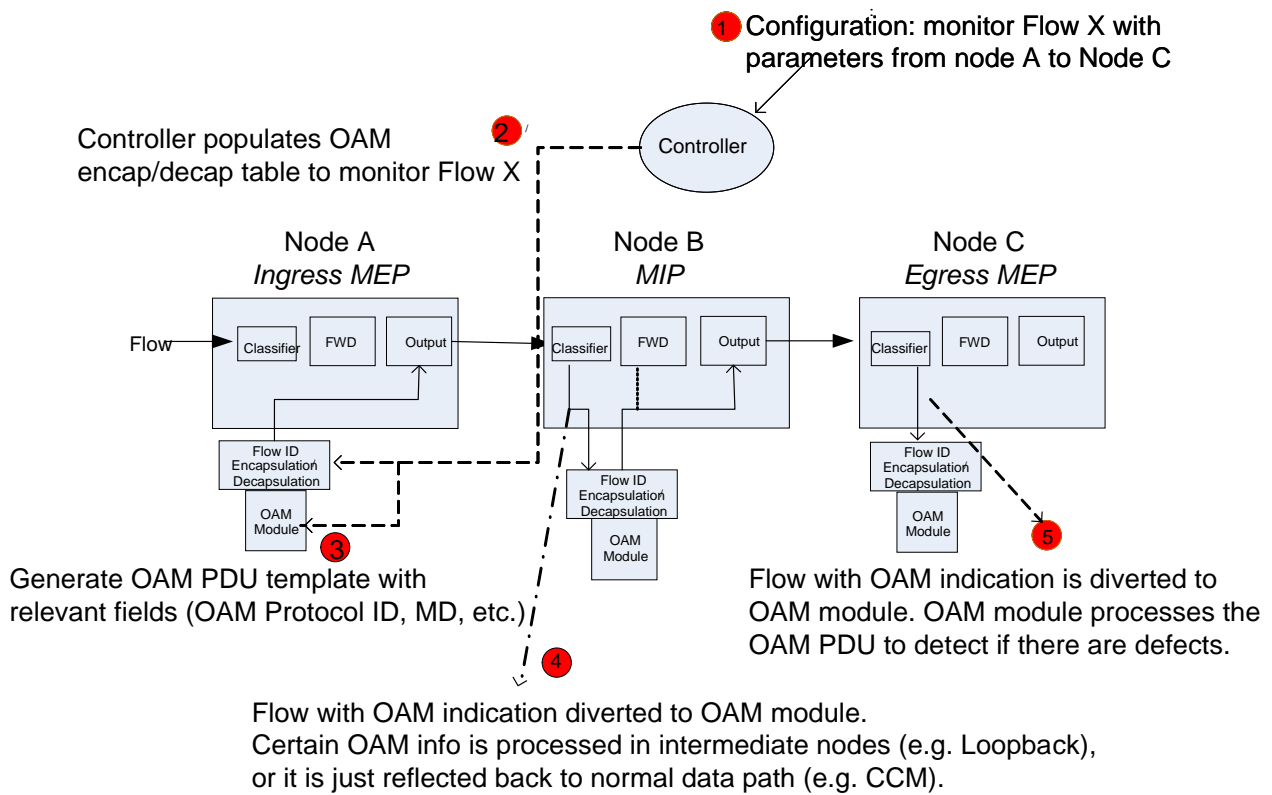

**Figure 33: Flow OAM Architecture**

Figure 33 contains a schematic overview of the proposed flow OAM, depicting the Flow ID module as well as the OAM module. To highlight the procedure of a typical OAM session including configuration, we list the necessary steps for an exemplary continuity check (CC) session:

1. The OAM configuration module in the controller receives a network management command to monitor traffic flow with certain parameters (Flow ID such as VLAN or MPLS label) from Node A to Node C. The OAM configuration module also manually or dynamically generates the corresponding MD (Maintenance Domain), MA (Maintenance Association), MEP and MIP.

2. The controller populates the Flow ID encapsulation/decapsulation tables of all nodes based on the traffic flow information, which is to be monitored; for example, if the traffic flow is MPLS, the OAM PDU must be encapsulated with the correct Flow ID (MPLS label).

3. The OAM module in Node A generates an OAM packet template and fills the fields of the OAM PDU, such as OAM Protocol ID, MD, MEP, etc. In this case, the OAM module will create a CCM OAM PDU. This OAM PDU is fed into the Flow ID module which encapsulates this OAM PDU with correct Flow IDs (e.g., Ethernet MAC, VLAN or MPLS label) based on the Flow ID encapsulation/decapsulation table.

4. When the intermediate Node B receives traffic flow from Node A, the traffic flow with OAM indication (e.g., OAM EtherType) will be guided to the Flow ID module. The Flow ID module will strip the Flow ID and send the OAM PDU to the OAM Module for further processing. In this case, OAM PDU is a CCM OAM PDU, thus the intermediate Node B will not process this OAM PDU. The OAM PDU will be sent back to the normal traffic flow path.

5. When MEP Node C receives traffic flow from Node B, the traffic flow will be matched against a default flow table which has a special flow entry to guide traffic flow with OAM Type (e.g., OAM EtherType) to the Flow ID module. The Flow ID module will strip the Flow ID and send the OAM PDU to the OAM Module for further processing. In this case, the OAM module recognizes it is the destination of this CCM OAM PDU. The OAM module finally processes the OAM PDU to detect if there is a problem.

This is a general Flow OAM mechanism for *SplitArchitecture*. In practice, we can reuse all the existing implementation in IEEE 802.1ag and Y.1731. For example, the OAM PDU can be based on Ethernet OAM, whereas the actual traffic can be MPLS or IP or Ethernet. However, it is necessary to define an identifier for OpenFlow OAM indication. We suggest reusing the Ethernet OAM EtherType 0x8902, but a new special Flow OAM type or other fields in the OpenFlow match structure might also be used. Furthermore, since OAM is not part of current OpenFlow implementations [19], the selected OAM identifier thus needs to be considered by the OpenFlow parser (i.e., classifier) within each datapath element. Besides configuration routines, this modification of the OpenFlow classifier is also the major extension to the OpenFlow specifications required in order to implement the proposed flow OAM mechanisms.

In order to support multi-service and/or multi-operator scenarios, the Flow ID module can be implemented as a service multiplex entity as in Figure 34. The Flow ID service multiplex entity has one General SAP (Service Access Point), and a number of multiplexed SAPs, each of them assigned to a single VLAN identifier, or a MPLS label, or a generic Flow Identifier depending on the configuration of the controller. Upon receiving an OAM packet from the General SAP, the Flow ID multiplex entity uses the identifier (VLAN ID, MPLS label, etc.) or a generic flow identifier to select one of its corresponding multiplexed SAPs to present the OAM PDU to the specific OAM module instance. Similarly, upon receiving an OAM PDU from a specific OAM module instance from the multiplexed SAP, the identifier associated with this SAP is added to the OAM PDU before going to the General SAP.

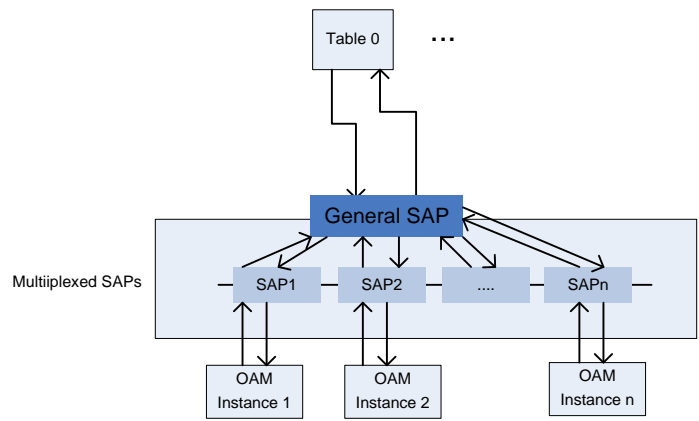

**Figure 34: Flow ID Module**

For technology-specific OAM requirements which an Ethernet OAM PDU cannot satisfy, we may define new PDU types to extend the functionality. Furthermore, it would also possible be possible to base OAM instances on MPLS BFD instead of Ethernet OAM.

The Flow OAM architecture in 5.3.4.1 assumes that an OAM indication is considered by the OpenFlow classifier, so that the OAM flow can be detected and diverted to the corresponding OAM instance in the common OAM module. In other words, the architecture proposed actually decouples OAM traffic from regular data traffic. For this reason, fate sharing between the OAM flow and the data flow, which is to be monitored, needs to be enforced. However, we have not yet given a detailed description of how to ensure fate sharing. We will therefore present two proposals of how to achieve fate sharing. One is to create a virtual data packet in datapath elements in order to test the forwarding engine. The other approach is to use the META data header in the internal flow processing of each node.

### 5.3.4.2     Fate sharing with virtual data packets

The idea is to test the forwarding engine with a virtual data packet created from the information carried in the payload of the actual OAM packet. The next hop for the OAM packet is selected based on the forwarding decision made for the virtual data packet. Before exiting at the output port, the internal OAM logic strips off the virtual header and puts back the OAM information into the payload of the original OAM packet.

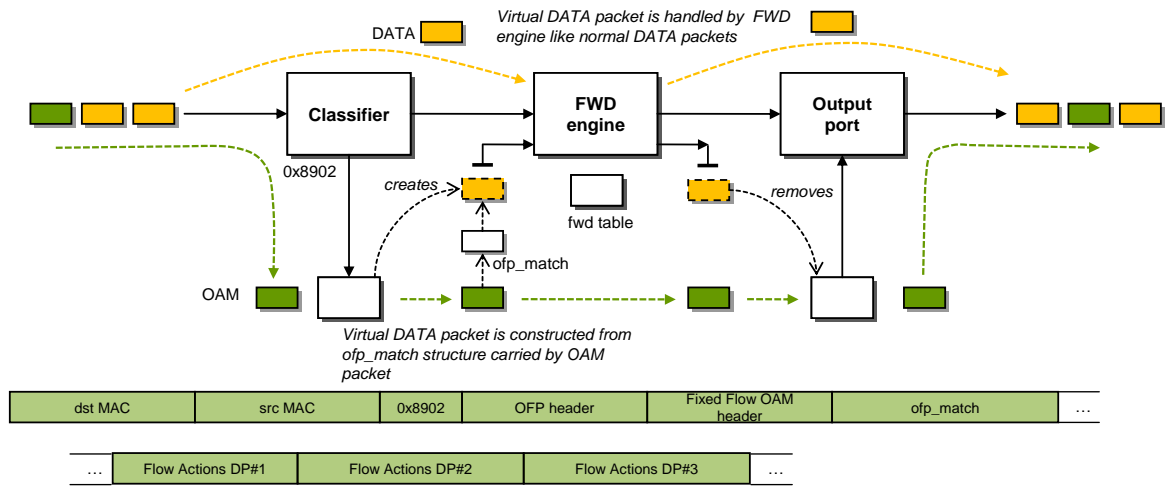

**Figure 35: Virtual Data Packet**

Virtual data packets enforce fate sharing, since OAM packets travel through the same links, the same node and the same filtering tables as the corresponding data traffic. We depict the process in Figure 35 and describe its details next: Figure 35

- OAM packets (highlighted in green) arrive at datapath elements interleaved with data packets (yellow) for the corresponding flow on a certain interface.

- The classifier, which has been updated to recognize OAM packets based on some ID (e.g., EtherType 0x8902 as used by Ethernet OAM) diverts the OAM packet to the Flow ID encap/decap module and the corresponding OAM module instance (simplified as a single OAM module box in the figure).

- The OAM packet payload carries information about the packet headers used in the corresponding data packets (e.g. in the form of an OpenFlow match structure). The OAM module uses this information in order to create a virtual data packet whose header matches those of regular data packets of the specific flow.

- The newly created virtual data packet is inserted into the forwarding engine like any regular data packet. Since it has an identical header as regular data packets, it will also match the same rules and actions as corresponding data packets from the monitored flow.

- The OAM module collects the virtual data packet before it is sent out at an output port. The header of the virtual data packet is adjusted in accordance with the actions performed by the OpenFlow forwarding engine. The virtual packet header is stripped off and the header information is stored in the OAM packets payload.

- The OAM packet is forwarded at an output port as chosen by the forwarding engine or the controller. Thus, the OAM packet is again interleaved with the data traffic of the monitored flow.

A flow OAM solution implementing the virtual data packet approach as depicted above would result in a single, generic OAM module in each datapath element, covering the entire flowspace as offered by OpenFlow, and at the same time would ensure fate sharing of OAM and data traffic. While this approach presents an initial concept for an OpenFlow OAM solution, we acknowledge that it has some limitations and is thus still the subject of ongoing discussions and future work. First of all, the separate handling of OAM packets at each datapath element could have undesired effects on OAM performance measurements (e.g., delay) or cause packet reordering. And second, this type of generic flow OAM works only within an OpenFlow domain, but does not provide compatibility with legacy domains (i.e., traditional OAM tools). However, legacy OAM needs to be taken into account in scenarios where an OpenFlow domain is adjacent to non-OpenFlow domains, as in the access/aggregation use case described in Deliverable D2.1.

#### 5.3.4.3        Fate sharing with metadata encapsulation

A different solution would be to add additional OpenFlow metadata of fixed length to each packet inside an OpenFlow domain. In order not to interfere with current standards, there are two options – either we add the additional metadata at the beginning of the packet or at the end of the packet. Since packets are of variable length, stripping bits off the packet may be easier at the beginning of the packet (i.e., before any protocol header). The OpenFlow metadata is always stripped by each OpenFlow datapath element for inspection by a module in the node, and added again to the packet on the line out. It is therefore not interfering with the switching in any way, but it enables additional management of flows without interfering with standardized protocols. OpenFlow switches which do not follow this OAM standard must perform the strip and add operation transparently (without processing the contents of the header).

Adding OpenFlow metadata to packets essentially corresponds to adding an extra encapsulation layer within OpenFlow domains, which would allow indication of special traffic (such as OAM or in-band controller traffic) without the need to interfere with the OpenFlow match structure. The disadvantages of this approach are that all switches must support the stripping operation. Furthermore, the OpenFlow domain would have a decreased effective MTU size. Advantages are simplicity and that all networked packets with the same headers (including OAM packets) follow the same path through the network, i.e., fate sharing is easy to ensure. Furthermore, all the processing of the metadata can be done in parallel to the switching operation.

For example, a 16-bit OpenFlow metadata field preceding packets in an OpenFlow domain could be specified as follows:

- Packet class (3 bits)
    - (000) Data packet
    - (001) OpenFlow control packet (e.g., for in-band OpenFlow)
    - (010) OAM packet
    - (011-111) currently undefined
- Protection class (3 bits)
    - (000) unprotected flow
    - (001) restored flow
    - (010) 1:1 protected flow
    - (011) 1:1 protected flow with restoration after dual failure
    - (100) 1+1 protected flow
    - (101) 1+1 protected flow with restoration after dual failure
    - (110-111) currently undefined
- Currently unused (9 bits)
- Parity check bit (1 bit)

## 5.4        Network Resiliency

Network resilience is the ability to provide and maintain an acceptable level of service in the presence of failures. Resilience is achieved in carrier-grade networks by first designing the network topology with failures in mind in order to provide alternate paths. The next step is adding the ability to detect failures and react to them using proper recovery mechanisms.

The resilience mechanisms that are currently used in carrier-grade networks are divided into two categories: reactive restoration and proactive protection. In the case of protection, redundant paths are preplanned and reserved before a failure occurs. Hence, when a failure occurs, no additional signaling is needed to establish the protected path and traffic can immediately be redirected. However, in the case of restoration, the recovery paths can be either preplanned or dynamically allocated, but the resources needed by the recovery paths are not reserved until a failure occurs. Thus, when a failure occurs, additional signaling is needed to establish the restoration path.

We divide this section into two parts: The first describes data plane failures managed by an out-of-band control plane; the second describes failure management when the failure is located in the control path.

### 5.4.1　　　　Data plane resiliency

The failure can be detected in OpenFlow by a Loss of Signal (LOS). This causes an OpenFlow port to change the state from *up* to *down*. The OpenFlow port is the port bounded to the OpenFlow instance to transmit and receive packets. This mechanism only detects link-local failures, for example, it says nothing about failures in forwarding engines on the path. As the link-local failures can be used to detect failures in restoration, the available detection method can be used in the restoration mechanism. However, since path protection requires an end-to-end failure detection, we require an additional method to detect those failures. As explained in Section 5.3.3, BFD can be used to detect end-to-end failures on a path in a network. The same protocol can be used to trigger protection in OpenFlow networks.

Fast restoration in OpenFlow networks can be implemented in the controller. It requires an immediate action from the controller after a notification of a change in a link status. Failure recovery can be performed by removing the existing flow entries affected by the failure and installing new entries in the affected switches as fast as possible following a failure notification. The restoration mechanism can be seen in Figure 36 A, which consists of the OpenFlow switches A, B, C, D and E. Assuming the controller knows the network topology, we can calculate a path from a source node to the destination node. In Figure 36 A, the controller first installs the path <ABC> by adding the flow entries in the switches A, B and C. Once the controller receives the failure notification message of the link BC, it calculates the new path, <ADEC>. For the OpenFlow switch A, as the flow in the flow entry for the working path <ABC> and restoration path <ADEC> is identical but the action is different for OpenFlow switch A (i.e., to forward to the switch B or D), the controller modifies the flow entry at A. In addition, for the restoration path <ADEC>, there are no flow entries installed in the nodes D, E, and C, so the controller adds these entries in the respective switches. The flow entry in C for the working path <ABC> and restoration path <ADEC> is likely different since the incoming port is assumed to be a part of the matching header in the flow entry. Once all the affected flows have been updated/installed in all the switches, the flow is recovered. After the immediate action of restoration, the controller can clean up the other nodes by deleting the flow entries at B and C related to the older path <ABC>.

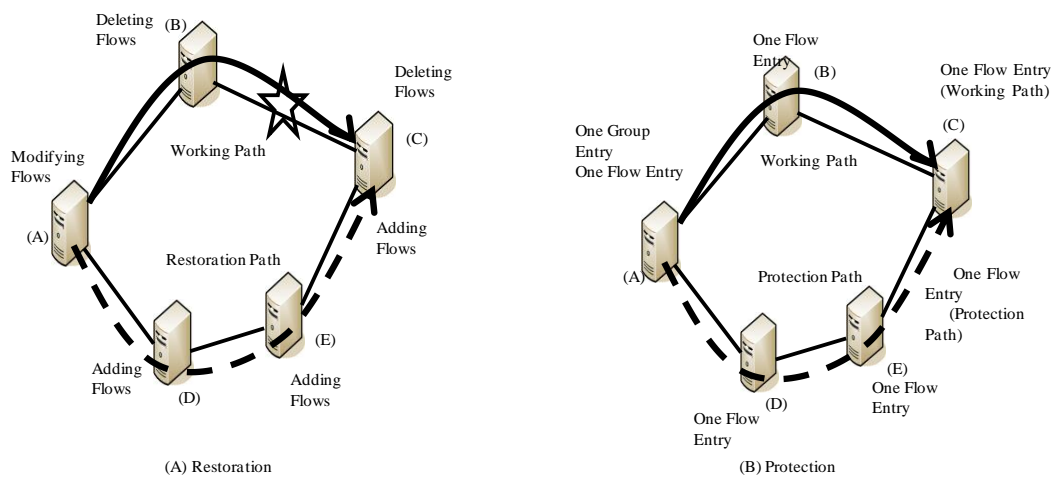

**Figure 36: Recovery mechanism for OpenFlow networks**

The total time to recover from failure using restoration depends on:

1. The amount of time it takes for a switch to detect a failure.
2. The propagation delay between the switch and the controller.
3. The time spent by the controller to calculate the new path.
4. The time spent by the controller to transmit the messages to modify/delete/add flow entries in the switches.
5. Time spent by the switches to modify/add the flow entries.

Our experiments carried out in WP5 Deliverable D5.2 have shown that the restoration time in such a recovery scenario depends on the number of flows to be restored via the alternative path. While the experiments showed that low-cost OpenFlow devices can restore traffic, its dependency on a centralized controller means that it will be hard to achieve 50 ms restoration in a large-scale carrier-grade network. In fact, the restoration time increases linearly with the number of flows, which indicates that the total recovery time is actually dominated by the flow specific messages and modifications listed above (i.e. list items 4. and 5.).

During the recovery time, i.e. the time between failure detection and complete restoration, packets may be lost. In order to reduce the packet loss resulting from a delay in executing the restoration action, we can switch over to the pre-established protection. Protection eliminates the need for the controller to update the datapath elements for modifying and adding new flow entries in order to establish an alternative path. This is accomplished by precomputing the protected path and provisioning it along with the working path. In this case, recovery is fast and the total recovery time is constant, i.e. it does not depend on the number of affected flows (as shown in WP5 Deliverabe D5.2). On the other hand, protection requires more overhead in terms of additional "idle" entries in the flow tables.

To implement a protection scheme, we can use the group table concept (specified for OpenFlow v1.1 ). Unlike a flow table, the group table consists of group entries which in turn contain a number of actions. To execute any specific entry in the group table, a flow entry forwards the packet to a group entry with a specific group ID. Each group entry consists of the group ID (which must be unique), a group type and a number of action buckets. An action bucket consists of an alive status (e.g., watch port and watch group in OpenFlow v1.1) and a set of actions that are to be executed if the associated alive status has a certain value. OpenFlow introduces the fast failover group type in order to perform fast failover without needing to involve the controller. This group type is important for our protection mechanism. Any group entry of this type consists of two or more action buckets with a well-defined order. A bucket is considered alive if its associated alive status is within a specific range (i.e., watch port or watch group is not equal to 0xffffffff). The first action bucket describes what to do with the packet under the normal condition. On the other hand, if this action bucket is declared as unavailable, for example due to a change in status of a bucket (i.e., 0xffffffff), the packet is treated according to a "next" bucket, until an available bucket is found. The status of the bucket can be changed by the OpenFlow switch by the monitored port going into "down state" or through other mechanisms, e.g., if a BFD session declared the bucket as unavailable. In our protection mechanism we used BFD to declare the bucket unavailable.

The protection mechanism for OpenFlow can be seen in Figure 36 B. When a packet arrives at the OpenFlow switch (A), the controller installs two disjoint paths in the OpenFlow network: one in <ABC> (working path) and the other one in <ADEC> (protected path). The OpenFlow switch (A) is the node that actually needs to take the switching action on the failure condition, i.e., to send the packet to B on the normal condition and to send the packet to D on the failure condition. For this particular flow of the packet, the group table concept can be applied at OpenFlow switch (A), which may contain two action buckets: one for output port B and the other for output port D. Thus, one entry can be added in the flow table of the OpenFlow switch (A) which points the packet to the above entry in the group table. Since the group entries do not require any flow information, it can be added proactively in the OpenFlow switch (A). For the other switches, B and C for the working path, and D, E and C for the protection path, only a normal flow entry can be added. Thus in our case, the switch in C contains two flow entries, one for the working path <ABC> and other for the protecting path <ADEC>. Once a failure is detected by BFD, the OpenFlow switch (A) can change the alive status of the group entry to make the specific bucket unavailable for the action. Thus, action related to the second bucket, i.e., whose output port is D, can be taken when the failure is detected in the working path. As the flow entries in D, E and C related to the <ADEC> path are already present; there is no need to establish a new path in these switches once the failure is detected.

### 5.4.2 Control channel resiliency

This subsection discusses recovery from a control channel failure, i.e., when the connection between the controller and the OpenFlow switches fails. Earlier, we considered the failure in data plane links, i.e., the links between the OpenFlow switches. However, because OpenFlow is a centralized architecture (relying on the controller to take action when a new flow is introduced in the network), reliability of the control plane is also an important issue. There are multiple options for control plane resiliency. One can provide two controllers, each on a separate control network, and when a connection to one controller is lost, the switch can switch over to the backup network. This is a very expensive solution. Control plane resiliency can also be obtained by having a redundant connection to the controller, where restoration and protection mechanisms can be applied in the out-of-band network. However, in some cases, it is difficult to provide redundant path for an out-of-band control network. Another option is to try to restore the connection to the controller by routing the control traffic over the data network, i.e., an in-band solution. When a switch loses connection to the OpenFlow controller, it can send its control traffic to a neighboring switch which forwards the traffic to the controller via its own control channel. This requires that the controller detects such messages and establishes flow entries for routing the control traffic through the neighboring switch. This solution is an intermediate step toward full in-band control. An effective scheme for control plane resiliency in carrier grade networks may be to implement out-of-band control until the failure occurs and switching to in-band control for switches that lose the controller connection after a failure. Thus, when the switch in out-of-band control network looses the connection, it can discover the controller via in-band control channel discovery, which is discussed in Section 5.5.1.

## 5.5 Control Channel Bootstrapping and Topology Discovery

Current OpenFlow specifications do not describe how initial address assignment and control channel setup are performed. In this section, we discuss a method that facilitates automatic bootstrapping of the control network for datapath elements. The bootstrapping procedure for newly connected datapath elements requires at least three configuration steps:

1. Establishment of a data control network which will implement the IP connectivity required by the OpenFlow protocols (i.e. OpenFlow and OF-config) between the datapath elements and the controllers.

2. Assignment of connection identifiers for connecting the datapath element to an OpenFlow controller (or alternatively to an OF-configuration point if the recent ONF model is considered). The connection identifiers required are at least the local address and the address of the OpenFlow controller. If non-default values are used, this may also include transport protocols and port numbers. Assuming an IP based control network, network address configuration can be done via DHCP.

3. Instantiation of an OpenFlow (or OF-config) session with the OpenFlow controller through the control network.

Once an OpenFlow or OF-config session is established, all further configuration and setup of the datapath element can be done remotely via the controller.

In the case of a dedicated out-of-band control network, the first step (1.) is automatically satisfied. When this network is implemented with "legacy" network control (e.g., spanning tree or IGP) the other two steps, namely the address auto-configuration (2.) and OpenFlow session establishment (3.) are realized with already standardized procedures and protocols. However, in the case of an in-band control network, the datapath elements need to be able to establish IP connectivity towards the network control in the absence of a node configuration protocol. In following subsections, we discuss bootstrapping of OpenFlow datapath elements in an in-band control network scenario. Furthermore, we will present our extensions to the topology discovery module that is implemented in the NOX controller.

### 5.5.1 Control-Channel Bootstrapping in an in-band OpenFlow network

In an in-band control network, the control-plane traffic is sent in the same communication channel used to transport the associated user data or management traffic. An example of the In-band OpenFlow topology is shown in Figure 37 where the control-plane traffic of the switches B or C passes the same connection as the data-plane traffic. The fundamental principle of in-band control in OpenFlow is that an OpenFlow switch must recognize the control traffic in the data plane traffic without involving the OpenFlow controller. In order to support in-band control, the Stanford reference switch implementation also includes an in-band control plane. In addition to in-band control, the reference switch implementation implements a discovery module in its local networking stack. This discovery module runs a DHCP client to configure the IP address of the LOCAL port and to discover the IP address of the controller.

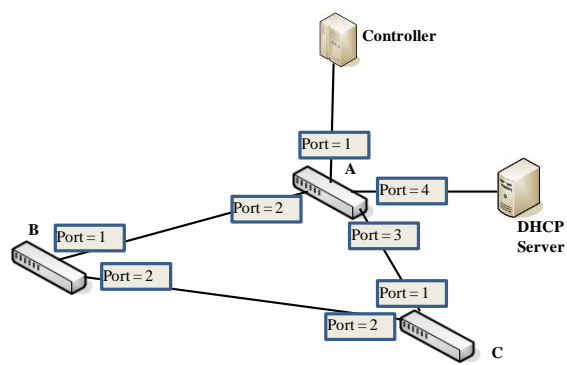

Figure 37: In-band Network topology

In-band control in a switch (with the reference 1.0 software) recognizes its own control messages but it does not recognize all the control messages from the other switches. In Figure 37, these messages are DHCP messages from the switches B and C, which are not recognized by in-band control of the switch A. Thus, when these messages reach to the switch A, these are forwarded to the controller for an action. Now if switch A connects to the controller via OpenFlow, the controller can reply the switch A for the action. We implement an application in the controller that can recognize these messages and can respond with an appropriate action. We call our implemented application as a bootstrapping application. In addition to this bootstrapping application, we modify in-band control of the switch such that in-band rules are applied to the packets only when the switch is not connected to the controller, using the reasoning that the controller should have complete control once it has established a connection with the controller.

We propose our solution because all the switches in the OpenFlow topology need a connection with the controller. The current solution in the OpenFlow reference implementation connects the switch A to the controller but it is not able to connect the switches B and C to the controller (in Figure 37). Section 5.5.1.1 describes working of in-band control in the OpenFlow reference software; Section 5.5.1.2 describes the controller connection to the switch A with the OpenFlow reference software; Section 5.5.1.3 describes our contribution to in-band control.

### 5.5.1.1　　In-band control module of the reference switch implementation

In-band control in the reference switch is an application on the OpenFlow switch. It is like a controller application that receives PACKET-IN event and replies with PACKET-OUT or FLOW-MOD messages. The OpenFlow switch communicates with this application before transmitting the packet to the controller. The packets are sent to the controller only when this application is not able to handle the packet. In-band control in the reference switch performs MAC learning on the source address. It handles the following packets when the switch does not have a matching flow-entry in the FlowTable.

(1) All the packets with the LOCAL port as the incoming port (in_port).

(2) All the packets with the destination MAC address (dl_dst) as the LOCAL port's MAC address.

(3) All the ARP packets from the controller i.e. the source MAC address (dl_src) as the controller's MAC address and destination MAC address as the broadcast address.

(4) All the packets that are sent to (or from) the controller. These are specifically TCP packets with dl_src as the controller MAC address and tp_src as 6633, or dl_dst as the controller's MAC address and tp_dst as 6633 (Controller TCP port = 6633). This case may happens when the controller traffic is to (or from) other switches. The switches own control traffic follows (1) or (2) case.

In the case (1) (above), in-band control floods the packet when the destination MAC address is the broadcast address. However, if the destination MAC address is not the broadcast address, it performs the MAC learning lookup. The performed action on packet is flood when MAC learning lookup does not know the output port for the destination. However, if it knows the output port then the flow-entry is added in the flow-table and packet is forwarded according to the flow-entry.

In the case (2), in-band control performs MAC learning on the source address and forwards the packet to the LOCAL port.

In the case (3), in-band control floods the packet.

In the cae (4), in-band control performs MAC learning and output port is decided by performing the MAC learning lookup. If the output port is found, the flow-entry is added, otherwise the packet is flooded.

### 5.5.1.2　　OpenFlow session establishment in the reference switch implementation

As explained above, the controller can establish a connection with the switch A in the topology shown in Figure 37. We describe the connection of the switch A with the controller in this section. The OpenFlow reference software implements a discovery module to establish a connection with the controller. This discovery module runs a DHCP client. The OpenFlow reference switch assumes that the DHCP server is present in the controller or it is connected to the one of the ports of the switch that directly connects the controller to the OpenFlow topology (e.g. shown in Figure 37).

The DHCP client and server interaction is shown in Figure 38. We describe exchange of messages between the DHCP client and the server together with in-band control processing in the switch A. In this description, the DHCP client runs in the switch A and the DHCP server is connected with the port 4 of the switch A (Figure 37).

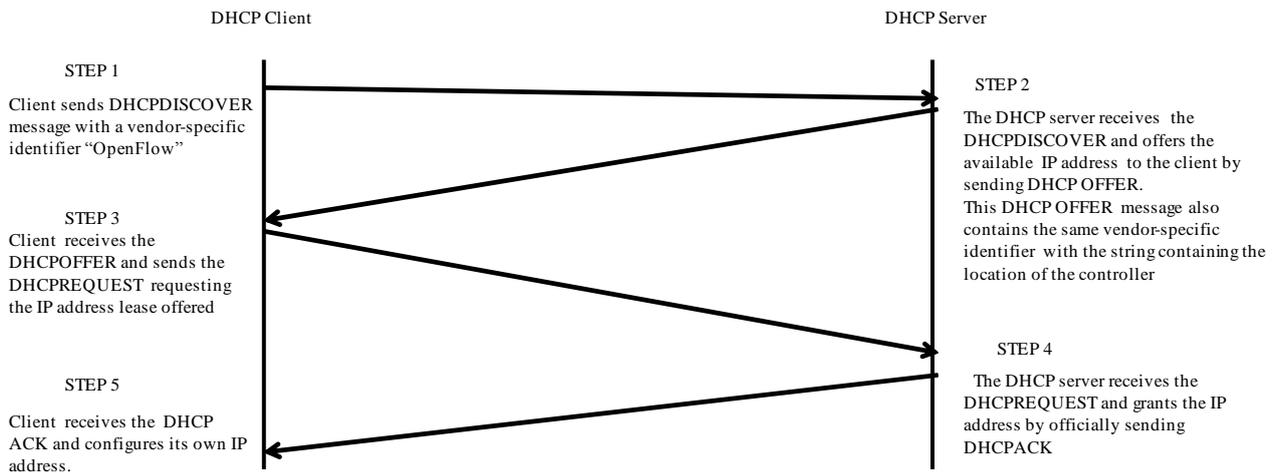

**Figure 38: DHCP client and server interaction**

In the first step, the DHCP client in A transmits the DHCPDISCOVER message on the LOCAL port, which contains a vendor specific identifier "OpenFlow" (STEP 1 in Figure 38). The DHCPDISCOVER message has the source MAC address as the MAC address of the LOCAL port of A, destination MAC address as the broadcast address, source UDP port as 68 and destination UDP port as 67. As the incoming port of the message is the LOCAL port (the case (1) of in-band control), in-band control handles this packet. Now, as the destination address of the DHCPDISCOVER message is the broadcast address, in-band control floods the message from each of the ports including Port 4 of A. The DHCP server receives this packet from the Port 4 of A and chooses the free IP address. Figure 39 shows the format of the configuration file (proposed by the OpenFlow reference implementation) of the DHCP server which chooses the IP address 192.168.1.20 through 192.168.1.30 to the DHCP clients that send the vendor specific identifier as "OpenFlow" in the DHCPDISCOVER message. In addition to the IP address for a DHCP client, the following configuration file also allows to send a string in the DHCPOFFER message containing the location of the controller i.e. tcp:192.31.1.1.

```
default-lease-time 600;

max-lease-time 7200;

option space openflow;

option openflow.controller-vconn code 1 = text;

class "OpenFlow" {

        match if option vendor-class-identifier = "OpenFlow";

        vendor-option-space openflow;

        option openflow.controller-vconn "tcp:192.31.1.1";

        option vendor-class-identifier "OpenFlow";

}

subnet 192.31.1.0 netmask 255.255.255.0 {

        pool {

                allow members of "OpenFlow";

                range 192.31.1.20 192.31.1.30;

        }

}
```

**Figure 39: Format of DHCP server configuration file**

Now the DHCP server sends the DHCPOFFER message to the DHCP client at the switch A (STEP 2 in Figure 38). The DHCPOFFER message has the source MAC address as the MAC address of the DHCP server, destination MAC address as the MAC address of the LOCAL port of A, UDP source port as 67 and UDP destination port as 68. The OpenFlow switch A receives this DHCPOFFER from the Port 4. Now as the destination address of the message is the LOCAL port of the switch A, in-band control handles this packet (due the case (2) of in-band control). Thereafter, in-band control transmits the packet to the LOCAL port. The LOCAL port handles this packet to the discovery module.

The discovery module parses the vendor specific identifier string (tcp:192.31.1.1) and sends the DHCPOFFER message to the DHCP client. The DHCP client receives this DHCPOFFER and sends the DHCPREQUEST to request the IP address offered (STEP 3 in Figure 38). Like the DHCPDISCOVER, the DHCPREQUEST message also contains the vendor specific identifier "OpenFlow". The DHCPREQUEST message has the source MAC address as the MAC address of the LOCAL port, destination MAC address as the broadcast address, source UDP port as 68 and destination UDP port as 67. As the incoming port of this DHCPREQUEST message is the LOCAL port, the message is handled by in-band control (case (1)). MAC learning in in-band control floods this message as the destination address of the DHCPREQUEST message is the broadcast address. The server receives the DHCPREQUEST from Port 4 and grants the IP address by sending the DHCPACK message (STEP 4 Figure 38). The DHCPACK message has the source MAC address as the MAC address of the DHCP server, destination MAC address as the MAC address of the LOCAL port of A, UDP source port as 67 and UDP destination port as 68. The OpenFlow switch A receives this DHCPACK from the Port 4. Now as the destination MAC address of the DHCPACK is the MAC address of the LOCAL port, in-band control forwards this message to the LOCAL port. The LOCAL port handles this to the discovery module which forwards this to the DHCP client and sets the IP address of the controller from the parsed vendor specific string (tcp:192.31.1.1). The DHCP client receives this message from the discovery module and configures the IP address of the LOCAL port (STEP 5 in Figure 38).

After STEP 5 of DHCP client/server interaction, the discovery module in the switch A transmits the ARP message to the controller (192.31.1.1) from the LOCAL port. The source MAC address of the ARP message is the MAC address of the LOCAL port. Now as the incoming port of the packet is the LOCAL port (case (1) of in-band control), in-band control handles this message. In-band control floods this packet as the destination MAC address of the packet is the broadcast address. In Figure 37, the Port 1 of the switch A connects to the controller. Thus, the packet reaches to the controller by the Port 1. The controller receives this message and sends the ARP reply. The OpenFlow switch A receives this ARP reply from the Port 1. Now, as the destination address of the ARP reply is the LOCAL port (the case (2) in in-band control), in-band control handles this packet. In-band control learns the MAC address of the controller and forwards the packet to the LOCAL port. The LOCAL port handles this to the discovery module which then performs the TCP 3 way handshake with the controller. Thus, now the OpenFlow switch A becomes connected with the controller.

### 5.5.1.3    SPARC contribution to in-band control channel bootstrapping

In parallel with the switch A in the OpenFlow topology shown in Figure 37, the DHCP client of the switch B and C also transmit the DHCP messages but these messages do not reach to the DHCP server. This is because when these messages reach to the switch A, in-band control of the switch A does not recognise this traffic (because these do not fall into case (1) to case (4) of in-band control). If switch A is not connected to the OpenFlow controller, these messages are dropped, otherwise, these are sent to the controller for the action. We implement an application in the controller so that the controller can decide the action of these messages. We call our application as a bootstrapping application. We also modify in-band control of the OpenFlow reference implementation such that a switch does not handle any control messages after it connects with the controller.

Our bootstrapping application works on the PACKET-IN and datapath-join event. It maintains the topology database (TD), a list of the datapath IDs that are connected to the controller (JOINED-IDs), tables that consist the MAC address of the LOCAL port verses datapath id (TABLE-1) and the IP address of the LOCAL port verses datapath id (TABLE-2), and a list (LIST) which contains JOINED and SENDER as two variables. The JOINED variable in LIST contains the MAC address of the LOCAL port of a switch such that if this switch is connected to the controller, the datapath id in the SENDER has to transmit LLDP packet from its ports. These LLDP packets are the probe packets which has the similar format as the LLDP packet. These are used to add or update links in the topology database. The topology database (TD) contains the ID of a node (OpenFlow switch, DHCP server and the controller) and the link between the nodes. The ID of the switches in our case is the MAC address of the LOCAL port. In the initial step, the bootstrapping application in the controller assigns the unique ID to the controller and the DHCP server, and initializes the topology database (TD) with the ID of the controller and the DHCP server, Other variables like JOINED-ID, TABLE-1, TABLE-2 and LIST are initialized with the NULL value, The pseudo code of our bootstrapping application on the packet-In event is shown in Figure 40.

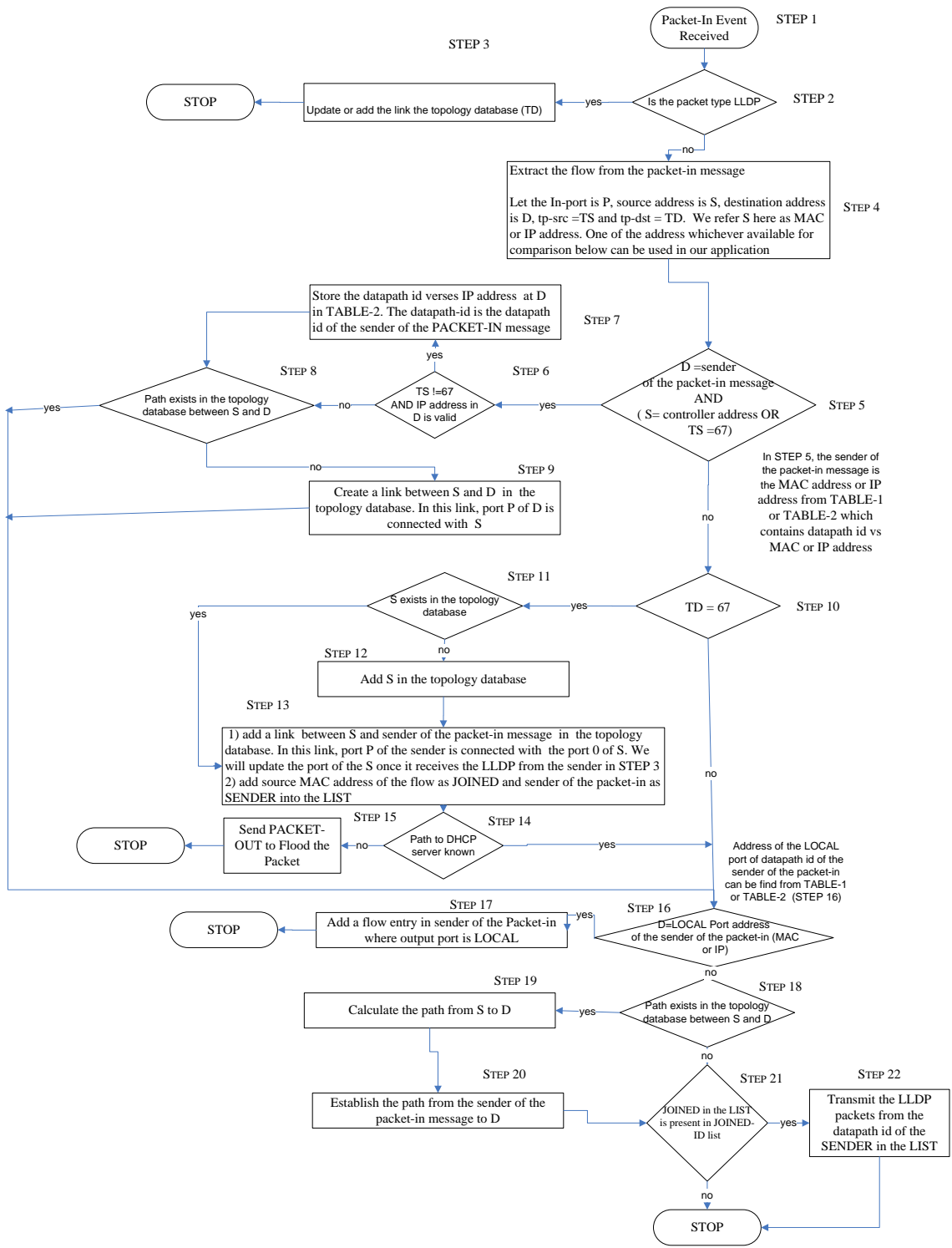

**Figure 40: Action of the bootstrapping application on the PACKET-IN event**

The action of the bootstrapping application on the datapath-join event is described in the Pseudo code written below:

In the datapath join event, the controller receives the FEATURE REPLY message from a switch. The FEATURE REPLY consists of the switch datapath ID and information about switch ports including LOCAL port. The information of the switch ports consists of the MAC address of the port which is important for our bootstrapping application. In the part of the datapath-event, our bootstrapping application adds the datapath ID into the JOINED-ID list, adds the datapath ID and the MAC address of the LOCAL port in the TABLE-1, and adds of the MAC address of the LOCAL port in the TD if it is not present. At this time, the bootstrapping application checks that if the MAC address of the LOCAL port is present in the JOINED variable of the LIST or not. If it is present then the bootstrapping application forwards the LLDP packet from the each of the ports of the switch whose datapath id is present in the SENDER variable of the LIST.

We explain working of above bootstrapping application together with the DHCP client and modified in-band control of the OpenFlow reference software. We take an example of the OpenFlow topology shown in Figure 37 to describe the bootstrapping of switches with our application. Our bootstrapping application recognises the DHCP messages. A message is recognised as a DHCP message if the UDP source or destination port of the message is 67.

The description of the controller connection with the OpenFlow switch A (Figure 37) in our modified in-band control is similar to the description with the unmodified in-band control (described above). Our bootstrapping application comes into picture after it receives the data-path join event of the switch A. In the part of the data-path join event of A, the bootstrapping application adds the datapath ID of A into the JOINED-ID list, adds the MAC address of the LOCAL port and datapath id of A in the TABLE-1, and adds the MAC address of the LOCAL port into TD.

When the switch is connected with the controller, the controller sends flow-remove message to the switch that triggers flushing of all the entries from the switch. As now the switch is connected and modified in-band control does not handle this message, the message is sent to the controller for the action. The controller receives this message as the PACKET-IN message (STEP 1 in Figure 40). As the packet type in the PACKET-IN is not the LLDP (STEP 2 in Figure 40), the flow is extracted from the packet-in message (STEP 4 in Figure 40). Let S and D are the source and destination MAC address of the flow. Now in our case, destination (D) of the packet is the sender of the packet-in message and source (S) is equal to the MAC address of the controller (STEP 5). The bootstrapping application adds the link between the controller and the switch A in its topology database (STEP 9) after following STEP 6, 7, 8 in Figure 40. Now, as the destination of the flow (STEP 16) is the LOCAL port of the switch A (sender of this packet in message) itself, the flow entry is added in the switch A where output port is LOCAL and thus now the acknowledgement of the flow-remove message is handled to the discovery module.

At present (when the switch A has established a connection to the controller), the switch B and C are in the initial phase of transmitting DHCPDISCOVER messages, because these messages do not reach to the DHCP server. The switch B and switch C continues transmitting DHCPDISCOVER messages until it does not receive the reply from the DHCP server.

Let us take the case when the switch A receives the DHCPDISCOVER message of the switch B from the Port 2 (after A has established a connection with the controller). The DHCPDISCOVER message of B has the source MAC address as the MAC address of the LOCAL port of B, destination MAC address as the broadcast address, source UDP port as 68 and destination UDP port as 67. The switch A sends this message to the controller as the PACKET-IN message. The controller receives the PACKET-IN message and generates the PACKET-IN event ((STEP 1 in Figure 40). As the packet is not LLDP one (STEP 2 in Figure 40), the flow is extracted from the packet-in message (STEP 4 in Figure 40). Now it follows the STEP 5 and reaches STEP 10 in Figure 40. As the destination port of the flow is 67, it reaches to STEP 11 in Figure 40. As S (switch B) is not present in TD (STEP 11), it is added in the topology database (TD) and the link is added in the topology database (TD) where port 2 of the switch A connects to the switch B (STEP 13). At this time, the bootstrapping application does not know the output port of the switch B that is connected to the port 2 of A. So, it adds the source MAC address of the flow in the JOINED variable of the LIST and datapath id of the sender of the PACKET-IN (switch A) into the SENDER variable of the LIST (STEP 13). Now as the bootstrapping application does not know the path to the DHCP server (STEP 14), it sends a packet-out message to the switch A to flood the DHCPDISCOVER message of switch B. Thereafter, the DHCPDISCOVER message reaches to the DHCP server via Port 4 of the switch A. The DHCP server receives this message and sends the DHCPOFFER to the switch B via the switch A. This DHCPOFFER message has the source MAC address as the MAC address of the DHCP server, destination MAC address as the MAC address of the LOCAL port of B, UDP source port as 67 and UDP destination port as 68. The switch A receives this packet and sends this to the controller as PACKET-IN message (STEP 1 in Figure 40) . Now the bootstrapping application reaches to the STEP 18 after following STEP 2, 4, 5, 10, 16. The destination the DHCPOFFER is the switch B. As there is a path to the switch B (LOCAL port of switch A) from the switch A in the TD, the controller calculates the path from A to B from TD (STEP 19), and it establishes the flow-entry in the switch A with the action Port 2 and forwards this DHCPOFFER to the switch B. Now switch B receives this and sends DHCPREQUEST.

After the exchange of the all DHCP messages and TCP-3 way handshake, the switch B connects to the controller. Thus the controller receives the datapath join event of the switch B. In the part of the data-path join event, the bootstrapping application adds the datapath ID of the switch B into the JOINED-ID list, adds the LOCAL port and datapath id in the TABLE-1. Now, as the MAC address of the LOCAL port of the switch B is present in the JOINED variable of LIST, the bootstrapping application forwards the LLDP packet from the each of the port of the datapath id in the SENDER variable of LIST ( i.e. switch A).  Now when the switch B receives this LLDP packet, it sends the LLDP packet to the controller as the PACKET-IN message (STEP 1). As the packet is LLDP (STEP 2), the bootstrapping application updates the link (in the TD) between switch B and switch A (STEP 3) with output port of B as 1. Thus, the bootstrapping also derives the topology of the network.

### 5.5.2　　SPARC Extension to the Topology Discovery Mechanism

Our topology discovery mechanism borrows the mechanism defined by the NOX original routing mechanism. The NOX routing mechanism implements three modules for routing a packet in an OpenFlow network. These modules are discovery, authentication and routing modules (Figure 41). The discovery module discovers links between the OpenFlow switches. It uses probe packets to discover the link which have similar as LLDP packet format. The authenticator module creates a Flow_in_event containing the source and destination access points, which the Routing module then listens for and uses to set up the flow's route through the network..

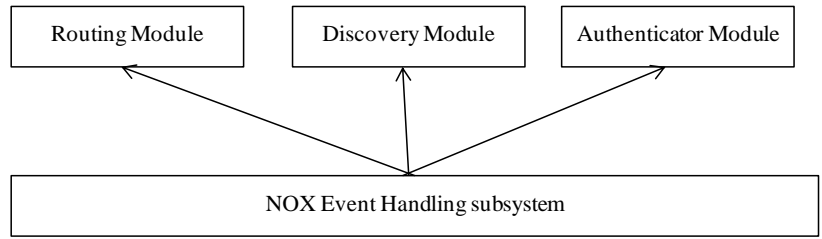

**Figure 41: The NOX Routing Mechanism**

The routing module discovers non-OpenFlow switches by MAC learning and route a packet to the destination non-OpenFlow switch. The MAC learning algorithm tracks source address of packets to discover the switches. It floods the packet if the destination is unknown. An OpenFlow network may have loops in its topology. Hence, the flooded packets may persist indefinitely or until TTL expire. Thus, MAC learning in the routing module may not function correctly since nodes may receive packets from multiple ports. The current solution in Ethernet networks to prevent loops is to draw a spanning tree and flood the packet around that spanning tree. We implement two algorithms to prevent loops in MAC learning. In the first algorithm, the controller performs MAC learning on each OpenFlow node and the packet is flooded along the spanning tree, and in the second algorithm, MAC learning is performed on the OpenFlow network and the packet is flooded outside of the OpenFlow network. Thus in first case, each OpenFlow node behaves like an Ethernet switch and in second case, an OpenFlow network controlled by a controller behaves like an Ethernet switch.

In the first algorithm where each OpenFlow node behaves like an Ethernet switch, the controller needs to draw a spanning tree of an OpenFlow network topology. The NOX discovery module learns the OpenFlow topology via the original topology discovery method (periodically sends out LLDP formatted probe packets to the node and waits for relaying back these packets by other switches). We used this topology to draw a spanning tree in the OpenFlow topology. We implemented Kruskal's Algorithm to draw a spanning tree. The sequence steps in our implementation are shown Figure 42A. The steps are shown after a packet-in event. The controller generates the packet-in event when it receives the packet-in message from the OpenFlow switch. If the packet contained in the packet-in message is not a LLDP packet (Figure 42A) and its destination is unknown, it is flooded along the spanning tree. On the other hand, if the packet is a LLDP packet then the NOX performs discovery. Furthermore, if the packet is not a LLDP packet and destination of the packet is known, the controller calculates the shortest path and establishes Flow Entries.

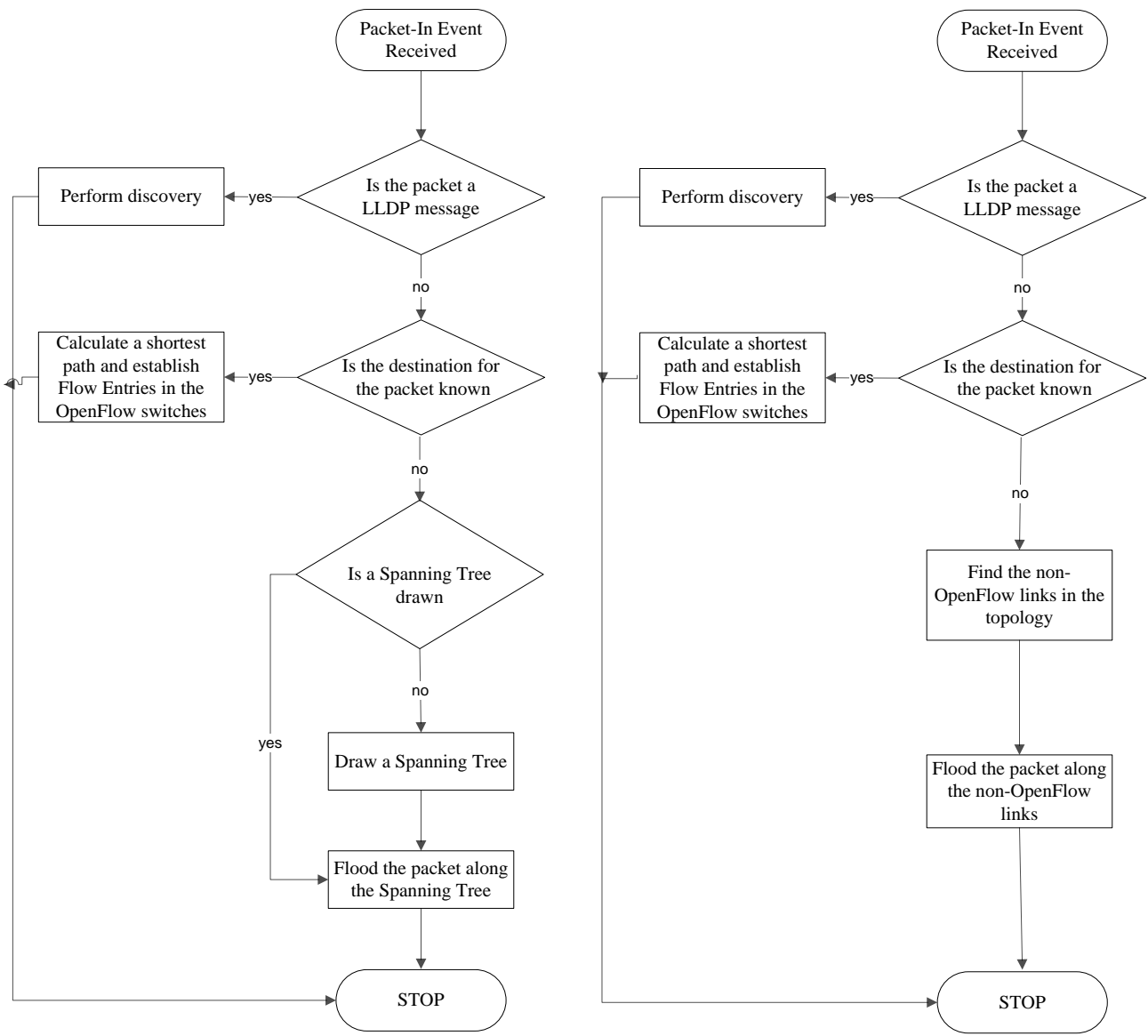

**Figure 42: A) NOX modified Mechanism (Spanning Tree solution)    (B) without Spanning Tree Creation**

In the second algorithm where an OpenFlow network behaves like an Ethernet switch, the controller needs to know the non-OpenFlow links to flood the packet outside of the OpenFlow network. The non-OpenFlow links are the links in the topology connected to the non-OpenFlow switches. In our implementation of this algorithm, reception or non-reception of a LLDP packet declares a link as an OpenFlow or a non-OpenFlow link. In the production network, there could be two situations (1) a non-OpenFlow switch does not transmit any LLDP packet (may be because it does not run its own LLDP protocol) (2) a non-OpenFlow switch transmits a LLDP packet to the OpenFlow network (may be because it runs its own LLDP protocol). We consider both the situations while declaring a link as a non-OpenFlow link. The controller in our implementation registers the MAC addresses of all the Ethernet interfaces of the OpenFlow switches when it receives the feature-reply message in the initial phase of connection. The controller declares a link as a non-OpenFlow link in one of the following situations (1) if it has not received any LLDP packet from the link (2) it has received the LLDP packet but the MAC address of the LLDP packet is not the registered MAC address of any of the Ethernet interface of the OpenFlow switches. The sequence of steps that are followed in this implementation is shown in Figure 42B. Like first algorithm, all the actions are taken upon packet-in event. The packet is first checked whether it is a LLDP packet or whether destination is known for the non-LLDP packet. Now if the destination is not known then the controller finds the non-OpenFlow links in the topology and floods the packet along the non-OpenFlow links (Figure 42B).

The OpenFlow switch can send a first few bytes of the packet or all the bytes of the packet to the controller in the packet-in message. It depends on configuration of the switch. In the second implemented algorithm, we transmitted all the bytes of the packet to the controller because the packet in packet-in message is flooded to the destination non-OpenFlow switch in this implementation

## 5.6        Service Creation

Service creation is a very general concept. In the telecommunication industry the concept describes service creation points (SCP) which are the points in the network where network functions with customer-specific or product-specific parameters needs to be configured. These are customer facing services, in network technologies often referred to as the service itself, and often referred to as a network service or a user session. For example, mobile networks provide adequate means for addressing various service slices via so called "Access Point Names" (APN) that tie a user session and its established PDP context (the tunnel actually connecting the user and the service creation gateway) to the service slice. Currently such a comprehensive architectural solution is lacking in fixed networks, e.g., in the architecture defined by the Broadband Forum for access aggregation domains. In this case a BRAS grants access to a user session which comprises only the default service slice (i.e., IP access) and does not allow differentiation between various service slices.

Besides the network service, there is also a transport function which typically aggregates multiple network services. One could say that network services are tunneled through transport services. For example, customers are using VLAN IDs to split between different services and in a certain forwarding engine; the VLAN IDs are mapped to MPLS labels in order to reduce configuration efforts since one only needs to configure a single transport service that can carry all the network services. Therefore, there might not be one-to-one mapping between a transport function, and the network services correlation with transport functions might not be given, or there may be a wish split transport and service (a service is a virtual connection between two points in this context, potentially stacked and transported within other virtual connections). Typical SCPs are located at the edge of the telecommunication network, for example the BRAS that is used to create residential services.

Service creation exists in various forms for different customer groups, e.g., an access service provided to residential customers is fundamentally different from the one provided to business customers. Typical examples for both cases are presented in the following section. Nonetheless, there are some requirements and process steps that are common to both cases and these will be explained first.

Another important aspect to be mentioned with regard to service creation is single point provisioning. Today, for certain protocols and their configurations, operators cannot cope with the growing complexity of the network. For example, the scalability of VLAN identifiers is limited to 4094 unique identifiers, but operators have many more users to organize in a typical network domain or segment. Technologies like Provider Bridge (Q-in-Q, IEEE 802.1ad) overcome this limitation, but require additional configuration efforts at the border of both technology variants. Therefore, the number of such provisioning points has to be minimized. This consequently requires good network design and/or qualified control plane protocols.

### 5.6.1        Service creation phases

Service creation denotes the process of connecting and granting access for a user and the creation of access to specific service slices, taking into account different limitations stemming from policy restrictions, which in turn are influenced by contractual constraints, etc. Service creation comprises several phases:

**Attachment phase**: A user must establish physical connectivity to the operator's network termination or authentication point. A network operator may adopt various options for establishing initial connectivity for the user, e.g., a legacy PPPoE/PPP session, a DHCP/IP-based solution, or an IPv6 link layer auto-configuration procedure. The network operator's policy will presumably demand prevention of user access to other service slices in this phase. Note that in this scenario there is a strict distinction between user identity and network attachment point. However, in existing solutions, this is not always the case, e.g., the line ID on a DSLAM access node identifies both the user ID and network attachment point. .

**Default session phase**: This phase is only applicable in cases where the user is not or could not be authenticated. e.g., the user does not have authentication information, but still requires network connectivity for emergency calls, resolving configuration/purchasing issues or to contact helpline services. Here special protocols or protocol configurations should be used such as PPPoE/PPP session establishment without authentication or DHCP-based IP configuration with limited access rights to a special service slice, e.g., a landing page or emergency VoIP system. If a user has sufficient authentication information, one should skip this phase and directly move to the authentication phase.

**Authentication phase**: Access to the user's service slices is granted based on proper user authentication, i.e., the network operator is able to identify the user and allocate a session for handling all management-related activities concerning service slice access. Again, a network operator should not be restricted in his selection of an authentication scheme and an authentication wrapper protocol for exchanging the necessary authentication PDUs – PPP, IEEE802.1x, PANA are some examples of authentication protocols that allow integration of various authentication schemes. Typically, this also includes some form of "binding" the user session between the network operator and user by deriving some keying material for message authentication and probably encryption.

**Authorization (and signaling) phase**: Once a user session has been established, various options for attaching the user to service slices exist: a) some form of dedicated signaling via a specific signaling protocol (resembling SIP-style signaling in mobile networks for accessing x-CSCF functions); b) some policy-driven automatic attachment based on user/operator contracts that correlates to the automatic attachment to the Internet service slice as done today in the existing access/aggregation architectures; c) some form of implicit signaling where the network performs some form of deep packet inspection in order to determine which is the appropriate service slice.

**Session phase**: User attachment is a complex process beyond the three phases discussed thus far. For attaching a user to a service slice, the management subsystem must ensure several constraints. Typically, a service slice consists of several service slice gateways and transport services between the user's network attachment point and the service slice gateway. It may be desirable to provide an adequate level of service quality as well. Furthermore, when IP connectivity is required for joining a service slice, compatibility of the addressing scheme adopted in the service slice and the addresses assigned during the initial attachment phase must be ensured. Today residential gateways (RGW) in fixed network environments are typically given a single IP address for accessing various service slices. A more advanced scheme may coordinate the assignment of different identifiers for different service slices on an RGW. The single IP address model also potentially stresses the service slice gateway with the need to do network address translation for providing services in a transparent manner, i.e., NAT helper applications may be required.

### 5.6.2 Relationship to requirements and OpenFlow 1.1

Overall, service creation relates to a multitude of the requirements detailed in D2.1. The list is as follows:

- R-1: A wide variety of services/service bundles should be supported.
- R-2: The Split Architecture should support multiple providers.
- R-3: The Split Architecture should allow sharing of a common infrastructure to enable multi-service or multi-provider operation.
- R-4: The Split Architecture should avoid interdependencies of administrative domains in a multi-provider scenario.
- R-8: The Split Architecture should support automatic transfer methods for distribution of customer profiles and policies in network devices.
- R-10: The Split Architecture should provide sufficient customer identification.
- R-21: The Split Architecture should monitor information required for management purposes.
- R-23: The Split Architecture should extract accounting information.
- R-29: It should be possible to define chains of processing functions to implement complex processing.
- R-30: The Split Architecture shall support deployment of legacy and future protocol/service-aware processing functions.

Obviously, R-1 – R-4 need to be covered by the service creation approach. Multiservice support is highly relevant both in current and future x-Play networks with business customer support. In multi-provider environments, it is important to distinguish between the various customers in relation to the appropriate provider. Here R-10 is demanded as well. In conjunction with customer identification, support for the configuration of customer-specific entities in edge devices (R-8) as service creation is required in telecommunication networks. In addition, R-21 and R-23 are relevant with the demand for management information in general and customer-specific terms. One of the most important requirements is the support of the deployment of legacy processing functions (like PPPoE or other tunneling protocols) as defined in R-30. In conjunction with PPPoE, for example, requirement R-29 is important as PPPoE is a mixture of different protocols and requires flexible decision logic (the complexity of PPP can be pointed out by 104 RFC's with "PPP" in the title).

Comparing the requirements in this section and the desired network functions from the previous section with the implemented OpenFlow switch specification version 1.1.0, it is rather difficult to outline specific missing features in the specification. However, the following existing features could be reused:

- Security mechanisms preventing unrestricted or authorized access could be implemented as a combination of default flow entries with appropriate counters and pointers to the authentication platform; counters must become more dynamic, allowing the submission of traps and notifications to network management systems.
- DDoS attacks could be prevented by appropriate counters for specific protocol requests like ARP requests; counters must become more dynamic allowing the submission of traps and notifications to network management systems.
- Spoofing could be prevented by appropriate flow entries after authentication.
- Accounting information collection mechanisms could be based on various counters.

- Management support is given to a limited extent with various "read state message" – here especially counters must become more dynamic allowing the submission of traps and notifications to network management systems.

- Authentication and related authorization of the datapath could be based on the setting of appropriate flow entries; however, the decision logic is currently beyond the scope of the OpenFlow specification.

- Forwarding to RADIUS entity, auto-configuration functions could be based on appropriate flow entries; however, the decision logic for these functions is currently beyond the scope of the OpenFlow specification.

- Configuration of profiles could be based on group tables; it is unclear how sophisticated the potential configuration options are.

- Limited support for IPv6 could be enabled by flow entries matching the IPv6 EtherType, but the specification lacks more advanced forwarding analytics.

Note that missing features do not have to be implemented in OpenFlow per se. However, it is necessary to point to functions and features which should be available to support service creation models. The essential missing features are:

- Modular decision logic for authentication/authorization and support for related protocols (depending on service creation models).

- Auto-configuration mechanism for customer services.

- Sophisticated management features like profile-based/policy-based network control and common protocol support like SNMP, potentially based on more dynamic counter mechanisms.

- Support for PPPoE matching and encapsulation/decapsulation.

- Support for IPv6 matching as well as modification actions.

### 5.6.3 Residential customer service creation (PPP and beyond)

For residential customers, service creation and single point provisioning requires a combination of different network functions like authentication, authorization and accounting, (auto-)configuration of layers 2 – 7, and security for the network infrastructure. In addition, it must be possible to trace and analyze problems and support today's current protocols. There are two typical protocol suites/models:

- The most common one is the Point-to-Point Protocol (PPP) in combination with RADIUS, in current deployments as a PPP over Ethernet (PPPoE) variant. PPP is a combination of a data plane encapsulation and protocol suite, which provides several functions like support for different data plane protocols, auto-configuration of layers 2 and 3, different authentication mechanisms and an interaction model with RADIUS. This interaction is implemented through a Broadband Remote Access Service (BRAS). The BRAS functionality is typically provided by a router and requires a high degree of processing and memory in the router compared to other operations. The BRAS is the mediation point between PPP and RADIUS. RADIUS provides AAA and configuration options as well as services between centralized user databases (like OSS and BSS systems) and the BRAS. The schematic diagram is illustrated in Figure 43.

- The second model is an adaptation of the Dynamic Host Configuration Protocol (DHCP) with several extensions, called DHCP++ in this deliverable. DHCP has its roots in LAN environments and provides auto-configuration support for layers 3-7. For carrier-grade support it requires additional protocols for AAA and security, and potentially auto-configuration for layer 2. Here a variety of options are discussed. Figure 43 only illustrates a model with AAA entry in the DSLAM.

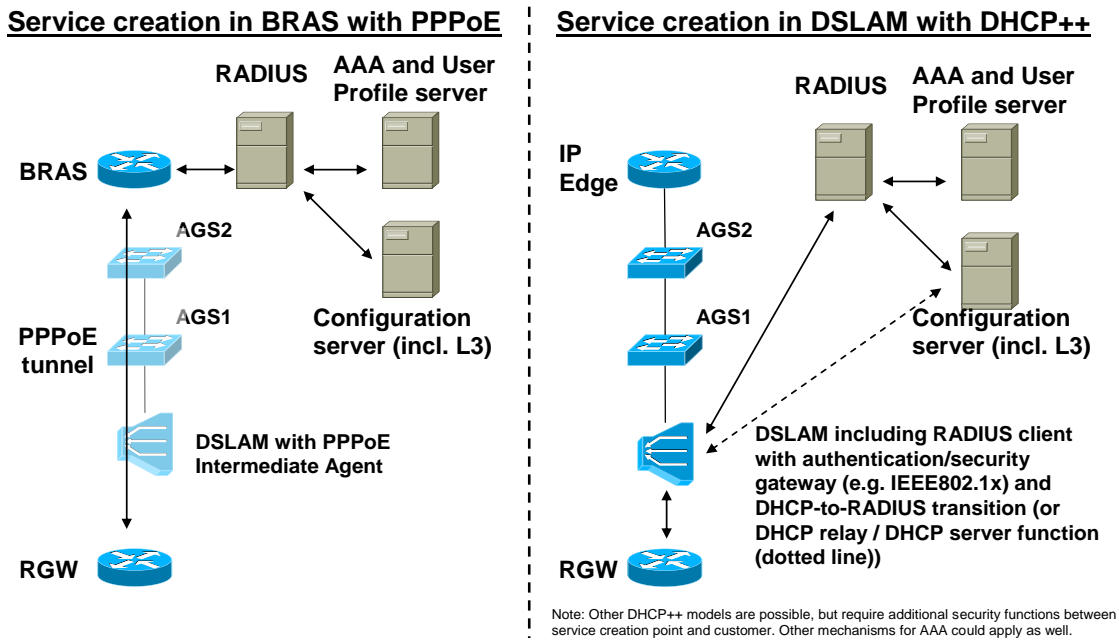

**Figure 43: Service creation models BRAS with PPPoE and DHCP++ based on DSLAM**

An important aspect is the transition from IPv4 to IPv6 in carrier networks. This aspect is not dedicated to service creation only, but it must be mentioned here explicitly. Several models are now available for the introduction of IPv6 in existing carrier environments like PPPv6, DHCPv6 or IPv6 route advertisements.

In summary, a service creation model has to cover the following aspects:

- Support for legacy protocols in general, PPPoE or DHCP-based IPv6 model as potential future production model.
- Support for common user databases like RADIUS.
- Security mechanisms
    - Authentication of users.
    - Authorization of users.
    - Security mechanisms preventing unrestricted or authorized access.
        - Discarding any data except authorization requests until successful authorization.
        - Restriction of access to resources associated to a user.
        - Protection of network infrastructure against any misbehavior like DDoS attacks (not necessarily with ill-effects) or spoofing.
- Collection of accounting information.
- Support for management.
- Auto-configuration of common L3-L7 functions like IP address, subnet mask, etc.

### 5.6.4 Business customer service creation based on MPLS pseudo-wires

Business customers typically require connectivity between different locations emulating a LAN environment over public telecommunication services. Service creation relies on the appropriate configuration of desired virtual networks. For example, a pseudo-wire is a generic point-to-point tunneling concept for emulating various services over packet-based networks. Several pseudo-wire-based services have been defined, e.g., TDM circuit emulation, ATM emulation, SONET/SDH emulation, and Ethernet emulation to provide business customers with sufficient options for interconnecting locations. Ethernet on MPLS pseudo-wires can be used to provide an emulated Ethernet connection also known as an E-Line service in MEF terminology. Typically, E-Line services are used to connect two branch offices at the Ethernet layer. Ethernet pseudo-wires are also often used to create Virtual Private LAN Services (VPLS) by establishing a mesh of MPLS LSPs between the provider edge routers and implementing an Ethernet to/from the LSP switch in these routers (similar to a normal Ethernet switch that handles the LSPs like ports) – in MEF terminology these are called E-LANs. MPLS-based VPLS services have become very popular, partly because of the possibility of performing extensive traffic engineering on MPLS LSPs, something that can be difficult with other packet-based networks.

There are several ways to create an E-Line/E-LAN service within an OpenFlow network. Since OpenFlow incorporates the Ethernet layer, one could imagine a network application running in the controller that simply installs Ethernet forwarding rules along a path in the OpenFlow network. However, since the MAC address space is flat, it would result in one flow table entry per OpenFlow switch for each MAC address that exists in the customers network, and that is something that does not scale very well. Additionally, there is a risk of MAC address collisions between E-Lines/E-LANs belonging to different customers (MAC addresses are not as unique as they should be, especially if they have been automatically generated and assigned to a virtual machine). Collisions between different E-Lines could be resolved through translation of the MAC addresses at the provider edge, however, this causes further complications in troubleshooting the network and might increase complexity if multiple controllers are involved, since this would have to be synchronized between the provider edge routers.

Pseudo-wire encapsulation resolves the scalability issues as well as the collision problem. Instead of requiring one flow entry per customer MAC address on all provider nodes, it requires two flow entries per E-Line service – one for each direction of the MPLS LSP carrying the pseudo-wire. Since the Ethernet frames are encapsulated, there is no risk of collisions – the customer MAC addresses are only "visible" at the provider edge nodes.

The overall pseudo-wire architecture is described in RFC3985, and the detailed specification for MPLS networks in RFC4385 and RFC4448. The first step when tunneling an Ethernet frame in MPLS PWE is to encapsulate the frame (source and destination MAC address, EtherType, and payload) with a PWE control word. The control word is used to provide various optional features, for example, it may contain a frame sequence number in order to enforce ordered delivery of the tunneled frames. This is optional when emulating Ethernet because Ethernet as such does not guarantee ordered delivery, therefore the control word can be left empty without violating the specification. Two MPLS labels are prepended on top of the control word, one for identifying the pseudo-wire tunnel and one for identifying the LSP. Finally, a new Ethernet header is added – this header does **not** contain any customer MAC addresses – instead it refers to the originating and next-hop provider router. Once the outer Ethernet header as been added, the frame is completed (see Figure 44).

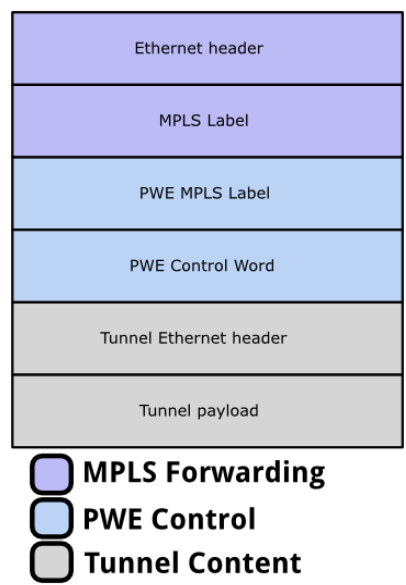

**Figure 44: Typical pseudo-wire frame structure for Ethernet emulation**

### 5.6.5　　　　　　　　Overall conclusions for service creation

This section summarizes the general missing aspects in the current OpenFlow specification for supporting service creation. They are:

- Detailed, complete service creation model based on OpenFlow for carrier environments.
- Sophisticated push operation of flow entries during bootstrap from controller to switch (access node, e.g., DSLAM) for initial protection of aggregation network.
- Profile-based/policy-based configuration possibilities are required.

- Potentially fine granular configuration of flow entries for protocols or user-specific authorized access to services and/or providers (the latter in the case of multi-provider forwarding).
- Legacy support, e.g., for PPPoE, MPLS Pseudo-Wire or other tunneling protocols, especially from business customer demands.
- General support for IPv6.

In Section 6.2 we will follow up on the issues raised here and present a detailed model for implementation of service creation for both residential and business customers in an OpenFlow-based *SplitArchitecture* environment.

## 5.7  Energy-Efficient Networking

The global concern about climate change on our planet is also influencing the ICT sector. Currently ICT accounts for 2 percent to 4 percent of carbon emissions worldwide. About 40 percent to 60 percent of these emissions can be attributed to energy consumption in the user phase, whereas the remainder originates in other life cycle phases (material extraction, production, transport, and end-of-life). By 2020 the share of ICT in worldwide carbon emissions is estimated to double in a "business as usual" scenario. Thus, an important goal for future networking is the reduction of its carbon footprint.

One way to gain higher energy efficiency in networking is the wider use of optical technologies, since optical signals consume less energy than electrical signals. Additionally, there has been an increase in research worldwide related to sustainable networking in recent years, with initiatives such as the current EU-funded Network of Excellence TREND [5], COST Action 804 [6], the GreenTouch consortium [7] and the CANARIE-funded GreenStar Network [8] which also has European members, including the SPARC members IBBT and Ericsson.

### 5.7.1  Current approaches to reducing network power consumption

#### 5.7.1.1  Network topology optimization

Different strategies to save power in networks are possible. At the highest level one can investigate whether *optimizations* are possible *in the network topology*. Currently networks are designed to handle peak loads. This means that when the loads are lower, there is overcapacity in the network. At night the traffic load can be only 25 percent to 50 percent of the load during the day. This lower load could allow for a more simplified network topology at night which in turn enables switching off certain links. Additionally, switching off these links allows for line cards to be switched off and thus leads to reduced node power consumption. A concept that implements this principle is MLTE (multilayer traffic engineering). The MLTE approach can lead to power savings of 50 percent during low load periods. However, many access networks are organized in tree structures, so shutting down links is not a feasible option. This means that dynamic topology optimization cannot be applied in all network scenarios, and is not feasible in access networks.

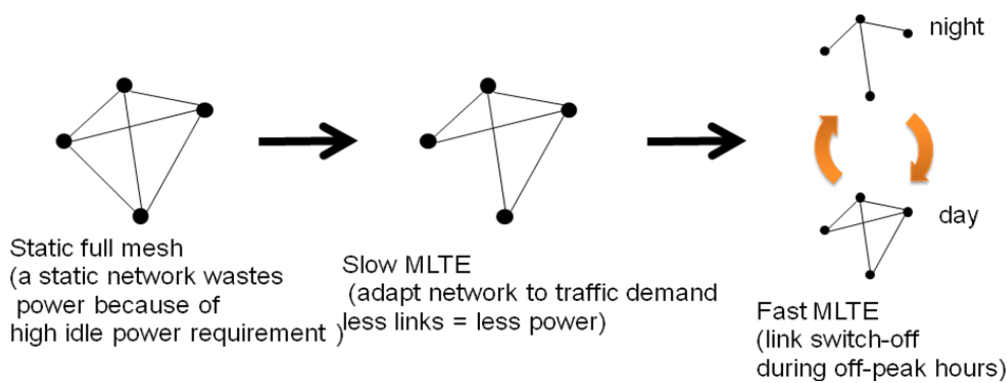

Figure 45: Multilayer Traffic Engineering (MLTE)

#### 5.7.1.2  Burst mode operation

Given a certain static network topology, further optimizations can be achieved by *burst mode* operation. In burst mode operation, packets are buffered in a network node and then sent over the link at the maximum rate. In between the bursts the line can be powered down. This strategy can be useful mainly in access networks due to the "burstiness" of the traffic. However, the difference in power consumption between different link rates is mainly manifested at the higher bit rates. Furthermore, burst mode operation works with very small time scales, so the number of components which can be switched off is limited. Finally, burst mode operation requires larger packet buffers which also need power. Hence it is yet unclear whether this strategy can lead to significant power optimization in reality.

#### 5.7.1.3 Adaptive link rate

Another approach for static network topologies is to use *adaptive link rates*. Adaptive link rate also exploits the "burstiness" of the traffic. The approach is based on the principle that lower link rates lead to lower power consumption in the network equipment. Saving energy is possible by estimating the average link rate required on a line and adapting the link rate to this level. However, similar to burst mode operation, adaptive link rate requires larger packet buffers, which might reduce some of the actual power savings.

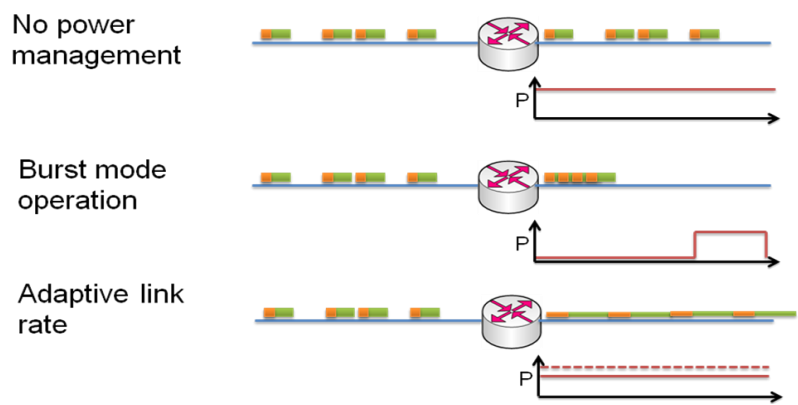

**Figure 46: Power Management**

### 5.7.2 Sustainable networking with OpenFlow

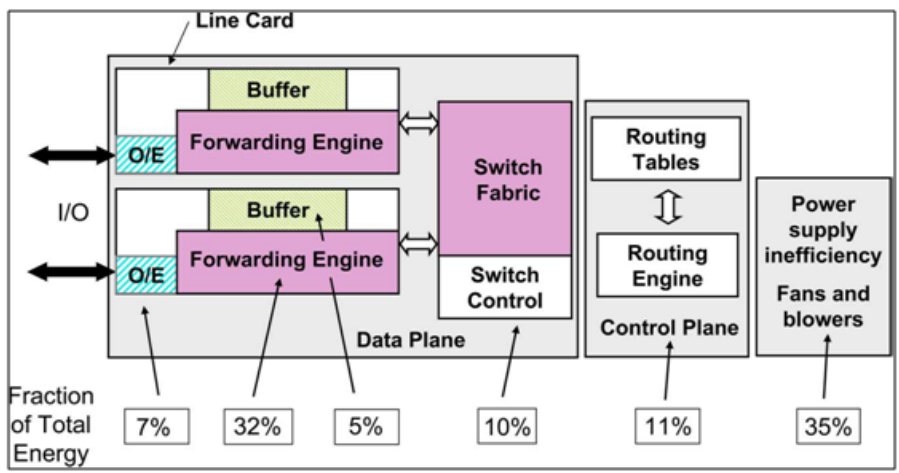

**Figure 47: Energy consumption across the functional blocks of a high-end core router [9]**

Centralizing the control software often means an automatic reduction in switch power consumption offset by consumption of a new element. As we can see from [9], most of the power consumption stems from the Forwarding Engine (32%) and the Cooling/Power Supply (35%). However, a rather small, but still significant amount (11%) of the power is consumed by the control plane. *SplitArchitecture* with OpenFlow enables us to reduce this part by moving the routing tables (RIB)/routing engine and control plane functionality to the controller and keeping only the forwarding engine (FIB) in the switch with a smaller OpenFlow software component and extra hardware for handling communication with the controller. However, the controller will consume more power due to additional computational loads. As a result, it is unclear if an OpenFlow architecture per se will actually reduce power consumption compared to conventional network architectures.

#### 5.7.2.1 OpenFlow protocol extensions for energy-efficient networking

The OpenFlow architecture enables us to implement *optimization of network topology*, e.g., MLTE operations as an application in the OpenFlow controller. In order to reap the maximum benefit, the controller should be able to power up/power down parts of the switch on demand as a function of the energy-efficient algorithms in the application. To enable power management for static network topologies as well, we need to add the *burst* and *adaptive link modes* in the switches and advertise them to the controller. On the controller side, it should be extended to allow control of such energy-efficient features.

This means that there are two sets of extensions designed within SPARC for energy efficiency:

- The first set of functions relates to port features: switching them on/off, enabling functions on the ports (Adaptive Line Rate / Burst Mode) and setting parameters for these functions (for instance, burst length for burst mode operation). We also need to disseminate these port capabilities to the controller.

- The second set relates to configuration and monitoring of components of the switch itself, which do not relate directly to forwarding, for instance internal power management and monitoring switch temperature.

To enable energy-efficient networking applications to run on the OpenFlow controller, there needs to be an interface towards the datapath elements to control the energy-efficiency related functions. Thus, we first proposed to add some extra messages to the OpenFlow specification which not only indicate the status of a port on the switch, but also allows us to control the individual ports. In D4.2 we will present details of our proposed extensions for dissemination of power management capabilities, for monitoring the related switch parameters and for controlling these capabilities.

However, with the recent addition of OF-config to the OpenFlow architecture (see Section 3.1.3) there is an additional interface available, dedicated for configuration tasks. Following the discussion in Section 4.2.6, it would make sense to include this configuration possibilites to the network management function in the control framework, since it helps to configure and steer the network in timely-fashion[6]. Due to the recency of OF-config, it is still unclear where the energy awareness features go in the updated ONF SDN architecture. As of yet, OF-config provides means for setting the following parameters for port configuration: no-receive, no-forward, no-packetin, admin-state. To support energy-efficient networking, the OF-config data model could be extended with a set of parameters allowing configuration of energy efficiency features, as listed above. The other extension set, relating to monitoring and management of the switch itself is also not unambiguously defined. Regarding additional switch capabilities, the OF-config specification readily support capability discovery, which could be extended in a straight-forward way to additionally discover capabilities designed for energy efficiency.

## 5.8 Quality of Service

To understand the QoS requirements, the different logical blocks of a 10 Gigabit Ethernet switch model with line cards and a backplane, is depicted in Figure 48 with the potential stress points highlighted. This model is from an IXIA white paper [34]. It addresses six typical stress points, which are designed to operate at line speed but under certain loads may become congested and cause increased latency, packet drops and other problems:

① Ingress packet buffer: The ingress packet buffer stores all received packets that are waiting to be processed. These buffers may overflow if upstream nodes are transmitting packets faster than the switch can processes them.

② Packet classification: The packet classifier uses the parsed header information incoming packets in order to classify them into different service classes. Depending on the design and the types of packets received, packet classification may not be able to operate fast enough. For example, complicated packets with multiple levels of headers (e.g., tunneled packets) may require multiple table lookups that increase the required amount of processing time per packet in order to classify them.

③ Traffic management: The traffic management modules are responsible for applying QoS through the mechanisms discussed below. During high load, these modules may be heavily stressed as queue management algorithms such as random early detection are activated in an attempt to reduce the load.

④ Crossbar switch and backplane interconnect: The crossbar switch and backplane interconnect are responsible for transferring packets between different connected line cards. Depending on the architecture of the particular switch, the backplane may cause blocking under certain traffic patterns. Advanced queuing and scheduling algorithms, combined with fast interconnects, can limit this problem or even completely resolve it.

⑤ Multicast replication and queues: Multicast replication is usually performed in two stages, one at the ingress line card in order to multicast the packet to different egress line cards, and another at the egress line card(s) in order to multicast the packet to multiple ports. Multicast packets compete for the same resources as the unicast packets and may be the cause of congestion both at the egress ports and when going through the crossbar and the backplane.

---

[6] Note that the extensions detailed in D4.2 were designed prior to the release of OF-config, thus only OpenFlow extensions have been considered so far.

⑥ Control plane: The rate at which the control plane is able to update the tables used by the switch, for example in forwarding tables, may cause problems in error conditions when a high number of changes needs to be performed.

As we can see, congestion can be caused by traffic consisting of complicated packets or multicast traffic. However, even simple unicast traffic may be the most obvious cause of congestion when multiple ports are trying to forward traffic through the same outgoing port and the combined packet rate/bit rate is higher than the line rate of the outgoing port. This may cause congestion first in the egress packet buffer, which in turn may cause blocking in the crossbar switch, which in turn may cause congestion in the incoming packet buffer.

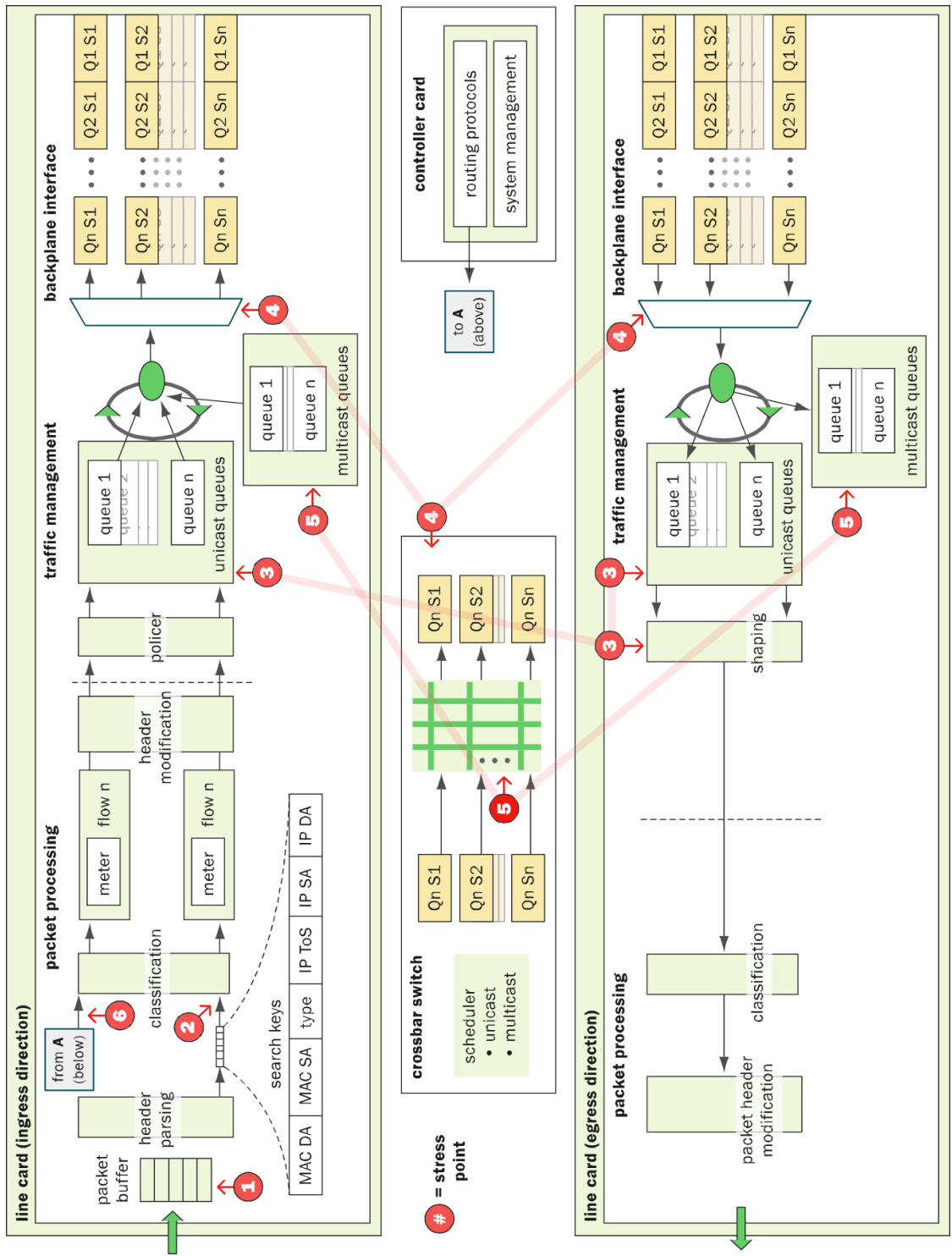

**Figure 48: Logical operation of a 10 Gigabit Ethernet switch with line cards, from [34].**

### 5.8.1 Queuing management, scheduling, traffic shaping

Typical quality of service (QoS) implementations in most routers and switches are constructed from a number of tools that are able to affect incoming packets in order to prioritize certain traffic, ensure proper bandwidth usage, smoothen traffic flows and reduce congestion in the network. Normally five tools are used: classification, traffic policing, traffic scheduling, traffic shaping, and packet rewriting. Figure 49 illustrates an example of QoS processing chaining; the order of the tools (or the tools used) in the figure are in no way canonical – different QoS goals may require different tools in a different order. Exactly which capabilities are needed by the different tools again depends on the requirements in a particular instance, as well as where they should be placed in the packet processing chain from incoming port to outgoing port. As shown in the figure, traditional devices may allow some QoS actions to take place before the packet enters the actual switching/routing stage, for example policing, or allow queue management actions on the incoming packet buffers. For a more detailed discussion of these subjects, see [35] and [36].

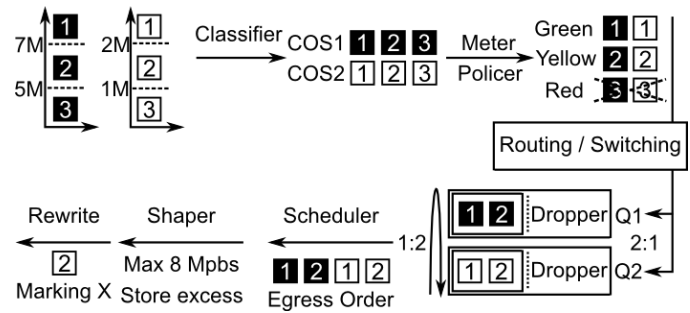

**Figure 49: QoS processing pipeline where packets from different flows are classified and metered before going through the routing/switching stage. The packets that were not dropped are then scheduled, shaped, and rewritten.**

**Classifier**

The classifier is responsible for recognizing different traffic flows based on a number of criteria, for example, Layer 2 and Layer 3 addresses, TCP/UDP port number, explicit packet marking (DSCP/802.1p), incoming physical port, etc. Classified packets are placed in a class of service that shares a common QoS treatment in the switch.

**Metering and coloring**

Traffic metering and coloring measures the packet rate and assigns colors to the incoming packets based on their arrival rate. Usually a dual-rate mechanism is employed: Packets not exceeding the first rate are colored green, packets exceeding the first rate but not the second are colored yellow, and finally packets that exceed both rates are colored red. Typical implementations may use token bucket-based algorithms such as the "two-rate three-color marker" [rfc2698] or "virtual scheduling" as used in ATM networks [37].

**Policing**

Traffic policing is used to enforce a maximum transmission rate based on the colors assigned to the packets. For example, red packets may be dropped immediately while yellow packets are candidates for dropping, whereas green packets should be allowed through without any restrictions.

**Shaping**

Traffic shaping is used to not only enforce a maximum transmission rate, but also to smoothen the traffic flow. This may be based on the same coloring used for policing, or new measurements may be made. Instead of dropping the offending packets, they are placed in a buffer until they may be transmitted without violating the defined packet rate or bit rate. However, if the shaping buffer fills up, packets may be dropped here as well.

**Scheduling**

Packet scheduling is responsible for selecting packets from queues representing different service classes and placing them in the output queue in a particular order. Many different scheduling/queue management algorithms exist, from a basic "round-robin" scheme to more complicated schemes such as "priority-based deficit-weighted round-robin" and congestion avoidance algorithms such as flow-based random early detection.

**Rewriting**

If one wishes to carry explicit service class or priority markers in the packets themselves (for use by other routers/switches), these are written to the packets by the packet rewriter which modifies the packets before transmission.

### 5.8.2 Improvement proposal

OpenFlow has basic support for some of the methods described above. Since OpenFlow is continuously evolving we will try to show the differences between the various versions (1.0, 1.1, and 1.3) and what has changed between them.

**Classification** can be performed by matching packets in the flow table(s).

In OpenFlow 1.0, multiple matching per packet is not possible since there is only one flow table and multiple passes through that table is not allowed. Thus it is not possible to decouple service class classification from flow classification (i.e., using one match for determining service class, and a different one for determining how to forward the packets). If one wishes to have a generic rule for a flow to perform forwarding but apply different QoS schemes, one needs to combine the forwarding and QoS rules, multiplying the number of flow table entries (e.g., with four different QoS service classes the number of entries becomes four times larger).

In OpenFlow 1.1 multiple matches are possible, so there is no need to combine the rules, and the number of entries may only increase with a constant number at the cost of an additional match. The metadata field available in OpenFlow 1.1 – which is associated with a packet while it traverses the processing pipeline – could be used to temporarily store the assigned service class for use in later matching stages if necessary.

The support for multiple lookups remains in OpenFlow 1.3, with the addition of "Extensible match support". Extensible matching replaces the previously static matching structure (i.e. which header fields could be matched and how matching was performed) with a more flexible TLV-based structure, that can be extended by vendors. This could potentially allow more advanced classification than was previously supported.

The major issue with classification is solved in OpenFlow 1.3 through multiple lookups and extensible matching for more advanced classification. However, OpenFlow still lacks a consistent way to color packets and change their behaviour (e.g. in queuing and scheduling) based on the this coloring. The metadata field is an ad-hoc way of carrying the color information but a more consistent way would be preferable, for example by adding a color field to the per-packet context data carried through the processing pipeline.

**Metering and coloring** is not available in ether OpenFlow versions 1.0 or 1.1. What is possible is to collect statistics on the number of matched packets and bytes per flow entry. However, these cannot affect how packets are processed within the switch but are purely for collecting statistics.

Our suggestion regarding improving metering and coloring would be to implement metering and coloring as a processing action that measures the packet rate or bit rate and writes the determined color to the metadata field. Following tables could then use the metadata to allow the packet to continue in the processing pipeline or to drop the packet. Since shaping uses similar mechanisms, it seems it should be as easy to implement as a processing action. However, since shaping requires buffering of packets that exceed the packet rate or bit rate, it becomes more complicated. This would require that the processing pipeline be capable of temporarily storing some packets somewhere in the pipeline, continue processing incoming packets, and then pick up processing of the stored packets mid-pipeline when the shaping mechanism allows it.

OpenFlow version 1.3 includes per-flow meter support. These meters are configured in a metering table and can be accessed from the flowtable on a per-flow basis. In the current state it support multiple bands per meter, meaning that different actions can be applied based on the packet rate. Currently only two actions are defined, drop and DSCP-remark. With the drop action packets can be dropped if they exceed the limit, making it possible to do **policing** with this type of meter band. DSCP-remark allows the meter to mark the IP header of a packet based on the current packet rate, which together with flexible matching could be used for decisions further along in the processing pipeline on the switch, or by other switches/routers in the network. This limited type of coloring is however only applicable to IP packets.

To improve the mechanism in 1.3 it would be natural to extend the existing metering framework to add support for more coloring methods that remark packets (e.g. using priority bits in Ethernet and CoS bits in MPLS). Additionally support for coloring that does not modify the packet would be useful, for example through the metadata field or a dedicated color field.

**Policing** before entering the flowtable is not supported by any of the OpenFlow versions, and only version 1.3 supports it within the flowtables with the mechanisms described for metering and coloring. However, this may leave the flowtables and associated processing vulnerable to congestion since it is not possible to limit the rate of packets entering the processing pipeline directly after packet classification has been performed (i.e. once the switch has parsed the packet, not classified in terms of QoS class of service assignment). One could imagine a solution to this where policing units (or meters) can be executed before the flowtables: However, these meters would need a similar matching structure to the one already available in the flowtables, which makes this solution would less elegant since we end up with a structure that is more or less identical to flowtables. However, in most situations where the flowtables are overwhelmed by a long complex processing pipeline it would be enough to do policing early in the pipeline and in that way reduce the load on the rest of it.

**Shaping** is supported by all OpenFlow version using the maximum-rate queues. Since packets can be implicitly marked with which queue they should go through it is possible to shape on both unicast and multicast flows, if somewhat awkwardly. It is not possible to shape a flow before its packets are sent to the port-queue, so for example shaping a flow and then spreading the packets over multiple outgoing ports is not possible; the shaping can only be performed after the packets have been sent to the ports. This makes it difficult to implement certain shaping schemes. For example, if one flow is spread over multiple ports using a Select group (which sends to one outgoing port based on a criterion, e.g. a hash of the packet header or by simple round robin) it is not possible to shape the actual flow. Rather, one is forced to shape each individual part of the flow that has been created by the Select group.

Shaping could be improved using the same mechanisms described for metering and coloring, e.g. by adding shaping support to the meter bands.

**Packet rewriting** is supported by all OpenFlow versions through built-in actions. OpenFlow version 1.0 supports modifying the Ethernet 802.1p priorities and IP ToS bits, this is extended in Version 1.1 by MPLS traffic class bits. In OpenFlow 1.3 the ToS bits of an IPv6 header can be modified as well. All these fields may also be used as matching targets in the flow table(s). Matching and modifying these fields makes it possible to interact with legacy systems on the QoS level, for example to utilize the QoS classification performed and written by legacy systems.

**Scheduling** has limited support in all versions, through queues that attach to a port. In OpenFlow 1.0 and 1.1 only a single type of queue is supported, namely the minimum guaranteed rate queue, which acts as a traffic shaper attached to an outgoing port. This queue allows to reserve a percentage of the outgoing bandwidth of a port, but it is not defined whether it should consider priorities, e.g., via the different packet marking mechanisms. This queuing concept could be extended with more advanced queuing mechanisms, supporting different kinds of queuing algorithms such as random early detection (RED). Additionally they could take into account not only the explicit packet markings (i.e., 802.1p, IP ToS or MPLS TC), but also use parts of the metadata field (only available from OpenFlow version 1.1) to allow packet prioritization without explicit marking. For example, the packet metering functions could implicitly modify bits 0-3 of the metadata, or these metadata bits could be modified by explicit metadata modification actions. OpenFlow version 1.3 introduces another type of queue, the maximum rate queue, which can be used to shape outgoing traffic.

One major issue with the current queuing mechanism is that the queues only attach to outgoing ports – it is not possible to chain queues with other queues. This makes it difficult to construct hierarchical queuing structures, something that highly simplifies construction of typical QoS schemes. Using hierarchical schemes one can easily design complex QoS setups like the one depicted on the left in Figure 50, where two different organizations share the bandwidth of a single link. Within their shares they have further subdivided the bandwidth between a number of services, some with low-latency requirements and some with less stringent latency requirements. The bandwidth values are the guaranteed values for these classes. However, if there is leftover bandwidth one level higher up in the hierarchy, this bandwidth may be used without violating any guarantees. For example, organization B may be using 75 Mbps of the total capacity if organization A is not currently utilizing all of its guaranteed bandwidth. Packets that do not match any of the defined QoS service classes may also utilize any leftover bandwidth, if no guarantees are broken. This type of sharing of the leftover bandwidth is difficult to manage in a fair way with a flat organization (on the right in Figure 50), whereas with a hierarchical organization it is easy to determine how the share of the leftover bandwidth should be distributed among the different service classes.

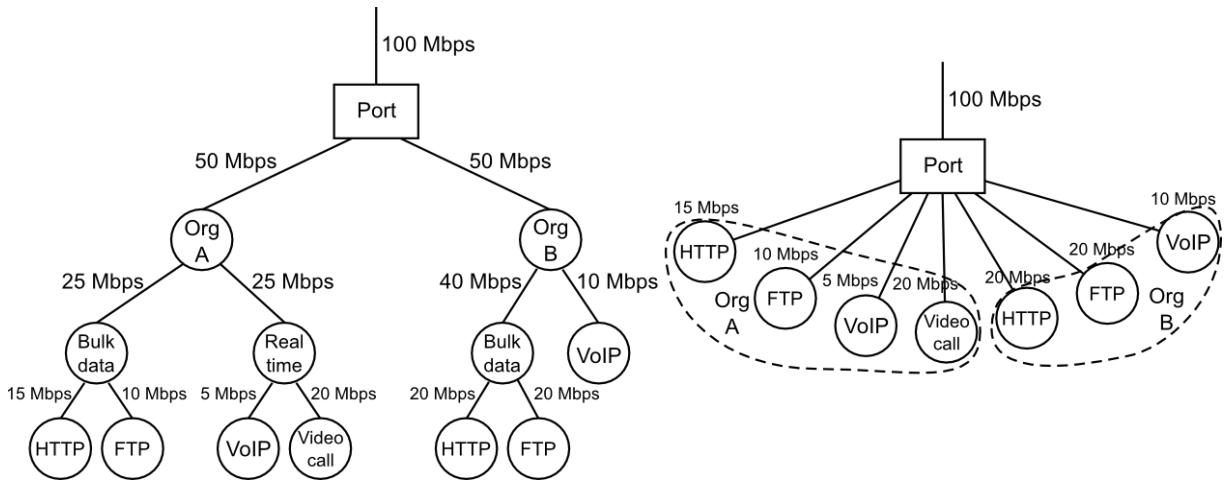

Figure 50: The hierarchical and the flat QoS model

A hierarchical QoS model would also fit very nicely in the virtualization concepts discussed in Section 5.2. In this case, the network operator could create and attach queues to the ports per virtual network. These queues would then be presented as physical ports to their respective virtual networks, which could attach their QoS classes directly to this queue (see Figure 52). This would not only ensure isolation between the different virtual networks and allows complex queuing setups within the virtual networks, but could also allow the creation of additional virtual networks within previously defined virtual networks (i.e., recursion).

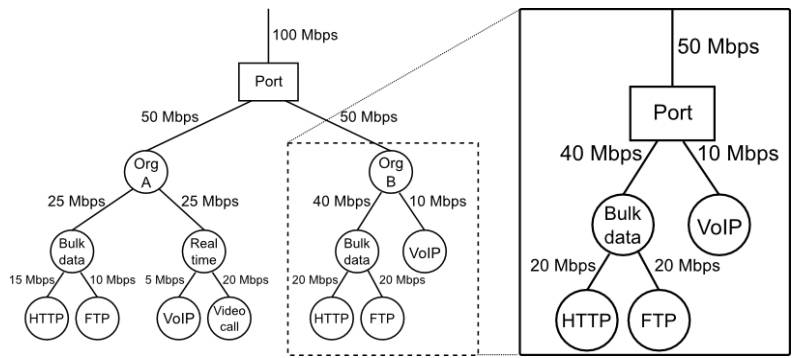

Figure 51: Representing a leaf of one QoS hierarchy as the root of another, virtualized one.

## 5.9  Multilayer Aspects: Packet-Optical Integration

The term multi-layer needs an up-front clarification: In GMPLS terminology (RFC5212), a *region* is defined by a switching technology. Multi-region nodes in exhibit at least two different interface switching capabilities (ISC) An ISC represents one out of six currently defined switching technologies for the interface: PSC (Packet Switch Capable), L2SC (Layer-2 Switch Capable), TDM capable, LSC (Lambda Switch Capable), FSC (Fiber Switch Capable), and DCSC (Data Channel Switch Capable). This section deals with the integration of optical networks and OpenFlow. Multiple layers may well be formed within a single switching technology, e.g., TDM, multiplexing fine-grained circuits into coarse-grained. The following considerations therefore consider multi-region nodes (which are always multi-layer, as well).

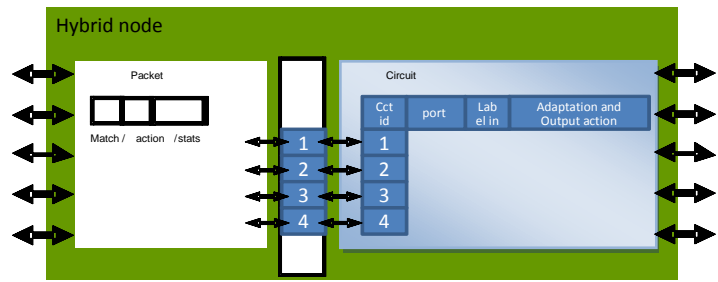

**Figure 52: GMPLS Multi-Region hybrid node, composed of a packet OpenFlow switch and a circuit (TDM or WDM) switch.**

Introduction of current generation optical networks into the domain controlled by OpenFlow is a tedious task because of the complexity of optical signal transmission:

- While an Ethernet **port** in a conventional OpenFlow switch will have little more status than being "UP" or "DOWN", WDM ports are characterized by the number of channels available, the nature of the transceivers (e.g., short/long range), available bit rate and in the near future even the size of the available spectrum. The transmission via Ethernet cable is assumed error-free (because the signal is terminated and regenerated at each switch). However, the **quality of the received signal** in an optical path varies as function of the length of the optical path, the number of hops without full regeneration, the number of active wavelengths, the modulation scheme, etc. This has the side effect that signal quality is attributed to an optical path (a flow) instead of to a port.

- While **actions** in OpenFlow are typically coded as packet header modifications, and output actions, which can be arbitrarily defined, optical crossconnects differ in the available switching capabilities. As an example, ROADMs can be fixed or colorless, directive or directionless, contention-prone or contention-less. These switching constraints are non-existent in Ethernet, and consequentially in OpenFlow switches. In addition, the number of transponders in a so-called "hybrid switch" (switch with non-homogeneous ports, see Section 4.2.1 of RFC5212) limits the throughput.

- While a **label** in OpenFlow can be directly matched on the packet headers, the label information for wavelength or OTN channels is indirect. Encoding for labels beyond packet transmission has been introduced along with GMPLS (RFC3471) and has been extended. All these labels, however, are only visible in the control plane and need to be mapped to a certain configuration of the switching matrix in an optical crossconnect.

### 5.9.1　　　　　GMPLS and OpenFlow

Most of the problems described above were tackled some five years ago in the IETF GMPLS working groups (ccamp) when defining GMPLS extensions for wavelength switched networks (WSON). There is no obvious reason to invent new encodings for ports, labels, adjustment capabilities and optical impairments [38].

| **Encodings** | **IETF GMPLS/WSON** | **OpenFlow 1.1.0 spec** |
|---|---|---|
| Port | RFC4202 (PSC,L2SC, TDM, LSC, FSC) RFC4206, RFC5212, RFC 6002 (DCSC) | Section A.2.1: `enum ofp_port_features` |
| Match (Flow label) | RFC 3471, RFC4606 (TDM), RFC6205(WDM), draft-farrkingel-ccamp-flexigrid-lambda-label-03 (FlexGrid) | Section A.2.3: `struct ofp_match` |
| Action (Adjustment capacity) | RFC 5212, RFC5339, RFC 6001 (IACD) | Section A3.6: Supported action bit field in `struct ofp_table_stats` |
| Flow stats (Impairment encoding) | RFC6566, draft-bernstein-wson-impairment-encode-01.txt | Section A3.6: `struct ofp_flow_stats` |

**Table 5: Options for the transport of encoded information in OpenFlow**

The question here is rather how and where these encodings can be introduced in OpenFlow. It turns out (see Table 5) that potentially all of the peculiarities of optical transmission can be architecturally covered by the existing OpenFlow standard 1.1.0. However, some of the mappings require a different understanding of the nature of a *flow* in optical networks. This is especially the case for the use of `ofp_flow_stats` for the encoding of optical impairments. In fact, it is the different nature of the optical flow that prohibits flow statistics relying on packet counts.

### 5.9.2    Virtual ports vs. adaptation actions

As discussed in Section 5.1.2, there would be a functional equivalence between virtual ports and potentially complex "processing actions" if the latter would be able to keep state information. For optical networks relying on circuit switching, one could create virtual ports that identify an established circuit, thereby hiding the optical details of this circuit and making the port appear again as a regular OpenFlow port.

All configuration of this virtual port would go through either the configuration protocol (OF-config) or OpenFlow itself. This would cover not only creating/updating/deleting the port, but also manage its internal behavior. The virtual port concept would split a multi-region hybrid node into two nodes (see Figure 52), a conventional (packet-switched) OpenFlow controlled part and another part that may or may not be under the control of OpenFlow.

Adaptation actions have the potential to pull the encapsulation/decapsulation into the domain controlled by OpenFlow, making the use of a separate configuration interface for virtual ports superfluous/obsolete.

Capabilities are bound to tables, starting from OpenFlow 1.1, which means that a specific encapsulation/decapsulation action like, e.g., the Generic Framing Procedure (GFP), could be called as an action on a flow. However, output actions are not yet port-specific, which means that a GFP-mapped TDM signal could end up in an Ethernet port by misconfiguration. It would be helpful to associate ports to tables for a clean configuration of hybrid nodes.

### 5.9.3    Three Levels of Integration

OpenFlow can be gradually introduced to control optical network equipment, leveraging on existing control plane implementations. This may be advantageous as it saves development cost and may as well protect IPR used in the Path Computation Elements [70] of vendors. The following three phases follow the implementation time line in the project OFELIA as presented in [39].

#### 5.9.3.1    Adaptation of Overlay model

One of the GMPLS architecture [69] considered routing models defines dedicated GMPLS control plane instances to the different transport technologies. These instances communicate with each other via a User-to-Network interface. In one possible implementation, a single abstract node represents the whole server domain. The optical transport plane takes care of the optical related attributes (ports, labels, actions) while the packet forwarding is configured via OpenFlow. This way the details of optical transport domainare hidden from the OpenFlow part and the optical domain appears as the backplane of a single Ethernet switch (Figure 53).

Operation is such that a `packet_in` message from one of the hybrid switches triggers the setup of a light path via the GMPLS UNI [71]. This means that the `flow_mod` entries that are generated by the OF controller for the hybrid switches are following the establishment of a light path between the transponders (appearing as virtual ports between the packet and the circuit switch part of the hybrid switches).

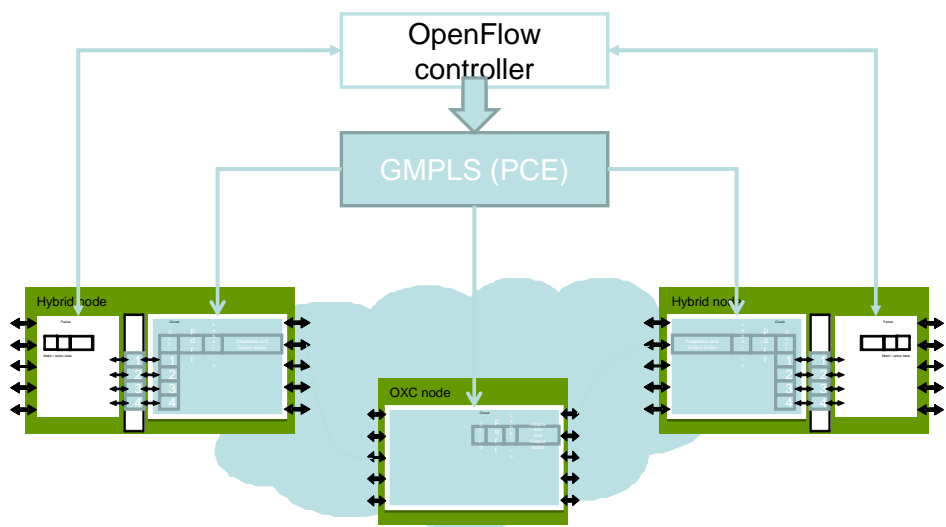

**Figure 53: Interworking of OpenFlow and GMPLS**

#### 5.9.3.2 Abstracted optical layer

A second step of integration is the OpenFlow control of optical nodes (Figure 54). This means that each network element has an agent translating the received OpenFlow messages into local (typically SNMP) control commands. A `flow_mod` from the controller is then used to configure an entry in the switching matrix.

Path computation will still be done in a GMPLS PCE, but all configuration of the network elements would now go through the OF controller.

While the optical routing is remaining in the GMPLS control plane with the PCE evaluating OSPF-TE messages from GMPLS nodes to create a view of the topology, OpenFlow replaces signaling. The controller requests an ERO (explicit routed object) from the PCE using PCEP, and then configures the nodes accordingly.

On the OpenFlow side this step will require encoding of label types and adaptation actions. Ports can still be considered abstract, as they only appear after a feasibility check

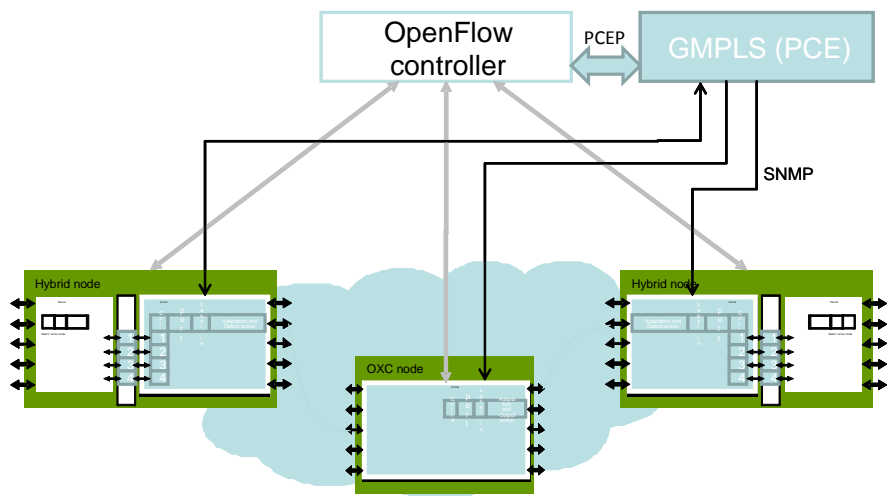

**Figure 54: Direct control of optical network elements by the OF controller. Path computation is still being done in a vendor's control plane.**

#### 5.9.3.3 Impairment-aware OpenFlow

The ultimate step integrating packet and optical transmission will be the mapping of the attached transport technologies (i.e., ISC) to flow tables. This will require association of ports to flow tables, the definition of impairment-annotated flow_stats and the path computation as an integral part of the multi-layer OpenFlow controller, as indicated in Figure 55.

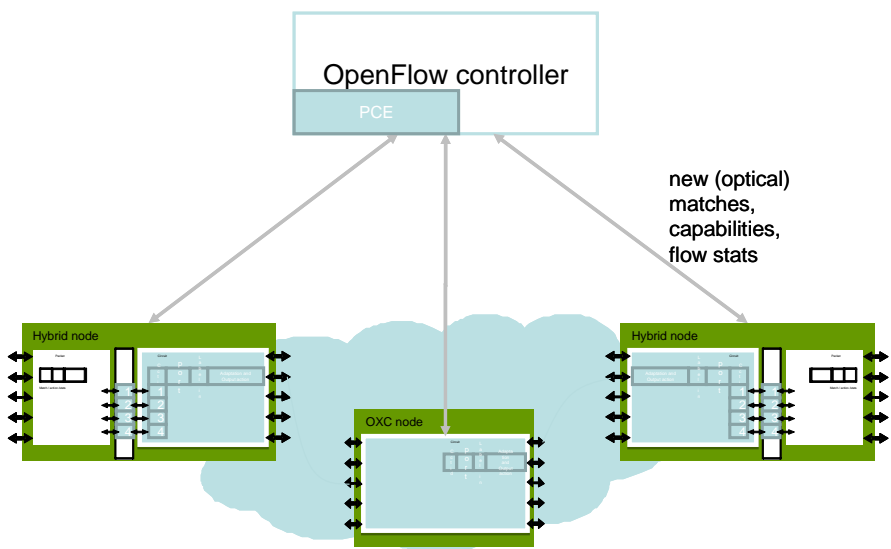

**Figure 55: Integration of a PCE into a multilayer OpenFlow controller**

Two main propositions exist in the literature on how to implement this step of full integrated circuit based transport networks to SDN. First we will briefly introduce the one proposed by the University of Stanford, and then we discuss the second solution, proposed by Ericsson.

**OpenFlow Circuit Switched Addendum by Stanford University:**

First, there is the OpenFlow Circuit Switched Addendum v.03 [52], which is an extension to OpenFlow 1.0. It consists of seven additions to the OpenFlow 1.0 specification. The Addendum proposes extensions to support certain transport layer technologies, in particular time division multiplexing (TDM). A basic circuit switched cross-connection table is defined inside the OpenFlow switch. This cross-connect table is to be kept separate from the usual OpenFlow packet flow table. The circuit switch flow table has four fields per input and output ports. These include the port, the lambda, the virtual port and the TDM signal and time-slots (starting time slot in the SONET/SDH encoding). Unfortunately, these extensions re-define from scratch the circuit resources used by the TDM and other transport technologies. This adds unnecessary complication to the OpenFlow protocol and compatibility issues with existing control planes (e.g. GMPLS).

**Ericsson's GMPLS-aware Multi-Layer/Multi-Region Extensions to OpenFlow:**

Second, Ericsson further enhanced the circuit switched addendum by reusing GMPLS encodings and the logical concept of GMPLS label switches paths (LSP) [53]. Compared to the former addendum, this solution removes the complexity of re-defining circuit and optical resource encodings and emulates LSP nesting features.

The Ericsson ML/MR proposal also contains a circuit flow table, which has however a different use than the current packet flow table; the former will only represent existing connections while the latter serves in a per packet lookup process. The fundamental difference between circuit switched and packet switched OpenFlow is therefore the fact that the circuit flow table is not used to lookup packets. The OpenFlow controller is responsible for setting up the circuits' cross-connections in the switch using the OpenFlow protocol and treating messages received from the switch regarding the current state of connections. The circuit cross-connections are established in a proactive way, i.e., no packet is forwarded to the controller for circuit flows. However, a packet sent to the controller can trigger the establishment of a new circuit cross-connect (e.g. pre-configured cross-connects, similar to virtual TE-links in GMPLS). The extensions proposed by Ericsson consider hybrid switches with both circuit based and packet based interfaces. This is not to be confused with the OpenFlow hybrid terminology defined by the ONF.

The Ericsson OpenFlow extensions partly rely on existing GMPLS features, specifically on GMPLS' way of provisioning new connections with the standardized label encodings. This implies that GMPLS routing function are taken over by the centralized OpenFlow controller, which furthermore contains traffic engineering (TE) and path computation (PCE) applications.

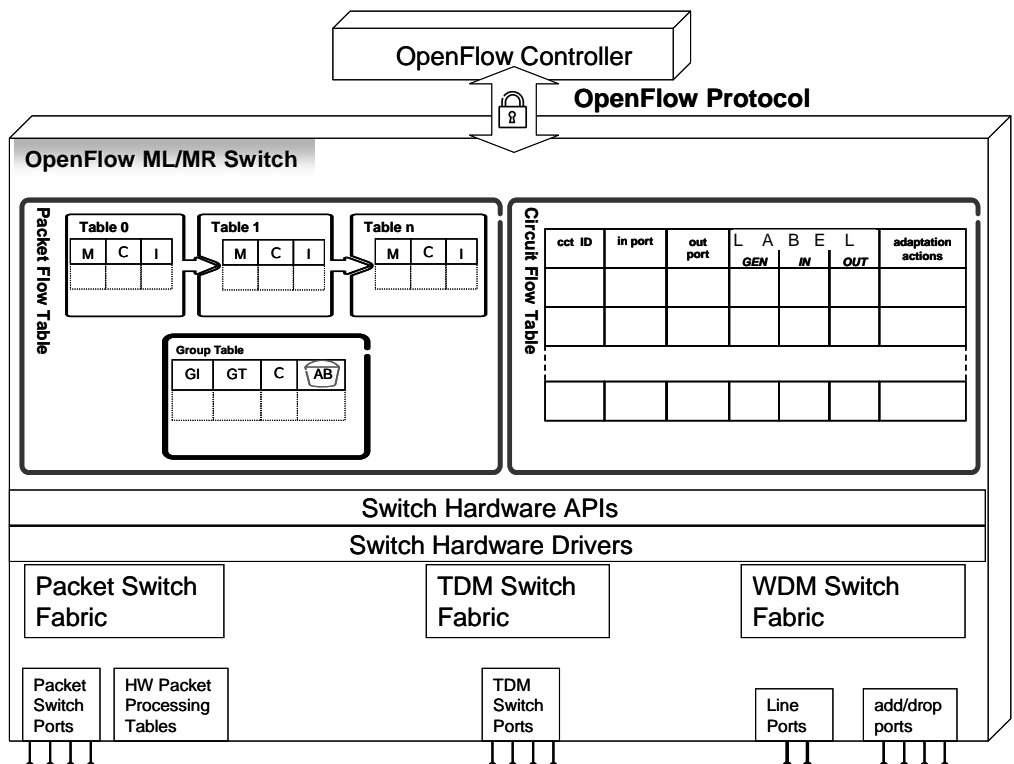

**Figure 56: Ericsson proposal for an OpenFlow multi-layer/multi-region switch architecture**

The proposed node architecture of an OpenFlow ML/MR switch is shown in Figure 56. Packet flow tables (left side of the figure) are consulted on the fly for each packet to determine its forwarding and required actions. Circuit flows (right side of the figure) represent existing physical circuits established by the switch. The circuit IDs serve as virtual ports to other flows. A circuit ID is a virtual port to which incoming packet flows can be forwarded. Other circuit flows can also point to a circuit ID and hence represent circuit hierarchies (the equivalent to GMPLS LSP nesting). The circuit flows do not affect the on the fly processing of packets. The proposed architecture is an extension to the OpenFlow 1.1 specification, the left side of Figure 56 is therefore left unmodified.

| CCT ID | in port | out port | Gen Label (encoding,ST,G-PID) | label in (e.g. TDM/WDM) | label out (e.g. TDM/WDM) | adaptation actions |
|--------|---------|----------|-------------------------------|-------------------------|--------------------------|--------------------|
|        |         |          |                               |                         |                          |                    |

**Figure 57: Circit flow table entry**

Figure 57 shows that each entry in the circuit flow table consists of a set of circuit identifiers and descriptive fields. Again, the circuit table is just an internal representation of existing cross-connects inside the switch and a new entry is added each time the controller signals the establishment of a new circuit. For the case of bidirectional circuits, the circuit will occupy two entries in the circuit flow table, as resources (In/Out Labels) may not be symmetrical in both directions. The flow table fields in Figure 57 are defined as follows:

- Circuit Identifier (CCT ID): a 32 bit unsigned integer represents the circuit flow and also corresponds to a virtual port to which other flows can be forwarded.

- In Port/Out Port: a 32 bit unsigned integer represents the incoming/outgoing port number between which the circuit cross-connects have been programmed.

- General Label (Signal Specification): a 32 bit unsigned integer represents the information required to fully characterize the cross-connect as part of an end-to-end circuit. It comprises of the following GMPLS specified attributes: encoding, switching type and payload identifier. The OpenFlow extension adopts the code point domains defined by IETF RFC 3473:

    o Encoding: an eight bit unsigned integer that designates the signal type of the connection within the transport technology class. For example, both SONET and OTN are denoted with TDM switching capabilities, but different encoding code points identify them.

    o Switch Type: an eight bit unsigned integer that designates the switching type used on the link. This is particularly important for hybrid switches, which has interfaces supporting more than one region.

    o G-PID: a sixteen bit unsigned integer that designates the payload of the client signal carried in the circuit.

- In Label/Out Label: a vector of 32 bit unsigned integer represents the incoming/outgoing label following GMPLS standardized labels per technology.

- Adaptation actions: the adaptation of the signal from the input towards the output port. This field can also be used in the future for specific technology related actions (e.g. related to optical technologies).

The establishment of a new circuit flow (and hence its addition to the circuit flow table) must carry enough information to allow the switch to program its cross-connections. To be able to signal the new circuit flow cross-connect, the controller first needs to know the features of the switch, its ports, and the available resources. The information stored in the circuit flow table comes from the controller and is sufficient for the switch to establish the circuit connection. To this end, the controller needs to keep an updated view of the switch's resources and state.

# 6      Implementing Carrier-Grade SplitArchitecture in an Operator Network

*SplitArchitecture* and SDN provide a significant degree of freedom to network planners and designers as they remove several constraints and boundaries found in legacy architectures. An operator network is typically structured in transport domains. Today the size and scope of such transport domains are fixed by manual network planning, as different datapath elements typically provide different functions. However, in *SplitArchitecture* the differences among datapath elements (e.g. switching and routing devices) start to disappear, as datapath elements are capable of providing both functions in parallel. This enables network operators to adapt a fixed transport domain and its boundaries based on dynamic conditions (e.g., load situation, etc.), of course within the limits defined by the physical deployment and wiring of datapath elements. We identified several types of integrating OpenFlow in carrier networks:

1. **Emulation of transport services**: As a first step, OpenFlow may be introduced in transport domains (e.g., Ethernet, MPLS, optics, etc.) by replacing legacy network devices with OpenFlow-compliant datapath elements and deploying a control plane that emulates behavior of the legacy transport technology in use, e.g., an Ethernet domain, an MPLS domain, etc. All nodes connected to such an OpenFlow enhanced transport domain still use legacy protocols for providing service and remain unaltered. OpenFlow in its versions 1.0 and 1.1 provide all the means to control Ethernet transport domains in such a scenario. However, support for enhanced Ethernet or MPLS services (e.g., those from the Metro Ethernet Forum), including OAM and reliability features, is beyond scope of OpenFlow 1.0/1.1.

2. **Enhanced emulation of transport services**: For a carrier-grade *SplitArchitecture*, a number of mandatory features and functions must be added to OpenFlow in order to fully comply with OAM requirements (among others), resiliency and scalability needs. OpenFlow lacks support for such advanced functions in versions 1.0 and 1.1 and must be extended accordingly to emulate carrier-grade transport services. Basic MPLS support was added to OpenFlow 1.1, but support (e.g., for MPLS-specific OAM schemes like BFD) is still lacking. We cover some of the necessary extensions to OpenFlow 1.0 and 1.1 in Section 5 of this deliverable, including OAM, advanced processing, interaction with legacy stacks, resiliency, and multilayer operation in OpenFlow. Again, all service nodes in this second integration scenario remain unaltered.

3. **Service node virtualization**: Thus far we have focused on ways of emulating legacy transport domains with OpenFlow. However, besides such basic transport services, carrier-grade operator networks provide a number of additional functional elements, e.g., for authentication and authorization, service creation, enforcing quality of service, etc. Most of these functions are today located on a limited set of network devices; the discussions in Deliverable D2.1 have documented the exposed position of the Broadband Router Access Service Gateway (BRAS) in carrier-grade operator networks according to the architecture defined by the Broadband Forum and deployed by most operators. A third integration level for a *SplitArchitecture* is virtualization of such service node functions inside OpenFlow. This involves control plane as well as datapath elements to cope with more advanced processing needs, as interfacing with more legacy protocol stacks must be supported. For the access/aggregation use case, we will showcase the virtualization of service nodes in OpenFlow in more detail in Section 3 based on an access/aggregation PPP/PPPoE UNI example.

4. **All-OpenFlow-Network**: Obviously, OpenFlow deployments may be pushed forward to other network domains as well, e.g., for controlling residential gateways (RGW) in customer premises networks or toward the operator's core domain. Controlling RGWs may simplify service slicing and service deployment in customer premises networks, but defines new constraints on an operator's control plane in a *SplitArchitecture*: Controlling customer-owned devices outside of the network operator's area of responsibility may impose additional security requirements. However, these security implications are beyond the scope of this deliverable.

Existing OpenFlow versions (1.x) suffice for integration of OpenFlow for pure emulation of transport services (type 1 listed above). Consequently, this deliverable is mainly targeting enhancements to the emulation of transport services (type 2) by extending the architecture and protocols as described in Sections 4 and 5. However, besides OpenFlow-based configuration of transport nodes, this deliverable also starts to look into service node virtualization (type 3), i.e. how service functionalities can be implemented through a centralized control architecture, discussed so far in Section 5.6. In the following sections, we will outline generic approaches of how to realize service creation in Access/aggregation network with OpenFlow. We will then present specific examples for residential and busienss service creation (BRAS, DHCP and PWE, respectively).

## 6.1 OpenFlow in Access/Aggregation Networks

One essential question in the design of the *SplitArchitecture* is how to integrate OpenFlow into the existing network architecture. There are three aspects to consider: The first aspect is dealing with the *level of the hierarchy* at which we introduce OpenFlow. The second aspect relates to the *number of hierarchy levels* controlled by OpenFlow. Finally, the third aspect focuses on *which functionalities* are configured through OpenFlow. Considering these aspects, we will present three evolutionary approaches for today's residential service creation as in Section 5.6.3.

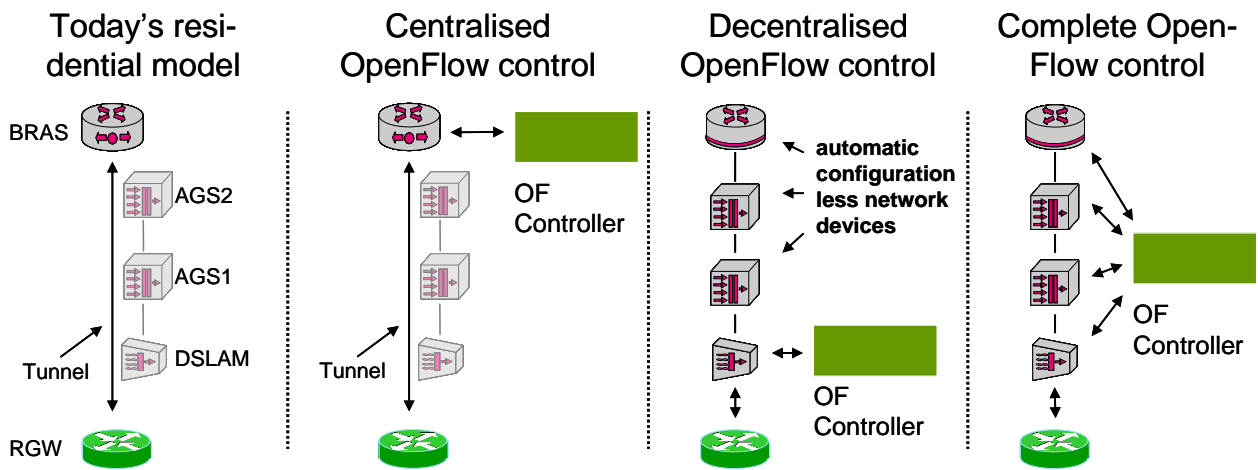

**Figure 58: Three models for attachment of OpenFlow in access/aggregation networks**

The "Centralized OpenFlow Control" model would be very similar to today's model with the exception that the OpenFlow controller would manage the central IP edge. This would result in a centralized network element in which potentially all customer traffic must be managed through OpenFlow with fine granular control – requiring powerful hardware. In addition, this model requires, similar to today's model, a mechanism to maintain routes to forward all packets from the RGW to the central element – the IP edge.

In the "Decentralized OpenFlow Control" model an OpenFlow controller would configure the DSLAMs. The difference to the previous model is that the DSLAMs control a significantly smaller amount of flows and bandwidth, but the OpenFlow controller needs to handle more connections to different DSLAMs than in the centralized model. The connection between the device and the OpenFlow controller, or between the OpenFlow controller and the network management system, requires additional network connectivity, which is implemented either with an out-of-band dedicated control plane network or multiplexed with the data links resulting in an in-band control network. Again, similar to the centralized model, the connection between DSLAM and the IP edge needs to be more automatic and the OpenFlow controller/subsystem needs to trigger the configuration of required transport connections.

In the "Complete OpenFlow Control" approach all devices, including the DSLAM, are controlled by an OpenFlow controller. Besides managing service configuration at the DSLAM, the controller is also responsible for provisioning transport functions of all forwarding devices along the network path to the IP edge. Note that, for performance and compliance reasons, mechanisms other than OpenFlow may be used for managing these transport functions, but how they are implemented depends on the capabilities of the OpenFlow controller

The decision favoring or rejecting a particular model is not part of the discussion in this section. Most likely there will be no clear general preference for a particular model in real network deployments, as this depends on several other aspects such as the size of the network, the availability of a hybrid OpenFlow switch model, which functionality needs be controlled, etc. The discussion of these models for the use of service creation continues in the following Section 6.2.

## 6.2 Implementing OpenFlow for Residential and Business Services in Carrier Environments

In Section 5.4 service creation was introduced and detailed for residential and business services. Aspects of OpenFlow and related implementations were presented. From a technical perspective, the level of detail was not sufficient. Thus a detailed proposal for residential and business customers is presented here.

### 6.2.1　　　　Residential customer service with OpenFlow

As described in Section 5.4, there are several solutions available with varying implementation options, see Figure 59 below for a general overview.

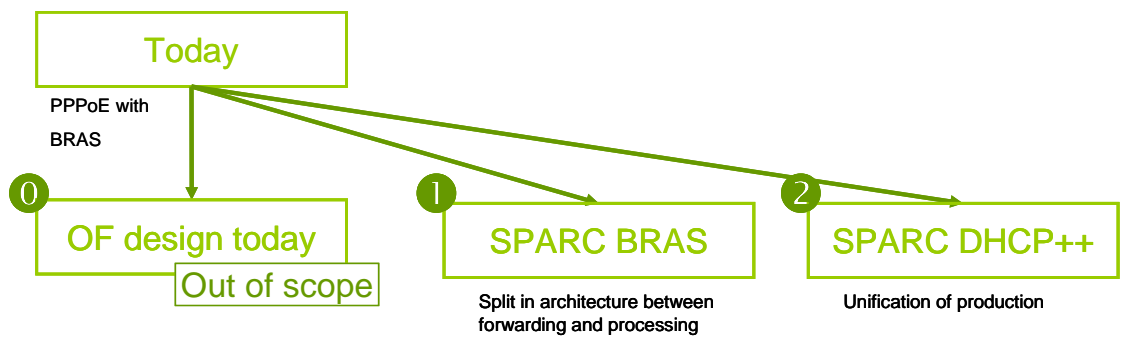

**Figure 59: General implementation options for service creation**

Option 0, "OF design today," follows the design rules and principles of the OpenFlow development done up to Version 1.1 – add any new protocol structure to the base protocol. As already discussed, this could result in an "explosion" of required protocol options, and hardware support might not be implemented in any case. In addition, the split between forwarding and processing could not be integrated. Therefore, this solution is considered out of scope.

Option 1, "SPARC BRAS," transforms current service creation design into the SPARC design principle of the split between forwarding and processing. The essential aspect in this solution is the required support for PPPoE processing, which could be hardware supported or emulated in various ways. Overall, several options are discussed in the following subsections.

Option 2, "SPARC DHCP++," uses the split of forwarding and processing, but essentially introduces a new set of protocols which needs to be supported in the OpenFlow environment. From a carrier's perspective, the model provides an integrative approach for the various service creation models used today. The target of this solution is to not introduce new required protocol support to the base specification, but to propose additions for required features in OpenFlow.

Beside the need for legacy support for PPPoE, this section details another important requirement on OpenFlow improvements as seen from SPARC: The decision logic for authentication and authorization. The authentication aspect is discussed a separate subsection in order to outline the requirements and the potential options in more detail.

#### 6.2.1.1　　　　Authentication

In Section 5.4, authentication has been identified as one of the requirements and important phases of service creation. Some more detailed information on the specific models of authentication for residential customers in the models based on PPP as used in the model SPARC BRAS and DHCP as used in the model SPARC DHCP++ is presented here.

The level of detail of authentication can differ depending on the desired level of information, ability of fine granular management and demands from legal aspects (which are omitted here). In general, there are two different levels: the authentication of a single user and the authentication of a connection per port. In the latter case, it could not be assumed that only one customer is authenticated. For example, fixed Internet dial-in environments (e.g., POTS, ISDN) were authenticating a single user only (and demanded the function for multiple users per dial-in connection) until the wide-spread deployment of residential gateways in xDSL environments, which typically uses only one set of user name/password for the whole group of users connected to them. Therefore, the most common model today is the authentication of a connection per port. Similar models exist in mobile environments as well, where the definition of connection/port differs, but uses the same principle. The authentication of a single customer could be required in different situations/cases:

- Each customer needs to be authenticated in the case of multiple customers per connection per port.
- Customer authentication in nomadism and a number of mobility cases.

Another important reason for authenticating on the port level is the configuration of a customer / service specific profile (parameters such as QoS profiles, nomadism, access to service platforms, etc.). In general the same reasoning applies for the authentication of a single user and therefore this model should be combined with the authentication of a port.

Port-level authentication is typically handled through the help of an (virtual) identifier (defined as Line ID) and a mechanism to include this identifier in the connection setup process. In principle, the Line ID could be anything that uniquely identifies the port (the process of creation/provision of the LineID is not yet standardized). Since the authentication is performed centrally at the BRAS node, a mechanism to pass the Line ID (however defined) to the BRAS is essential. The mechanism to include the Line ID in the connection setup process depends on the protocol used in the service creation. For PPPoE, this is done by the PPPoE Intermediate Agent. In DHCP a similar function exists with option 82 (DHCP Relay Agent Information Option) function, integrated into the DSLAM.

With OpenFlow authentication can be a rather simple or extremely complex processes depending on the two outlined authentication targets (single user or connection/port). Both mechanisms in today's carrier environments (PPPoE Intermediate Agent and DHCP option 82) could be supported easily with the right filtering mechanisms in the DSLAM and forwarding to appropriate processing, or through the use of the hybrid mode and ignoring the authentication messages from the OpenFlow-configured part of the switch. On the other hand, the port information is integrated in any OpenFlow packet-in message sent to the controller. Therefore, the information about the port could be used in any controller application that in turn could, e.g., fetch it from a database attached to the controller. This approach would require OpenFlow support in the DSLAM or any other network device terminating the customer's access line. Otherwise appropriate port information would not be available for the controller or would have to be added to the request (packet-in message) with appropriate mechanisms like adding a virtual identifier (e.g., VLAN ID) at the ingress of the DSLAM (the port) and the correlation of VLAN ID with the respective DSLAM port. Figure 60 depicts the principle of the PPPoE model and the potential application of the split control functions. Resource configuration is an essential part of service creation, but will be documented in an upcoming SPARC deliverable – it is included in the illustration in order to show the difference between the two models. Today service creation is performed centrally in the BRAS and the configuration information is transmitted via RADIUS. In an OpenFlow environment, an AAA application has to take over the task, or at least needs a link to the resource configuration system/function via the OpenFlow controller.

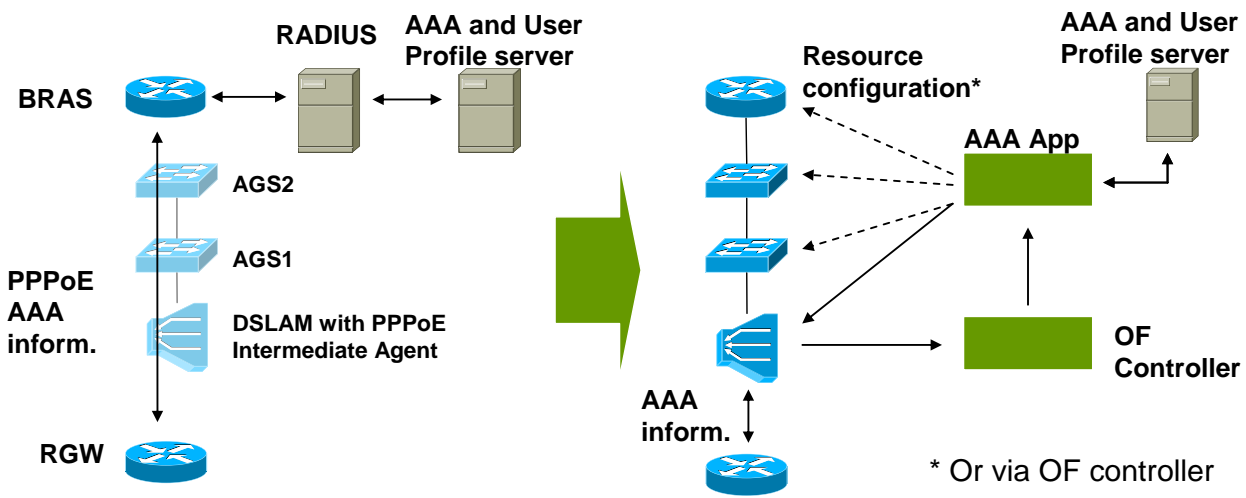

**Figure 60: AAA integration in PPPoE and OpenFlow**

Authorization is the step after authentication and could be handled in a simple fashion with OpenFlow too. The controller (or rather, an authorization application running on top of it) can simple send appropriate replies in "Packet-out" messages and configure the forwarding and processing policies at the DSLAM using e.g. flow modification messages.

### 6.2.1.2    Example: SPARC BRAS

We depict three deployment scenarios of a split BRAS (Broadband Remote Access Server) in Figure 61. Note that here we consider a BRAS to be a software component deployed on a Broadband Remote Access Router (BB-RAR) or Broadband Network Gateway (BNG) as defined in BBF TR-101. Therefore, BRAS and BNG are used synonymously within this document:

a) The first scenario covers a typical legacy BRAS deployment. The BRAS is deployed on a BNG and the latter acts as a termination point towards the Ethernet based access/aggregation domain. The BRAS is a central point containing all functions defined within TR-101 [61].

b) The second scenario replaces the BRAS/BNG device with an SDN based solution consisting of an extended datapath element and a control plane implementation that emulates the behavior of a legacy BRAS device. This scenario allows a smooth deployment of SDN-based elements within a legacy deployment and defines a migration path. Necessary extensions to the datapath element must provide the requested protocol encapsulation services, policy enforcement, and OAM functions.

c) In the third scenario, the BNG functions are distributed on different devices within the access/aggregation domain. The BNG provides VLAN support, QoS enforcement, hierarchical scheduling, user traffic isolation, IP forwarding services, multicast support, ARP processing, DHCP specific functions, security, and OAM support. Contrary to the centralized scenario (b), SDN allows distribution of these functions among all nodes within the access/aggregation domain, thus leading to load sharing, relieving the centralized BRAS device.

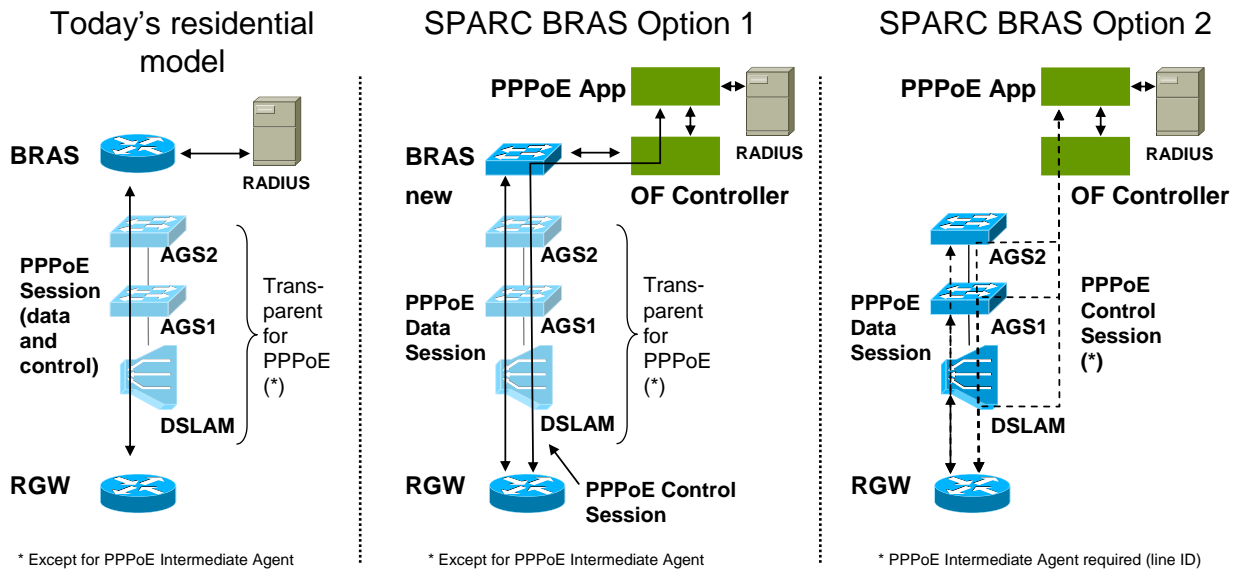

**Figure 61: SPARC BRAS options in contrast to today's residential model**

We have seen from the discussion in the previous section that *SplitArchitecture* allows network operators the freedom to distribute control and data plane functions across several network elements. In a legacy access/aggregation domain the BRAS, as a special purpose service node, provides all authentication and authorization services for specific service types and traffic shaping in a centralized manner. With SDN techniques, a network operator may virtualize all these functions in the control plane and may enforce them in the data plane at different locations.

SDN allows an operator to define a smooth migration strategy: in a first phase legacy network elements (e.g. BRAS) may be replaced with SDN enabled ones. An SDN enabled BRAS must support the same set of protocols and network functions as its legacy counterpart in order to act as a drop-in replacement. Besides BRAS compliant devices, a network operator may also replace the aggregation switching devices and/or access nodes with SDN enabled versions. After replacing all legacy devices, a network operator may decide to maintain the old set of protocol suits on all these devices. This in fact results in emulating a legacy domain and all network functions at their intended locations, respectively. As an alternative, a network operator may also decide to distribute or relocate functions inside the aggregation domain and thus, changing the network protocol stacks deployed on individual network elements.

In either case the SDN framework must provide adequate means to emulate all network functions defined by the network operator for his aggregation domain, e.g. as defined by the Broadband Forums TR-101 specification [61]. For an exhaustive list of requirements for a revised BRAS (named Broadband Network Gateway or BNG for short), please refer to the aforementioned document. We limit our discussion on mapping these functions to an OpenFlow-based SDN framework.

TR101 Issue 2 chapter 5 defines the requirements for a Broadband Network Gateway (BNG) for the following categories:

- VLAN support
- Quality of Service and Hierarchical Scheduling
- Multicast
- ARP processing
- DHCP relaying
- OAM
- Security Functions

We compare the requested protocols with the current version of OpenFlow at the time of writing, i.e. OpenFlow v1.3.

---

**High level Architectural Requirements**

**R-10** The Broadband Network Gateway MUST be able to terminate the Ethernet layer and corresponding encapsulation protocols.

**R-11** The Broadband Network Gateway MUST be able to implement the counterpart of the functions added to the Access Node for access loop identification, Ethernet-based QoS, security and OAM.

**R-12** Following TR-059 QoS principles, the Broadband Network Gateway SHOULD be able to extend its QoS and congestion management logic (e.g. hierarchical scheduler) to address over-subscribed Ethernet-based topologies.

---

- *Supported* ➔ In principle a BRAS/BNG acts as an IP routing device and forwards packets between the operator's core network and the access/aggregation domain and vice versa. Since version 1.1, OpenFlow datapath elements provide support for IPv4 based layer 3 forwarding and since OF 1.2, also for IPv6 datagrams. Also supported: TTL decrement, IP ECN and DSCP fields for matching and queueing.

- *Supported* ➔ An IP router terminates L2 transport domains and requires Ethernet transport endpoints in each of these domains. Thus, MAC address rewriting is required for replacing source and destination MAC addresses in the Ethernet header.

- *Unsupported* ➔ OpenFlow 1.3 lacks PPPoE/PPP support, although some preliminary extensions for PPP termination have been defined.

---

**VLAN support**

**R-190** The Broadband Network Gateway MUST be able to attach a single S-Tag to untagged frames in the downstream direction.

**R-191** The Broadband Network Gateway MUST be able to double-tag frames (S-C-VID pair) in the downstream direction.

**R-192** The Broadband Network Gateway MUST be capable of associating one or more VLAN identifications with a physical Ethernet aggregation port. These may be S-VIDs or S-C-VID pairs.

**R-193** The Broadband Network Gateway MUST support a one-to-one mapping between an S-VID or S-C-VID pair and a user PPPoE or IPoE session.

**R-194** The Broadband Network Gateway MUST support a one-to-many mapping between an S-VID or S-C-VID pair and a user PPPoE or IPoE sessions, where multiple PPPoE and/or IPoE sessions from the same user are within the same S-VID or S-C-VID pair.

**R-195** The Broadband Network Gateway MUST support a one-to-many mapping between a S-VID or S-C-VID pair and users sessions, where multiple PPPoE and/or IPoE sessions from multiple users are within the same S-VID or S-C-VID pair.

---

- *Supported* ➔ OpenFlow supports IEEE 802.1ad style Q-in-Q and IEEE 802.1aq MAC-in-MAC virtual LANs.

---

**QoS – Hierarchical Scheduling / Policing**

**R-196** The Broadband Network Gateway MUST be able to perform at least 3-level HS towards the Ethernet aggregation network.

**R-197** The Broadband Network Gateway SHOULD be able to perform 4-level HS towards the Ethernet aggregation network.

**R-198** The Broadband Network Gateway MUST be able to identify the root level by a single physical port.

**R-199** The Broadband Network Gateway SHOULD be able to identify the root level by a group of physical ports.

**R-200** The Broadband Network Gateway MUST be able to identify the second level (and potentially the third) by either 1 or 2 below.

**R-201** The Broadband Network Gateway SHOULD identify the second level (and potentially the third) by a combination of 1 and 2 above.

**R-202** The Broadband Network Gateway MUST identify the access loop by: a single C-VID, S-VID or S-C-VID pair, or by the User Line Identification (described in Section 3.9).

**R-203** The Broadband Network Gateway MUST identify the logical port or session by a C-VID, S-VID or S-C-VID pair, by the User Line Identification, by IP address, or by PPPoE session.

**R-204** The Broadband Network Gateway MUST be able to map between IP traffic classes and the Ethernet priority field.

**R-205** The Broadband Network Gateway MUST support marking Ethernet drop precedence within at least 2 traffic classes and MUST support configurable mapping from both the classes as well as drop precedence to the 8 possible values of the Ethernet priority field.

**R-206** The Broadband Network Gateway MUST support marking Ethernet direct indication of drop precedence within all supported traffic classes based on setting the DEI bit value of the S-Tag header.

**R-207** The Broadband Network Gateway, when receiving information about Broadband line rate parameters through PPP or DHCP, MUST NOT apply the information in an additive fashion when multiple sessions are active on the same Broadband line (the underlying rate is shared by all the sessions on a given line although each session will report the rate independently).

**R-208** The Broadband Network Gateway MUST support the application of ingress policing on a per user basis.

**R-209** The Broadband Network Gateway MUST support the application of Ingress policing on a per C-VID, S-VID or S-C-VID pair basis.

**R-210** The Broadband Network Gateway SHOULD support the application of ingress policing of a group of sessions or flows for a given user.

- *Out-of-scope*→Implementing a scheduling policy is out of scope of the OpenFlow specification. OpenFlow defines the SetQueue action that can be used for mapping specific flows to QoS enhanced queues on an outgoing port. The queue-id namespace is a 32-bit number which should be sufficient for implementing the set of required queues.

**Multicast**

**R-261** The Broadband Network Gateway MUST support multicast routing capabilities per TR-092 Appendix A "Multicast Support."

**R-262** The Broadband Network Gateway MUST support IGMPv3. Note: IGMP v3 includes support for endpoints using earlier IGMP versions.

**R-263** The Broadband Network Gateway MUST support IGMPv2 group to source address mapping for IGMP v2 to PIM/SSM compatibility.

**R-264** The Broadband Network Gateway MUST provide the following statistics: … (see TR 101 Issue 2 for details)

**R-265** The Broadband Network Gateway MUST support forwarding the multicast traffic on the same Layer 2 interface on which it receives the IGMP joins.

**R-266** The Broadband Network Gateway MUST support the following configurable parameters per port (i.e. physical or logical port (VLAN), but not per end user). This allows the Broadband Network Gateway to enforce service level agreements in real-time.

**R-267** The Broadband Network Gateway MUST support IGMP immediate leave as part of the IGMP router function.

> **R-268** The Broadband Network Gateway MUST immediately send Group Specific Queries out of an interface if it receives an IGMP query solicitation message (i.e. a Group Leave for group „0.0.0.0").

- *Supported* ➔ All multicast related requirements can be fulfilled by the functions defined for datapath elements in the OpenFlow specification.
- *Supported in slow path* ➔ All IGMP specific functions will be dealt with in the slow path, i.e. in the control plane.

> **IGMP processing and Hierarchical Scheduling**
>
> **R-276** A Broadband Network Gateway supporting hierarchical scheduling MUST support dynamic adjustment of the user-facing QoS shapers to reflect changes in the number of multicast groups joined by a user. (These adjustments would be inclusive of all levels of the hierarchy).
>
> **R-277** A Broadband Network Gateway supporting hierarchical scheduling MUST be able to trigger dynamic adjustment of the user-facing QoS shapers based on the tracking of IGMP messages received on both regular user-facing interfaces as well as on the appropriate multicast VLAN, and also based on local knowledge of the peak-rate of multicast streams. The correlation mechanism to identify the proper scheduler node with an associated multicast group or groups is an implementation option of the BNG. The PPPoE session VLAN and IPoE multicast VLAN may or may not be the same.
>
> **R-278** A Broadband Network Gateway supporting hierarchical scheduling SHOULD debit the amount of traffic offered by a given multicast group from the user-facing QoS shapers based on a provisioned association between a multicast group and a peak information rate.
>
> **R-279** A Broadband Network Gateway supporting hierarchical scheduling MAY debit on a packet by packet basis the amount of traffic offered by a given multicast group from the user-facing QoS shapers on a real time basis.

> **ARP processing**
>
> **R-211** For a given IP interface (say in subnet Z), the Broadband Network Gateway MUST be able to work in „Local Proxy ARP" mode: routing IP packets received from host X on this interface to host Y (X and Y are in subnet Z) back via the same interface. Any ICMP redirect messages that are usually sent on such occasions MUST be suppressed.
>
> **R-212** The Broadband Network Gateway MUST respond to ARP requests received on this interface for IP addresses in subnet Z with its own MAC address. This requirement refers to both N:1 VLANs as well as to several 1:1 VLANs sharing the same IP interface on the BNG.

- *Supported* ➔ OpenFlow supports ARP as a native protocol in the data plane and can emulate MAC addresses in its action lists or sets.

> **DHCP relay**
>
> **R-213** The Broadband Network Gateway MUST be able to function as a DHCP Relay Agent as described in RFC 951 "BOOTP", RFC 2131"DHCP" and RFC 3046 "DHCP Relay Agent Information Option" on selected untrusted interfaces.
>
> **R-214** The Broadband Network Gateway MUST be able to disable the DHCP Relay Agent on selected interfaces.
>
> **R-215** The Broadband Network Gateway MUST be able to function as a DHCP relay agent on selected trusted interfaces, from which it does not discard packets arriving with option-82 already present, and does not add or replace option-82 in these packets.
>
> **R-216** The Broadband Network Gateway MUST be able to function as a DHCP relay agent on selected trusted interfaces and MUST NOT strip out option-82 from the corresponding server-originated packets it relays downstream.

> **R-217** The Broadband Network Gateway, when functioning as a DHCP Relay Agent, MUST discard any DHCP packets with non-zero „giaddr" in the DHCP request from the client.
>
> **R-218** The Broadband Network Gateway, when functioning as a DHCP Relay Agent, MUST send the DHCP packets downstream as Layer 2 unicast or Layer 2 broadcast, according to the broadcast bit in the request.
>
> **R-219** The Broadband Network Gateway, when functioning as a DHCP relay agent, MUST be able to transparently forward any DHCP option information other than for option 82.

- *Supported on slow path* ➔ DHCP is not supported by OpenFlow. All requirements defined in R-213 – R-219 must be implemented in the control plane (i.e. in the slow path). As the number of DHCP packets is typically low, this does not seem to be a critical limitation.

> **OAM ➔ Short Intra-Carrier Maintenance Level**
>
> **R-354** The BNG MUST support an outward-facing Maintenance association End Point (MEP) on a per user-facing port and per S-VLAN basis.
>
> **R-355** The BNG MUST support initiating a Loopback Message (LBM) towards its peer MEPs and receiving the associated Loopback Reply (LBR), for the MEP(s) on the user-facing port.
>
> **R-356** The BNG MUST support receiving a Loopback Message (LBM) from its peer MEPs and initiating the associated Loopback Reply (LBR), for the MEP(s) on the user-facing port.
>
> **R-357** The BNG MUST support initiating a Link Trace Message (LTM) towards its peer MEPs and receiving the associated Link Trace Reply (LTR) messages, for the MEP(s) on the user-facing port.
>
> **R-358** The BNG MUST support receiving a Link Trace Message (LTM) from its peer MEPs and initiating the associated Link Trace Reply (LTR), for the MEP(s) on the user-facing port.
>
> **R-359** For business customers and/or premium customers requiring proactive monitoring, the BNG SHOULD support generating Continuity Check Messages (CCMs) towards its peer MEPs for the MEP(s) on the user-facing port.
>
> **R-360** The BNG MUST support turning off sending CCMs for the MEP(s) on the user-facing port, while keeping the associated MEP active.
>
> **R-361** The BNG SHOULD be configurable to assume continuity exists from a remote MEP while not receiving CCMs from this MEP.
>
> **R-362** The BNG MUST support receiving AIS messages on the MEP(s) on the user-facing port.
>
> **R-363** The BNG SHOULD trigger the appropriate alarms for Loss of Continuity.

A TR 101 compliant access/aggregation domain utilizes Ethernet like OAM mechanisms based on IEEE 802.1ag-2007 and ITU Y.1731. In a hybrid environment, SDN enabled devices must emulate legacy OAM behavior for interacting with non-SDN legacy devices. There has been no native OAM support defined in OpenFlow so far, although some basic ingredients for brewing an OAM solution exist (group tables, buckets, liveness of buckets, etc.). In Section 5.3, we introduced and discussed three proposals for defining a flow OAM solution for OpenFlow in order to address the specific OAM needs of SDN frameworks. A brief summary:

- Carrier-grade OAM solutions require usually high precision timers (e.g. < 50ms for protecting voice carrying connectivity) for detecting loss of connectivity. An OAM endpoint's core logic must detect a loss of connectivity on a primary path within a specific period of time, typically by exchanging echo messages with its peer, and maintain a second alternative path in case of a failure. OpenFlow provides the basic functionality in terms of group table entries and buckets for this task. However, the SDN control channel induces additional delay, OAM endpoints should not be instantiated within the slow path (=control plane) and must be deployed directly in the user plane.

- In case of a failure event, the OAM endpoint's core logic must notify the control plane about this incident any further countermeasures. OpenFlow lacks specific notification functions for signaling such error conditions currently, but either OpenFlow's ERROR message or an experimental message may be used for sending such indications to the control plane.

- Usually, OAM frameworks define specifically tailored OAM packet formats for achieving their monitoring goal. We use specific virtual ports for creating BFD OAM packets within our demo implementation, but a more capable framework for defining arbitrary packet creation may be defined later within OpenFlow (see the packetC discussion in Section 5.1). We assume that creation, configuration, and destruction of virtual ports for OAM packet management happen via a proprietary management interface of the datapath element.
- **Final conclusion ➔ An OpenFlow based SDN framework can provide the necessary OAM functionality, but additional logic and an additional interface for managing OAM endpoints and signaling state and events is required.**

---

**OAM ➔ Carrier Maintenance Level**

**R-364** For 1:1 VLANs, the BNG MUST support using a Multicast LBM towards its peer MEP.

---

**OAM ➔ Customer Maintenance Level**

**R-365** The BNG MUST support receiving a unicast or multicast Loopback Message (LBM) from its peer MEPs and initiating the associated Loopback Reply (LBR), for the MEP(s) on the user-facing port.

**R-366** The BNG SHOULD be able to be configured to assume continuity exists from a remote RG MEP while not receiving CCMs from this MEP.

**R-367** The BNG MUST be able to be configured to discard all incoming LTMs on a per user-facing port, per S-VLAN and per C-VLAN basis.

**R-368** The BNG MUST support rate limiting of received CFM Ethernet OAM messages arriving on a per user-facing port.

---

**Security Functions (Source IP spoofing)**

**R-220** The Broadband Network Gateway MUST only respond to user ARP requests when they originate with the proper IP source address and are received on the appropriate 802.1q VLAN, or 802.1ad stacked VLAN.

**R-221** The Broadband Network Gateway MUST be able to detect and discard ARP requests and reply messages with „sender protocol address" other than the one assigned (i.e. spoofed). Specifically, the Broadband Network Gateway MUST NOT update its ARP table entries based on received ARP requests.

**R-222** The DHCP relay agent in the Broadband Network Gateway MUST inspect downstream DHCP ACK packets, discover mapping of IP address to MAC address and populate its ARP table accordingly.

**R-223** The DHCP relay agent in the Broadband Network Gateway SHOULD follow the lease time and lease renewal negotiation, and be able to terminate any user sessions and remove the corresponding ARP table entry when the lease time has expired.

**R-224** Having the knowledge of MAC to IP mapping (achievable by following R-222 and R-223), the Broadband Network Gateway MUST NOT send broadcast ARP requests to untrusted devices (i.e. RGs).

---

- *Supported in slow path* ➔ These requirements cannot be fulfilled on a datapath element within the OpenFlow specification. However, an implementation may be moved in the control plane.

**Mandatory datapath element extensions**

- Provide support for PPP-over-Ethernet encapsulation/decapsulation according to RFC 2516. The current OpenFlow specification (v1.3 at time of writing) supports flexible matches for new protocols. However, push and pop operations for PPPoE and PPP must be added to the OpenFlow specification.
- The SET-FIELD action adopts the OpenFlow Extensible Match (OXM) TLVs and can be used for setting protocol header fields within PPPoE and PPP. For both PPPoE and PPP the following OXM TLVs should be defined:

| Header field | Description |
|---|---|
| PPPoE session id | 16bit session id → identifies jointly with Access Concentrator MAC address and user MAC address uniquely the user session |
| PPPoE code | Identifies the PPPoE frame type during 4-way handshake, i.e. PADI, PADO, PADR, PADS, and PADT |
| PPPoE version | PPPoE version, defined as value "1" by RFC 2516 |
| PPPoE type | PPPoE type, defined as value "1" by RFC 2516 |
| PPP protocol | The PPP protocol field (one byte or two bytes long) defined according to RFC 1661 |

- An Ethernet based access/aggregation domain adopts IEEE 802.1ag and Y.1731 for OAM. In TR-101 the authors have defined four levels of OAM: customer, carrier, intra-carrier, and access link. The endpoints of these maintenance domains depend on the intended model (broadband access vs. wholesale service, etc.). However, a datapath element used for emulating BNG services must provide support for Ethernet connectivity fault management. Refer to Section 5.3 for a description of the CFM OAM toolset. Beginning with version 1.1 OpenFlow provides basic support for implementing OAM services using so-called groups. Groups contain buckets and ActionLists and allow implementation of fast-failover strategies in the case of a network port failure.

- For the extended IEEE 802.1ad based model of TR-101 OpenFlow provides necessary functions since version 1.1. Implementers may decide to support an arbitrary depth of chained VLAN tags and provide push/pop operations in order to insert or remove tags accordingly.

- For IP forwarding services, a datapath element must be enabled to decrement TTL values in the IPv4 header. This functionality has been added to the OpenFlow specification since version 1.1.

- Policy enforcement and QoS support as a 3-level hierarchical scheduler must be available (see the final paragraphs of Section 5.8 for details on hierachical QoS models). QoS management is out of scope of the base OpenFlow specification and is expected to be configured via a third-party proprietary interface. However, OpenFlow provides means to queue packets to specific CoS queues on a network interface (see action ActionSetQueue). OpenFlow lacks efficient means for implementing traffic shaping strategies. However, we assume that such functionality is defined outside of the OpenFlow datapath element via a proprietary interface.

- Since version 1.2 of the OpenFlow specification, a precise definition of a datapath element's classifier has been omitted. For supporting PPPoE/PPP, a datapath must be enabled to parse PPPoE and PPP protocol headers and to match these against the above defined OXM TLVs. With OpenFlow version 1.3, a control plane developer may define an arbitrary sequence of push operations, i.e. it is up to the control plane to ensure that the final sequence of headers (PPPoE within VLAN within VLAN within Ethernet) is useful. The datapath may not check validity of the defined order of push commands.

**Modules for emulating a BRAS/BNG in the control plane**

The following list provides a set of mandatory functions for a BRAS/BNG emulation in the control plane. An architecture for implementing such a control plane is detailed in deliverable D4.3.

- A module for providing an emulated PPPoE access concentrator functionality bound to an operator defined MAC address must be available.

- Functionality for hosting a PPPoE session must be available.

- Functionality for hosting a PPP session must be available. The PPP module must be enabled to connect to AAA interfaces for session management, e.g. Radius or Diameter.

- An IP routing module must be available capable of defining Flow-Mod entries for L3 based forwarding.

- The IP routing module must take into account the session's authentication state. Non-authenticated sessions must not be forwarded (except L2TP encapsulated towards a remote BRAS/BNG for wholesale services).

- Functionality for hosting IP-over-Ethernet or IP-over-MPLS should be available.

- A management function for observing and configuring OAM state must be available.

- A traffic shaping control module must be available.

- A VLAN control module must be available supporting IEEE 802.1ad.
- A DHCP relay control module must be available.
- Multicast support must be provided.

### 6.2.1.3 Example SPARC DHCP++

This example is splitting data and control plane of the SPARC BRAS model Option 1 while leveraging some advantages of this separation. Fundamentally, it changes nothing of the requirements as detailed in Section 5.6 and is relative similar to Option 2 of the SPARC BRAS example detailed in the previous section. The key change is that, instead of PPP(oE) and its integrated protocols, DHCP is used in combination with other protocols and mechanisms. Typically, the required protocols are already available at customer's residential gateways (RGW), but commonly not used for the WAN side / interface. Therefore, it would require some modifications of the RGW, but this could be done by firmware updates and would require no hardware upgrades. How the legacy devices could be updated or how firmware will be deployed is out of scope of the SPARC project and thus not covered in this deliverable.

As stated before, the DHCP++ proposal splits between data and control plane. Furthermore, it requires support of forwarding decisions in terms of forwarding entries in the FIB of the devices (potentially DSLAM, AGS1, AGS2, edge router) and implementation of QoS profiles in scheduler, shaper or police engine in the devices (see Section 5.8).

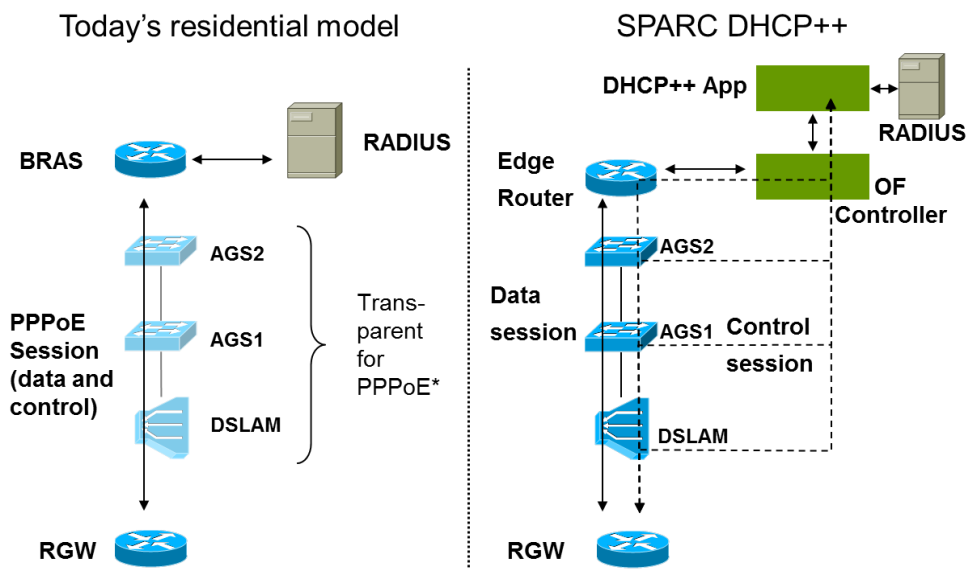

**Figure 62: SPARC DHCP++ in contrast to today's residential model**

The current model is already explained in detail in Section 5.6.3. The important aspect is that the data and control part of the PPPoE session is encapsulated and sent within the same session, presented in Figure 62 left side.

In SPARC DHCP++, a DHCP++ app is introduced on top of the OpenFlow controller which is connected to either the edge of the network (edge router) or to any other device in the data plane chain from customer to Internet (more specifically: DSLAM, AGS1 or AGS2 as depicted with dotted lines in Figure 62 right side). Unfortunately, different implementation options with different requirements will exist concurrently. In the following paragraphs, we concentrate our analysis first on design options related to authentication, and then briefly discuss also the required routing function.

First, it is important to acknowledge that the desired authentication target is important (see Section 6.2.1.1 for details). In current fixed access networks, a router on customer premises (RGW) performs the authentication by sending credentials to the network, thus applies a per port-based authentication scheme. If more than one user needs to be authenticated (referred to as customer authentication in Section 6.2.1.1), the model becomes more complex and requires more advanced models (e.g. see Protocol for Carrying Authentication for Network Access (PANA) defined in [62]). Such a model is out of scope for this document and subject to future work. This deliverable will concentrate on the per-port based authentication.

Again, different models might be implementable, shown in Figure 63 below. In the following, two general models with different implementation options are discussed. Essential is the support of DHCP Option 82, the DHCP Relay Agent Information Option (standardized in [63]) in one or the other way. For now this function is integrated in the DSLAM and adds a Line Identifier based on the port which could be used for authentication (see Section 6.2.1.1 for details) and correlation of customer profiles. In OpenFlow networks, one could use the port identifier directly.

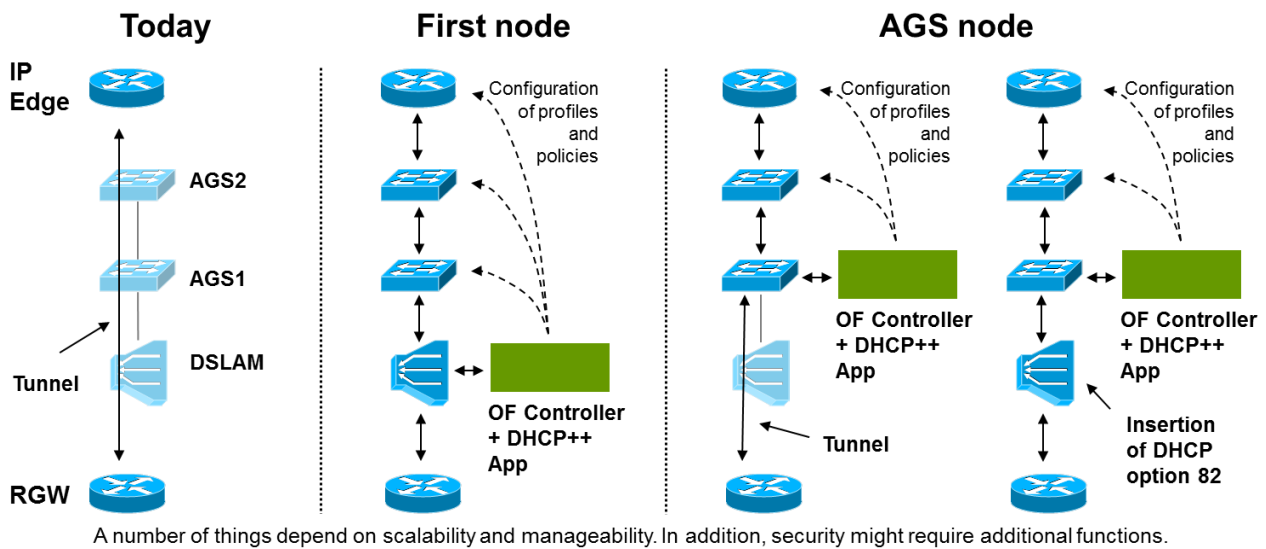

**Figure 63: SPARC DHCP++ integration options for first node (DSLAM) and AGS node**

The first possible model is the attachment of the OpenFlow controller with the DHCP++ app (and thus connectivity to RADIUS) at the first node (the DSLAM) as shown in the middle of Figure 63. The DSLAM is OpenFlow enabled. Here forwarding entries are installed in a way that any traffic despite DHCP messages for the attachment of the client is dropped (similar to the IEEE 802.1X Port Access Entity [64]). After a successful DHCP ACK message (and related configuration of network parameters in the client e.g. address, subnet mask, default gateway, etc), the forwarding entries are updated so that Internet access is granted. This initial set of forwarding rules must add some port information. As detailed previously, here one has two options. In option one, the DHCP option 82 is used (which would result in some kind of hybrid node) and the DHCP++ app forwards this ID to the RADIUS server in order to receive appropriate profile and policing information. In option two, the DSLAM will forward the incoming packet with a port information and based on some additional data base information, e.g. the port information in an OpenFlow Packet-In message. Here the DHCP++ app forwards the Line ID to the RADIUS (like in the DHCP option 82 mechanism) or it forwards information like the port number and the DSLAM ID to the RADIUS and the RADIUS itself must figure out the client based on this information. Optional is the configuration of profiles and policies in the other devices. This could be done during the procurement of the customer or during the connection setup. In general, the implications are similar to the considerations discussed in the evolutionary approaches described in Section 6.1, and each operator has to decide which model is the best fit for its respective requirements.

The second possible model is the attachment of the OpenFlow controller with the DHCP++ app at the OpenFlow enabled AGS1, shown on the right side of Figure 63. In principle, the concept is similar to the previous model, but requires smaller modifications. The general problem is the notification of the port. Again, this could be done in two different ways. First, at the DSLAM, a unique identifier for the datapath (e.g. VLAN ID) representing a tunnel is attached per port in each packet. Based on the ingress port of the AGS1 and this VLAN ID, the DHCP++ app can then identify the incoming port. The other option is to use the DHCP option 82 in the DSLAM and appropriate processing by the DHCP++ app. Other aspects like configuration of profile and policies are similar to the aspects covered by first node integration option. Another possibility is the integration at the AGS2 node. Here the model would be the same like for the integration at AGS1 node.

A second major aspect beside the authentication target is the integration of the routing function into the service creation architecture. This is important for the following reasons:

- Client requires a default gateway, e.g. for addressing ARP requests
- Scalability / security demand might require some termination of the broadcast domain
- Scalability of the OpenFlow platform might require some independent network segments
- Organization of backbone networks and related routing environment might require some subnetting

In the SPARC BRAS model, these issues could be handled in a rather simple manner because of the termination of the PPPoE sessions at the BRAS and the required routing functionality in the PPPoE function. Therefore, the BRAS could be changed from being a router (with PPPoE support to customer side) to a switch. In the SPARC DHCP++ model, this transformation could not be applied easily. Again several options for the integration are possible and will depend on the existing platform (attachment to legacy) or the desired targets of the platform to be built. Therefore, we give only some hints for possible options in this deliverable.

- Several options are detailed for the integration of MPLS environments in Section 6.3, similar models could be applied
- Integration / interworking of routing function (e.g. RouteFlow [65]) into / with the DHCP++ app
- Shift of routing function to some core networks, e.g. Label Edge Router

### 6.2.2　Business customer services based on MPLS pseudo-wires with OpenFlow

Business customers use a number of different technologies to interconnect locations, and carriers provide different options for service creation of these services. In order to analyze the impact of OpenFlow on the service creation, and to have a meaningful relation to the work done in the demonstrator development in SPARC WP4, it was restricted to Ethernet services and MPLS first.

In Figure 64 a provider edge router providing pseudo-wires is illustrated; while this example deals with Ethernet over MPLS, the same general architecture is used for multiple types of pseudo-wires. Frames are received from a customer edge switch/router and are first processed in the "Native Service Processing" (NSP) module – this refers to Ethernet-specific processing such as modifying VLAN tags, priority marking, bridging between different pseudo-wires, etc. Once through the NSP module, the "PW Termination" module is responsible for maintaining the pseudo-wire by, e.g., encapsulating/decapsulating frames and performing any necessary processing such as buffering in case of ordered delivery. Finally the packets are delivered to the "MPLS Tunnel" for MPLS encapsulation and transmission across the network.

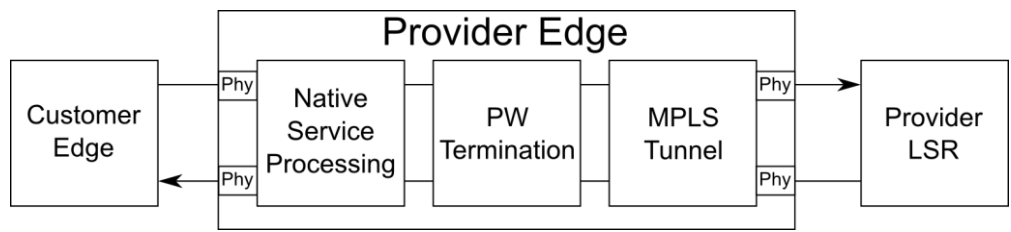

**Figure 64: Pseudo-wire processing in Ethernet over MPLS pseudo-wires (based on Fig.3 of RFC4448)**

In OpenFlow Version 1.1 this model can be followed, for example, by having one flow table per module, with the encapsulation/decapsulation actions implemented as vendor specific actions, and any additional processing managed either through process actions or virtual ports (as discussed in Section 5.1.2). If we focus on MPLS pseudo-wires with Ethernet emulation (and for the moment ignoring other types of pseudo-wires), we need a number of new actions corresponding to the different steps described above. First, at the ingress of the pseudo-wire, we need an action that takes an incoming Ethernet frame and turns it into the payload of a new frame. Once the new frame has been created, we need to add the control word. From this point on we can use existing actions to push MPLS labels, set the correct EtherType, and add correct MAC source and destination addresses. At the egress of the tunnel this should be mirrored by the inverse actions, in particular one to strip the outer headers and convert the payload back into an Ethernet frame.

## 6.3　Split Control of Transport Networks

### 6.3.1　Cooperation with legacy control planes

As discussed in the introduction to this section (Section 6), we identified four types of introducing OpenFlow to an access/aggregation network. In the first two types, OpenFlow-based control solutions configure the transport nodes, while in the third type the service functionalities are implemented through an OpenFlow-based centralized control scheme. In these three types, some nodes and/or some functions of those nodes are configured with protocols other than OpenFlow; only the fourth type considers OpenFlow to be the sole forwarding node configuration protocol.

There are several OpenFlow integration options (as discussed in Section 6.1) for the third type: the centralized and the distributed OpenFlow control. In both options, the intermediate transport nodes are still not configured via OpenFlow, but provisioned through the management plane or by making use of legacy control protocols. For example, an Ethernet-based aggregation switch can be configured using SNMP. Therefore, the OpenFlow controller should be able to interact with a management/control entity responsible for provisioning the transport connections. For this purpose, an interface between the legacy management/control entity and the OpenFlow controller is essential. This raises the demand for cooperation with legacy management or control planes. In our above example, the OpenFlow controller must be able to communicate with an entity responsible for managing Ethernet transport and thus for provisioning connectivity between the service nodes (e.g., DSLAM, BRAS).

The fourth integration type, in which all aspects of all aggregation nodes are configured via OpenFlow, does not require such "vertical" cooperation with non-OpenFlow control entities. Even in this latter case, the whole network, as shown in Figure 2, is still not under the control of OpenFlow due to these aspects:

- Mature control planes are deployed in some network segments and any gain of substituting them with OpenFlow is not clear, and
- Covering a whole network of thousands of nodes with a single controller entity raises scalability issues.

In any of the discussed implementation cases, the OpenFlow controller must be able to cooperate with the other domains of the operators network. It is important to emphasize that any kind of control solution (centralized, distributed, static, dynamic, etc.) can be deployed in those domains. Therefore, the controller must be able to cooperate with the control function of those domains; this is referred to as horizontal interworking or peering and raises two major issues.

The various control functions use different resource description models. For instance, the MPLS control plane was designed as a protocol running between physical nodes. Hence, the internal structure of routers is less relevant and information about the internal details of the nodes is not disclosed as a simplified view only. This view encodes the router with its interfaces and capacity information, assigned only to these interfaces. GMPLS control follows similar abstractions even in multilayer cases: All information is tied/bound to the interfaces of the nodes. In the case of WSON, this abstraction level is changed by adding further details of the node's internal capabilities, but it still uses a generic model. On the other hand, an OpenFlow-based transport controller has much more detailed information about the managed domain. For horizontal interworking with legacy domains controllers via MPLS, GMPLS, etc., that information will be essentially filtered: A virtualized view of the managed domain is derived and provided toward the peering control entities.

Furthermore, the control plane entities may allow for different control plane network implementations: For example, MPLS supports an in-band control plane, where the protocol messages travel together with the regular data traffic. GMPLS is also able to operate with in-band control channels, but it also supports use of the out-of-band control plane. While the *SplitArchitecture* inherently supports the out-of-band control network, it can provide in-band options as well: The controller is able to instruct the data plane nodes to inject the protocol messages into the data stream toward the peer control node and to demultiplex the protocol messages received from the peer node.

### 6.3.2   Semi-centralized control plane for MPLS access/aggregation/core networks

Considering typical network structure as shown in Figure 2, the network consists of two parts: The access/aggregation using various forwarding technologies (e.g., Ethernet or MPLS), whereas IP/MPLS is the predominant technology in the core network segments. This also implies the control plane used in the core. This means that the controller, which manages the transport connections in the access/aggregation network segment, must be able to exchange IP/MPLS control protocol messages with the distributed IP/MPLS control plane of the core. The IP/MPLS control plane has both of the major issues discussed above.

As IP/MPLS uses link state routing protocols (either OSPF or ISIS) to keep the topology databases synchronized at the protocol speakers, a very simple network model is used. According to this model each protocol speaker advertises only its identifiers to the attached network and the detected adjacent routers, and by default does not report any information about its internal capabilities or structure. It additionally uses signaling protocols, e.g., LDP, to provision end-to-end MPLS label switched paths. However, the label distribution mechanism allows the adjacent nodes to agree on the used label value, but it does not instruct any node about how to configure its internal elements. This means that the controller must implement an IP/MPLS control-compliant view of the managed domain and a mapping mechanism between the physical network and the logical representation. As discussed, the IP/MPLS control plane is an in-band control plane, so the OpenFlow controller must be aware of that.

An extension to the link state routing protocols allows the assignment of further opaque attributes to the link. These additional attributes are also disseminated to other protocol speakers, although they do not carry any relevant information for the routing protocol. Assigning link characteristics such as available bandwidth, delay, etc., as opaque attributes supports implementation of traffic engineering in IP/MPLS networks. However, to provision traffic-engineered LSPs, an additional protocol is used: RSVP-TE. The IGP protocol must implement additional structures to advertise the opaque attributes as well, and such extended protocols are referred to as IGP-TE (e.g., OSPF-TE). These extensions convert the link database to a Traffic Engineering Database (TEDB).

#### 6.3.2.1 Options for connecting the controller to the IP/MPLS control plane

The dissemination areas of the link state IGPs (OSPF/ISIS) define a structure for the IP/MPLS control plane. Adding all protocol speakers to a common dissemination area will result in an accurate view of the network at all speakers, allowing proper calculation of the LSPs. The drawback of such a solution is the scalability because the number of nodes of the same dissemination area increases. Several documents (see [40]) report that the while the core network can be covered by a single dissemination area, the whole network cannot. Therefore, in this section we discuss two alternatives of how a controller managing an access/aggregation domain could be attached to the distributed IP/MPLS control plane of the core.

A possible implementation is when the controllers act as simple IP/MPLS protocol speakers and they are attached directly to the core network's control plane, just like the simple core routers. Then the controllers and the core routers share the same dissemination area (OSPF area) as shown in Figure 65.

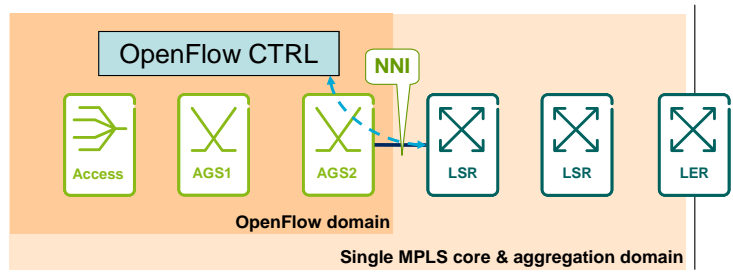

**Figure 65: Single dissemination area option**

Consequently, each controller has the same topology database as all the other core routers and controllers. Based on the shared database, every controller or router can initiate an LSP configuration to all other routers or controllers. The signaling protocols, LDP and the RSVP-TE are used to manage the LSPs.

In some cases, the IP/MPLS network splits into multiple dissemination areas. Area Border Routers (ABRs) reside at the border of the dissemination areas. Thus the controller can be part of either the backbone dissemination area or any of the stub/attached areas.

In the former option, the ABR can be considered part of the OpenFlow controller domain (as shown on Figure 66). This option is roughly similar to the single dissemination area option because the controller communicates with all nodes in the backbone using OSPF-TE, LDP or RSVP-TE for dissemination. Relying only on these protocols, it is possible to create LSPs in the core area only. Spanning LSPs covering multiple dissemination areas require additional protocols:

- One possibility is to introduce MP-BGP as described in the seamless MPLS concept [40]. This means that the controller must support the MP-BGP protocol as well as extended router and LSP redistribution mechanisms.

- Another option is to introduce some parts of the GMPLS' multi-domain extensions based on signaling extensions of RSVP-TE. Since such an LSP spans multiple dissemination areas, the source node has an accurate view of the local area only, and it has connectivity information only for the remote one. This affects the path calculation mechanisms used. It is possible to calculate the path domain-by-domain, but it will not be optimal. A better solution is to adapt the Backward Recursive Path Computation (BRPC) algorithm and use the PCEP to synchronize the calculated path fragments. To support this option the controller must implement the extended version of RSVP-TE as well as PCEP.

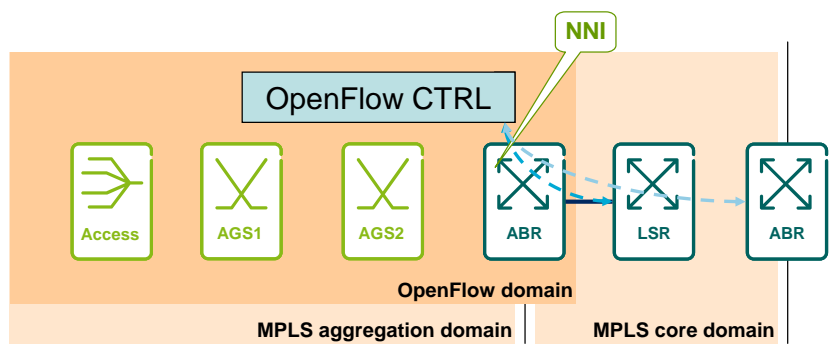

**Figure 66: ABR is under OpenFlow control**

In the latter option (see Figure 67) the controller is part of an attached dissemination area. Just like the other multi-area option, it implements the basic protocols to configure the LSPs within the area and applies BGP or multi-domain RSVP-TE for configuring the end-to-end path. The significant difference is that the controllers are not directly involved in the core network configuration, and they have a limited view. The pros and cons of the two multi-area options are discussed as part of the scalability evaluation in Section 6.4.

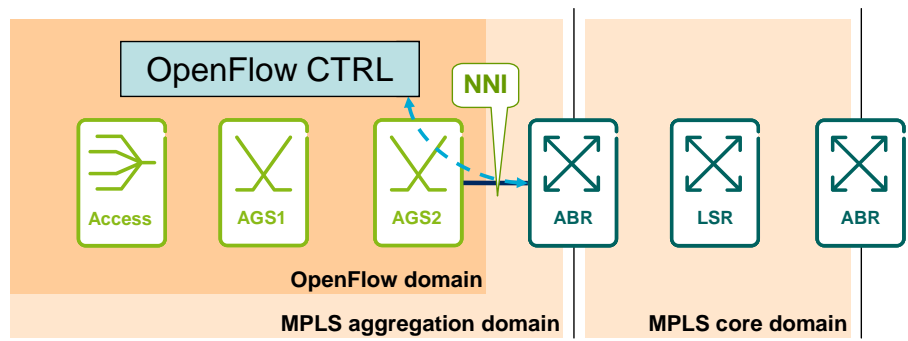

**Figure 67: ABR is not under OpenFlow control**

Based on the implementation cases discussed, the NNI interface running between the controller and the IP/MPLS domain shall implement the following protocols:

- A link state IGP protocol (either OSPF or ISIS) to share connectivity information.
- TE extensions of the above IGP protocols to share TE attributes and/or provide detailed information about the internal structure of the OpenFlow domain.
- To signal intra-domain MPLS connections, LDP can be used for best effort, RSVP-TE for TE-enabled connections.
- To signal inter-domain MPLS connections, MP-BGP can be used for best effort, RSVP-TE (RFC5151) with optional PCEP support for TE-enabled connections.
- To support multicast, mLDP or RSVP-TE (RFC4875) can be used.

### 6.3.2.2 Domain representation model

The IP/MPLS control plane design assumes that the protocol speakers are actually routers, and they could consider all other speakers as routers. As a result, that information model focuses on the links running between these routers, and very limited information is disclosed about the internals of the routers. This model was kept when TE capabilities were added: The TE attributes were tied/bound to the link descriptors (see Traffic Engineering Link, TE Link concept). One alternative is to keep the IP/MPLS model as is, and the controller is then developed with functions to support the existing models. As an alternative the IP/MPLS information model may be extended, similar to the WSON extensions for GMPLS [27]. However, it will violate the assumption of not making any updates to the core network. Therefore, this latter alternative is not discussed here.

A trivial consequence is that the OpenFlow controller must implement an appropriate mapping function between the controller's internal model and the IP/MPLS information model because the IP/MPLS model is not sufficient to describe all aspects of the OpenFlow domain. This mapping function can be implemented in many ways.

A possible realization is the emulation of the control plane (see for instance QuagFlow [28]). The OpenFlow domain switches are replicated as emulated routers running the IP/MPLS control plane. This creates a logical view of the OpenFlow domain topology and all control plane actions are emulated. If more routers run in the emulated environment, they will synchronize their state even though they are running in the same controller.

Instead of replicating the whole OpenFlow domain with emulated IP/MPLS routers, we propose representing the whole OpenFlow domain as a single IP/MPLS router. This single virtual router is considered during the interaction with the legacy IP/MPLS control plane. Its identifiers and virtual interfaces and are advertised in OSPF, and its content determines the sent signaling messages (LDP or RSVP-TE) as well. Upon reception of any signaling messages its content will be uploaded. Besides eliminating the unnecessary domain internal state synchronizations, this approach has scalability advantages as well: The rest of the IP/MPLS does not need to take care of the internal structure of the OpenFlow domain.

### 6.3.2.3　　　Updated controller architecture

The general control plane architecture described in Section 4.1 does not cope with all the requirements dictated by the considered MPLS-based access/aggregation scenario. That hierarchical organization of controller layers does not consider peering control plane entities at the same control layer. To support such peering control plane entities, the control plane model must be extended as described below.

According to the above examples, the peering control planes may use different approaches to manage their supervised network segments. Therefore, direct exchange of the internal data models of the network segments is impossible without any agreed translation functions. Even if such translation functions were defined, implementing would not be recommended due to scalability and privacy issues. For example, one operator may not want to disclose all internal information to other operators. Furthermore, sharing the information describing databases may place unacceptable processing burdens on the interoperating control planes: One controller would process each and every change in the databases of other controllers.

Router virtualization is an obvious choice to alleviate both problems. In this case, the peering controller plane must support a common information model and associated procedure set. The virtualization models used here are roughly the same as those used in hierarchical interworking, i.e., the managed network domain can be represented as a switch, or a set of switches with a physical/emulated topology. The associated procedures sit on the top of this virtualization model and are implemented by protocols. In order to enhance flexibility, the virtualization model and the associated procedures are detached, i.e., the different sets of protocols may use the same virtualization model, and the protocols actually used may be configured. The resulting updated control architecture is depicted in Figure 68.

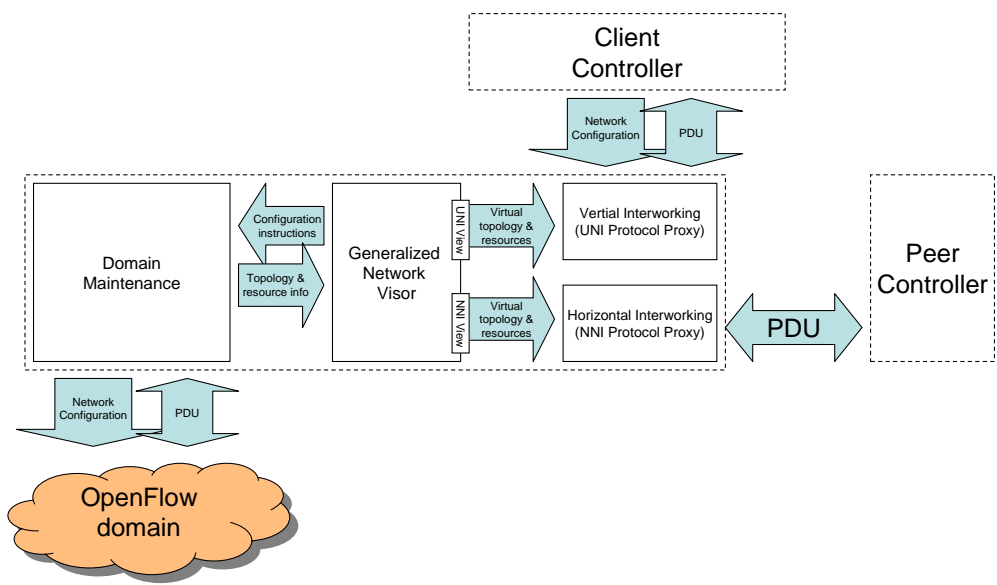

**Figure 68: Revised controller architecture**

Here the controller is comprised of three major elements:

- The Domain Maintenance module manages the OpenFlow domain, deploys flows, reacts to topology changes etc.

- A Generalized Network Visor realizes the virtualization feature by managing virtual (emulated) topologies. The namespace and resource management is also implemented here.

- Interworking modules implement the protocol and functional modules to communicate with other control plane entities. These modules operate on top of the virtualized topologies provided by the generalized visor, which controls the access of the interworking modules to the OpenFlow switches.

With these extensions a controller implementing the split transport control plane will comprise several key modules.

The Domain Maintenance module is responsible for managing the OpenFlow domain. It authenticates the connected data plane switches, and it maintains an appropriate view of the topology of the available resources of the managed network domain. It also provisions all desired aspects of flows within the OpenFlow domain: calculating paths fulfilling the traffic engineering objectives, deploying monitoring endpoints and the protection infrastructure. It also provides an interface to other modules. Through this interface the other modules can post configuration requests to the Domain Maintenance module and they can receive reports of events in the managed domain.

The Generalized Network Visor supports the various virtualization models and implements a virtual router model. The essential transformation functions between the virtual model and the managed topology are realized by the visor as well. For example, a configuration request in the virtual router may trigger a request for a flow establishment from the Domain Maintenance module. The result of a successful deployment of such a flow may result in installation of a new forwarding entry in the virtual router.

The virtual router is used by the NNI protocol proxy that steers the communication with the IP/MPLS protocol control plane. For example, it can integrate all relevant legacy protocols (OSPF-TE, LDP, BGP, etc.) that run as part of the controller. Another option is to make use of external protocol implementations. In this case the controller hosts a proxy or kernel part of the protocol stack and uses stack-specific protocols (e.g., the zebra protocol) to communicate with the protocol implementation. As a third option, external protocol stacks are used, but the virtual router model is exported and the controller acts as a switch by providing standard switch forwarding configuration interfaces (e.g., SNMP).

## 6.4 Scalability Characteristics of Access/Aggregation Networks

This section presents our scalability investigations regarding the access/aggregation use case. First we provide a high level introduction and then we present the results of our numerical analysis before finally showing the simulation results.

### 6.4.1 Introduction to the scalability study

There are many aspects of scalability, e.g., the scalability of an architecture, of a protocol, of a given scenario. We are focusing on the generic *SplitArchitecture* concept and the main use-case of the project.

#### 6.4.1.1 Deployment – network topology

Our deployment scenario is based on D2.1 and serves as a basis for the scalability study. Figure 69 shows the common view of the project partners in the schematic view of the access/aggregation network area.

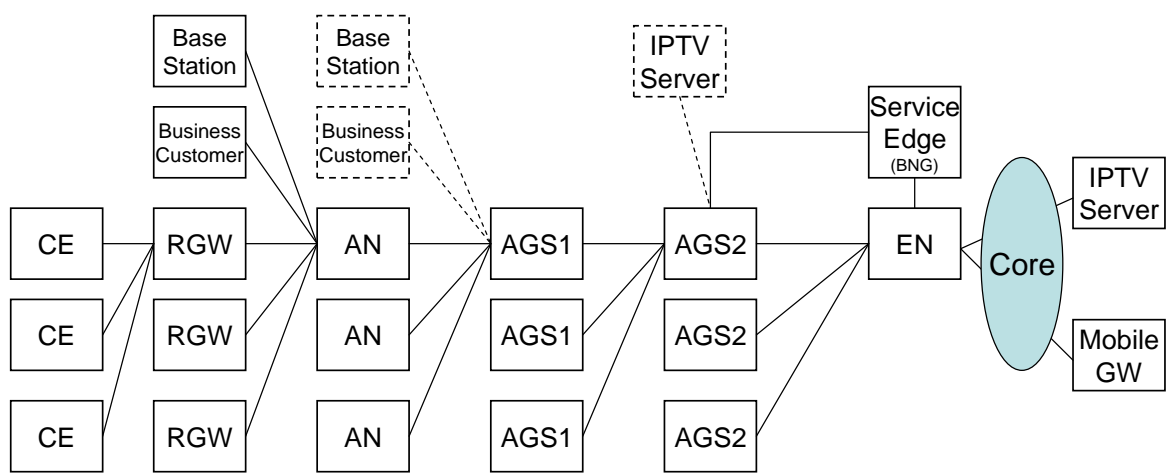

**Figure 69: Deployment scenario for scalability studies**

The OpenFlow domain consists of Access Nodes (AN), first-level aggregation switches (AGS1), second-level aggregation switches (AGS2) and edge nodes (EN). All other entities, both on the client side (Customer Equipment, Residential Gateway, Business Customer [BC], mobile Base Station [BS]) and the server side (Service Edge, IPTV server, Mobile GW) are not part of the OpenFlow domain. Note that multiple access/aggregation domains can be connected to the core network and they can be OpenFlow-based, but they have no direct (OpenFlow) interaction.

Based on D2.1 we have three scenarios that cover current and future deployment numbers.

- In the "Today" scenario there is 1 PoP location (corresponding to our EN) for 500,000 households (which are represented by the RGWs). This covers about 1 million inhabitants.

- In the "Future" scenario 1 PoP location will cover around 2,000,000 customers.

- In the "Long-term" scenario 1 PoP location will serve 4,000,000 customers.

In all scenarios the number of devices will relate to each other as: customers devices (CE) >> customer edge (RGW) >> access nodes (AN) >> edge nodes (EN).

Additionally, the sum of access nodes (AN) relates to the sum of aggregation nodes (AGS1+AGS2) and edge node (EN) as 10:1.

Based on these assumptions, network topologies can be drawn for the different scenarios.

### 6.4.1.2　Deployment - services

Regarding the services provided by using this network, we have made the following assumptions based on D2.1.

There are residential services, namely the simple service (e.g., PPPoE) for Internet access and IPTV. The simple service provides bidirectional connectivity between the RGW and the service edge. The IPTV service provides bidirectional connectivity between the RGW and the IPTV server, where the direction from the IPTV server dominates, and typically the same data is simultaneously sent to multiple RGWs.

There are business services, namely the simple service (as for residential customers), the Point-to-Point (PtP) and the Multipoint to Multipoint (MPtMP). The PtP service connects two business customers, while the MPtMP connects multiple business customers. All of these services are bidirectional.

The mobile backhaul service connects a base station to a mobile GW in a bidirectional way.

### 6.4.1.3　Scalability concerns

There are theoretical constraints due to protocol limitations, e.g., the maximum number of fields or objects, maximum length or value of fields or objects, maximum size of packets, etc. The OpenFlow protocol is relevant in our case. Other protocols that may be affected are the distributed control plane protocols, e.g., MPLS (OSPF, LDP, RSVP). Our focus is not on the analysis of a given protocol, but rather on a more generic approach to the split control plane. Protocol-specific issues are beyond our scope.

Computational resource limits may be a bottleneck at the central controller (in such a case these calculations may be made in a decentralized way, e.g., single central logical view vs. distributed physical view). The complexity of the algorithms used can be checked here, e.g., NP hardness. However, the time budgeted for running an algorithm is important – from this point of view the P complexity class could be still too difficult in some practical cases.

The control network's capacity limits the overall control traffic, thus introducing upper limitations on the frequency and size of node configuration commands. This can have an effect on the possible network size (e.g., the number of nodes/ports/links) or on the reaction time of the overall network.

The control network introduces non-negligible delays due to the geographical distance between switches and the controller. This results in a lower limit on reaction time even if the controller had infinite computational power.

## 6.4.2　Numerical model

This section describes our numerical model and the results of the of the study based on this model.

### 6.4.2.1　High-level description of the model

The model is based on Section 6.4.1 and it includes the network nodes, services and scenarios identified there, with the following assumptions. Clients (RGW, BC, BS) are connected only to the AN, not to higher level aggregation nodes. SE is not directly interconnected to an AGS2. The MPtMP business service is not modeled. The UNI interface is on the left side of the AN, while the NNI is on the right side of the EN, see Figure 71.

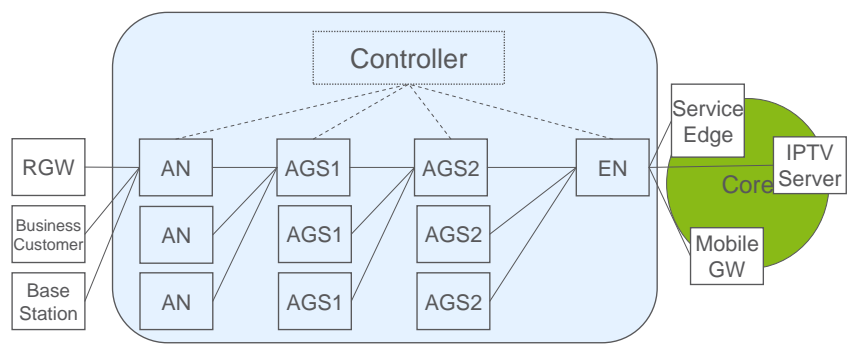

**Figure 70: Simplified domain view**

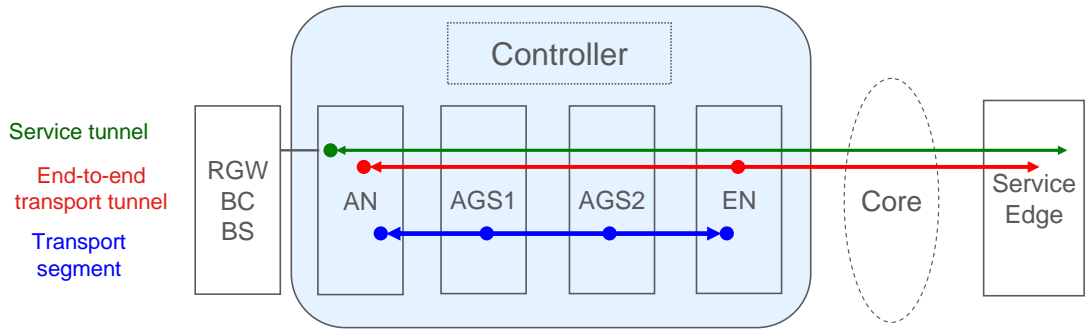

**Figure 71: Tunnels considered**

To transport the above services appropriately, the transport connection architecture is assumed as depicted in Figure 71. The blue connections are OpenFlow domain internal transport segment tunnels. There are multiple options for organizing these tunnels. There can be a segment tunnel connecting each AN to the EN as shown in the figure.

End-to-end tunnels (shown as a red line) are defined between nodes, e.g., the AN and the SE. Such tunnels overlap OpenFlow and non-OpenFlow domains.

Service tunnels (shown as a green line) within the E2E tunnels identify the service, e.g., the user in the case of an AN where multiple RGWs are connected. The exception is IPTV, which is treated as a special E2E tunnel in the model.

This triple-layer connectivity set provides compatibility with the IP/MPLS core, flexibility within the OpenFlow domain, and it also supports scalability (by aggregating connections).

#### 6.4.2.2　　Static requirements from the model

The figure below (Figure 72) shows the amount of equipment units to be managed by the controller according to the main scenarios based on our model. Two alternatives are offered for each scenario, both fulfilling the requirements. Note the we will later concentrate on one alternative only because the results of the alternative topologies have the same order of magnitude results in each case.

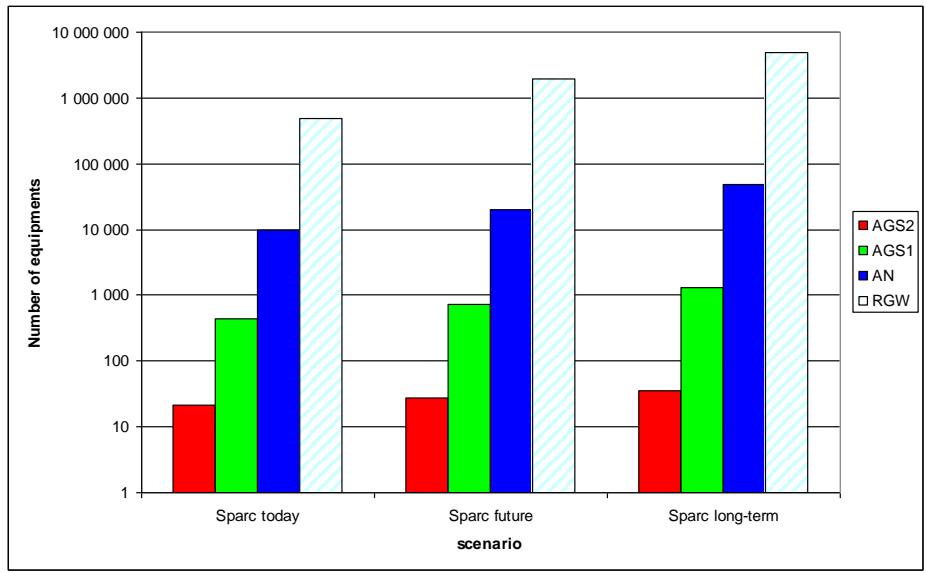

**Figure 72: Number of equipment units**

Figure 72 shows that the number of OpenFlow switches in an OpenFlow domain is around 10,000 / 20,000 / 50,000 for the specific scenarios.

Figure 73 shows the number of tunnels terminating or crossing a given type of node.

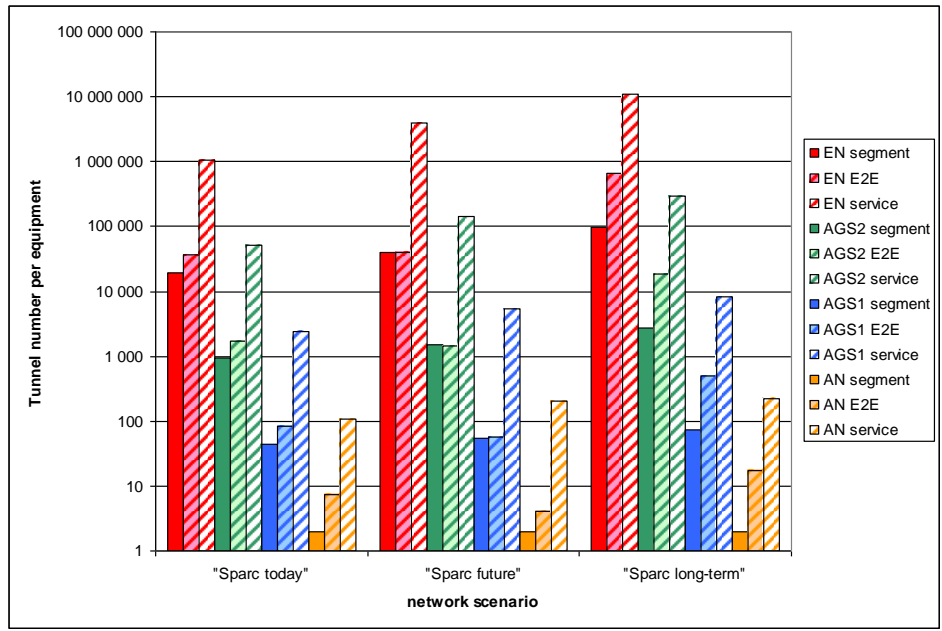

**Figure 73: Number of tunnels**

Note that a flow entry isn't needed for each type of tunnel in each type of node – for example, an aggregation switch is unaware of the service tunnels passing by because they are encapsulated in E2E tunnels which are encapsulated in transport segment tunnels. This figure illustrates why encapsulation is needed here and why automatic advertising of all interfaces (e.g., via LDP) is not recommended for use here.

Figure 74 below shows the number of flow entries to be supported by the network entities in our access/aggregation use case for the different scenarios.

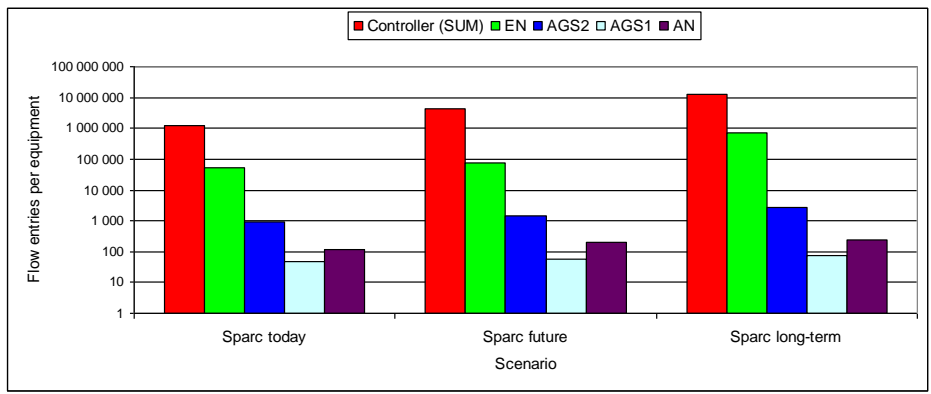

**Figure 74: Number of flow entries**

We can see that for the lower part of the aggregation about one hundred, while for the upper part of the aggregation thousands of flow entries have to be supported by the switches. The controller has to manage from one to ten million flow entries overall in the whole domain.

Based on the static analysis the following KPIs have been calculated for the scenarios (today/future/long-term) for a given OpenFlow access/aggregation domain:

- Number of OF switches in an OF domain: 10,000 / 20,000 / 50,000
- Number of UNI interfaces: 500,000 / 2,000,000 / 5,000,000
- Number of NNI interfaces: 1
- Number of flow entries in a switch, depending on aggregation level: hundreds – thousands – tens of thousands
- Number of actions of a flow entry: 1 – 2

### 6.4.2.3 Dynamic results from the model

To address the dynamic behavior of the OpenFlow domain, we considered the effects of two type of events.

The first one was a link going down an AN-AGS1, an AGS1-AGS2 or an AGS2-EN link, to be handled by the controller. We assume redundancy in the topology, so all traffic can be rerouted to a backup path. Regarding the reconfiguration of the flow entries, we calculated two values for each link-down:

- The best case, where there is a direct alternative link between the same two nodes, resulting in limiting the modifications only to those two nodes.
- The worst case, where the alternative link is connected to a different (but same aggregation level) node, with the most disjoint path to the EN compared to the original failed one. In this case the flow reconfigurations are not limited to the nodes directly affected, and higher aggregation level nodes will also be reconfigured.

Figure 75 shows how a link-down affects the various tunnels according to the model.

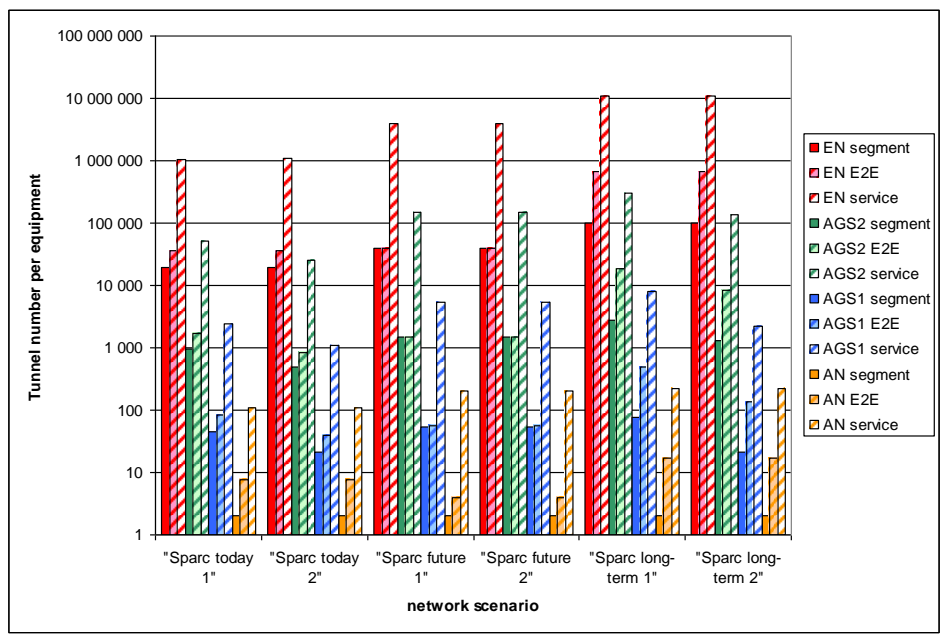

**Figure 75: The effect of a link down: tunnels**

From the best case we can derive requirements for the switches, while from the worst case we can derive requirements for the controller.

Figure 76 shows the required flow modification commands for restoring the tunnels via the controller.

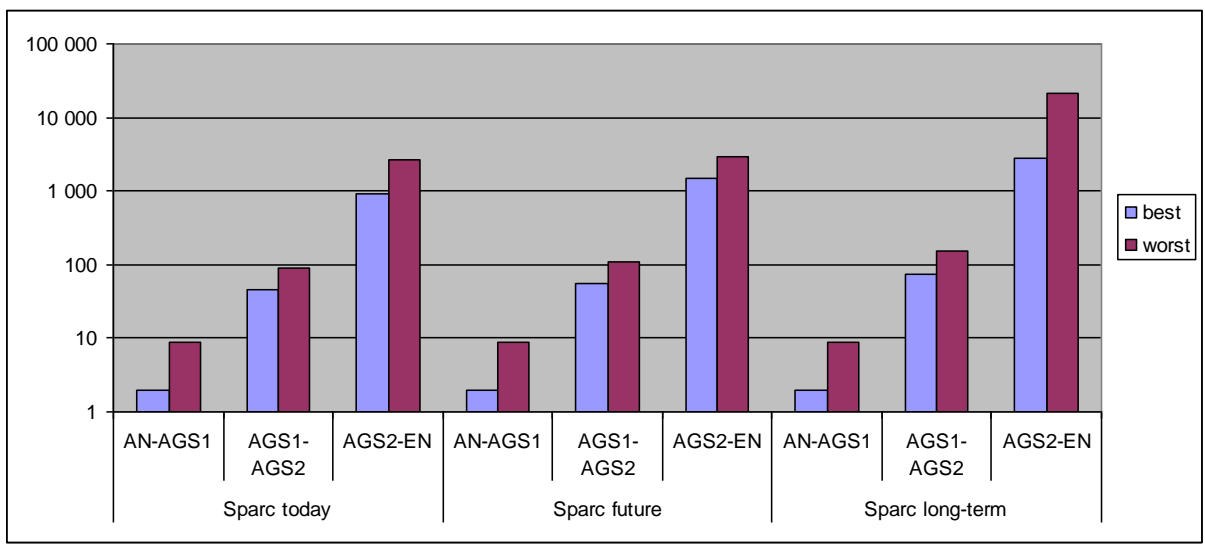

**Figure 76: The effect of a link-down: flow mods**

Flow_mods to be handled by a switch in the case of a non-AGS2-EN link failure are in the magnitude range of tens and a few hundreds. For the controller, the magnitude range is in the hundreds of flow entries.

In the case of an AGS2-EN link failure, the switch has to be able to handle a few thousand of flow_mods, while the controller magnitude range is a thousand or a few ten thousands.

Taking into account the time available for sending these changes results in Figure 77. These diagrams can be used to derive requirements, e.g., if 1 sec. can be used for the controller to send the configuration messages (see horizontal axis), then this results in a given flow mod/sec controller requirement (vertical axis).

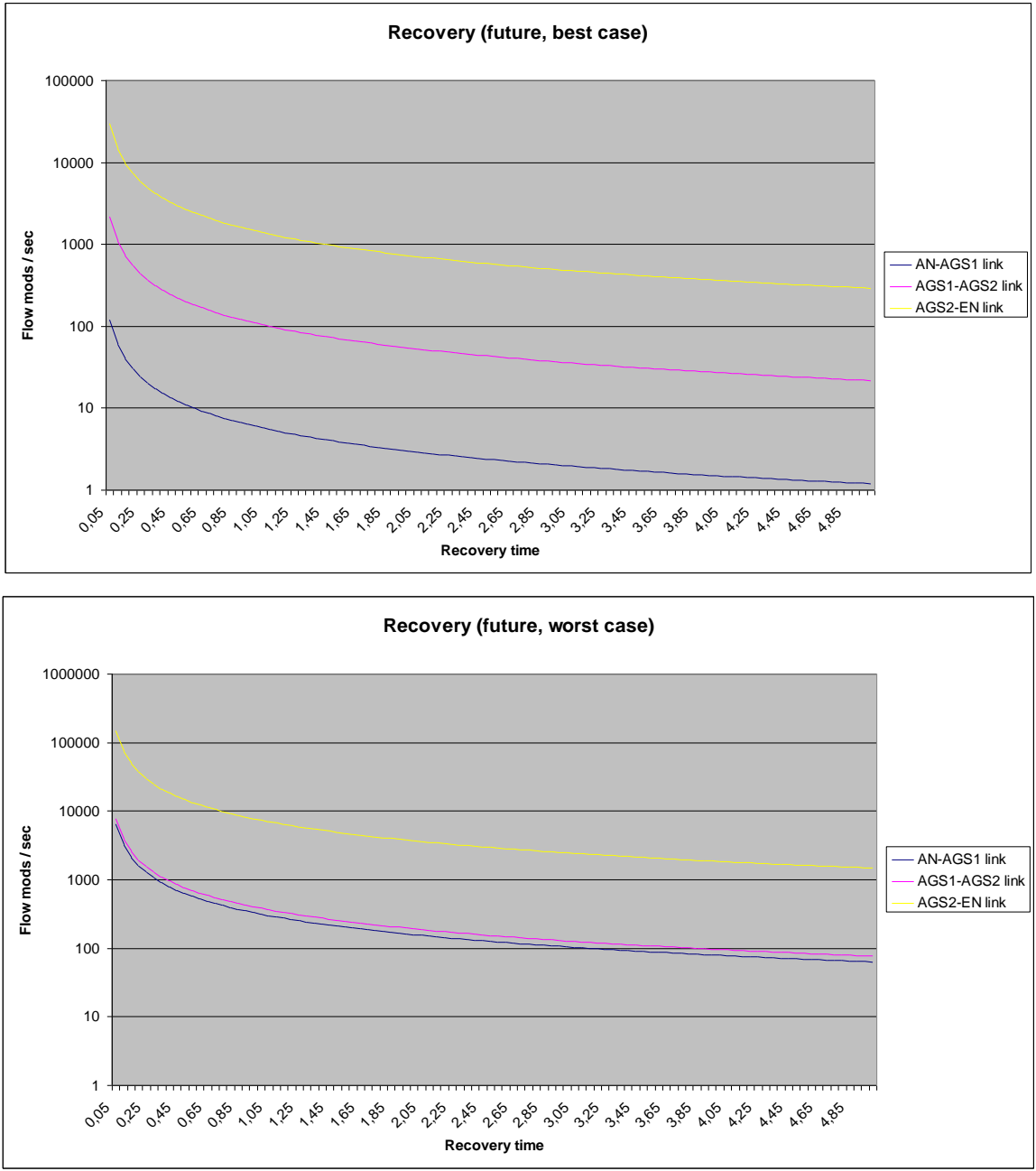

**Figure 77: Recovery times**

Note that the time refers only to sending the flow_mod messages – other aspects are not shown here.

The second type of event we considered is a burst of IPTV channel changes. This could happen if a significant percent of the users change channels, for example in the case of the beginning of a sport event or the evening news. The figure below shows the number of flow_mod messages the controller has to send. The best case refers to the situation where the video stream is already at the AN, whereas in the worst case the path for the stream has to be configured from the core for each request (which is slightly above the realistic scenario, but definitely an upper limit). The result is shown in Figure 78.

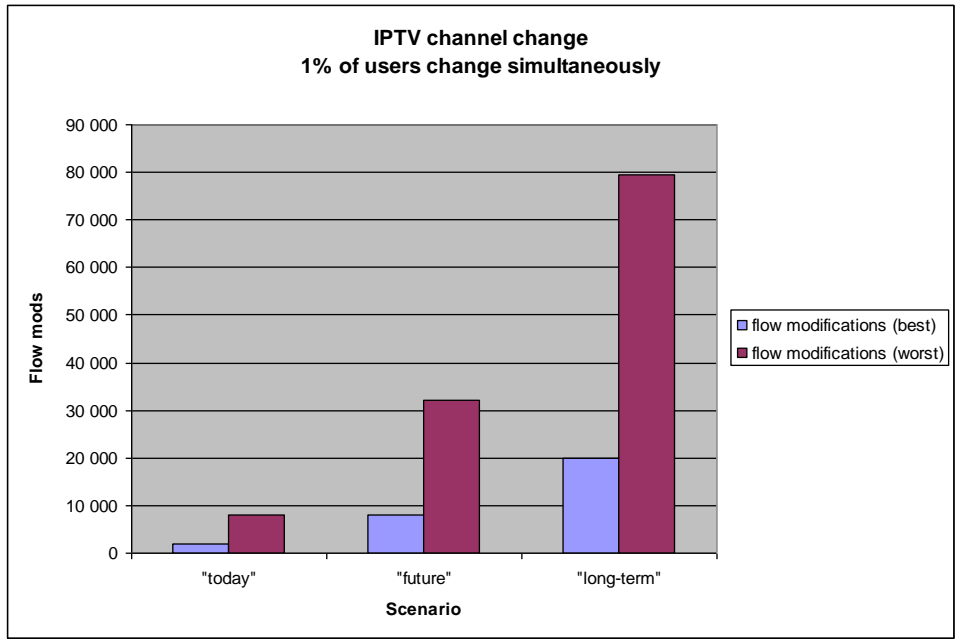

**Figure 78: IPTV channel change**

While the best case is a requirement for the AN and a lower limit for the controller, the worst case is a pessimistic upper limit requirement for the controller. Note: 1 percent of IPTV users change channels simultaneously in this example, but since it is scaled linearly, it is easy make calculations for other numbers.

Taking into account the time available for sending these changes results in Figure 79. Note that the time refers only to sending the flow_mod messages – other aspects are not shown here.

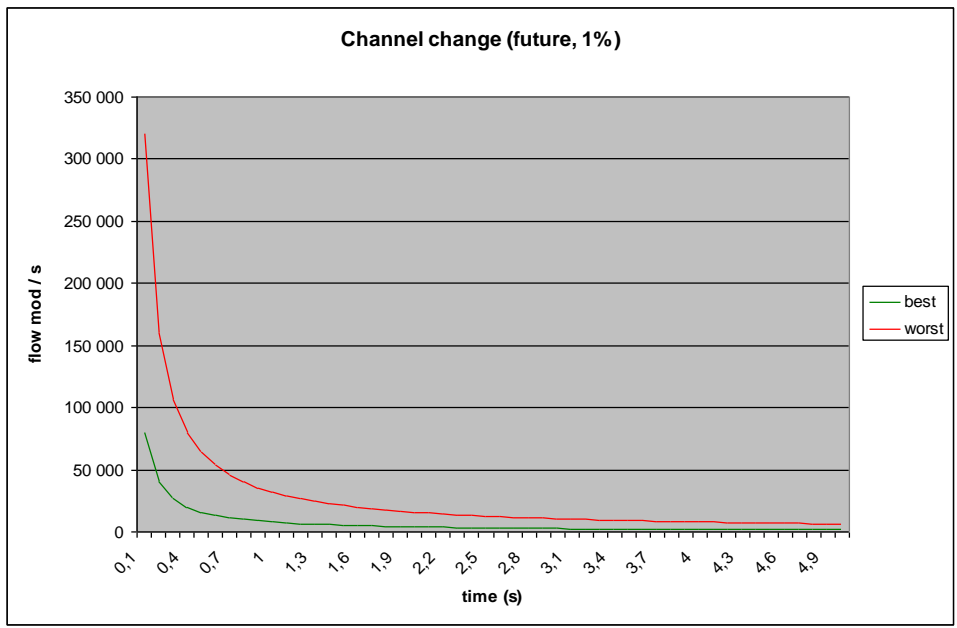

**Figure 79: IPTV channel change time**

If we assume 1 second response time for the channel change request and 1 percent of the users change, then the flow mod rate per second to be issued by the controller is in the best case 2,000 / 8,.000 / 20,000, while in the worst case 8,000 / 32,000 / 80,000.

#### 6.4.2.4 Comparison – existing switch and controller performance

We found the following related numbers in the DevoFlow [10] paper, which discusses a different use case and setup, however it provides numbers for expectable switch and controller performance, regardless of the actual setup:

> "We found that the switch completes roughly 275 flow setups per second. This number is in line with what others have reported [11]."

This performance is far below that needed to provide carrier-grade controller-based resiliency during a link-down event, however, it may be suitable for best effort type of service. Regarding the IPTV channel change event, this switch performance seems to suit the lower parts of the access/aggregation network.

> "[12] report that one NOX controller can handle 'at least 30K new flow installs per second while maintaining a sub-10 ms flow install time ... The controller's CPU is the bottleneck.'"

> "Maestro [13] is a multi-threaded controller that can install about twice as many flows per second as NOX, without additional latency."

This reported controller performance seems to be in the order of magnitude suitable for handling link-down effects in the network in a carrier-grade sense. For the IPTV channel change, it depends on the specific requirements.

#### 6.4.2.5 Updates to the initial topology and tunnel structure

The high flow_mod values in the case of link failure restoration close to the EN are the result of the many tunnels between the ANs and the EN. To decrease this high number of flow_mod messages, those tunnels could be aggregated, thus having only a single transport segment tunnel between neighboring nodes. In the case of restoration, an alternative segment tunnel can be built between the same two nodes over other multiple nodes. This increases the static number of flow entries per forwarding device and the traffic volume. However, it drastically decreases the message number in the restoration case. Figure 82 shows the comparison of a tree (snowflake) topology (Figure 70), a tree + ring combination topology (Figure 80), and the same topology with the single hop transport segments (Figure 81). Order of magnitude differences are evident.

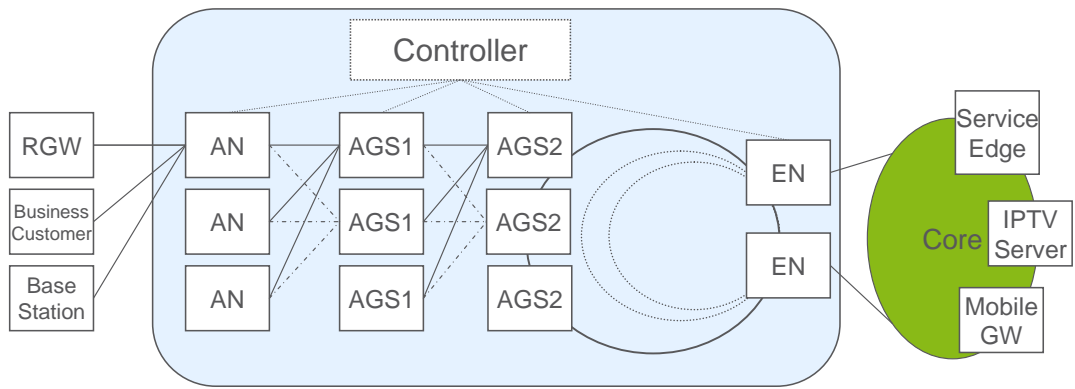

**Figure 80: Ring topology**

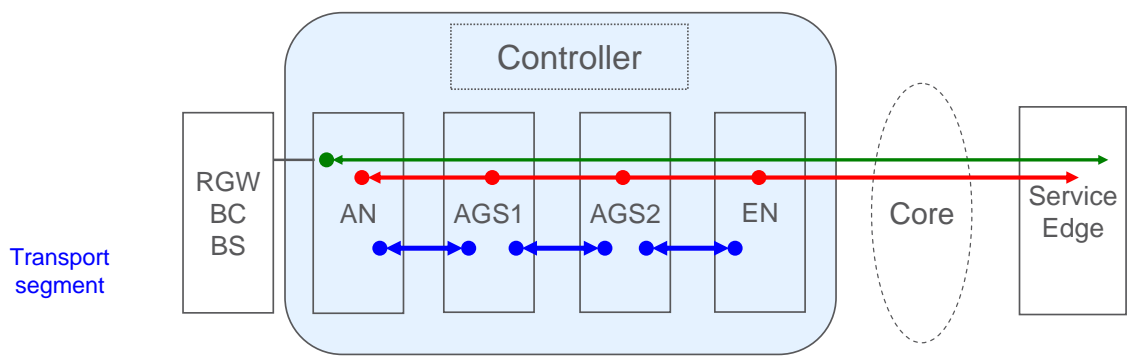

**Figure 81: Alternative OpenFlow domain internal tunnel structure**

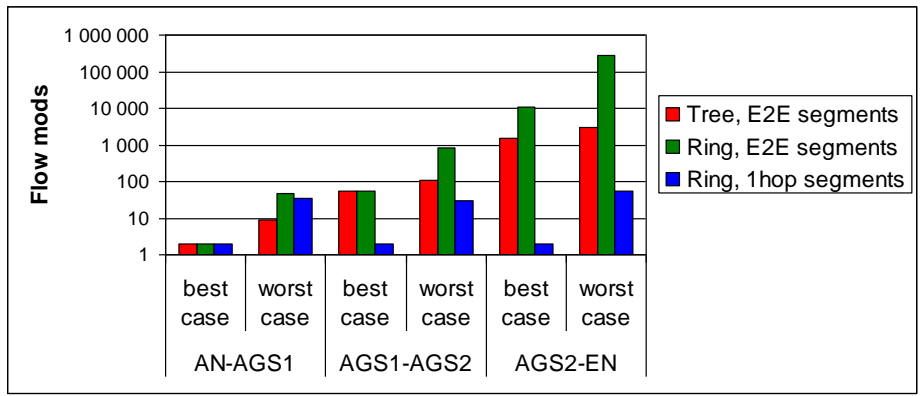

**Figure 82: Topology and domain internal tunnel structure effects**

However, reducing the required flow_mod messages has specific disadvantages as well, such as more static flow entries or increased traffic volume in the case of restoration. A detailed analysis is beyond the scope of this deliverable.

### 6.4.2.6  Conclusions from the scalability study

Looking at the static behavior, requirements from the model seem to be in the order of magnitude or below the capabilities of existing controllers and switches.

Regarding the dynamic behavior, there are scalability concerns, especially when strict time constraints are given. We showed that by changing the connection structures scalability can be significantly increased.

# 7 Conclusions

This deliverable defined the proposed carrier-grade *SplitArchitecture*, applying the concept of Software Defined Networking (SDN) to operator networks. Taking the use cases and requirements defined in WP2 into account, we presented our proposal and evaluated technical issues against certain architectural trade-offs. First, we considered a control and management architecture, consisting of a hierarchical, recursive control plane that enables operators to deploy several control planes with minimal interference and assign flows dynamically based on given policies. Additionally, we outlined an initial proposal of how to integrate network management and OpenFlow control with the flexibility to choose for each management function whether to integrate it into an SDN controller or place it in a separate network management system. Next, we discussed the required extensions to the OpenFlow protocol for supporting the carrier-grade *SplitArchitecture*. These include general extensions for openness & extensibility, virtualization support, OAM solutions, resiliency measures, bootstrapping and topology discovery issues, service creation solutions, energy-efficient networking approaches, QoS aspects, and multilayer considerations. In addition, we outlined selected deployment and adoption scenarios faced by modern operator networks. Specifically, we discussed procedures for scenarios such as service creation, general access/aggregation network scenarios and peering aspects, i.e., how to interconnect with legacy networks. Finally, a numerical scalability study indicates the feasibility of the *SplitArchitecture* approach in access/aggregation network scenarios in terms of scalability.

As overarching goals, the SPARC project defined

i) A *SplitArchitecture* blueprint for carrier-grade networks, and

ii) Required extensions to the OpenFlow protocol.

We are confident that this deliverable provides a comprehensive blueprint for control-, data- and management planes, as well as protocol aspects, as summarized above. Here, we would like to highlight two specific requirements with respect to carrier-grade networks, namely *scalability* and *availability*.

- Regarding *scalability*, our goal for the *SplitArchitecture* blueprint was to meet requirements to support large-scale deployments for carrier-grade networks. Specifically, the *SplitArchitecture* device shall be able to control forwarding devices that could count in the order of hundreds.

  In fact, the targeted use case determined the size of a network even in the order of thousands of nodes (see deliverable D2.1). This confirmed the requirement for a controller device to be able to control forwarding devices that could count in the order of thousands. We conclude that practical software architectures do not preclude maintaining OpenFlow connections toward such number of equipment. However, the rate of configuration data exchanged between the controller and the switches may raise scalability concerns. Our numerical scalability study showed that the control traffic required to proactively deploy transport connections with OpenFlow will not cause scalability limitations in the considered network sizes. On the other hand, reactive operation impairs the scalability of the *SplitArchitecture*. The measurements performed in deliverable D5.2 confirmed that tasks requiring strict time constraints, such as 50 milliseconds in the case of recovery or fast detection of link failures, cannot be realized via current OpenFlow based split control. This observation led to OpenFlow switch function proposals, together with OpenFlow protocol extensions for providing configuration support for these functions, such as OAM and protection, as discussed in this deliverable.

- Regarding *availability*, our goal for the *SplitArchitecture* blueprint was to ensure that the availability of networking services shall be equivalent to that of traditional technologies.

  We conclude that the availability of applications and services is dependent on the reliability of the server hosting them and the ability of the network to connect the user to the service. Server reliability itself is not affected by introducing *SplitArchitecture*, and therefore the main impact is by facilitating the implementation of custom protection and restoration per flow. As discussed in this deliverable and shown in deliverable D5.2, the proposed extensions to OpenFlow can enable sub-50 milliseconds protection for high-availability services while providing restoration for others. We must note that restoration in *SplitArchitecture* is slower than in traditional networks due to the centralized nature of the controller. However, OpenFlow can enable sub-second restoration in networks that previously had no such capability in a very cost-efficient solution.

In SPARC, we applied the concept of SDN/*SplitArchitecture* to the network operator domain with the promise of improving network design and operation in large-scale networks with multi-million customers, high availability and high automation. Furthermore, *SplitArchitecture* should open the field for new market players by lowering the entry barriers that exist in the complexity of individual components. From the technical point of view, we have been able to present and to demonstrate an architecture that can fulfill scalability and availability requirements, as well as many other specific requirements such as network management and OAM, virtualization and QoS support. From the business point of view, two things should be noted. First, our techno-economic analysis in WP2 confirms that there exists a potential for optimization of costs in terms of capital (CAPEX) as well as operational expenditures (OPEX) for carrier networks. Second, our analysis of the business environment, including a questionnaire and an analysis of standardization organizations, indicates that major players already present in today's telecommunication industry might be the leaders in a software defined telecommunication network environment as well.

Hence, the conclusion of the SPARC project is that it is both technically feasible and economically beneficial to apply SDN/*SplitArchitecture* to the carrier domain. Furthermore, the results of this project clearly indicate that the promises made by this novel architecture paradigm in terms of simplified and automated operation are still valid and definitely deserve further attention not only by academia, but also the network industry in general and telecom operators in particular.

The positive conclusions of the SPARC project also manifest itself in many ongoing and planned contributions to the main SDN standardization body, the ONF, by the SPARC ONF members (Ericsson and DTAG). During the initial period of the project, the SPARC project was concentrating on the setup of an ONF working group dealing with questions regarding a general architecture and use cases, which was eventually successful with the installation of the Architecture and Framework WG (including a SPARC member in the design team). Currently, also the Configuration & Management WG receives much input from SPARC partners based on the content on network management, OAM and bootstrapping as provided in this deliverable (Sections 4.2, 5.3 and 5.5). While these efforts will continue, the SPARC ideas on service creation (Sections 5.6 and 6.2), as well as the concepts for a controller architecture and openness and extensibility (presented in Sections 4.1 and 5.1) are planned to be discussed within the newly founded ONF Architecture and Framework WG. Additionally, SPARC members contribute actively to the New Transport Discussion Group within the ONF with the results of the multilayer discussions in Section 5.9, specifically the GMPLS-aware extensions for OpenFlow (Section 5.9).

As the next steps for carrier-grade SDN solutions, the SPARC project suggests continuing research towards full service node virtualization. SPARC already considered first SDN scenarios for service creation (e.g. split BRAS or SPARC DHCP), but mostly focused on an enhanced emulation of transport services in order to comply with legacy network technologies (e.g. support for MPLS and related OAM). However, the concept of split service creation could evolve to fast and flexible service deployment, ranging from connectivity services for residential business and backhaul, but also more sophisticated ones such as VoIP and IPTV. This would require study of programmable network datapath elements with generic hardware abstraction, enabling support for these flexible services. Furthermore, evolving the control plane for flexible service creation requires support for unified service and transport control along with additional necessary protocol extensions. Functions to consider include support for mobility, AAA and advanced service OAM.

# Abbreviations

| | | | |
|---|---|---|---|
| 3GPP | Third-generation partnership program | CPU | Central Processing Unit |
| ADSL | Asymmetric Digital Subscriber Line | CRC | Cyclic Redundancy Check |
| AES | Advanced Encryption Standard | CR-LDP | Constraint-based LDP |
| AGS | Aggregation Switch | CRUD | Create, read, update, delete |
| ANDSF | Access Network Discovery Selection Function | CSCF | Call Session Control Function |
| AP | Access Point | DCF | Distributed Coordination Function |
| API | Application Programming Interface | DHCP | Dynamic Host Configuration Protocol |
| ARP | Address Resolution Protocol | DHT | Distributed Hash Table |
| AS | Autonomous System | DiffServ | Differentiated Services, IETF |
| ATM | Asynchronous Transfer Mode | DNS | Domain Name Server |
| AWG | Arrayed-waveguide Grating | DOCSIS | Data Over Cable Service Interface Specification |
| BB | Broadband | DPI | Deep Packet Inspection |
| BBA | Broadband Access | DRR | Deficit Round Robin |
| BBF | Broadband Forum | DS | Differentiated Services |
| BB-RAR | Broadband Remote-Access-Router (SCP for Fast Internet) | DSCP | Diff Serve Code Point |
| BE | Best-Effort | DSL | Digital Subscriber Line |
| BFD | Bidirectional Forwarding Detection | DSLAM | Digital Subscriber Line Access Multiplexer (network side of ADSL line) |
| BG | Broadband Aggregation Gateway | DWDM | (Dense) Wave-Division-Multiplex |
| BGP | Border Gateway Protocol; Distance Vector Routing protocol of IETF (EGP) | dWRED | Distributed WRED |
| BRAS | Broadband Remote Access Server / Service | DXC | Digital Cross-Connect |
| BRPC | Backward Recursive Path Computation | ECMP | Equal Cost Multi-Path |
| BSS | Basic Service Set | ECN | Explicit Congestion Notification |
| CAR | Committed Access Rate | ECR | Egress Committed Rate |
| CBO | Class-Based Queuing | EGP | Exterior Gateway Protocol |
| CBWFQ | Class-Based Weighted Fair Queuing | EIGRP | Enhanced IGRP |
| CCM | Continuity Check Message | EN | Edge Node |
| CDMA | Code Division Multiple Access | ePDG | Evolved Packet Data Network Gateway |
| CE | Control Element | ESS | Extended Service Set |
| CHG | Customer HG; HG in customer site | FE | Forwarding Element |
| CIDR | Classless Inter-Domain Routing | FEC | Forwarding Equivalence Class |
| CIPM | Cisco IP Manager | FEC | Forward Error Correction |
| CIR | Committed Information Rate | FIB | Forwarding Information Base |
| CLI | Command Line Interface | FMC | Fixed Mobile Convergence |
| CLNC | Customer LNS, LNS in customer site | ForCES | Forwarding and Control Element Separation |
| CORBA | Common Object Request Broker Architecture | FPGA | Field Programmable Gate Array |
| CoS | Class of Service | FSC | Fiber Switching |
| CP | Connectivity Provider | FTTCab | Fiber to the Cabinet |
| CPE | Customer Premise Equipment | FTTH | Fiber to the Home |
| FTTLEx | Fiber to the Local Exchange | LAC | L2TP Access Concentrator |
| FW | Firewall | LAN | Local Area Network |
| GbE | Gigabit Ethernet | LDP | Label Distribution Protocol |
| GFP | Generic Framing Procedure | LER | Label Edge Router; MPLS-based router with MPLS, IP-VPN and QoS edge support |
| GLONASS | Globalnaja Nawigazionnaja Sputnikowaja Sistema (Russian satellite system) | LER-BB | Broadband LER; LER for DS and higher |
| GMPLS | Generalized Multi-Protocol Label Switching | L-GW | Local Gateway |
| GNSS | Global Navigation Satellite System | LLDP | Link Layer Discovery Protocol |
| GPON | Gigabit Passive Optical Network | L-LSP | Label-inferred LSP |
| GPS | Global Positioning System | LMP | Link Management Protocol |
| GRE | Generic Route Encapsulation | LNS | L2TP Network Server |
| GTS | Generic Traffic Shaping | LSP | Label Switch Path |
| GUI | Graphical User Interface | LSR | Label Switch Router; MPLS-based router in the inner IP network. Only IGP knowledge. |
| HCF | Hybrid Coordination Function | LTE | Long Term Evolution |
| HDLC | High-level Data Link Control | MAC | Media Access Control |
| HG | Home Gateway | MAN | Metropolitan Area Network |
| HIP | Host Identify Protocol | MEF | Metro Ethernet Forum |
| IACD | Interface Adjustment Capability Descriptor | MGW | Media Gateway |
| ICR | Ingress Committed Rate | MIB | Management Information Base |
| ICT | Information and Communication Technology | MLD | Multicast Listener Discovery |
| IEEE | Institute of Electrical and Electronics Engineers | MLTE | Multilayer Traffic Engineering |
| IETF | Internet Engineering Task Force (www.ietf.org) | MME | Mobility Management Entity |
| | | MPLS | Multi-Protocol Label Switching |
| IF | Interface | MPLS-TP | MPLS Transport Profile |
| ISC | Interface Switching Capabilities | MSC | Mobile Switch Controller |
| IGMP | Internet Group Management Protocol | MTU | Maximum Transmission Unit |
| IGP | Interior Gateway Protocol | NAT | Network Address Translation |
| IGRP | Interior Gateway Routing Protocol | NGN | Next-Generation Network |
| IntServ | Integrated Services, IETF | NIC | Network Interface Controller |
| IP | Internet Protocol | NMS | Network Management System |
| IPTV | IP television | NNI | Network-to-Network Interface |
| ISCD | Interface Switching Capability Descriptor | NP | Network Provider |
| ISDN | Integrated Services Digital Network | NSP | Native Service Processing |
| IS-IS | Intermediate System - Intermediate System; Link State Routing Protocol from OSI (IGP) | NTP | Network Time Protocol |
| | | OAM | "Operation, Administration and Maintenance" or "Operations and Maintenance" |
| ISO | International Organization for Standardization | ODU | Optical Data Unit |
| ISP | Internet Service Provider | OF | OpenFlow |
| ITIL | IT Infrastructure Library | OFDM | Orthogonal Frequency Division Multiplexing |
| ITU | International Telecommunication Union | OLT | Optical Line Termination |
| L2F | Layer 2 Forwarding | OSI | Open Systems Interconnection |
| L2TP | Layer 2 Tunnel Protocol | | |
| OSNR | Optical Signal-to-Noise Ratio | RTT | Round Trip Time |
| OSPF | Open Shortest Path First; Link State Routing Protocol from IETF (IGP) | SAE | System Architecture Evolution |
| | | SAP | Service Access Point |
| OTN | Optical Transport Network | SBC | Session Border Controller |
| OTU | Optical Transport Unit | SCP | Service Creation Point |
| OXC | Optical Cross-Connect | SDH | Synchronous Digital Hierarchy |
| PANA | Protocol for carrying Authentication for Network Access | RSVP (-TE) | ReSource reserVation Protocol (-Traffic Engineering) |
| PBB-TE | Provider Backbone Bridge Traffic Engineering | SDN | Software Defined Networking |
| | | SDU | Service Data Unit |
| PCEP | Path Computation Element Communication Protocol | SE | Service Edge |
| | | SGW | Serving Gateway |
| PDU | Protocol Data Unit | SHG | Separate HG, separate HG device with virtual HG |
| PE | Provider Edge; Service Creation Point for IP-VPN | | |
| | | SIP | Session Initiation Protocol |
| PER | Provider Edge Router | SIPTO | Selective IP Traffic Offload |
| PGW | Packet Data Network Gateway | SLA | Service Level Agreement |
| PIM | Protocol Independent Multicast | SLNS | Separate LNS, separate LNS device with virtual LNS |
| PIP | Physical Infrastructure Provider | | |
| PMIP | Proxy Mobile IP | SMS | Service Management System |
| PoP | Point of Presence | SNMP | Simple Network Management Protocol |
| POTS | Plain Old Telephony Service | SONET | Synchronous Optical Network |
| PPP | Point-to-Point Protocol | SP | Service Provider |
| PPPoE | PPP over Ethernet | SPARC | Split Architecture for carrier-grade networks |
| PSTN | Public Switched Telephone Network | | |
| PVC | Permanent Virtual Circuit (permanent L2 connection, e.g., Frame Relay, ATM) | SSID | Service Set Identifier |
| | | SSM | Source Specific Multicast |
| PWE | PseudoWire Emulation | STA | Station |
| QoE | Quality of Experience | STM | Synchronous Transfer Module (STM-1: 155 Mbit/s, STM-4: 622 Mbit/s, STM-16: 2.5 Gbit/s; STM-64: 10 Gbit/s) |
| QoS | Quality of Service; general for differentiated quality of services or absolute quality of services. | | |
| | | TCAM | Ternary Content Addressable Memory |
| QPSK | Quadrature Phase-Shift Keying | TCM | Tandem Connection Monitoring |
| RADIUS | Remote Authentication Dial-In User Service | TCP | Transmission Control Protocol |
| RAR | Remote Access Router (SCP for OCN) | TDM | Time Division Multiplexing |
| RARP | Reverse ARP | TE | Traffic Engineering |
| RFC | Request for Comment (in IETF) | TKIP | Temporal Key Integrity Protocol |
| RGW | Residential Gateway | ToR | Top of the Rack |
| RIB | Routing Information Bases | ToS | Type of Service |
| RIP | Routing Information Protocol; Distance Vector Routing Protocol from IETF (EGP) | TR | Technical Report (from BBF) |
| | | TTL | Time to live |
| ROADM | Reconfigurable Optical Add-Drop Multiplexer | UDP | User Datagram Protocol |
| | | UNI | User Network Interface |
| RR | Route Reflector for BGP/MP-BGP | | |
| RTP | Real Time Protocol | | |
| VEB | Virtual Ethernet Bridges | VSI | Virtual Station Interface |
| VEPA | Virtual Ethernet Port Aggregator | WAN | Wide Area Network |
| VLAN | Virtual LAN | WDM | Wavelength Division Multiplexing |
| VM | Virtual Machine | WEP | Wired Equivalent Privacy |
| vNIC | Virtual NIC | WFQ | Weighted Fair Queuing |
| VoIP | Voice over IP | WP | Work Package |
| VPLS | Virtual Private LAN Service | WSON | Wavelength Switched Optical Network |
| VPN | Virtual Private Network | WT | Working Text (from BBF) |